\documentclass[11pt,notitlepage]{article}

\usepackage[affil-it]{authblk}
\usepackage{amsmath,amssymb,amsfonts,amscd,mathtools}
\usepackage{mathabx}
\usepackage[bbgreekl]{mathbbol}
\usepackage{cite}
\usepackage[colorlinks,allcolors=blue]{hyperref}
\usepackage[none]{hyphenat}
\usepackage[utf8]{inputenc}
\usepackage{array}
\usepackage{multirow,bigdelim}
\usepackage[stable]{footmisc} 
\usepackage{stmaryrd}

\usepackage{tikz-cd}   
\usepackage{tikz}
\usetikzlibrary{knots,shapes.misc,decorations.markings,intersections,patterns,decorations.pathreplacing}

\font\teneurm=eurm10 \font\seveneurm=eurm7 \font\fiveeurm=eurm5
\newfam\eurmfam
\textfont\eurmfam=\teneurm \scriptfont\eurmfam=\seveneurm
\scriptscriptfont\eurmfam=\fiveeurm

 \font\teneusm=eusm10 \font\seveneusm=eusm7 \font\fiveeusm=eusm5
\newfam\eusmfam
\textfont\eusmfam=\teneusm \scriptfont\eusmfam=\seveneusm
\scriptscriptfont\eusmfam=\fiveeusm
\def\eusm#1{{\fam\eusmfam\relax#1}}
\font\tencmmib=cmmib10 \skewchar\tencmmib='177
\font\sevencmmib=cmmib7 \skewchar\sevencmmib='177
\font\fivecmmib=cmmib5 \skewchar\fivecmmib='177
\newfam\cmmibfam
\textfont\cmmibfam=\tencmmib \scriptfont\cmmibfam=\sevencmmib
\scriptscriptfont\cmmibfam=\fivecmmib

\def \x{{\eusm{X}}}
\def \y{{\eusm{Y}}}

\newcommand{\kbf}[1]{{\boldsymbol{#1}}}

\def \CL{\mathcal{L}}
\def \CD{\mathcal{D}}
\def \CC{\mathcal{C}}
\def \CA{\mathcal{A}}

\def \GG{\mathfrak{G}}
\def \fd{{\text{fd}}}
\def \ch{{\mathfrak{c}}}

\def \Q{\mathbb{Q}}
\def \Z{\mathbb{Z}}
\def \R{\mathbb{R}}
\def \C{\mathbb{C}}

\def \Im{\mathrm{Im}\,}
\def \re{\mathrm{Re}\,}
\def \im{\mathrm{Im}\,}
\def \Tr{\mathrm{Tr}\,}
\def \CS{\mathrm{CS}}

\def \deg{\mathrm{deg}\,}
\def \pert{\mathrm{pert}}
\def \Crit{\mathrm{Crit}\,}
\def \CritVal{\mathrm{CritVal}\,}
\def \Hom{\mathrm{Hom}}
\def \Res{\mathop{\mathrm{Res}}}

\def \tA{\widetilde{\mathcal{A}}}
\def \tilde{\widetilde}
\def \hat{\widehat}
\def \check{\widecheck}

\newcommand{\pto}{{\,{\propto}\,}}

\usepackage{tikz-cd}                     

\usepackage[left=3cm,top=4cm,right=3cm,bottom=3cm]{geometry}

\title{On categorification of Stokes coefficients in Chern-Simons~theory}
\author[1,2]{Sergei Gukov}
\author[3]{Pavel Putrov}
\affil[1]{Merkin Center for Pure and Applied Mathematics, California Institute of Technology,
1200 E. California Blvd., Pasadena, CA 91125, USA}
\affil[2]{Dublin Institute for Advanced Studies, 10 Burlington Rd, Dublin, Ireland}
\affil[3]{ICTP, Strada Costiera 11, Trieste 34151, Italy}
\date{}

\begin{document}

\maketitle
\begin{abstract}
  We consider a finite-dimensional oscillatory integral which provides a ``finite-dimensional model'' for analytically continued $SU(2)$ Chern-Simons theory on closed 3-manifolds that are described by plumbing trees. This model allows an efficient description of Stokes phenomenon for perturbative expansions in Chern-Simons theory around classical solutions -- $SL(2,\mathbb{C})$ flat connections. Moreover, the Stokes coefficients can be  categorified, i.e. promoted to graded vector spaces, in terms of this finite-dimensional model. At least naively, the categorification gives BPS spectrum of 5d maximally supersymmetric Yang-Mills theory on the 3-manifold times a line with appropriate boundary conditions. We also comment on necessity of taking into account ``flat connections at infinity'' to capture Stokes phenomenon for certain 3-manifolds.
\end{abstract}

\tableofcontents

\section{Introduction and summary}

Homological invariants of knots and 3-manifolds play an important role in the low-dimensional topology. Not only do they often provide strong 3-dimensional invariants, but due to their functoriality properties they can be of use in 4-dimensional topology. From the physics point of view, homological invariants can be usually interpreted as the BPS spectrum ($Q$-cohomology) of certain twisted theories, where the knot / 3-manifold plays the role of a part of the space, with the time being transversal to it
\cite{Gukov:2004hz,Gukov:2007ck,witten2011fivebranes}. One of the most well-known homological invariant of knots is the Khovanov homology \cite{khovanov2000} (and its higher-rank generalization \cite{khovanov2008matrix}). The important examples of homological invariants of 3-manifolds are Heegaard Floer \cite{ozsvath2004holomorphic}, Monopole Floer \cite{Kronheimer_Mrowka_2007}, Instanton Floer \cite{floer1988instanton}, and Embedded Contact Homology \cite{hutchings2009embedded}, which are all closely related. These homology theories also have versions for 3-manifolds with a knot.   

The homological invariants can be understood as a categorification of numerical integer-valued invariants, which are recovered by the Euler characteristic of the homology. For example, the ($\Z^2$-graded) Khovanov homology categorifies the Jones polynomial (with one of the gradings corresponding to powers of the variable in the polynomial) and Instanton Floer homology categorifies the Casson invariant.

In this paper we consider a physically-motivated homology theory of a 3-manifold $Y$ that categorifies Stokes coefficients in analytically continued $SU(2)$ Chern-Simons (CS) theory on 3-manifolds. The Stokes coefficients are integers that describe the wall-crossing of the Borel resummations of perturbative expansions in CS theory around different flat $SL(2,\C)$ connections on $Y$. Informally they can be also understood as counts of gradient flows between flat connections with respect to the real part of the Chern-Simons functional (multiplied by an extra phase) on the space of all $SL(2,\C)$ connections \cite{Witten:2010cx}. The gradient flow equations are known to be equivalent to 4-dimensional Kapustin-Witten equations on $Y\times \R$. Solutions of Kapustin-Witten equations, in turn, can be understood as the $\R$-invariant solutions of 5d Haydys-Witten equations on $Y\times \R\times \R$ \cite{witten2011fivebranes,Haydys:2010dv}. (Note, this is different from 5d Haydys-Witten equations on $Y\times S^1 \times \R$ or 6d fivebrane theory on $Y\times T^2 \times \R$.) Therefore, at least naively, the categorification of the Stokes coefficients can be provided by the Floer-type homology based on counting solutions of Haydys-Witten equations on $Y \times \mathbb \R^2$. Physically, it is given by the BPS spectrum of the 5d $\mathcal{N}=2$ $SU(2)$ Super-Yang-Mills theory topologically twisted along $Y$.

Alternatively, considering a Heegaard splitting of the 3-manifold along a Riemann surface $\Sigma$, one can consider a Floer-type theory based on counting solutions to 3d Fueter equations with the complex symplectic target being the $SL(2,\C)$ character variety of $\pi_1(\Sigma)$ \cite{doan2022holomorphic,Bousseau:2022jaf}. Physically it should correspond to the BPS spectrum of an A-twisted 3d $\mathcal{N}=4$ sigma-model with the aforementioned target. However, both of these approaches neither have been made mathematically rigorous, nor allow systematic calculations at the level of rigor of theoretical physics.

In this work we consider a special class of 3-manifolds -- plumbed (or graph), for which we formulate a different approach to categorification of Stokes coefficients. It is based on the existence of a finite-dimensional oscillatory integral that provides an analytic continuation of the Witten-Reshetikhin-Turaev (WRT) invariant of $Y$ with respect to the level parameter. It therefore captures at least some sector\footnote{The precise meaning of this is explained in Section \ref{sec:single-thimble-info}.} in the analytically-continued CS theory on $Y$. We refer to this integral as the ``finite-dimensional model'' of the analytically continued CS theory. The categorification can be described combinatorially in terms of the data defining the finite-dimensional oscillatory integral. It can be understood as the Floer homology of pairs of Lagrangian submanifolds in the space of integration.

During our analysis, we find that in some examples, in order to fully capture the Stokes phenomenon, it is not sufficient to consider perturbative expansions around the ordinary $SL(2,\C)$ flat connections. One has to also take into account certain ``flat connections at infinity'', which are not part of the standard moduli space $\Hom(\pi_1(Y),SL(2,\C))/SL(2,\C)$.

The rest of the paper is organized as follows. In Section \ref{sec:stokes-plumbed} we first provide a brief review of resurgence theory and Stokes phenomenon in analytically continued Chern-Simons theory. We then describe the finite-dimensional model of the analytically continued CS theory on plumbed 3-manifold and show how to systematically calculate Stokes coefficients from it. In Section \ref{sec:flat-at-infinity} we point out the necessity of consideration of ``flat connections at infinity'' to capture the Stokes phenomenon in CS theory on certain 3-manifolds. In Section \ref{sec:categorification} we go into the details of  categorification of the Stokes coefficients in Chern-Simons theory. In Section \ref{sec:hemisphere} we provide an interpretation of the finite-diemensional model for plumbed 3-manifold as a hemisphere partition function of a 2d $\mathcal{N}=(2,2)$ quantum field theory preserving $A$-type of $B$-type supersymmetry. 

\section{Stokes coefficients for plumbed 3-manifolds.}
\label{sec:stokes-plumbed}

\subsection{Review of resurgence and Stokes phenomenon in Chern-Simons theory}
\label{sec:CS-resurgence-review}

Consider a closed oriented connected 3-manifold $Y$. In the analytically continued Chern-Simons theory \cite{Witten:2010cx,Kontsevich} one formally considers integrals over middle-dimensional contours $\Gamma$ in the space $\tA$ of complex connection 1-forms $A\in \Omega^1(Y)\otimes \mathfrak{sl}(2,\C)$ on a principle $SL(2,\C)$ bundle over $Y$ modulo gauge transformations isotopic to identity:
\begin{equation}
    I^\Gamma(k)=\int_{\Gamma} \CD [A] \, e^{2\pi ik\CS([A])}
    \label{CS-general-integral}
\end{equation}
where $k\in \C\setminus \{0\}$ is a complex parameter -- complexified level of Chern-Simons theory. The Chern-Simons functional,
\begin{equation}
\begin{array}{rrcl}
    \CS: & \tA & \longrightarrow & \C,  \\
    & [A]& \longmapsto & \frac{1}{8\pi^2}\int_{Y} AdA+\frac{2}{3}A^3,
\end{array}
\end{equation}
is considered as a holomorphic function on $\tA$ and $\CD [A]$ is the natural ``holomorphic volume form'' on $\tA$. The quotient is performed only over gauge transformations isotopic to identity so that the Chern-Simons functional $\CS$ is a well-defined function on the quotient with values in $\C$, and not just in $\C/\Z$. Moreover, it is often beneficial to consider the $SL(2,\C)$ principle bundle over $Y$ to be framed at a point $\ast\in Y$, so that the action of the group of the gauge transformations (consisting of smooth maps $f: Y\rightarrow SL(2,\C),\;f|_\ast =1$) is free and the resulting quotient space does not have singularities (nor needs to be considered as a stack). We will follow this approach by default. The space $\tA$ is simply connected and has a free action by the group of all gauge transformations modulo gauge transformations isotopic to identity, which can be identified with integers $\pi_3(SL(2,\C))\cong \Z$. The action, if denoted by $(\,\cdot\,)^{(n)}:\tA\rightarrow \tA,\;n\in\Z$ is such that $\CS([A]^{(n)})=\CS([A])+n$.

The countour $\Gamma$ should be chosen such that $\exp\,2\pi ik\,\CS[A]$ vanishes as $A$ tends to infinitity along $\Gamma$. Without loss of generality, one can assume a stronger condition -- that $ik\CS[A]$ tends to $-\infty$ along the real axis, as $A$ goes to infinity along the contour.  Because of the holomorphicity of the integrand, $I_\Gamma(k)$ depends on the class of $\Gamma$ in the appropriate relative homology group, which formally can be written as $H_{\dim_{\C}\tA}(\tA,\CS^{-1}(i\infty/k);\Z)$. For a generic value of $k\in \C\setminus \{0\}$ this group of homology classes of admissible contours of integration, which we denote by $\GG$, has a natural basis given by Lefschetz thimbles for the holomorphic function $\CS:\tA\rightarrow \C$. To be precise, one needs a generalized version of the standard Lefschetz thimbles that takes into account that the function is not Morse. This is ordinarily the case for the Chern-Simons functional on a 3-manifold. The critical set is the space of flat connections modulo isotopic to identity gauge transformations of the $SL(2,\C)$ bundle framed at a point is the following\footnote{As we will see in Section \ref{sec:flat-at-infinity}, in general, to capture completely the Stokes phenomenon, one needs to complete $\tA$ by including certain connections at infinity that have finite values of $\CS$. This may result in having extra components in the critical set.}:
\begin{equation}
    \tA\;\;\supset\;\; \Crit \CS\;\cong\; \Hom(\pi_1(Y),SL(2,\C))\times \Z
\end{equation}
so that the free $\Z$ action on $\tA$ acts on the $\Z$-component by integral shifts. Therefore, unless $\pi_1(Y)\cong \Z_2^N$, the critical set contains non-isolated points.

In the case when the function is Morse and critical points are isolated or, more generally, the function is Morse-Bott and each connected component of the critical set has a form of the total space of the cotangent bundle over a compact real manifold, there is a single Lefschetz thimble associated with each connected component. In general, however, one may need to consider multiple (generalized) Lefschetz thimbles associated with a single connected component. Following the convention of \cite{Gukov:2016njj} we use blackboard bold Greek letters $\bbalpha,\bbbeta,\bbgamma,\ldots$ for the indices labeling different Lefschetz thimbles. In the case when the function is Morse-Bott, for a connected component $C\subset \mathrm{Crit}\,\CS\subset \tA$, one has a Lefschetz thimble for each free generator of $H_{\dim_\C C}(C)$. They can be defined as the union of steepest descent flows with respect to $\re (2\pi ik\CS)$ starting from a middle-dimensional compact cycle in $C$. Since $\CS$ is shifted by a constant under the free action of $\Z$ on $\tA$, this action induces a free action of $\Z$ on $\mathrm{Crit}\,\CS$ such that each critical point is taken into a different connected component. One can choose the basis Lefschetz thimbles in a way that respects this $\Z$-action, i.e. so that we have an induced free $\Z$ action on the set of all Lefschetz thimbles. Following again the convention of \cite{Gukov:2016njj}, we use a non-bold Greek letter (e.g. $\alpha$) to denote the $\Z$-orbits of  Lefschetz thimbles (e.g. $\bbalpha$). Note that although the number of thimbles is infinite, the number of orbits is finite (assuming $Y$ has a finitely represented fundamental group).

One can consider an integral over a particular thimble:
\begin{equation}
    I^\bbalpha(k)= \int_{\bbalpha} \CD [A] \, e^{2\pi ik\CS([A])}.
    \label{CS-thimble-integral}
\end{equation}
Because Lefschetz thimbles generate (over integers)  all admissible contours of integration, an integral over an arbitrary contour can for generic value of $k$ be expressed as an integral linear combination of $I^\bbalpha(k)$. However, since the Lefschetz thimbles depend on $\arg k$, the coefficients of the decomposition may jump as one changes $k$. This is known as the Stokes phenomemon, and is the main focus of this paper.

Note that although the functional integrals in (\ref{CS-general-integral}) and (\ref{CS-thimble-integral}) are not mathematically well-defined, the perturbative expansion of $(\ref{CS-thimble-integral})$ at $k\rightarrow \infty$ is nevertheless well-defined in many cases. In general it is expected to have the following form \cite{Gukov:2003na,Dimofte:2009yn} similar to that of Chern-Simons with compact gauge group \cite{Witten:1988hf} (the same when the thimble is associated with a cycle homologous to a one contained in the subspace of $SU(2)$ connections in $\mathrm{Crit}\,\CS$):
\begin{equation}
    I^{\bbalpha}(k) \stackrel{\text{asympt}}{\cong} e^{2\pi i k\CS_\bbalpha}
    \sum_{n\geq 0}\frac{a^{(\alpha)}_n}{k^{n+\delta_\alpha}},\qquad k\rightarrow\infty,
    \label{thimble-asympt}
\end{equation}
where $\CS_\bbalpha$ is the value of the Chern-Simons functional on any connection from the connected component $\Crit \CS$ with which the thimble $\bbalpha$ is associated. The shift $\delta_\alpha$ is expected to be a half-integer, and under certain assumptions is explicitly given by the formula \cite{Freed:1991wd,Jeffrey:1992tk,ohtsuki2002problems,Gukov:2006ze} $\delta_\alpha=(\dim H^0(Y,d+A)-\dim H^1(Y,d+A))/2$ where $A$ is a connection 1-form from the connected component of the critical set. Here and below we assume that indeed $\delta_\alpha \in \frac{1}{2}\Z$ and that $a_0^{(\alpha)}\neq 0$ (which, unless all $a^{(\alpha)}_n$ simultaneously vanish, can always be achieved by choosing an appropriate $\delta_\alpha$).
Note that $\delta_\alpha$ and the coefficients $a^{(\alpha)}_n$ depend only on orbit $\alpha$, but not on a particular lift $\bbalpha\in\alpha$.

Under certain assumptions on the properties of the connected component of the critical set, one can define the coefficients $a^{(\alpha)}_n$ independently in terms of certain finite-dimensional integrals over $Y$ corresponding to Feynman graphs of the perturbative expansion in Chern-Simons theory \cite{Axelrod:1991vq,Axelrod:1993wr,kontsevich1994feynman,Dimofte:2012qj}. For the trivial flat connection one can also define the coefficients via Ohtsuki invariant \cite{ohtsuki1996polynomial}. Assuming we are in a setting where $a^{(\alpha)}_n$ can be mathematically defined, we then can define the formal power series in $1/k$: 
\begin{equation}
    I^{\bbalpha}_\pert(k) :=  e^{2\pi i k\CS_\bbalpha}
    \sum_{n\geq 0}\frac{a^{(\alpha)}_n}{k^{n+\delta_\alpha}},
    \label{thimble-asympt-formal}
\end{equation}
without a direct use of mathematically ill-defined functional integrals. The series are in general divergent for any $k$ and make sense only as formal power series. 

To deal with the divergence, consider the Borel transform of the series (\ref{thimble-asympt-formal}) defined as
\begin{equation}
    B^\bbalpha(\xi):= \sum_{n\geq 0}\frac{(-2\pi i)^{n+\delta_\alpha}\,a_n^{(\alpha)}\,(\xi-\CS_\bbalpha)^{n+\delta_\alpha-1}}{\Gamma(n+\delta_\alpha)}
\end{equation}
where the powers of $(-2\pi i)$ were included for later convenience. Conjecturally, the series $B_\bbalpha(\xi)$ has a finite radius of convergence and can be analytically continued to a cover of $\C\setminus \{\CS_\bbbeta\}_\bbbeta$. That is, it can have singularities (including branch points) only at other critical values of the Chern-Simons functional.

The series (\ref{thimble-asympt-formal}) then can be recovered as the asymptotic expansion of a generalization of a Laplace transform of $B^\bbalpha(\xi)$:
\begin{equation}
    I^\bbalpha_\pert(k)\stackrel{\text{asympt}}{\cong} 
    \int_{\CS_{\bbalpha}}^{\infty} d\xi\, B^\bbalpha(\xi)\,e^{2\pi ik\xi}
\end{equation}
where the integral is taken over any contour starting from $\CS_\bbalpha$  and going to infinity in the direction where $e^{2\pi ik\xi}\rightarrow 0$, avoiding the possible singularities at $\xi_*=\CS_\bbbeta$ for some other $\bbbeta$. Note that when $\delta_\alpha\leq 0$ the integral must be regularized. There is a standard way to do it, by considering instead a Hankel-type contour going around the original contour  and modifying $B^\bbalpha(\xi)$ appropriately (cf. \cite{Gukov:2016njj}):
\begin{multline}
    B^\bbalpha(\xi) \rightsquigarrow
    \\
    \widetilde{B}^\bbalpha(\xi):=
    \sum_{n\geq 0}a_n^{(\alpha)}(-2\pi i)^{n+\delta_\alpha}\,(\xi-\CS_\bbalpha)^{n+\delta_\alpha-1}\cdot
    \left\{
        \begin{array}{ll}
             -\frac{1}{2\Gamma(n+\delta_\alpha)},
             & \delta_\alpha\in \frac{1}{2}+\Z, \\ \frac{\log (\xi-\CS_\bbalpha)}{2\pi i\Gamma(n+\delta_\alpha)}, & \delta_\alpha\in \Z,\;n+\delta_\alpha\geq 1,  \\
        \frac{(-1)^{n+\delta_\alpha}\,(-n-\delta_\alpha)!}{2\pi i}, & \delta_\alpha\in \Z,\;n+\delta_\alpha\leq 0.  \\
        \end{array}
    \right.
\end{multline}

\begin{figure}
\centering
\begin{tikzpicture}
    \draw[->] (0,-3) -- (0,4) node[right=2] {$\im \xi$};
     \draw[->] (-4,0) -- (4,0) node[above=2] {$\re \xi$};
    
    \draw plot[only marks,mark=x,mark size=4pt,mark options={draw=red}] coordinates {(1,1) (-1,-2) (-2,2) (3,2)};

    \draw[ultra thick, 
        decoration={markings, mark=at position 0.5 with {\arrow{>}}},
        postaction={decorate}
        ] (1,1) node[below=8] {$\CS_\bbalpha$} -- (4,4);

    \draw[blue, ultra thick, 
        decoration={markings, mark=at position 0.4 with {\arrow{>}}},
        postaction={decorate}
        ] (3.7,4) -- (1,1.3) .. controls ++(-135:0.7) and ++(-135:0.7) .. (1.3,1) -- (4,3.7);

    \draw (1,1) -- (3,1);

    \draw ([shift=(0:1.5)]1,1) node[above right] {$\mathrm{arg}\,\frac{i}{k}$} arc (0:45:1.5);
\end{tikzpicture}
\caption{The black ray originating at $\xi=\CS_\mathbb{\alpha}$ shows the contour in (\ref{thimble-integral-borel}) corresponding to the Lefschetz thimble, valid for the case $\delta_\bbalpha\geq 0$. The blue contour shows the Hankel-type contour needed instead in the case $\delta_\bbalpha<0$, with $B^\bbalpha(\xi)$ replaced by $\widetilde{B}^\bbalpha(\xi)$ in the integrand.  The red crosses mark singularities at critical values of the Chern-Simons functional.}

\label{fig:Borel-thimble-contour}
\end{figure}
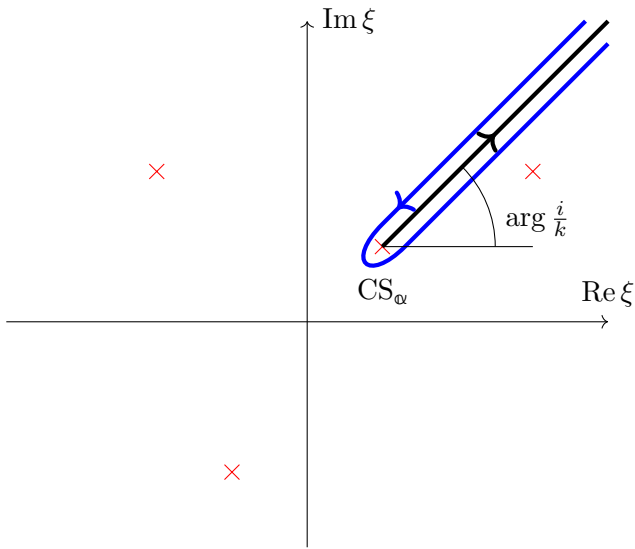

For a generic $k$, there is a canonical choice of the contour: the straight ray determined by the condition $-ik(\xi-\CS_\bbalpha)\in \R_+$. For finite-dimensional oscillatory integrals, the integral over such a contour recovers the value of the original integral over the corresponding Lefschetz thimble \cite{pham1983}. Therefore one can use it to \textit{define}
\begin{equation}
    I^\bbalpha(k) := \int\limits_{\CS_\bbalpha+\frac{i}{k}\R_+} 
     d\xi\, B^\bbalpha(\xi)\,e^{2\pi ik\xi}
     \label{thimble-integral-borel}
\end{equation}
without directly using an ill-defined functional integration as in (\ref{CS-thimble-integral}). Here again if $\delta_\alpha<0$, the integral should be modified appropriately: the contour should be changed to the Hankel contour surrounding the original integration ray $\CS_\bbalpha+\frac{i}{k}\R_+$ and $B^\bbalpha(\xi)$ should be replaced by $\widetilde{B}^\bbalpha(\xi)$ (see Figure \ref{fig:Borel-thimble-contour}).

Conjecturally, the singularity structure of $B^\bbalpha(\xi)$ has the following form:
\begin{equation}
    B^\bbalpha(\xi)\stackrel{\xi\rightarrow \xi_\ast}{\sim}
    \sum_{\bbbeta:\,\CS_\bbbeta=\xi_*} m^\bbalpha_\bbbeta\,\widetilde{B}^\bbbeta (\xi)+\text{regular}.
    \label{B-singularity}
\end{equation}
The coefficients of the singular part, $m^\bbalpha_\bbbeta$, known as Stokes coefficients, is conjecturally an integer. They can be formally interpreted as the monodromy coefficients of the middle-dimensional cycles in the holomorphic fibration $\CS:\tA\rightarrow \C$. Namely, assume the Morse-Bott situation. Then, as described before, each thimble $\bbalpha$ corresponds to a compact middle-dimensional cycle in $\Crit \CS$. For $\xi\in\C$ in the vicinity of the critical value $\CS_\bbalpha\in \C$ one then consider a middle dimensional cycle $\Xi^\bbalpha$ in the fiber $\CS^{-1}(\xi)$ which degenerates to the cycle in $\Crit \CS$ corresponding to $\bbalpha$, as $\xi\rightarrow \CS_\bbalpha$. Then one can first transport $\Xi^\alpha$ from the vicinity of $\CS_\bbalpha\in\C$ to the vicinity of $\xi_*$, and then transport it along a small circle surrounding  $\xi_*$. The latter part of the process will result in a monodromy of the corresponding homology classes:
\begin{equation}
    [\Xi^\bbalpha] \;\longmapsto\; [\Xi^\bbalpha]+\sum_{\bbbeta:\,\CS_\bbbeta=\xi_*} m^\bbalpha_\bbbeta\,[\Xi^\bbbeta].
\end{equation}

The coefficients $m^\bbalpha_\bbbeta$ in principle depend on the homotopy class of the path in the $\xi$-plane from $\CS_\bbalpha$ to $\xi_*$ along which the analytic continuation of $B^\bbalpha(\xi)$ was performed. The naive canonical prescription is to do it along a straight line. However, if there are singularities along this line this prescription has to be refined: a choice from which side we ``dodge'' the singularities along the path has to be made. 

\begin{figure}
\centering
\begin{tikzpicture}
    \draw[->] (0,-1) -- (0,5) node[right=2] {$\im \xi$};
     \draw[->] (-1,0) -- (5,0) node[above=2] {$\re \xi$};
    
    \draw plot[only marks,mark=x,mark size=4pt,mark options={draw=red}] coordinates {(1,1) (2,2) (3,3) (4,4)};

    \draw[blue, ultra thick]
        (1.2,1.2) -- (1.8,1.8);

    \draw[blue, ultra thick]
        (2.2,2.2) -- (2.8,2.8);

    \draw[blue, ultra thick,->]
        (3.2,3.2) -- (3.8,3.8);

    \draw (1,1) node[below=7] {$\CS_\bbalpha$};

    \draw (4,4) node[right=7] {$\CS_\bbbeta$};
        
     \draw[blue, ultra thick] (1.8,1.8) arc (-135:45:0.283);

     \draw[blue, ultra thick] (2.8,2.8) arc (-135:45:0.283);
\end{tikzpicture}
\caption{The path of analytic continuation in the $\xi$-plane from the neighborhood of $\CS_\bbalpha$ to the neighborhood of $\CS_\bbbeta$, in the case when there are other critical values of the Chern-Simons functional (shown as red crosses) on the straight line connecting $\CS_\bbalpha$ with $\CS_\bbbeta$.}

\label{fig:analytic-continuation-path}
\end{figure}
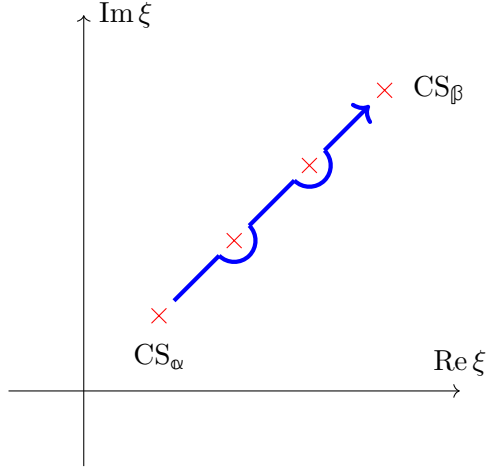

There is, in a sense, a canonical choice of the path: to go from $\CS_\bbalpha$ to $\xi_\ast$ along a segment infinitesimally shifted to the right (with respect to the direction of movement) from the straight segment connecting $\CS_\bbalpha$ with $\xi_\ast$ (see Figure \ref{fig:analytic-continuation-path}).

This prescription is in agreement with the standard interpretation of the Stokes coefficients as the coefficients describing the jumps of the integrals over the Lefschetz thimbles occurring when $k\in \C\setminus \{0\}$ passes through a ``wall'' determined by the condition $k(\CS_\bbbeta-\CS_\bbalpha)\in i\R_+$. Namely, from (\ref{thimble-integral-borel}) and (\ref{B-singularity}) it follows that:
\begin{equation}
    \lim_{\epsilon\rightarrow 0+} 
    I^\bbalpha(ke^{i\epsilon}) =
    \lim_{\epsilon\rightarrow 0-} 
\left(I^\bbalpha(ke^{i\epsilon})+
\sum_{\bbbeta:\;k(\CS_\bbbeta-\CS_\bbalpha)\in i\R_+\setminus\{0\}}  
m^{\bbalpha}_\bbbeta I^\bbbeta(ke^{i\epsilon})\right).
\label{thimble-jump}
\end{equation}

Assuming the conjectures mentioned above are true, and using the described ``canonical'' prescription for $m^\bbalpha_\bbbeta$, we can then define the following $\tilde{q}$-series labeled by a pair of $\Z$-orbits of thimbles:
\begin{equation}
    I_\beta^\alpha(\tilde{q}) :=
    \sum_{\bbbeta\in \beta}m^\bbalpha_\bbbeta \,\tilde{q}^{\CS_\bbalpha-\CS_\bbbeta},\qquad \in \tilde{q}^{\Delta_{\alpha}^{\beta}}\Z[[\tilde{q},\tilde{q}^{-1}]],
    \label{Stokes-series-definition}
\end{equation}
where $\bbalpha$ is any lift of $\alpha$. The overall power shift $\Delta_\alpha^\beta\in[0,1)$ is such that $\Delta_\alpha^\beta=\CS_\bbalpha-\CS_\bbbeta \mod 1$. Note that the result of the summation in the right-hand side of (\ref{Stokes-series-definition}) is independent of the choice of $\bbalpha$. This is because $m^\bbalpha_\bbbeta$ is invariant under simultaneous action of $\Z$ on $\bbalpha$ and $\bbbeta$. To take into account the possibility of coinciding orbits (i.e. $\alpha=\beta$), we introduce the ``self-monodromy'' coefficient:
\begin{equation}
    m^\bbalpha_\bbalpha:=\left\{
\begin{array}{cl}
     1, & \delta_\alpha\in \Z,  \\
     -1, & \delta_\alpha\in \frac{1}{2}+\Z.
\end{array}
    \right.
\end{equation}

Note that so far $\tilde{q}$ has appeared as just a formal variable in the definition of formal power series (\ref{Stokes-series-definition}). However, as we will see shortly, when one considers a ``total monodromy'', or equivalently ``total jump'' of Lefschetz thimbles, it is natural to identify $\tilde{q}=e^{-2\pi ik}$. Let us start with $I^\bbalpha(k)$ for some generic value of $k$. It is locally an analytic function in $k$ and can be analytically continued. Consider its analytic continuation with $\arg(1/k)$ starting from $\epsilon$ and increasing to $\pi+\epsilon$ (for some sufficiently small $\epsilon>0$). Applying the formula (\ref{thimble-jump}) for all the rays in the walls in the corresponding sector in the $k$-plane, the analytic continuation can be related to a linear combination of Lefschetz thimbles (cf. \cite{Garoufalidis:2020nut,Garoufalidis:2020xec,Garoufalidis:2021osl,Wheeler:2023cht}):
\begin{equation}
    I^\bbalpha(k)\;\longmapsto \;
    \sum_{\bbbeta} \prescript{+}{}{m}^\bbalpha_\bbbeta \,I^\bbbeta(k)
    \label{total-jump-plus}
\end{equation}
with some integer coefficients $\prescript{+}{}{m}^\bbalpha_\bbbeta$ (assuming that the infinite sum over $\bbbeta$ is convergent). Then defining
\begin{equation}
    \widetilde{I}^{\alpha}(k):=e^{-2\pi ik\CS_\bbalpha}\,I^\bbalpha(k),
    \label{CS-I-tilde-def}
\end{equation}
one can rewrite (\ref{total-jump-plus}) in terms of series (\ref{Stokes-series-definition}) and a \textit{finite} sum over the orbits:
\begin{equation}
     \widetilde{I}^\alpha(k)\;\longmapsto \;
    \sum_{\beta} \prescript{+}{}{I}^\alpha_\beta(e^{-2\pi ik}) \,\widetilde{I}^\beta(k),
\end{equation}
where
\begin{equation}
    \prescript{+}{}{I}^\alpha_\beta(\tilde{q}):=
    \sum_{\bbbeta\in \beta}\prescript{+}{}{m}^\bbalpha_\bbbeta \,\tilde{q}^{\CS_\bbalpha-\CS_\bbbeta}.
\end{equation}
By construction, $\prescript{+}{}{I}^\alpha_\beta(\tilde{q})$ are series in non-negative powers of $\tilde{q}$, as $\re\CS_\bbalpha>\re\CS_\bbbeta$ for non-zero terms. Therefore it is natural to expect series to be convergent for $|\tilde{q}|<1$, which indeed happens in all the known examples.

Similarly, one can consider analytic continuation with $\arg(1/k)$ starting from $\pi+\epsilon$ and increasing to $2\pi+\epsilon$:
\begin{equation}
    I^\bbalpha(k)\;\longmapsto \;
    \sum_{\bbbeta} \prescript{-}{}{m}^\bbalpha_\bbbeta \,I^\bbbeta(k)
    \label{total-jump-minus}
\end{equation}
with some integer coefficients $\prescript{-}{}{m}^\bbalpha_\bbbeta$. As before, one can rewrite (\ref{total-jump-plus}) as a  sum over the orbits:
\begin{equation}
     \widetilde{I}^\alpha(k)\;\longmapsto \;
    \sum_{\beta} \prescript{-}{}{I}^\alpha_\beta(e^{-2\pi ik}) \,\widetilde{I}^\beta(k),
\end{equation}
where
\begin{equation}
    \prescript{-}{}{I}^\alpha_\beta(\tilde{q}):=
    \sum_{\bbbeta\in \beta}\prescript{-}{}{m}^\bbalpha_\bbbeta \,\tilde{q}^{\CS_\bbalpha-\CS_\bbbeta}.
\end{equation}
By construction, $\prescript{-}{}{I}^\alpha_\beta(\tilde{q})$ are series in non-positive powers of $\tilde{q}$ and it is natural to expect them to converge for $|\tilde{q}|>1$. As we will see later, in the case of weakly negative-definite plumbed manifolds the Stokes jump happens across a single wall $k\in -i\R_+$. Therefore in this case $\prescript{-}{}{I}^\alpha_\beta(\tilde{q})\equiv 0$ and $\prescript{+}{}{I}^\alpha_\beta(\tilde{q})\equiv {I}^\alpha_\beta(\tilde{q})$ (up to a constant self-monodromy term for $\beta=\alpha$, which is not included in  $\prescript{+}{}{I}^\alpha_\beta(\tilde{q})$).

The goal of the rest of the Section \ref{sec:stokes-plumbed} is to provide an explicit computation algorithm for $I^\alpha_\beta(\tilde{q})$ in the case of plumbed 3-manifolds. 

Before we proceed, let us also note that as explained in \cite{Witten:2010cx}, one can interpret $m^\bbalpha_\bbbeta$ as the count of solutions of Kapusin-Witten equations on $Y\times \R$, which are known to be equivalent to gradient flow equations of CS functional in the space of $SL(2,\C)$ connections on $Y$. More precisely, assuming that $B^\bbalpha(\xi)$ has no singularities inside the straight interval connecting $\CS_\bbalpha$ and $\CS_\bbbeta$, the coefficient $m^\bbalpha_\bbbeta$ can be interpreted as counting flows starting from the compact middle-dimensional cycle corresponding to $\bbalpha$ and terminating at a certain (in general, non-compact) middle-dimensional cycle in $\Crit \CS$ which is ``dual'' in a certain sense to the cycle corresponding to $\bbbeta$. In the case when the cycle $\bbbeta$ is in the connected component which has a form of the total space of the cotangent bundle over a compact real manifold, $\bbbeta$ itself must be homologous to the base of the fibration, and the dual cycle can be chosen to be a fiber at some point.

\subsection{Review of plumbed 3-manifolds}
\label{sec:review-plumbed}

In this section we provide a brief review of \textit{plumbed manifolds} and fix notations relevant to the rest of the paper. A plumbed manifold is a 3-manifold associated with a \textit{plumbing graph} $\Gamma$, a graph with a certain additional data \cite{neumann1981calculus}. We restrict our attention to the case when the plumbing graph is a tree and the plumbing is \textit{orientable} and \textit{closed} (in the terminology of \cite{neumann1981calculus}), although the latter condition will be relaxed at some point later in the paper. In this case, to each vertex $i$ of the graph one assigns a non-negative integer $g_i\in\Z_{\geq 0}$ and an arbitrary integer $f_i\in\Z$ (see Figure \ref{fig:contours-original}). The manifold $Y_\Gamma$ that one associates with the graph can be constructed from basic building blocks associated with the vertices $V$. Namely, to each vertex $i\in V$ one associates a circle fibration over a genus $g_i$ Riemann surface $\Sigma_i$ with Euler class $f_i$. An edge between a pair of vertices $i,j$ then corresponds to the following operation. First, for each of the two fibrations, one removes the restriction of the bundle to a sufficiently small disk in the base (the disks corresponding to different edges should not overlap). This creates a torus boundary in the total space for each of the two fibrations. Then one glues them along the boundaries with the map that identifies the boundary circle in the base of one fibration with that in the fiber of the other, and vice-versa. 

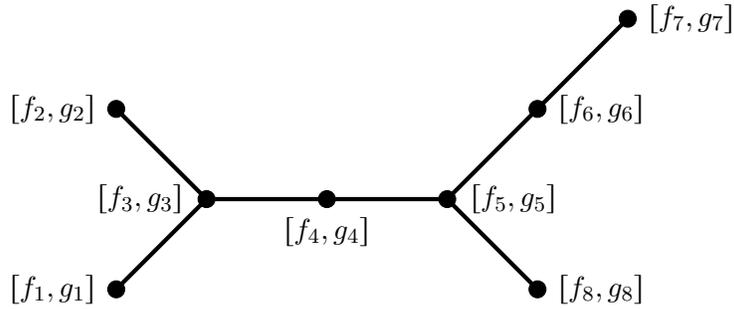
\begin{figure}
\centering
\begin{tikzpicture}[scale=0.4]

\draw[ultra thick] (5,0)  -- (9,0);
  \draw[ultra thick] (9,0) -- (13,0);
  \draw[ultra thick] (13,0) -- (16,3);
  \draw[ultra thick] (16,3) -- (19,6);
   \draw[ultra thick] (13,0) -- (16,-3);
    \draw[ultra thick] (5,0) -- (2,-3);
     \draw[ultra thick] (5,0) -- (2,3);
    \filldraw[black] (5,0) circle (8pt) node[left=5] {$[f_3,g_3]$};
    \filldraw[black] (2,3) circle (8pt) node[left=4] {$[f_2,g_2]$};
    \filldraw[black] (2,-3) circle (8pt) node[left=4] {$[f_1,g_1]$};
    \filldraw[black] (9,0) circle (8pt) node[below=3] {$[f_4,g_4]$};
    \filldraw[black] (13,0) circle (8pt) node[right=5] {$[f_5,g_5]$};
    \filldraw[black] (16,3) circle (8pt) node[right=4] {$[f_6,g_6]$};
    \filldraw[black] (19,6) circle (8pt) node[right=4] {$[f_7,g_7]$};
    \filldraw[black] (16,-3) circle (8pt) node[right=4] {$[f_8,g_8]$};
\end{tikzpicture}
\caption{An example of a plumbing graph.}
\label{fig:plumbing-graph}
\end{figure}

Some basic topological invariants of the resulting closed 3-manifold $Y_\Gamma$ can be expressed in terms of the $|V|\times |V|$ matrix $B$ which off the diagonal coincides with the adjacency matrix of the graph $\Gamma$ and on the diagonal is given by $B_{ii}=f_i$.

\begin{figure}
\centering
\begin{tikzpicture}[scale=0.4]
\begin{knot}[end tolerance=1pt]
\strand[ultra thick] (2, 0) 
  .. controls ++(90:1.5) and ++(0:-1) .. (5,2) node[pos=1,above] {$f_3$}
  .. controls ++(0:1) and ++(90:1.5) .. 
  (8,0)
  .. controls ++(90:-1.5) and ++(0:1) .. (5,-2)
  .. controls ++(0:-1) and ++(90:-1.5) .. (2, 0);
  \strand[ultra thick] (6, 0) 
  .. controls ++(90:1.5) and ++(0:-1) .. (9,2) node[pos=1,above] {$f_4$}
  .. controls ++(0:1) and ++(90:1.5) .. (12,0)
  .. controls ++(90:-1.5) and ++(0:1) .. (9,-2)
  .. controls ++(0:-1) and ++(90:-1.5) .. (6, 0);
 \strand[ultra thick] (10, 0) 
  .. controls ++(90:1.5) and ++(0:-1) .. (13,2) node[pos=1,above] {$f_5$}
  .. controls ++(0:1) and ++(90:1.5) .. (16,0)
  .. controls ++(90:-1.5) and ++(0:1) .. (13,-2)
  .. controls ++(0:-1) and ++(90:-1.5) .. (10, 0);
   \strand[ultra thick] (14, 1) 
  .. controls ++(135:1) and ++(45:-1) .. (15,4) 
  .. controls ++(45:1) and ++(135:1) .. (18,5)
  .. controls ++(135:-1) and ++(45:1) .. (17,2) node[pos=1,right] {$f_6$}
  .. controls ++(45:-1) and ++(135:-1) .. (14, 1);
     \strand[ultra thick] (14, -1) 
  .. controls ++(-135:1) and ++(-45:-1) .. (15,-4) 
  .. controls ++(-45:1) and ++(-135:1) .. (18,-5)
  .. controls ++(-135:-1) and ++(-45:1) .. (17,-2) node[pos=1,right] {$f_8$}
  .. controls ++(-45:-1) and ++(-135:-1) .. (14, -1);
    \strand[ultra thick] (17, 4) 
  .. controls ++(135:1) and ++(45:-1) .. (18,7) 
  .. controls ++(45:1) and ++(135:1) .. (21,8)
  .. controls ++(135:-1) and ++(45:1) .. (20,5) node[pos=1,right] {$f_7$}
  .. controls ++(45:-1) and ++(135:-1) .. (17, 4);
    \strand[ultra thick] (0, -5) 
  .. controls ++(135:1) and ++(45:-1) .. (1,-2) node[pos=1,left] {$f_1$}
  .. controls ++(45:1) and ++(135:1) .. (4,-1)
  .. controls ++(135:-1) and ++(45:1) .. (3,-4)
  .. controls ++(45:-1) and ++(135:-1) .. (0, -5);
      \strand[ultra thick] (0, 5) 
  .. controls ++(-135:1) and ++(-45:-1) .. (1,2) node[pos=1,left] {$f_2$}
  .. controls ++(-45:1) and ++(-135:1) .. (4,1)
  .. controls ++(-135:-1) and ++(-45:1) .. (3,4)
  .. controls ++(-45:-1) and ++(-135:-1) .. (0, 5);

   \flipcrossings{5,4,1,7,9,12,13}
\end{knot}
\end{tikzpicture}
\caption{Dehn surgery diagram corresponding to the plumbing graph shown in Figure \ref{fig:plumbing-graph}, with $g_i=0,\,\forall i$.}
\label{fig:plumbing-dehn-surgery}
\end{figure}
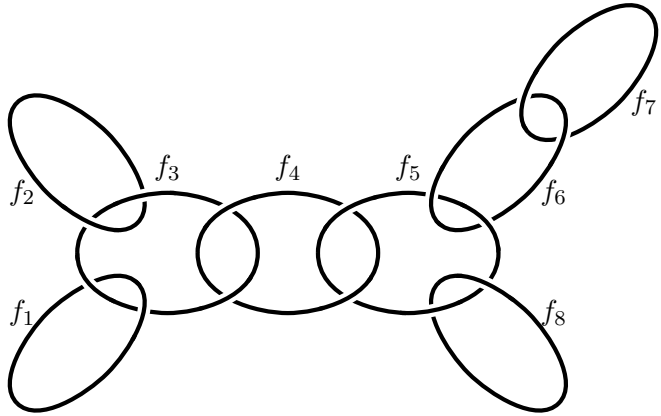

In particular, the 1st homology group of the 3-manifold is given by 
\begin{equation}
    H_1(Y_\Gamma)=\Z^{2\sum_i g_i}\oplus \mathrm{Coker}\,B.
\end{equation}
The manifold is a rational homology sphere (i.e. $b_1=0$) if and only if all the fibrations are over spheres (i.e. $g_i=0,\forall i$) and $\det{B}\neq 0$. When all the genera are zero, the manifold also admits a simple representation in terms of Dehn surgery on a framed link associated with the plumbing graph in a natural way (see Figure \ref{fig:plumbing-dehn-surgery}). The link has $|V(\Gamma)|$ components. For a vertex $i\in V(\Gamma)$ of the graph one associates a copy of an unknot with framing $f_i$. A pair of unknots associated with a pair of vertices forms a Hopf link if the vertices are connected by an edge and forms an unlink otherwise.

Of course, as in the case of Dehn surgery representation of a 3-manifold, different plumbing graphs can produce diffeomorphic 3-manifolds. The analogue of the 3-dimensional Kirby moves are the so-called Neumann moves on the plumbing graphs, listed in \cite{neumann1981calculus}. A pair of plumbings $\Gamma$ and $\Gamma'$ produce diffeomorphic manifolds if and only if they can be related by a sequence of Neumann moves.

\subsection{Information from a single thimble}
\label{sec:single-thimble-info}

Assume we have only the knowledge of the contribution of the ``zero'' Lefschetz thimble $I^{\bbalpha_0}(k)$ corresponding to the trivial flat connection with zero value of the Chern-Simons functional.\footnote{A reader mostly interested in computational aspects can skip this subsection.} Instead of $\bbalpha_0$ one could also consider any other distinguished thimble, the general analysis in this subsection would remain the same. One could ask a question: what information about the monodromy coefficients $m^\bbalpha_\bbbeta$ and other $I^{\bbalpha}(k)$ can be recovered from just $I^{\bbalpha_0}(k)$?

The answer, which is rather tautological, can be formulated as follows. As was reviewed in Section \ref{sec:CS-resurgence-review}, the space $\GG=\oplus_{\bbalpha}\Z\bbalpha$ of all admissible contours is a free abelian group generated by the Lefschetz thimbles $\bbalpha$. In addition, it is equipped with:
\begin{enumerate}
    \item Direct sum decomposition \begin{equation}
        \GG=\bigoplus_{S\in \CritVal\CS} \GG_S
        \label{crit-val-decomposition}
    \end{equation} indexed by the set $\CritVal\CS$ of the critical values of the Chern-Simons functional, so that a Lefschetz thimble $\bbalpha\in\GG_{\CS_\bbalpha}$.
    
    \item The free $\Z$-action on $\GG$ that shifts the components: 
    \begin{equation}
        n:\GG_S\stackrel{\cong}{\longrightarrow}\GG_{S+n},\qquad n\in\Z ,
        \label{crit-value-decomp-Z-action}
    \end{equation}
    and acts on the thimbles as was described in Section \ref{sec:CS-resurgence-review}: $\bbalpha\rightarrow \bbalpha^{(n)}$.
    \item For each critical value $S\in \CritVal\CS$ there is a Stokes automorphism of the group $\GG$:
    \begin{equation}
        \begin{array}{rrcl}
            \mathcal{S}_S: & \GG & \longrightarrow & \GG,  \\
             & \GG_{S'} & \longrightarrow & \GG_{S'}\oplus \GG_{S},  \\
            & \bbalpha & \longmapsto & \bbalpha +\sum\limits_{\bbbeta:\,\CS_\bbbeta=S}m^\bbalpha_\bbbeta\,\bbbeta.
        \end{array}
    \end{equation}
\end{enumerate}

 Let $\GG^0\subset \GG$ be the minimal subgroup containing the zero Lefschetz thimble $\bbalpha_0$ and closed under the action of all Stokes automorphisms $\mathcal{S}_S$, and projections on the components $\GG_S$. (The latter condition is actually redundant.) Let $\widehat{\GG}^0$ be the orbit of $\GG^0$ under $\Z$-action. Since $\GG$ was a free abelian group, the subgroups $\GG^0\subset \widehat{\GG}^0\subset \GG$ are also free abelian. We can choose a basis in $\widehat{\GG}^0$ which contains $\bbalpha_0$ and respects both the decomposition (\ref{crit-val-decomposition}) and the $\Z$-action so that we have an induced free $\Z$-action on the basis elements of $\widehat{\GG}^0$. We denote the basis in the subgroup $\widehat{\GG}^0\subset \GG$ by $\bblambda,\bbmu,\bbnu,\ldots$ to emphasize that it is in general different from the basis elements in $\GG$ which we denote by $\bbalpha,\bbbeta,\bbgamma,\ldots$

 By construction, we have an induced action of the Stokes automorphisms on $\widehat{\GG}^0$:
    \begin{equation}
        \begin{array}{rrcl}
            \mathcal{S}_S: & \widehat{\GG}^0 & \longrightarrow & \widehat{\GG}^0,  \\
            & \bblambda & \longmapsto & \bblambda +\sum\limits_{\bbmu\in \GG_S^0}m^\bblambda_\bbmu\,\bbmu.
        \end{array}
    \end{equation}
More explicitly, if the basis in $\widehat{\GG}^0$ are expressed through the basis in $\GG$ as $\bblambda=\sum_{\bbalpha}C_\bbalpha^\bblambda\bbalpha$, the coefficients $m^\bblambda_\bbmu$ can be determined from the original Stokes coefficients $m^\bbalpha_\bbbeta$ by the following linear system of equations:
\begin{equation}
    \sum_{\bbalpha\in \GG_S}C^\bblambda_\bbalpha\,m^\bbalpha_\bbbeta =
    \sum_{\bbmu\in \GG_S^0}m^\bblambda_\bbmu\,C^\bbmu_\bbbeta.
\end{equation}
By construction of $\widehat{\GG}^0$, the solution for $m^\bblambda_\bbmu$ exists and is unique.

Then, starting from $I^{\bbalpha_0}(k)$, and applying Stokes jumps and the $\Z$-action realized as multiplication by $e^{2\pi ik}$, one can recover all $I^{\bblambda}(k)\equiv \sum_{\bbalpha} C^{\bblambda}_\bbalpha I^{\bbalpha}(k)$ and $m^\bblambda_\bbmu$ for all $\bblambda, \bbmu \in \widehat{\GG}^0$. 
 
Although in simple cases one has $\widehat{\GG}^0=\GG$, as we will see later in the paper, there are indications that this is not true in general, cf. ``phantom saddles'' in \cite{Costin:2023kla}. However, it is still plausible that always $\widehat{\GG}^0\cap \GG_S\neq 0,\;\forall S$. That is, all critical values of the Chern-Simons functional can be detected from just $I^{\bbalpha_0}(k)$. Based on the current evidence, one can also conjecture that a contour $\Gamma_{SU(2)}$ for which (\ref{CS-general-integral}) in the limit $k\rightarrow r\in \Z_{\geq 2}$ recovers the WRT invariant $\tau_r(Y)$ of the 3-manifold can be chosen to be inside this subgroup: $\Gamma_{SU(2)}\in \widehat{\GG}^0$.

\subsection{A finite-dimensional model}
\label{sec:finite-dimensional-model}

In this section we argue that for a plumbed manifold one can find a ``finite-dimensional model'' for the functional integral (\ref{CS-general-integral}) of the complex Chern-Simons theory. In particular, for the integral over the Lefschetz thimble of the lift $\bbalpha_0$ of the trivial flat connection $\alpha_0$ with $\CS_\bbalpha=0$, we obtain
\begin{equation}
    I^{\bbalpha_0}(k)=c\,k^{a/2}e^{\frac{2\pi ib}{k}}\int\limits_{\gamma^0\subset \C^n} d^nv\,R(v)\, e^{2\pi ik S(v)}
    \label{zero-thimble-model}
\end{equation}
where  $R(v)$ is a $k$-independent meromorphic function in $v\in \C^n$, given explicitly as a composition of elementary functions,  $S(v)$ is a $k$-independent homogeneous quadratic polynomial, and  $a,b,c$ are $k$-independen constants: $a$ is an integer, and $b$ is a rational number. The contour $\gamma^0$ is the Lefschetz thimble for the holomorphic fibration  $S:\C^n\rightarrow \C$ with a unique Morse critical point at the origin $v=0$.  Using this expression one can then obtain analytically the Stokes coefficients in the group $\widehat{\GG}^0$ which is generated by all Stokes automorphisms and $\Z$-action starting from $\bbalpha_0$, as explained in Section \ref{sec:single-thimble-info}. 

The basic idea behind this is the following. The integral
\begin{equation}
    \int\limits_{\gamma\subset \C^n} d^nv\,R(v)\, e^{2\pi ik S(v)}
    \label{finte-dimensional-integral-general-contour}
\end{equation}
has its own free abelian group $\GG_\fd$ generated by admissible contours $\gamma$. As we describe explicitly in what follows, one can choose a basis for this group analogous to the Lefschetz thimble basis in the sense that applying a Borel transform to the $1/k$ asymptotic expansion of the integral over such a contour and then applying directional Laplace transform to it recovers exactly the original integral, cf. (\ref{thimble-integral-borel}). Such contours will then exhibit Stokes jump phenomenon. Similarly to $\GG^0$ considered in Section \ref{sec:single-thimble-info}, one can then consider the minimal subgroup $\GG^0_\fd\subset \GG_\fd$ which contains $\gamma^0$ and is closed under Stokes jumps. As reviewed in Section \ref{sec:CS-resurgence-review}, the Stokes jumps are completely determined by the structure of branchpoints of the Borel transform of the perturbative expansion, which is not affected by the coefficient in front of the integral in (\ref{zero-thimble-model}), see Appendix \ref{app:borel}. It then follows that one must have $\GG^0\cong \GG^0_\fd$, equivariantly with respect to the Stokes automorphisms. That is, $\mathcal{S}_S:\GG^0\rightarrow \GG^0$ can be identified with the Stokes automorphisms of $\GG^0_\fd$. The latter can be determined by analysis of the finite-dimensional integrals (\ref{finte-dimensional-integral-general-contour}). Applying the $\Z$-action, one can then determine the Stokes automorphism action on $\widehat{\GG}^0\subset \GG$, the minimal subgroup containing the Lefschetz thimble $\bbalpha_0$ associated to the trivial flat connection, and closed under Stokes automorphisms and $\Z$-action.

Note, however, that it is \textit{not true} that $\GG_\fd\cong \GG$. Moreover, there is no $\Z$-action on $\GG_\fd$. Yet, in general there are pairs of elements from $\GG_\fd^0\cong \GG^0$ that can be transformed into one another by the action of a certain $\pm n\in \Z$. And, as we shall see below, in simple examples one can have
$$
\GG_\fd^{\Z_2}=\GG_\fd^0\cong \GG^0.
$$
Where $\GG_\fd^{\Z_2}$ is the subgroup of $\GG_\fd$ invariant under $\Z_2$ action corresponding to the symmetry $v\mapsto -v$ in the integral. However, in general $\GG_\fd$ cannot be identified with any subgroup of $\GG$.

\subsection{Weakly negative-definite plumbings}
\label{sec:stokes-weakly-negative}

Consider a 3-manifold $Y$ associated to a plumbing tree $\Gamma$, as in Section \ref{sec:review-plumbed}. For simplicity, we initially assume that all circle fibrations associated with the vertices of the plumbing graph are of genus zero: $g_i=0$ and, moreover that $|\det B|=1$. In particular, this implies that $Y$ is an integer homology sphere. 

Let $H,L\subset V$ be subsets of vertices of degrees $>2$ and $<2$ respectively. We refer to them as \textit{high-valency} and \textit{low-valency} vertices respectively. Furthermore, let 
\begin{align}
    C:= & (B^{-1})_{HH}, \label{C-block-def}\\
    D:= & (B^{-1})_{HL}, \label{D-block-def} \\
    A:= & (B^{-1})_{LL}, \label{A-block-def}
\end{align}
be the corresponding blocks in the matrix $B^{-1}$, the inverse to the linking matrix. Another initial assumption we make is that $C$ is negative-definite. The plumbings satisfying this condition are called \textit{weakly negative-definite} \cite{Gukov:2019mnk}. By $\sigma$ and $b_+$ we denote, respectively, the signature and the number of positive eigenvalues of $B$.

Other technical assumptions will be introduced at later stages, and in Section \ref{sec:more-general-plumbings} we explain how various assumptions can be relaxed.

Define a sequence of half-integer numbers $F_{i,\ell}\in\frac{1}{2}\Z,\,i\in V,\,\ell\in\Z$ as coefficients of the following symmetric expansion (a formal Laurent series in $z$ given by the half-sum of the expansions at $z=0$ and $z=\infty$):
\begin{equation}
    \sum_{n\in\Z}F_{i,n}z^n  =
    \left.\frac{1}{2}\,(z-1/z)^{2-\deg(i)}\right|_{\substack{\text{expansion at} \\ z=0}}+
    \left.\frac{1}{2}\,(z-1/z)^{2-\deg(i)}\right|_{\substack{\text{expansion at} \\ z=\infty}}.
    \label{vertex-half-sum-expansion}
\end{equation}

The following formula for the WRT invariant \cite{Witten:1988hf,reshetikhin1991invariants} $\tau_r(Y),\,r\in \Z_{\geq 2}$ was conjectured in \cite{Gukov:2017kmk} (generalizing \cite{Gopakumar:1998ii,lawrence1999modular,deHaro:2004id,deHaro:2004wn,hikami2005quantum,Blau:2006gh,hikami2006quantum,hikami2011decomposition,Gukov:2016gkn}) and later proved in \cite{Murakami:2023oam,Murakami:2023csv}:
\begin{equation}
   \sqrt{\frac{2}{r}}\,\sin\frac{\pi}{r}\,\tau_r(Y)
    =
    \lim_{\epsilon\rightarrow 0+} Z(r-i\epsilon),
    \label{WRT-from-limit}
\end{equation}
where for $\im k <0$, with $q\equiv e^{\frac{2\pi i}{k}}$ ($|q|<0$), 
\begin{equation}
Z(k):=\frac{(-1)^{b_+}\,q^\frac{3\sigma-\Tr B}{4}}{2\sqrt{2k}}\sum_{n\in \Z^V}\prod_{i\in V} F_{i,n_i}\,q^{-\frac{n^TB^{-1}n}{4}},
\label{Z-hat-from-F}
\end{equation}
and, to emphasize the distinction between continuous (complex) and discrete (integer) values of the ``level'' we denote the fomer by $k$ and the latter by $r$.

In \cite{Gukov:2017kmk,Gukov:2019mnk} it was argued that $Z(k)$ is invariant under a subset of Neumann moves on the plumbing graphs that preserve the above-mentioned conditions on the plumbings.\footnote{Note that this does not immediately imply that $Z(k)$ is the same for any pair of plumbings $\Gamma$ and $\Gamma'$ that satisfy the specified conditions and realize the same 3-manifold up to a diffeomorphism, as in principle there can be a sequence on Neumann moves relating $\Gamma$ and $\Gamma'$ that contained intermediate plumbings that do not satisfy these conditions. See \cite{Ri:2022bxf} for a related discussion. } Here we assume that the WRT invariant $\tau_r(Y)$ has a standard normalization such that $\tau_r(S^3)=1$. The prefactor on the left-hand side of (\ref{WRT-from-limit}) modifies it in such a way that it equals to 1 for $Y=S^2\times S^1$ instead. This is a more natural normalization from the point of view of identifying the left-hand side of (\ref{WRT-from-limit}) with the value of the corresponding 3d WRT TQFT and also naively interpreting it as the value of the functional integral in level $r$ Chern-Simons theory. Therefore one can interpret $Z(k)$ as a certain analytic continuation of the WRT invariant in the TQFT normalization, with respect to the level. As we show below, it also has the expected properties of an analytic continuation in the sense of \cite{Witten:2010cx,Kontsevich}, which is supposed to be realized by a functional integral of the form (\ref{CS-general-integral}) over a special (intrinsically non-unique) contour designed to recover WRT invariant in the limit when $k$ tends to an integer.

\begin{figure}
\centering
\begin{tikzpicture}
    \draw[->] (0,-4) -- (0,4) node[right=2] {$\re v_I$};
     \draw[->] (2,0) -- (-2,0) node[above=2] {$\im v_I$};
    \draw plot[only marks,mark=x,mark size=4pt,mark options={draw=red}] coordinates {(0,1) (0,2) (0,3) (0,-1) (0,-2) (0,-3)};

    \draw[blue,ultra thick, 
        decoration={markings, mark=at position 0.6 with {\arrow{>}}},
        postaction={decorate}
        ] (0.5,-3.5) -- (0.5,3.5) node[midway,above right=5] {$\frac{1}{2}$};

    \draw[blue,ultra thick, 
        decoration={markings, mark=at position 0.6 with {\arrow{>}}},
        postaction={decorate}
        ] (-0.5,-3.5) -- (-0.5,3.5) node[midway,above left=5] {$\frac{1}{2}$} node[below left = 15] {$\gamma_I$};

\end{tikzpicture}
\qquad
\begin{tikzpicture}
    \draw[->] (0,-4) -- (0,4) node[right=2] {$\re v_I$};
     \draw[->] (2,0) -- (-2,0) node[above=2] {$\im v_I$};
    \draw plot[only marks,mark=x,mark size=4pt,mark options={draw=red}] coordinates {(0,1) (0,2) (0,3) (0,-1) (0,-2) (0,-3)};

     \draw[blue,ultra thick, 
        decoration={markings, mark=at position 0.7 with {\arrow{>}}},
        postaction={decorate}
        ] (-2,-3.5) -- (2,3.5);

     \draw[blue,ultra thick, 
        decoration={markings, mark=at position 0.7 with {\arrow{<}}},
        postaction={decorate}
        ] (0,1) circle (0.4)  node[left=9] {$\frac{1}{2}$};

\draw[blue,ultra thick, 
        decoration={markings, mark=at position 0.7 with {\arrow{<}}},
        postaction={decorate}
        ] (0,2) circle (0.4)  node[left=9] {$\frac{1}{2}$};

\draw[blue,ultra thick, 
        decoration={markings, mark=at position 0.7 with {\arrow{<}}},
        postaction={decorate}
        ] (0,3) circle (0.4)  node[left=9] {$\frac{1}{2}$} node[left = 30] {$\gamma_I$};

     \draw[blue,ultra thick, 
        decoration={markings, mark=at position 0.7 with {\arrow{>}}},
        postaction={decorate}
        ] (0,-1) circle (0.4)  node[right=9] {$\frac{1}{2}$};

\draw[blue,ultra thick, 
        decoration={markings, mark=at position 0.7 with {\arrow{>}}},
        postaction={decorate}
        ] (0,-2) circle (0.4)  node[right=9] {$\frac{1}{2}$};

\draw[blue,ultra thick, 
        decoration={markings, mark=at position 0.7 with {\arrow{>}}},
        postaction={decorate}
        ] (0,-3) circle (0.4)  node[right=9] {$\frac{1}{2}$};
        
  \draw (1,0) arc (0:60:1) node[midway,right] {$\frac{1}{2}\mathrm{arg}\frac{i}{k}$};

\end{tikzpicture}
\caption{Left: the contour $\gamma_I$ in the $v_I$-plane, such that $\gamma=\bigtimes_{I\in H} \gamma_I$ in (\ref{Z-hat-to-LG-model}). Right: a homologically equivalent contour. The red crosses schematically denote the positions poles in $v_I$ of the integrand in (\ref{Z-hat-to-LG-model}), located potentially at integer values.}
\label{fig:contours-original}
\end{figure}
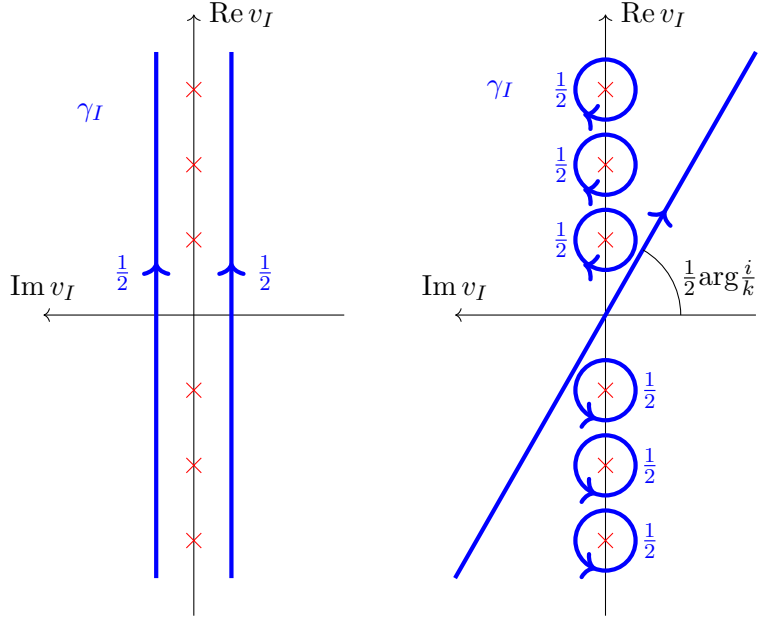

Next, we show that $Z(k)$ can be rewritten in the integral form a la (\ref{finte-dimensional-integral-general-contour}). In what follows $\pto$ is used to denote equality up to a factor of the form $c\cdot k^{\frac{a}{2}}\cdot q^{\frac{2\pi ib}{k}}$ for some $k$-independent constants $a,b,c$, with non-zero $c$, integral $a$ and rational $b$. We have
\begin{multline}
Z(k)
    \pto
    \sum_{n\in\Z^V}
\prod_{i\in V} {F_{i,n_i}} \, 
q^{-\frac{n^TB^{-1}n}{4}} =\\
\sum_{m\in \Z^H}\sum_{s\in \{\pm 1\}^L} \prod_{I\in H} {F_{I,m_I}} \prod_{a\in L} s_a\,
q^{-\frac{m^TCm}{4}-\frac{m^TDs}{2}-\frac{s^TAs}{4}} =\\
\sum_{m\in \Z^H}\sum_{s\in \{\pm 1\}^L} \prod_{I\in H} {F_{I,m_I}} \prod_{a\in L} s_a\,
q^{-\frac{(m+C^{-1}Ds)^TC(m+C^{-1}Ds)}{4}-\frac{s^T(A-D^TC^{-1}D)s}{4}} \pto \\
\sum_{m\in \Z^H}\sum_{s\in \{\pm 1\}^L} \prod_{I\in H} {F_{I,m_I}} \prod_{a\in L} s_a\,
q^{-\frac{(m+C^{-1}Ds)^TC(m+C^{-1}Ds)}{4}}
\pto\\
\sum_{m\in \Z^H}\sum_{s\in \{\pm 1\}^L} \prod_{I\in H} {c_{I,m_I}} \prod_{a\in L} s_a \int\limits_{\R^{|H|}} d{v^{|H|}}
e^{\pi i(m+C^{-1}Ds)^Tv}\,e^{2\pi ik\,\frac{v^TC^{-1}v}{4}}\pto\\
\int\limits_{\gamma} d{v^{|H|}}
\,
\frac{ \prod_{a\in L} \sin \pi (v^TC^{-1}D)_a}{\prod_{I\in H}(\sin\pi v_I)^{\deg(I)-2}}
\,e^{2\pi ik\,\frac{v^TC^{-1}v}{4}}.
\label{Z-hat-to-LG-model}
\end{multline}
Let us elaborate on the steps in this chain of relations. To transition from the first line to the second, we use the fact that $F_{I,\ell}=\delta_{\ell,0}$ and $F_{I,\ell}=\sum_{s\in \{\pm 1\}}s\delta_{\ell,s}$ for vertices of degree 2 and 1, respectively. Therefore, the sum over $\Z^V$ can be reduced to a sum over $\Z^H$. The transition from the second to the third line is simply a completion of squares with respect to $m\in\Z^H$. To arrive to the fourth line we then use the fact that the matrix
\begin{equation}
    A-D^TC^{-1}D
\end{equation}
is diagonal\footnote{We thank Cagri Karakurt for explaining the proof of this property.}. This implies that the term $s^T(A-D^TC^{-1}D)s$ is actually independent of $s$. The expression in the fifth line is related to the one on the fourth line via term-by-term Gaussian integration. Finally, to obtain the final expression, one exchanges the infinite summation over $m$ with the integration. In order to do that, one first needs to deform the contour of integration to ensure the absolute convergence of the sum over $m$ inside the integral. Namely, for each integration variable $v_I$ one needs to shift integration contour $\R$ in positive (resp. negative) imaginary direction to make convergent the part of the sum over $n_I$ corresponding to the expansion at $z=0$ (resp. at $z=\infty$) in (\ref{vertex-half-sum-expansion}). The deformation preserves the Cartesian product structure of the contour, i.e. the new contour of integration $\gamma\subset\C^H$ is still a product of contours inside complex $v_I$-planes: $\gamma=\bigtimes_{I\in H}\gamma_I$. The contours $\gamma_I$ are shown in Figure \ref{fig:contours-original} on the left. 

Thus, as claimed in (\ref{finte-dimensional-integral-general-contour}), we obtain an expression for $Z(k)$ representing an analytically continued partition function of Chern-Simons theory, with $n=|H|$,
\begin{equation}
    R(v)=\frac{ \prod_{a\in L} \sin \pi (v^TC^{-1}D)_a}{\prod_{I\in H}(\sin\pi v_I)^{\deg(I)-2}},
    \label{R-expression-raw}
\end{equation}
and
\begin{equation}
    S(v)= \frac{1}{4}\,v^TC^{-1}v.
    \label{S-expression}
\end{equation}
Alternatively, one could arrive at this final result by directly manipulating the integral representation for ${Z}(k)$ from \cite{Gukov:2017kmk}, see Appendix \ref{app:another-derivation}.

\begin{figure}
\centering
\begin{tikzpicture}[scale=0.4]

\draw[ultra thick] (2,0)  -- (10,0);
\draw[ultra thick] (16,0)  -- (25,0);
    \draw[ultra thick] (5,0) -- (2,-3);
     \draw[ultra thick] (5,0) -- (2,3);
    \filldraw[black] (5,0) circle (8pt) node[below=8] {$h(a)$};
    \filldraw[black] (9,0) circle (8pt) node[above=5] {$[f_{i_s},0]$} node[below=5] {$i_s$};

\draw (13,0) node {$\cdots$};

\filldraw[black] (17,0) circle (8pt) node[above=5] {$[f_{i_2},0]$} node[below=5] {$i_2$};
    \filldraw[black] (21,0) circle (8pt) node[above=5] {$[f_{i_1},0]$} node[below=5] {$i_1$};
   \filldraw[black] (25,0) circle (8pt) node[above=5] {$[f_a,0]$} node[below=5] {$a$};
  
\end{tikzpicture}
\caption{A high-valency vertex $h(a)$ associated to a low-valency vertex $a$ in a plumbing graph.}
\label{fig:plumbing-leg}
\end{figure}
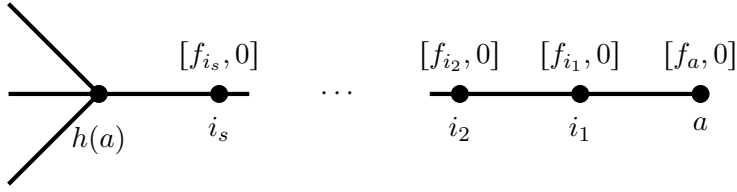

The arguments of the sines in the numerator of (\ref{R-expression-raw}) can be made more explicit: 
\begin{equation}
    (v^TC^{-1}D)_a=\frac{v_{h(a)}}{P_{a}},\qquad P_a\in\Z,
\end{equation}
where $h(a)\in H$ is the high-valency vertrex most adjacent to the low-valency vertex $a\in L$, as illustrated in Figure \ref{fig:plumbing-leg}. In other words, $a$ and $h(a)$ are connected by a sequence of edges $(a,i_1)$, $(i_2,i_3)$, ..., $(i_s,h(a))$ such that all vertices in this chain have valency two, i.e. $\deg (i_j)=2,\forall j$. The integer number $P_a$ then can be determined as the numerator of the following continuous fraction corresponding to the framings in the sequence of vertices between $a$ and $h(a)$:
\begin{equation}
    f_{a}-\cfrac{1}{f_{i_1}-\cfrac{1}{f_{i_2}-\cfrac{\ldots}{\ldots -\cfrac{1}{f_{i_s}}}}}=-\frac{P_a}{Q_a}.
\end{equation}
As usual, the integers $P_a$ and $Q_a$ are assumed to be coprime. This allows to write a neater expression for $R(v)$:
\begin{equation}
    R(v)=\frac{ \prod_{a\in L} \sin \cfrac{\pi v_{h(a)}}{P_a}}{\prod_{I\in H}(\sin\pi v_I)^{\deg(I)-2}}.
    \label{R-expression}
\end{equation}

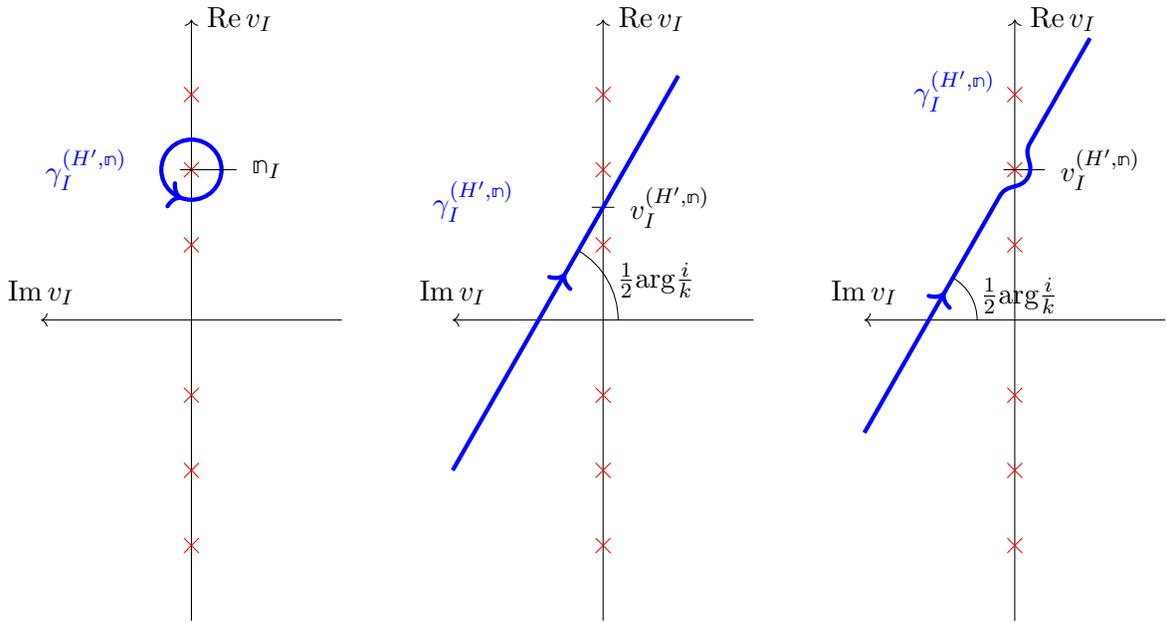
\begin{figure}
\centering
\begin{tikzpicture}
    \draw[->] (0,-4) -- (0,4) node[right=2] {$\re v_I$};
     \draw[->] (2,0) -- (-2,0) node[above=2] {$\im v_I$};
    \draw plot[only marks,mark=x,mark size=4pt,mark options={draw=red}] coordinates {(0,1) (0,2) (0,3) (0,-1) (0,-2) (0,-3)};

\draw[blue,ultra thick, 
        decoration={markings, mark=at position 0.7 with {\arrow{>}}},
        postaction={decorate}
        ] (0,2) circle (0.4)  node[left=20] {$\gamma_I^{(H',\mathbb{n})}$};

  \draw (-0.15,2) -- (0.6,2) node[right=2] {$\mathbb{n}_I$};

\end{tikzpicture}
\qquad
\begin{tikzpicture}
    \draw[->] (0,-4) -- (0,4) node[right=2] {$\re v_I$};
     \draw[->] (2,0) -- (-2,0) node[above=2] {$\im v_I$};
    \draw plot[only marks,mark=x,mark size=4pt,mark options={draw=red}] coordinates {(0,1) (0,2) (0,3) (0,-1) (0,-2) (0,-3)};
    
    \draw (-0.15,1.5) -- (0.15,1.5) node[right=2] {$v_I^{(H',\mathbb{n})}$};

     \draw[blue,ultra thick, 
        decoration={markings, mark=at position 0.5 with {\arrow{>}}},
        postaction={decorate}
        ] (-2,-2) -- (1,3.25) node[midway,above left=15] {$\gamma_I^{(H',\mathbb{n})}$};
   \draw (0.2,0) arc (0:60:1.07) node[midway,right] {$\frac{1}{2}\mathrm{arg}\frac{i}{k}$};
  
\end{tikzpicture}
\qquad
\begin{tikzpicture}
    \draw[->] (0,-4) -- (0,4) node[right=2] {$\re v_I$};
     \draw[->] (2,0) -- (-2,0) node[above=2] {$\im v_I$};
    \draw plot[only marks,mark=x,mark size=4pt,mark options={draw=red}] coordinates {(0,1) (0,2) (0,3) (0,-1) (0,-2) (0,-3)};
    
    \draw (-0.15,2) -- (0.4,2) node[right=2] {$v_I^{(H',\mathbb{n})}$};

     \draw[blue,ultra thick, 
        decoration={markings, mark=at position 0.35 with {\arrow{>}}},
        postaction={decorate}
        ] (-2,-1.5) -- (-0.2,1.65) .. controls ++(0.1,0.175) and ++(-0.1,-0.175) .. (0.175,1.9) .. controls ++(0.1,0.175) and ++(-0.1,-0.175) .. (0.2,2.35) -- (1,3.75) node[midway,left=20] {$\gamma_I^{(H',\mathbb{n})}$};

   \draw (-0.5,0) arc (0:60:0.65) node[midway,right] {$\frac{1}{2}\mathrm{arg}\frac{i}{k}$};
\end{tikzpicture}

\caption{A factor $\gamma_I^{(H',\mathbb{n})}$ in the Cartesian product decomposition of a basis ``Lefschetz thimble'' contour $\gamma^{(H',\mathbb{n})}=\bigtimes_{I\in H}\gamma_I^{(H',\mathbb{n})}$ corresponding to a subset $H'\subset H$ and $\mathbb{n}\in \Z^{H'}$, such that $\mathbb{n}_I$ is a pole in $v_I$ of $R(v)$ (\ref{R-expression}). Left: the case with $I\in H'$. Center: generic case when $I\notin H'$, with $v_I^{(H',\mathbb{n})}$ not being at a pole of $R(v)$. Right: a special case when $v_I^{(H',\mathbb{n})}$ coincides with a pole of $R(v)$.}
\label{fig:contours-basis}
\end{figure}

When $k\notin -i\R_+$, the contours $\gamma_I$ can be further deformed into contours shown on the left panel of Figure \ref{fig:contours-original}. After such a deformation, the contour can then be decomposed into a sum of the ``Lefschetz thimble'' contours of the finite-dimensional model. The latter can be explicitly classified as follows. Consider a subset $H'\subset H$ of the high-valency vertices and let $H'':=H\setminus H'$ be its complement. We have the corresponding block decomposition of the matrix $C^{-1}\in GL(\Q^{H})$:
\begin{equation}
  C^{-1}=
  \begin{array}{ccccc}
  & & {}_{H'} & {}_{H''} & \\
{}_{H'} & \ldelim({2}{4mm} & \CA_{(H')} & \CD_{(H')}^T & \rdelim){2}{4mm} \\
{}_{H''} &  & \CD_{(H')} & \CC_{(H')} &  
\end{array}.
\end{equation}
A ``Lefschetz thimble'' contour $\gamma^{(H',\mathbb{n})}$, labeled by $H'\subset H$ and an $|H'|$-tuple of integers\footnote{When $H'=\emptyset$ we omit $\mathbb{n}$ from the pair $(H',\mathbb{n})$ and write simply $(\emptyset)$, e.g. $\gamma^{(\emptyset)}$.} $\mathbb{n}\in \Z^{H'}$, can be defined as the cartertesian product of contours $\gamma^{(H',\mathbb{n})}_I$ in complex planes of variables $v_I,\;I\in H$, shown in Figure \ref{fig:contours-basis}. For $I\in H'$ the contour $\gamma^{(H',\mathbb{n})}_I$ is a small circle going around a pole at $v_I=\mathbb{n}_I$. The integration over such factors then results in taking corresponding residues:
\begin{equation}
    \int\limits_{\gamma^{(H',\mathbb{n})}}d^{|H|}v\,R(v)\,e^{2\pi ikS(v)}
    = \int\limits_{\bigtimes\limits_{I\in H''}\gamma_I^{(H',\mathbb{n})}}d^{|H''|}v''\,R^{(H',\mathbb{n})}(v'',k)\,e^{2\pi ikS^{(H',\mathbb{n})}(v'')}
    \label{integral-tuple-reduction}
\end{equation}
where 
\begin{equation}
    S^{(H',\mathbb{n})}(v''):=S(v)|_{v_I=\mathbb{n}_I,\,I\in H'}=\frac{1}{4}\,{v''}^T\CA_{(H')}{v''}+\frac{1}{2}\,{v''}^T\CD_{(H')}\mathbb{n}+\frac{1}{4}\,\mathbb{n}^T\CA_{(H')}\mathbb{n}
    \label{S-restricted}
\end{equation}
is a quadratic --- in general, non-homogeneous --- function in $v''\in \C^{H''}$ obtained from $S(v)$ by restricting it to $v_I=\mathbb{n}_I,\;I\in H'$, and
\begin{equation}
    R^{(H',\mathbb{n})}(v'',k) := e^{-2\pi ikS^{(H',\mathbb{n})}(v'')}
    \Res_{\{v_I=\mathbb{n}_I\}_{I\in H'}}R(v)\,e^{2\pi ikS(v)}\,
\end{equation}
is a meromorphic function in $v''$ which is polynomial in $k$. In the remaining directions $v_I''\equiv v_I,\,I\in H''$ we then take $\gamma^{(H',\mathbb{n})}$ to be the standard Lesfschetz thimble contour, which (due to the fact that $C<0$) is a product of straight lines $\gamma_I^{(H',\mathbb{n})}$ with slopes determined by $k$ (see Figure \ref{fig:contours-basis}) and passing through the extremum of $S^{(H',\mathbb{n})}(v'')$ at
\begin{equation}
    v''=v^{(H',\mathbb{n})}:=
    -\CA_{(H')}^{-1}\CD_{(H')}\,\mathbb{n}\qquad\in \Q^{H''}.
\end{equation}
The corresponding critical value of the quadratic function $S^{(H',\mathbb{n})}(v'')$ is
\begin{equation}
    S^{(H',\mathbb{n})}_\ast:=
    \frac{1}{4}\mathbb{n}^T\left(\CA_{(H')}
    -\CD^T_{(H')}\CC_{(H')}^{-1}\CD_{(H')}
    \right)\mathbb{n}.
    \label{fd-S-extremal-value}
\end{equation}
Note that since $C^{-1}<0$, the quadratic form in the above expression is also negative-definite.  The $k\rightarrow \infty$ asymptotics of the integral over $\gamma^{(H',\mathbb{n})}$ then can be obtained by expanding $R^{(H'',\mathbb{n})}(v'',k)$ in (\ref{integral-tuple-reduction}) in series in $(v''-v^{(H',\mathbb{n})})$ around the extremum. The result is either identically zero, or of the following form:
\begin{equation}
    \int\limits_{\gamma^{(H',\mathbb{n})}}d^{|H|}v\,R(v)\,e^{2\pi ikS(v)}
    = e^{2\pi ikS^{(H',\mathbb{n})}_\ast}\,\sum_{n\geq 0}\frac{a_n^{(H',\mathbb{n})}}{k^{n+\delta_{(H',n)}}}\,
    \label{fd-thimbe-expansion}
\end{equation}
for some $\delta_{(H',n)}\in \frac{1}{2}\Z$ and $a^{(H',n)}_0\neq 0$. The former case means that the contour $\gamma^{(H',\mathbb{n})}$ for such a particular pair $(H',\mathbb{n}\in\Z^{H'})$ is actually trivial and can be ignored. By construction, the exact integral over $\gamma^{(H',\mathbb{n})}$ can be recovered by performing the Borel resummation along a ray as in Figure \ref{fig:Borel-thimble-contour}. It is also clear that the contours $\gamma^{(H',\mathbb{n})}$ generate all admissible contours, and, in particular, include the original contour of integration shown in Figure \ref{fig:contours-original}.

If the plumbing graph is such\footnote{See Figure (\ref{fig:plumbing-leg}) to recall the definition of $h:L\rightarrow H$, the map from low- to high-valency vertices.} that $|h^{-1}(I)|\geq \deg(I)-2$, the meromorphic function has no poles at $v_I=0$. In that case one can assume $\mathbb{n}_I\neq 0,\forall I\in H'$. Then, due to the non-degeneracy of the quadratic form in (\ref{fd-S-extremal-value}), the only pair $(H',\mathbb{n}\in \Z^{H'})$ that has the vanishing critical value of the action, i.e. $S^{(H',\mathbb{n})}_*=0$, is $(\emptyset)$. Assume it is also true that the only flat connection with zero value of the Chern-Simons functional on the plumbed 3-manifold $Y$ is the trivial flat connection, and that the asymptotic expansion conjecture for the WRT invariant holds.  It then follows that $\gamma^{(\emptyset)}$ thimble in the finite-dimensional model corresponds to the thimble associated to the lift $\bbalpha_0$ of the trivial flat connection $\alpha_0$ with $\CS_\bbalpha=0$, as was anticipated in (\ref{zero-thimble-model}):
\begin{equation}
    I^{\bbalpha_0}(k)\pto \int_{\gamma^{(\emptyset)}}d^{|H|}v\,R(v)\,e^{2\pi ikS(v)},
    \label{CS-fd-zero-thimble-correspondence}
\end{equation}
where ``$\pto$'' has the same meaning as before. Since both the contour $\gamma^{(\emptyset)}$ in the finite-dimensional model and the trivial flat connection Lefschetz thimble in Chern-Simons theory are very distinguished contours it is tempting to conjecture that this relation holds even if the condition $|h^{-1}(I)|\geq \deg(I)-2$ on the graph is not satisfied. When $|H|=1$, i.e. there is a single high-valency vertex, the plumbed manifold is a Seifert fibration over the sphere. The formula (\ref{CS-fd-zero-thimble-correspondence}) then becomes the formula provided by Lawrence and Rozansky in \cite{lawrence1999witten} (see also \cite{Marino:2002fk,Beasley:2005vf}).

This implies that Stokes phenomena in the finite-dimensional model (i.e. jumps of the thimble contours $\gamma^{(H',\mathbb{n})}$ as one changes $k\in \C\setminus \{0\}$) capture the Stokes phenomena \textit{at least} in the minimal subgroup of all admissible contours in Chern-Simons theory that contains the Lefschetz thimble of the trivial flat connection and is closed under the monodromies in the Borel plane and the $\Z$-action (large gauge transformations). See Section \ref{sec:single-thimble-info} for details. In simple cases this subgroup contains all admissible contours.

\begin{figure}
\centering

\begin{tikzpicture}[baseline=(current  bounding  box.center)]
    \draw[->] (0,-4) -- (0,4) node[right=2] {$\re v_I$};
     \draw[->] (3,0) -- (-2.5,0) node[above=2] {$\im v_I$};
    \draw plot[only marks,mark=x,mark size=4pt,mark options={draw=red}] coordinates {(0,1) (0,2) (0,3) (0,-1) (0,-2) (0,-3)};

     \draw[blue,ultra thick, 
        decoration={markings, mark=at position 0.35 with {\arrow{>}}},
        postaction={decorate}
        ] (0,1.5) -- ++(30:3) node[above] {$\gamma_I^{(H',\mathbb{n})}$};
     \draw[blue,ultra thick] (0,1.5) -- ++(210:2.3);

    \draw (2,0) arc (0:30:4.6) node[midway,right] {$\frac{1}{2}\mathrm{arg}\frac{i}{k}$};

      \draw (-0.15,1.5) -- (0.4,1.5) node[right] {$v_I^{(H',\mathbb{n})}$};
\end{tikzpicture}
$\stackrel{ \substack{ \arg(1/k)\rightarrow\\ \arg(1/k)+2\pi \\ ~}}{\longmapsto}$
\begin{tikzpicture}[baseline=(current  bounding  box.center)]
    \draw[->] (0,-4) -- (0,4) node[right=2] {$\re v_I$};
     \draw[->] (3,0) -- (-2.5,0) node[above=2] {$\im v_I$};
    \draw plot[only marks,mark=x,mark size=4pt,mark options={draw=red}] coordinates {(0,1) (0,2) (0,3) (0,-1) (0,-2) (0,-3)};

     \draw[blue,ultra thick, 
        decoration={markings, mark=at position 0.35 with {\arrow{>}}},
        postaction={decorate}
        ] (0,1.5) -- ++(210:2.3) ;
     \draw[blue,ultra thick] (0,1.5) -- ++(30:3) node[above] {$-\gamma_I^{(H',\mathbb{n})}$};

    \draw (2,0) arc (0:30:4.6) node[midway,right] {$\frac{1}{2}\mathrm{arg}\frac{i}{k}$};

    \draw[blue,ultra thick, 
        decoration={markings, mark=at position 0.7 with {\arrow{>}}},
        postaction={decorate}
        ] (0,2) circle (0.3);

     \draw[blue,ultra thick, 
        decoration={markings, mark=at position 0.7 with {\arrow{>}}},
        postaction={decorate}
        ] (0,3) circle (0.3) node[left=10] {$\sigma_I^{\ell_I}$};
        
          \draw[blue,ultra thick, 
        decoration={markings, mark=at position 0.7 with {\arrow{<}}},
        postaction={decorate}
        ] (0,1) circle (0.3);
          \draw[blue,ultra thick, 
        decoration={markings, mark=at position 0.7 with {\arrow{<}}},
        postaction={decorate}
        ] (0,-1) circle (0.3);
          \draw[blue,ultra thick, 
        decoration={markings, mark=at position 0.7 with {\arrow{<}}},
        postaction={decorate}
        ] (0,-2) circle (0.3) node[left=10] {$-\sigma_I^{\ell_I}$};;
          \draw[blue,ultra thick, 
        decoration={markings, mark=at position 0.7 with {\arrow{<}}},
        postaction={decorate}
        ] (0,-3) circle (0.3);

          \draw (-0.15,1.5) -- (0.4,1.5) node[right] {$v_I^{(H',\mathbb{n})}$};

    \draw (-0.15,3) -- (0.4,3) node[right] {$\ell_I$};

    \draw (-0.15,-2) -- (0.4,-2) node[right] {$\ell_I$};
    
\end{tikzpicture}

\caption{Monodromy of a factor $\gamma_I^{(H',\mathbb{n})}$, $I\in H''$, in the Cartesian product decomposition of a basis ``Lefschetz thimble'' contour $\gamma^{(H',\mathbb{n})}=\bigtimes_{I\in H}\gamma_I^{(H',\mathbb{n})}$, as $\arg(1/k)$ increases by $2\pi$. The line $\gamma_I^{(H',\mathbb{n})}$ rotates by $\pi$ and absorbs the contours $\sigma_I^{(\ell_I)}$ encircling poles at $\ell_I$ with plus or minus sign, depending on whether the pole was above or below the line.}
\label{fig:contour-monodromy}
\end{figure}
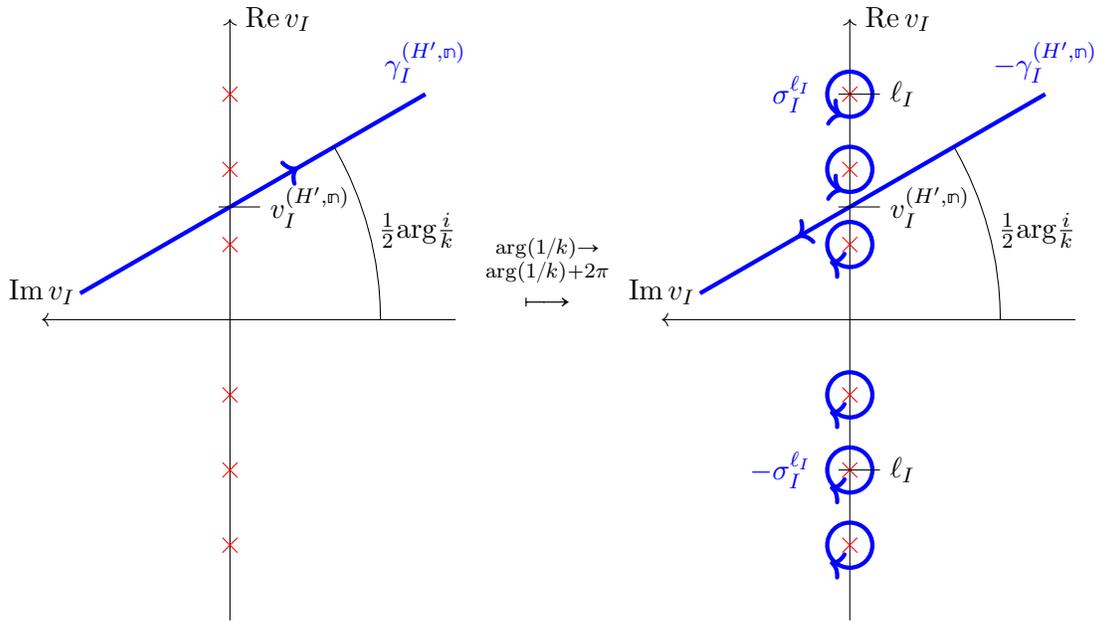

The Stokes jumps in the finite-dimensional model can be combinatorially described as follows. Consider a finite dimensional ``thimble'' contour $\gamma^{(H,\mathbb{n})}$ (Figure \ref{fig:contours-basis}). As one increases $\arg\frac{1}{k}$, the circle factors $\gamma_I^{(H,\mathbb{n})}$ always remain the same. The line factors, however, rotate counter-clockwise, with half the rate of the rotation of $k$. As $k$ makes a full (clockwise) rotation, the lines in the $v_I$-planes make a half counter-clockwise turn, illustrated in Figure \ref{fig:contour-monodromy}. As $k$ passes through the ray of the negative imaginary direction, the line overlaps with the real axis and absorbs the small circle contours going around poles, with positive/negative signs depending on whether the poles were above/below the line along the real axis. 

\begin{figure}
\centering
\begin{tikzpicture}[baseline=(current  bounding  box.center)]
    \draw[->] (0,-4) -- (0,4) node[right=2] {$\re v_I$};
     \draw[->] (3,0) -- (-3,0) node[above=2] {$\im v_I$};
    \draw plot[only marks,mark=x,mark size=4pt,mark options={draw=red}] coordinates {(0,1) (0,2) (0,3) (0,-1) (0,-2) (0,-3)};

     \draw[blue,ultra thick, 
        decoration={markings, mark=at position 0.35 with {\arrow{>}}},
        postaction={decorate}
        ] (0,0.5) -- ++(30:3) node[midway,right=10] {$\gamma_I^{(H',\mathbb{n})}$};
     \draw[blue,ultra thick] (0,0.5) -- ++(210:3);

         \draw (-0.4,0.5) -- (0.15,0.5) node[left=15] {$v_I^{(H',\mathbb{n})}$};

\end{tikzpicture}
\;\;$\simeq$\;\;
\begin{tikzpicture}[baseline=(current  bounding  box.center)]
    \draw[->] (0,-4) -- (0,4) node[right=2] {$\re v_I$};
     \draw[->] (3,0) -- (-3,0) node[above=2] {$\im v_I$};
    \draw plot[only marks,mark=x,mark size=4pt,mark options={draw=red}] coordinates {(0,1) (0,2) (0,3) (0,-1) (0,-2) (0,-3)};

     \draw[blue,ultra thick, 
        decoration={markings, mark=at position 0.35 with {\arrow{>}}},
        postaction={decorate}
        ] (0,2.5) -- ++(30:3) node[midway,right=10] {$\gamma_I^{(\tilde{H}',\tilde{\mathbb{n}})}$};
     \draw[blue,ultra thick] (0,2.5) -- ++(210:3);

         \draw (-0.4,2.5) -- (0.15,2.5) node[left=15] {$v_I^{(\tilde{H}',\tilde{\mathbb{n}})}$};

    \draw[blue,ultra thick, 
        decoration={markings, mark=at position 0.7 with {\arrow{>}}},
        postaction={decorate}
        ] (0,2) circle (0.3);

     \draw[blue,ultra thick, 
        decoration={markings, mark=at position 0.7 with {\arrow{>}}},
        postaction={decorate}
        ] (0,1) circle (0.3);

\end{tikzpicture}

\caption{Shifting a line contour factor $\gamma_I^{(H',\mathbb{n})}$ to $\gamma_I^{(\tilde{H}',\tilde{\mathbb{n}})}$, for $H'\subset  \tilde{H}'$ and $\tilde{\mathbb{n}}=\mathbb{n}\oplus \ldots\in \Z^{\tilde{H}'}$.}
\label{fig:contour-shift}
\end{figure}
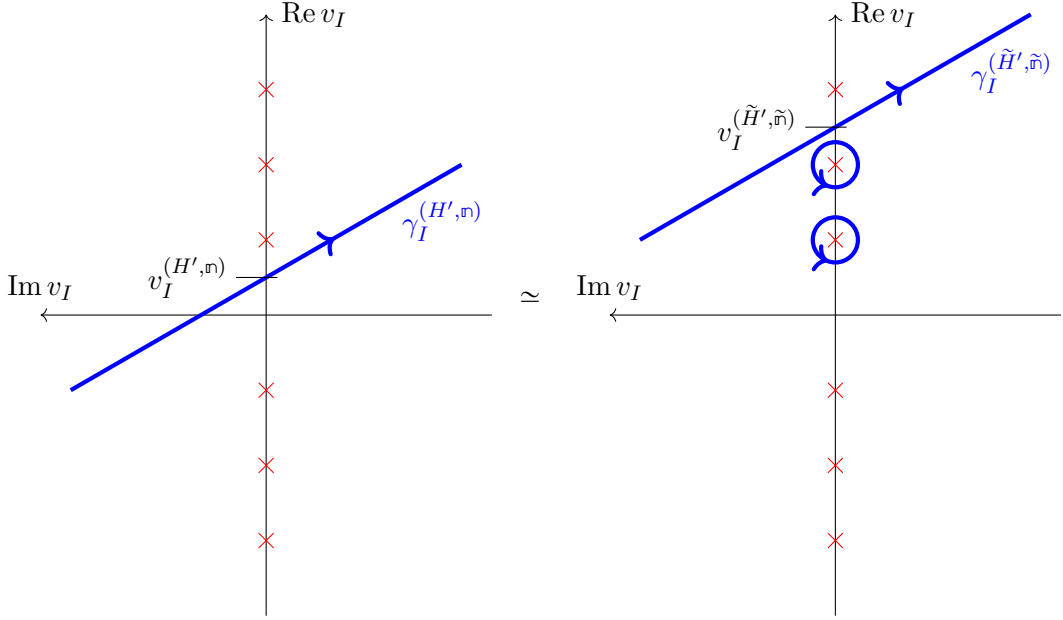

The action of the global monodromy on the finite-dimensional ``thimbles'' can be written as follows:
\begin{equation}
    \gamma^{(H',\mathbb{n})}\equiv 
    \bigtimes\limits_{I\in H} \gamma_I^{(H',\mathbb{n})}
    \longmapsto
   \bigtimes\limits_{I\in H'} \gamma_I^{(H',\mathbb{n})}\times
    \bigtimes\limits_{I\in H''} 
    \left(-\gamma_I^{(H',\mathbb{n})}+\sum_{\ell_I\geq v_I^{(H',\mathbb{n})}}\sigma_{I}^{(\ell_I)}-\sum_{\ell_I<v_I^{(H',\mathbb{n})}}\sigma_{I}^{(\ell_I)}\right)
    \label{fd-monodromy-first-step}
\end{equation}
where $\sigma_I^{(\ell_I)}$ is a small circle contour in $v_I$-plane around a pole at $v_I=\ell_I\in\Z$.
After applying the distributive law to all the $I\in H''$ factors, the right-hand side of (\ref{fd-monodromy-first-step}) can be expressed as a linear combination of the contours of the form
\begin{equation}
    \bigtimes_{I\in H\setminus K}\gamma_I^{(H',\mathbb{n})}\times
    \bigtimes_{I\in K}\sigma_I^{(\ell_I)}
\end{equation}
for a subset $K\subset H''\subset H$ and $\ell\in \Z^K$. Such a product, in general, is not equal to one of the thimble contours, as the ``line factors'' are not necessarily positioned at extrema. However, it can be transformed into a linear combination of the thimble contours, by applying recursively the following homological equivalence:
\begin{multline}
    \bigtimes_{I\in H\setminus K}\gamma_I^{(H',\mathbb{n})}\times
    \bigtimes_{I\in K\subset H''}\sigma_I^{(\ell_I)}
    \simeq \\
    \bigtimes_{I\in H' \cup K}\gamma_I^{(H'\cup K,\mathbb{n}\oplus\ell)}\times
    \bigtimes_{I\in H''\setminus K}\left(\gamma_I^{(H'\cup K,\mathbb{n}\oplus \ell)}\pm\sum_{\substack{\ell_I'\text{ is between}\\v_I^{(H',\mathbb{n})}\text{ and }v_I^{(H'\cup K,\mathbb{n}\oplus \ell)}}}\sigma_I^{(\ell_I')}\right)
    \label{contour-shift-transformation}
\end{multline}
with the plus sign in front of the sum when $v_I^{(H',\mathbb{n})}<v_I^{(H'\cup K,\mathbb{n}\oplus \ell)}$ and minus otherwise. The sum over integer $\ell'_I$ includes the lower bound (if it coincides with the pole, as in the scenario shown in Figure \ref{fig:contours-basis} on the right) but not the upper bound. 
The transformation (\ref{contour-shift-transformation}) is realized by interpreting the circle contours $\sigma^{(\ell_I)}_I$ on the left-hand side as the thimble factors $\gamma^{(\tilde{H}',\tilde{\mathbb{n}})}_I,\;I\in \tilde{H}'$ for a new pair $(\tilde{H}',\tilde{\mathbb{n}})=(H'\cup K,\mathbb{n}\oplus \ell)$ and then shifting the line factors $\gamma_I^{(H',\mathbb{n})},\;I\in H''\setminus K=\tilde{H}''$ accordingly to $\gamma^{(\tilde{H}',\tilde{\mathbb{n}})}_I$, at the cost of absorption of new circle contours $\sigma_I^{(\ell_I')}$, as shown in the Figure \ref{fig:contour-shift}. The recursion must stop after a finite number of steps (at most $|H|$), since each transformation makes the subset $H''$ smaller (as $\tilde{H}''=H''\setminus K$).

Applying such transformation rules to the right-hand side of (\ref{fd-monodromy-first-step}) one obtains at the end the total monodromy transformation purely in terms of thimbles:
\begin{equation}
    \gamma_I^{(H',\mathbb{n})}
\longmapsto 
\sum_{(K,\mathbb{m})}M_{(K,\mathbb{m})}^{(H',\mathbb{n})}\,\gamma_I^{(K,\mathbb{m})}
\label{fd-monodromy-coefficients}
\end{equation}
with $M_{(K,\mathbb{m})}^{(H',\mathbb{n})}\in\Z$ determined by the algorithmic procedure described above. It is clear that $M^{(H',\mathbb{n})}_{(K,\mathbb{m})}$ can only be non-zero if $H'\subset K$ with $\mathbb{m}=\mathbb{n}\oplus \ldots \in \Z^{K}=\Z^{H'}\oplus \ldots$

Now, let us assume that in Chern-Simons theory there is a single thimble $\bbalpha$ for a given critical value $\CS_\bbalpha\in \C$ of the Chern-Simons functional (as we earlier assumed for the zero CS value and the thimble $\bbalpha_0$ associated to the trivial flat connection). The modifications required to avoid such assumptions will be discussed in Section \ref{sec:more-general-plumbings}. We can then use the critical CS values to label different thimbles in Chern-Simons theory: $\bbalpha\equiv (\CS_\bbalpha)$.

Then, similarly to (\ref{zero-thimble-model}), we have:
\begin{equation}
    \frac{N_{(H',\mathbb{n})}}{2}\,I^{(S_\ast^{(H',\mathbb{n})})}(k)\pto \int_{\gamma^{(H,\mathbb{n})}}d^{|H|}v\,R(v)\,e^{2\pi ikS(v)},\qquad H'\neq \emptyset,
    \label{CS-fd-thimble-correspondence}
\end{equation}
with some multiplicity factors $N_{(H',\mathbb{n})}$, assuming the same proportionality coefficient as in (\ref{CS-fd-zero-thimble-correspondence}). The factor $1/2$ on the left-hand side of (\ref{CS-fd-thimble-correspondence}) is introduced to take into account the redundancy 
\begin{equation}
\int_{\gamma^{(H,\mathbb{n})}}d^{|H|}v\,R(v)\,e^{2\pi ikS(v)}\equiv
    \int_{\gamma^{(H,-\mathbb{n})}}d^{|H|}v\,R(v)\,e^{2\pi ikS(v)},\qquad H'\neq \emptyset,
\end{equation}
due to the symmetry $v\leftrightarrow -v$ of the integrand. The symmetry has its origin in the $\Z_2$ Weyl group of the $\mathfrak{sl}(2,\C)$ algebra.

Since the monodromies both in the finite-dimensional model (\ref{fd-monodromy-coefficients}) and in the infinite-dimensional setting of Chern-Simons theory (\ref{thimble-jump}) are integers, the multiplicities $N_{(H',\mathbb{n})}$ in the relation (\ref{CS-fd-thimble-correspondence}) must be rational numbers. In simple, generic enough cases one can actually expect that $N^{(H',\mathbb{n})}\in \Z$,  with at least one being $\pm 1$ among all $N^{(H',\mathbb{n})}$ for fixed CS value $S_*^{(H',\mathbb{n})}\mod 1$. Then, up to an overall (depending only on the CS value modulo 1) sign,   one can determine the multiplicities from the leading coefficients of the perturbative expansions (\ref{fd-thimbe-expansion}) in the finite-dimensional model as follows:
\begin{equation}
    N_{(H',\mathbb{n})}=\frac{a_0^{(H',\mathbb{n})}}{\pm a_0^{(H'_0,\mathbb{n}_0)}} 
    \label{CS-fd-multiplicities-a0}
\end{equation}
where $(H'_0,\mathbb{n}_0)$ minimizes $|a_0^{(H',\mathbb{n})}|$ for all $(H',\mathbb{n})$ with fixed $S_\ast^{(H',\mathbb{n})}\mod 1$.

More generally, without any knowledge of the perturbative expansions in CS theory around non-trivial flat connections, one can fix the multiplicities $N_{(H',\mathbb{n})}$ as follows\footnote{Alternatively, one can proceed as described in Section \ref{sec:more-general-plumbings}.}. Up to a rational factor depending only on CS value modulo 1, we still have (\ref{CS-fd-multiplicities-a0}). By rescaling  $I^{(S_\ast^{(H',\mathbb{n})})}$ by factors depending only on $S_\ast^{(H',\mathbb{n})}\mod 1$ (which implies a rescaling of the corresponding multiplicities by the inverse factor), one can then achieve that
\begin{equation}
    \frac{N_{(H'\cup \{h\},\mathbb{n}\oplus m)}}{N_{(H',\mathbb{n})}}\in \Z,\qquad\forall (H',\mathbb{n}),\;h\in H,\;m\in \Z,
    \label{CS-fd-multiplicities-normalization-conditions}
\end{equation}
recursively, starting with $N_{(\emptyset)}=1$ and increasing $H'$ by a single element. At each step the rescaling is required to be ``minimal'' in the sense that any further rescaling $I^{(H'\cup \{h\},\mathbb{n}\oplus m)}\rightarrow p\cdot I^{(H'\cup \{h\},\mathbb{n}\oplus m)}$, $N_{(H'\cup \{h\},\mathbb{n}\oplus m)}\rightarrow  N_{(H'\cup \{h\},\mathbb{n}\oplus m)}/p$ for an integer $p\neq \pm 1$ depending only on $S_\ast^{(H'\cup \{h\},\mathbb{n}\oplus m)}\mod 1$ violates this condition. Then, starting with the ``naive'' multiplicities given by (\ref{CS-fd-multiplicities-a0}) and applying recursively the rescaling procedure dictated by the normalization conditions (\ref{CS-fd-multiplicities-normalization-conditions}) one determines $N_{(H',\mathbb{n})}$, up to an overall sign depending only on CS value modulo 1. The sign ambiguity can be fixed, non-canonically, by choosing the sign in (\ref{CS-fd-multiplicities-a0}) such that $\arg(\pm a_0^{(H'_0,\mathbb{n}_0)})\in [0,\pi)$. By construction, we have 
\begin{equation}
    \frac{N_{(K,\mathbb{m})}}{N_{(H',\mathbb{n})}}\in\Z,\qquad\forall K\supset H,\;\mathbb{m}=\mathbb{n}\oplus \ldots\in \Z^K.
\end{equation}

The monodromy coefficients in CS theory can be related to the ones in the finite-dimensional model as follows:
\begin{equation}
    m^{(S)}_{(S')}=
   (-1)^{|H|-1} \sum_{(K,\mathbb{m}):\,S_\ast^{(K,\mathbb{m})}=S'}M^{(H',\mathbb{n})}_{(K,\mathbb{m})}\cdot \frac{N_{(K,\mathbb{m})}}{N_{(H',\mathbb{n})}}\times\left\{
    \begin{array}{rl}
        \frac{1}{2}, & H'=\emptyset,\,K\neq \emptyset, \\
        1, & \text{otherwise},
    \end{array}
    \right.
        \label{CS-fd-monodromies}
\end{equation}
where $S=S_\ast^{(H',\mathbb{n})}$ and the sign $(-1)^{|H|-1}$ takes into account the monodromy of the proportionality coefficient in (\ref{CS-fd-zero-thimble-correspondence}) and (\ref{CS-fd-thimble-correspondence}), which contains $k^{\frac{|H|-1}{2}}$. With the algorithmic procedures to determine $M^{(H',\mathbb{n})}_{(K,\mathbb{m})}$  and $N_{(H',\mathbb{n})}$ this gives us a method to calculate the monodromy coefficients in CS theory (with Lefschetz thimbles possibly replaced by their multiples, with factors depending only on the CS value modulo 1). 

The expected resurgence structure in CS theory reviewed in Section \ref{sec:CS-resurgence-review} implies certain non-trivial properties of the entries on the right-hand side of (\ref{CS-fd-monodromies}). First of all, the result should be independent of the choice of $(H',\mathbb{n})$ that gives $S_\ast^{(H',\mathbb{n})}$. Second, we must have the invariance under $\Z$-action (large gauge transformations):
\begin{equation}
m^{(S+f)}_{(S'+f)}=m^{(S)}_{(S')},\qquad f\in\Z , 
\label{CS-stokes-coefs-Z-invariance}
\end{equation}
so that, in particular, one has a well-defined $\tilde{q}$-series (\ref{Stokes-series-definition}):
\begin{equation}
    I^{(S\mod 1)}_{(S'\mod 1)}(\tilde{q})=
    \sum_{S'\mod 1\text{ is fixed}}m^{(S)}_{(S')} \,\tilde{q}^{S-S'}.
    \label{Stokes-series-from-fd}
\end{equation}
We verified in many examples that these properties are indeed satisfied. A nice exercise for future work would be to prove this as a general Theorem in the class of plumbings considered here (and their generalizations described in Section \ref{sec:more-general-plumbings}).

\subsection{A concrete example}
\label{sec:a-specific-example}

\begin{figure}
    \centering
  \begin{tikzpicture}[scale=0.4]

\draw[ultra thick] (5,0)  -- (9,0);
  \draw[ultra thick] (9,0) -- (12,3);
   \draw[ultra thick] (9,0) -- (12,-3);
    \draw[ultra thick] (5,0) -- (2,-3);
     \draw[ultra thick] (5,0) -- (2,3);
    \filldraw[black] (5,0) circle (8pt) node[left=5] {$[-1,0]$};
     \draw[blue] (5,0) node[above right] {$v_1$};
    \filldraw[black] (2,3) circle (8pt) node[left=4] {$[-3,0]$};
    \filldraw[black] (2,-3) circle (8pt) node[left=4] {$[2,0]$};
    \filldraw[black] (9,0) circle (8pt) node[right=5] {$[-1,0]$};
    \draw[blue] (9,0) node[above left] {$v_2$};
    \filldraw[black] (12,3) circle (8pt) node[right=4] {$[-3,0]$};
    \filldraw[black] (12,-3) circle (8pt) node[right=4] {$[5,0]$};
\end{tikzpicture}
    \caption{An example of a weakly negative-definite plumbed integer homology sphere. The variables $v_{1,2}$ corresponding to the high-valency vertices are shown in blue and labeled with $v_1$ and $v_2$.}
    \label{fig:plumbing-example-2m3-5m3}
\end{figure}
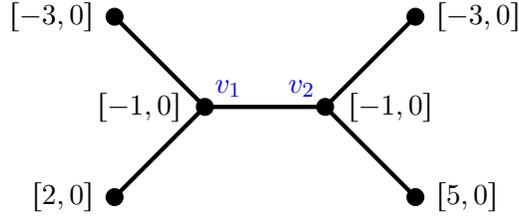

Consider the plumbing shown in Figure \ref{fig:plumbing-example-2m3-5m3}. The linking matrix, with its block decomposition corresponding to high- and low-valency vertices is the following:
\begin{equation}
    B=
\begin{array}{cccc|ccccc}
& & \multicolumn{2}{c}{H} & \multicolumn{4}{c}{L} & \\
\multirow{2}{*}{$H$} & \ldelim({6}{4mm} &-1 & 1 & 1 & 1 & 0 & 0 & \rdelim){6}{4mm} \\
 & &1 & -1 & 0 & 0 & 1 & 1 & \\
 \cline{3-8}
\multirow{4}{*}{$L$} & &1 & 0 & 2 & 0 & 0 & 0 & \\
 & &1 & 0 & 0 & -3 & 0 & 0 & \\
 & &0 & 1 & 0 & 0 & -3 & 0 & \\
 & &0 & 1 & 0 & 0 & 0 & 5 & \\
\end{array}.
\end{equation}
Its inverse is
\begin{equation}
    B^{-1}=\begin{array}{cccc|ccccc}
& & \multicolumn{2}{c}{H} & \multicolumn{4}{c}{L} & \\
\multirow{2}{*}{$H$} & \ldelim({6}{4mm} 
 & -78 & -90 & 39 & -26 & -30 & 18 & \rdelim){6}{4mm} \\
 & &-90 & -105 & 45 & -30 & -35 & 21 & \\
 \cline{3-8}
\multirow{4}{*}{$L$} & & 39 & 45 & -19 & 13 & 15 & -9 & \\
 & &-26 & -30 & 13 & -9 & -10 & 6 & \\
 & &-30 & -35 & 15 & -10 & -12 & 7 & \\
& & 18 & 21 & -9 & 6 & 7 & -4 & \\
\end{array}
\end{equation}
with the blocks given by the matrices $C$, $D$, $D^T$ and $A$ (\ref{A-block-def}-\ref{C-block-def}). With the labeling of high-valency vertices by 1, 2 as in Figure (\ref{fig:plumbing-example-2m3-5m3}), the meromorphic function (\ref{R-expression}) in the integrand of the finite-dimensional model reads
\begin{equation}
    R(v)=\frac{\sin \left(\frac{\pi  v_1}{3}\right) \sin \left(\frac{\pi  v_1}{2}\right) \sin \left(\frac{\pi  v_2}{5}\right) \sin \left(\frac{\pi  v_2}{3}\right)}{\sin \left(\pi  v_1\right) \sin \left(\pi  v_2\right)}.
\end{equation}
The action (\ref{S-expression}) is
\begin{equation}
    S(v)=-\frac{7v_1^2}{24}+\frac{v_2 v_1}{2}-\frac{13 v_2^2}{60}.
\end{equation}
The sets of critical values of the action $S_*^{(H',\mathbb{n})}\mod 1$ corresponding to different subsets $H'\subset H$ are the following:
\begin{equation}
    H'=\emptyset: \qquad S_*^{(H',\mathbb{n})} = 0 \mod 1,
\end{equation}
\begin{equation}
    H'=\{1\}: \quad S_*^{(H',\mathbb{n})} \in \left\{\frac{23}{312},\frac{95}{312},\frac{191}{312},\frac{263}{312},\frac{287}{312},\frac{311}{312}\right\} \mod 1,
\end{equation}
\begin{multline}
    H'=\{2\}:\; S_*^{(H',\mathbb{n})} \in
    \left\{\frac{59}{420},\frac{26}{105},\frac{131}{420},\frac{41}{105},\frac{59}{105},\frac{251}{420},\frac{299}{420},\frac{311}{420},\frac{89}{105},\frac{101}{105},\frac{104}{105},\frac{419}{420}\right\} \\ \mod 1,
\end{multline}
\begin{equation}
    H'=\{1,2\}:\quad S_*^{(H',\mathbb{n})} \in \left\{\frac{29}{120},\frac{71}{120},\frac{101}{120},\frac{119}{120}\right\} \mod 1.
\end{equation}
Applying the algorithms described in Section \ref{sec:finite-dimensional-model} we can obtain the monodromy coefficients $M_{(K,\mathbb{m})}^{(H',\mathbb{n})}$ of the finite dimensional model and the multiplicites $N_{(H',\mathbb{n})}$, all equal to $\pm 1$ in this case. Using the relation (\ref{CS-fd-monodromies}) we then can obtain the generating function of the the Stokes coefficients (\ref{Stokes-series-from-fd}). The non-trivial ones are the following:
\begin{equation}
\footnotesize
\begin{array}{rcl}
  I_{\left(\frac{23}{312}\right)}^{(0)} & = & \tilde{q}^{-\frac{23}{312}}\,\left(-\tilde{q}-\tilde{q}^4-\tilde{q}^6-\tilde{q}^{12}+\tilde{q}^{29}+\tilde{q}^{41}+\tilde{q}^{47}+\tilde{q}^{62}-\tilde{q}^{96}-\tilde{q}^{117}-\tilde{q}^{127}+\ldots\right) \\
 I_{\left(\frac{95}{312}\right)}^{(0)} & = & \tilde{q}^{-\frac{95}{312}}\,\left(-\tilde{q}^2-\tilde{q}^3-\tilde{q}^8-\tilde{q}^{10}+\tilde{q}^{33}+\tilde{q}^{37}+\tilde{q}^{52}+\tilde{q}^{57}-\tilde{q}^{103}-\tilde{q}^{110}-\tilde{q}^{135}-\tilde{q}^{143}+\ldots\right) \\
 I_{\left(\frac{191}{312}\right)}^{(0)} & = & \tilde{q}^{-\frac{191}{312}}\,\left(\tilde{q}+\tilde{q}^5+\tilde{q}^6+\tilde{q}^{15}-\tilde{q}^{26}-\tilde{q}^{43}-\tilde{q}^{46}-\tilde{q}^{68}+\tilde{q}^{90}+\tilde{q}^{120}+\tilde{q}^{125}+\ldots\right) \\
 I_{\left(\frac{263}{312}\right)}^{(0)} & = & \tilde{q}^{-\frac{263}{312}}\,\left(\tilde{q}-\tilde{q}^2-\tilde{q}^{12}+\tilde{q}^{17}-\tilde{q}^{24}+\tilde{q}^{31}+\tilde{q}^{61}-\tilde{q}^{72}+\tilde{q}^{86}-\tilde{q}^{99}-\tilde{q}^{149}+\ldots\right) \\
 I_{\left(\frac{287}{312}\right)}^{(0)} & = & \tilde{q}^{-\frac{287}{312}}\,\left(\tilde{q}+\tilde{q}^4+\tilde{q}^8+\tilde{q}^{18}-\tilde{q}^{23}-\tilde{q}^{39}-\tilde{q}^{51}-\tilde{q}^{74}+\tilde{q}^{84}+\tilde{q}^{113}+\tilde{q}^{133}+\ldots\right) \\
 I_{\left(\frac{311}{312}\right)}^{(0)} & = & \tilde{q}^{-\frac{311}{312}}\,\left(-\tilde{q}+\tilde{q}^3+\tilde{q}^{10}-\tilde{q}^{20}+\tilde{q}^{21}-\tilde{q}^{35}-\tilde{q}^{56}+\tilde{q}^{78}-\tilde{q}^{80}+\tilde{q}^{106}+\tilde{q}^{141}+\ldots\right) \\
\end{array}
\end{equation}

\begin{equation}
\footnotesize
\begin{array}{rcl}
 I_{\left(\frac{59}{420}\right)}^{(0)} & = & \tilde{q}^{-\frac{59}{420}}\,\left(\tilde{q}-\tilde{q}^9-\tilde{q}^{15}-\tilde{q}^{19}+\tilde{q}^{35}+\tilde{q}^{41}+\tilde{q}^{53}-\tilde{q}^{87}+\tilde{q}^{125}+\ldots\right) \\
 I_{\left(\frac{26}{105}\right)}^{(0)} & = & \tilde{q}^{-\frac{26}{105}}\,\left(\tilde{q}^3-\tilde{q}^{10}-\tilde{q}^{14}-\tilde{q}^{26}+\tilde{q}^{27}+\tilde{q}^{43}+\tilde{q}^{51}-\tilde{q}^{74}+\tilde{q}^{142}+\ldots\right) \\
 I_{\left(\frac{131}{420}\right)}^{(0)} & = & \tilde{q}^{-\frac{131}{420}}\,\left(-\tilde{q}-\tilde{q}^7-\tilde{q}^{11}-\tilde{q}^{13}+\tilde{q}^{45}+\tilde{q}^{49}+\tilde{q}^{59}+\tilde{q}^{89}-\tilde{q}^{123}+\ldots\right) \\
 I_{\left(\frac{41}{105}\right)}^{(0)} & = & \tilde{q}^{-\frac{41}{105}}\,\left(\tilde{q}+\tilde{q}^2+\tilde{q}^5-\tilde{q}^{18}+\tilde{q}^{37}-\tilde{q}^{66}-\tilde{q}^{81}-\tilde{q}^{90}+\tilde{q}^{122}+\tilde{q}^{133}+\ldots\right) \\
 I_{\left(\frac{59}{105}\right)}^{(0)} & = & \tilde{q}^{-\frac{59}{105}}\,\left(\tilde{q}^3+\tilde{q}^4+\tilde{q}^7+\tilde{q}^{19}-\tilde{q}^{36}-\tilde{q}^{60}-\tilde{q}^{71}-\tilde{q}^{76}+\tilde{q}^{140}+\tilde{q}^{147}+\ldots\right) \\
 I_{\left(\frac{251}{420}\right)}^{(0)} & = & \tilde{q}^{-\frac{251}{420}}\,\left(-\tilde{q}+\tilde{q}^5+\tilde{q}^{17}+\tilde{q}^{23}-\tilde{q}^{31}-\tilde{q}^{39}-\tilde{q}^{67}+\tilde{q}^{93}-\tilde{q}^{119}+\ldots\right) \\
 I_{\left(\frac{299}{420}\right)}^{(0)} & = & \tilde{q}^{-\frac{299}{420}}\,\left(-\tilde{q}-\tilde{q}^3-\tilde{q}^9+\tilde{q}^{25}-\tilde{q}^{29}+\tilde{q}^{55}+\tilde{q}^{77}+\tilde{q}^{95}-\tilde{q}^{117}-\tilde{q}^{139}+\ldots\right) \\
 I_{\left(\frac{311}{420}\right)}^{(0)} & = & \tilde{q}^{-\frac{311}{420}}\,\left(-\tilde{q}^2-\tilde{q}^4-\tilde{q}^6-\tilde{q}^{26}+\tilde{q}^{28}+\tilde{q}^{64}+\tilde{q}^{72}+\tilde{q}^{84}-\tilde{q}^{130}-\tilde{q}^{146}+\ldots\right) \\
 I_{\left(\frac{89}{105}\right)}^{(0)} & = & \tilde{q}^{-\frac{89}{105}}\,\left(-\tilde{q}-\tilde{q}^2-\tilde{q}^{10}+\tilde{q}^{21}-\tilde{q}^{34}+\tilde{q}^{53}+\tilde{q}^{85}+\tilde{q}^{98}-\tilde{q}^{114}-\tilde{q}^{129}+\ldots\right) \\
 I_{\left(\frac{101}{105}\right)}^{(0)} & = & \tilde{q}^{-\frac{101}{105}}\,\left(-\tilde{q}+\tilde{q}^6+\tilde{q}^{14}+\tilde{q}^{22}-\tilde{q}^{33}-\tilde{q}^{45}-\tilde{q}^{65}+\tilde{q}^{102}-\tilde{q}^{110}+\ldots\right) \\
 I_{\left(\frac{104}{105}\right)}^{(0)} & = & \tilde{q}^{-\frac{104}{105}}\,\left(\tilde{q}+\tilde{q}^9+\tilde{q}^{12}+\tilde{q}^{17}-\tilde{q}^{40}-\tilde{q}^{49}-\tilde{q}^{56}-\tilde{q}^{104}+\tilde{q}^{108}+\ldots\right) \\
 I_{\left(\frac{419}{420}\right)}^{(0)} & = & \tilde{q}^{-\frac{419}{420}}\,\left(\tilde{q}+\tilde{q}^3+\tilde{q}^5-\tilde{q}^{13}+\tilde{q}^{47}-\tilde{q}^{69}-\tilde{q}^{79}-\tilde{q}^{105}+\tilde{q}^{107}+\tilde{q}^{137}+\ldots\right) \\
 \end{array}
\end{equation}

\begin{equation}
\footnotesize
\begin{array}{rcl}
 I_{\left(\frac{29}{120}\right)}^{(0)} & = & \tilde{q}^{-\frac{29}{120}}\,\left(\tilde{q}^6+\tilde{q}^{10}-\tilde{q}^{21}-\tilde{q}^{27}-\tilde{q}^{32}-\tilde{q}^{37}-\tilde{q}^{43}-\tilde{q}^{50}-\tilde{q}^{57}+\tilde{q}^{61}-\tilde{q}^{64}+\tilde{q}^{66}+\tilde{q}^{74}+\tilde{q}^{79}\right.\\
 & &\left.
 +\tilde{q}^{85}+\tilde{q}^{90}+\tilde{q}^{92}+\tilde{q}^{103}+\tilde{q}^{106}+\tilde{q}^{110}+\tilde{q}^{120}-\tilde{q}^{122}+\tilde{q}^{126}-\tilde{q}^{133}+\tilde{q}^{143}-\tilde{q}^{147}+\ldots\right) \\
 I_{\left(\frac{71}{120}\right)}^{(0)} & = & \tilde{q}^{-\frac{71}{120}}\,\left(-\tilde{q}^5-\tilde{q}^{15}+\tilde{q}^{26}+\tilde{q}^{31}+\tilde{q}^{36}+\tilde{q}^{39}+\tilde{q}^{43}+\tilde{q}^{44}+\tilde{q}^{50}-\tilde{q}^{77}-\tilde{q}^{84}-\tilde{q}^{85}-\tilde{q}^{90}-\tilde{q}^{93}
 \right.\\
 & &\left.
 -\tilde{q}^{97}-\tilde{q}^{99}-\tilde{q}^{106}-2 \tilde{q}^{113}-\tilde{q}^{120}-\tilde{q}^{127}-\tilde{q}^{137}+\tilde{q}^{144}+\ldots\right) \\
 I_{\left(\frac{101}{120}\right)}^{(0)} & = & \tilde{q}^{-\frac{101}{120}}\,\left(\tilde{q}^2-\tilde{q}^3+\tilde{q}^{14}-\tilde{q}^{17}+\tilde{q}^{19}-\tilde{q}^{21}+\tilde{q}^{23}-\tilde{q}^{31}-\tilde{q}^{42}-\tilde{q}^{48}-\tilde{q}^{53}+\tilde{q}^{56}-\tilde{q}^{57}-\tilde{q}^{64}
 \right.\\
 & &\left. +\tilde{q}^{72}+\tilde{q}^{80}+\tilde{q}^{94}+\tilde{q}^{99}+\tilde{q}^{103}+\tilde{q}^{104}+\tilde{q}^{114}+\tilde{q}^{116}+\tilde{q}^{117}+\tilde{q}^{126}+\tilde{q}^{129}+\tilde{q}^{136}-\tilde{q}^{141}+\ldots\right) \\
 I_{\left(\frac{119}{120}\right)}^{(0)} & = & \tilde{q}^{-\frac{119}{120}}\,\left(\tilde{q}^2-\tilde{q}^3-\tilde{q}^{11}+\tilde{q}^{13}-\tilde{q}^{19}+\tilde{q}^{20}-\tilde{q}^{33}+\tilde{q}^{35}+\tilde{q}^{57}+\tilde{q}^{62}-\tilde{q}^{63}+\tilde{q}^{69}-\tilde{q}^{78}
 \right.\\
 & &\left.
 +\tilde{q}^{80}-\tilde{q}^{89}-\tilde{q}^{94}-\tilde{q}^{106}-\tilde{q}^{107}-\tilde{q}^{116}+\tilde{q}^{118}-\tilde{q}^{123}-\tilde{q}^{126}+\tilde{q}^{134}-\tilde{q}^{137}-\tilde{q}^{148}+\ldots\right) \\
 \end{array}
\end{equation}

\begin{equation}
\footnotesize
\begin{array}{rcl}
 I_{\left(\frac{29}{120}\right)}^{\left(\frac{23}{312}\right)} & = & \tilde{q}^{-\frac{131}{780}}\,\left(-\tilde{q}^3-\tilde{q}^7-\tilde{q}^9-\tilde{q}^{45}+\tilde{q}^{53}+\tilde{q}^{121}+\tilde{q}^{129}+\ldots\right) \\
 I_{\left(\frac{71}{120}\right)}^{\left(\frac{23}{312}\right)} & = & \tilde{q}^{-\frac{101}{195}}\,\left(\tilde{q}^2+\tilde{q}^3+\tilde{q}^{10}-\tilde{q}^{35}+\tilde{q}^{66}-\tilde{q}^{119}+\ldots\right) \\
 I_{\left(\frac{101}{120}\right)}^{\left(\frac{23}{312}\right)} & = & \tilde{q}^{-\frac{599}{780}}\,\left(-\tilde{q}^2+\tilde{q}^{16}+\tilde{q}^{30}+\tilde{q}^{34}-\tilde{q}^{68}-\tilde{q}^{74}-\tilde{q}^{102}+\ldots\right) \\
 I_{\left(\frac{119}{120}\right)}^{\left(\frac{23}{312}\right)} & = & \tilde{q}^{-\frac{179}{195}}\,\left(-\tilde{q}-\tilde{q}^{17}-\tilde{q}^{20}-\tilde{q}^{29}+\tilde{q}^{76}+\tilde{q}^{93}+\tilde{q}^{100}+\ldots\right) \\
  \end{array}
\end{equation}

\begin{equation}
\footnotesize
\begin{array}{rcl}
 I_{\left(\frac{29}{120}\right)}^{\left(\frac{95}{312}\right)} & = & \tilde{q}^{\frac{49}{780}}\,\left(-1+\tilde{q}^{12}+\tilde{q}^{24}+\tilde{q}^{34}-\tilde{q}^{66}-\tilde{q}^{82}-\tilde{q}^{110}+\ldots\right) \\
 I_{\left(\frac{71}{120}\right)}^{\left(\frac{95}{312}\right)} & = & \tilde{q}^{-\frac{56}{195}}\,\left(\tilde{q}^3-\tilde{q}^{20}-\tilde{q}^{24}-\tilde{q}^{40}+\tilde{q}^{59}+\tilde{q}^{83}+\tilde{q}^{91}+\ldots\right) \\
 I_{\left(\frac{101}{120}\right)}^{\left(\frac{95}{312}\right)} & = & \tilde{q}^{-\frac{419}{780}}\,\left(-\tilde{q}-\tilde{q}^5-\tilde{q}^7+\tilde{q}^{29}-\tilde{q}^{75}+\tilde{q}^{131}+\tilde{q}^{141}+\ldots\right) \\
 I_{\left(\frac{119}{120}\right)}^{\left(\frac{95}{312}\right)} & = & \tilde{q}^{-\frac{134}{195}}\,\left(\tilde{q}^2+\tilde{q}^5+\tilde{q}^{13}-\tilde{q}^{46}+\tilde{q}^{53}-\tilde{q}^{110}-\tilde{q}^{142}+\ldots\right) \\
  \end{array}
\end{equation}

\begin{equation}
\footnotesize
\begin{array}{rcl}
 I_{\left(\frac{29}{120}\right)}^{\left(\frac{191}{312}\right)} & = & \tilde{q}^{\frac{289}{780}}\,\left(1+\tilde{q}^2+\tilde{q}^{16}-\tilde{q}^{38}+\tilde{q}^{60}-\tilde{q}^{98}+\ldots\right) \\
 I_{\left(\frac{71}{120}\right)}^{\left(\frac{191}{312}\right)} & = & \tilde{q}^{\frac{4}{195}}\,\left(-1-\tilde{q}^4-\tilde{q}^7+\tilde{q}^{23}-\tilde{q}^{84}+\tilde{q}^{128}+\tilde{q}^{143}+\ldots\right) \\
 I_{\left(\frac{101}{120}\right)}^{\left(\frac{191}{312}\right)} & = & \tilde{q}^{-\frac{179}{780}}\,\left(\tilde{q}^5-\tilde{q}^{19}-\tilde{q}^{25}-\tilde{q}^{47}+\tilde{q}^{51}+\tilde{q}^{81}+\tilde{q}^{93}-\tilde{q}^{139}+\ldots\right) \\
 I_{\left(\frac{119}{120}\right)}^{\left(\frac{191}{312}\right)} & = & \tilde{q}^{-\frac{74}{195}}\,\left(-\tilde{q}+\tilde{q}^9+\tilde{q}^{30}+\tilde{q}^{41}-\tilde{q}^{58}-\tilde{q}^{73}-\tilde{q}^{122}+\ldots\right) \\
  \end{array}
\end{equation}

\begin{equation}
\footnotesize
\begin{array}{rcl}
 I_{\left(\frac{29}{120}\right)}^{\left(\frac{263}{312}\right)} & = & \tilde{q}^{\frac{469}{780}}\,\left(-\tilde{q}^3-\tilde{q}^7-\tilde{q}^{13}-\tilde{q}^{31}+\tilde{q}^{69}+\tilde{q}^{105}+\tilde{q}^{125}+\tilde{q}^{145}+\ldots\right) \\
 I_{\left(\frac{71}{120}\right)}^{\left(\frac{263}{312}\right)} & = & \tilde{q}^{\frac{49}{195}}\,\left(1+\tilde{q}^5+\tilde{q}^{17}-\tilde{q}^{48}+\tilde{q}^{49}-\tilde{q}^{96}-\tilde{q}^{136}+\ldots\right) \\
 I_{\left(\frac{101}{120}\right)}^{\left(\frac{263}{312}\right)} & = & \tilde{q}^{\frac{1}{780}}\,\left(-1+\tilde{q}^8+\tilde{q}^{22}+\tilde{q}^{42}-\tilde{q}^{56}-\tilde{q}^{86}-\tilde{q}^{124}+\ldots\right) \\
 I_{\left(\frac{119}{120}\right)}^{\left(\frac{263}{312}\right)} & = & \tilde{q}^{-\frac{29}{195}}\,\left(-\tilde{q}^2-\tilde{q}^{11}-\tilde{q}^{18}-\tilde{q}^{26}+\tilde{q}^{79}+\tilde{q}^{95}+\tilde{q}^{114}+\ldots\right) \\
  \end{array}
\end{equation}

\begin{equation}
\footnotesize
\begin{array}{rcl}
 I_{\left(\frac{29}{120}\right)}^{\left(\frac{287}{312}\right)} & = & \tilde{q}^{\frac{529}{780}}\,\left(-1-\tilde{q}^{14}-\tilde{q}^{20}-\tilde{q}^{22}+\tilde{q}^{84}+\tilde{q}^{88}+\tilde{q}^{102}+\ldots\right) \\
 I_{\left(\frac{71}{120}\right)}^{\left(\frac{287}{312}\right)} & = & \tilde{q}^{\frac{64}{195}}\,\left(-1+\tilde{q}^{11}+\tilde{q}^{27}+\tilde{q}^{35}-\tilde{q}^{64}-\tilde{q}^{76}-\tilde{q}^{112}+\ldots\right) \\
 I_{\left(\frac{101}{120}\right)}^{\left(\frac{287}{312}\right)} & = & \tilde{q}^{\frac{61}{780}}\,\left(\tilde{q}+\tilde{q}^3+\tilde{q}^{13}-\tilde{q}^{41}+\tilde{q}^{57}-\tilde{q}^{107}-\tilde{q}^{149}+\ldots\right) \\
 I_{\left(\frac{119}{120}\right)}^{\left(\frac{287}{312}\right)} & = & \tilde{q}^{-\frac{14}{195}}\,\left(-\tilde{q}^5-\tilde{q}^6-\tilde{q}^{10}-\tilde{q}^{38}+\tilde{q}^{61}+\tilde{q}^{117}+\tilde{q}^{133}+\tilde{q}^{138}+\ldots\right) \\
  \end{array}
\end{equation}

\begin{equation}
\footnotesize
\begin{array}{rcl}
 I_{\left(\frac{29}{120}\right)}^{\left(\frac{311}{312}\right)} & = & \tilde{q}^{\frac{589}{780}}\,\left(\tilde{q}-\tilde{q}^5-\tilde{q}^{35}-\tilde{q}^{47}+\tilde{q}^{49}+\tilde{q}^{63}+\tilde{q}^{133}+\ldots\right) \\
 I_{\left(\frac{71}{120}\right)}^{\left(\frac{311}{312}\right)} & = & \tilde{q}^{\frac{79}{195}}\,\left(-\tilde{q}^7+\tilde{q}^{14}+\tilde{q}^{30}-\tilde{q}^{43}+\tilde{q}^{54}-\tilde{q}^{71}-\tilde{q}^{103}+\tilde{q}^{126}+\ldots\right) \\
 I_{\left(\frac{101}{120}\right)}^{\left(\frac{311}{312}\right)} & = & \tilde{q}^{\frac{121}{780}}\,\left(-1+\tilde{q}^2+\tilde{q}^{10}-\tilde{q}^{18}+\tilde{q}^{94}-\tilde{q}^{116}+\ldots\right) \\
 I_{\left(\frac{119}{120}\right)}^{\left(\frac{311}{312}\right)} & = & \tilde{q}^{\frac{1}{195}}\,\left(-1-\tilde{q}-\tilde{q}^{21}+\tilde{q}^{32}-\tilde{q}^{69}+\tilde{q}^{88}+\ldots\right) \\
 \end{array}
\end{equation}

\begin{equation}
\footnotesize
\begin{array}{rcl}
  I_{\left(\frac{119}{120}\right)}^{\left(\frac{59}{420}\right)} & = & \tilde{q}^{-\frac{143}{168}}\,\left(-\tilde{q}-\tilde{q}^3-\tilde{q}^4-\tilde{q}^9+\tilde{q}^{14}+\tilde{q}^{23}+\tilde{q}^{26}+\tilde{q}^{38}-\tilde{q}^{48}-\tilde{q}^{64}-\tilde{q}^{69}-\tilde{q}^{88}+\tilde{q}^{103}+\tilde{q}^{126}+\tilde{q}^{133}+\ldots\right) \\
 I_{\left(\frac{29}{120}\right)}^{\left(\frac{26}{105}\right)} & = & \tilde{q}^{\frac{1}{168}}\,\left(-1+\tilde{q}+\tilde{q}^5-\tilde{q}^{10}+\tilde{q}^{11}-\tilde{q}^{18}-\tilde{q}^{30}+\tilde{q}^{41}-\tilde{q}^{43}+\tilde{q}^{56}+\tilde{q}^{76}-\tilde{q}^{93}+\tilde{q}^{96}-\tilde{q}^{115}-\tilde{q}^{143}+\ldots\right) \\
 I_{\left(\frac{71}{120}\right)}^{\left(\frac{131}{420}\right)} & = & \tilde{q}^{-\frac{47}{168}}\,\left(-\tilde{q}-\tilde{q}^2-\tilde{q}^4-\tilde{q}^6+\tilde{q}^{17}+\tilde{q}^{21}+\tilde{q}^{27}+\tilde{q}^{32}-\tilde{q}^{54}-\tilde{q}^{61}-\tilde{q}^{71}-\tilde{q}^{79}+\tilde{q}^{112}+\tilde{q}^{122}+\tilde{q}^{136}+\tilde{q}^{147}+\ldots\right) \\
 I_{\left(\frac{29}{120}\right)}^{\left(\frac{41}{105}\right)} & = & \tilde{q}^{\frac{25}{168}}\,\left(-1-\tilde{q}^2-\tilde{q}^3-\tilde{q}^8+\tilde{q}^{13}+\tilde{q}^{22}+\tilde{q}^{25}+\tilde{q}^{37}-\tilde{q}^{47}-\tilde{q}^{63}-\tilde{q}^{68}-\tilde{q}^{87}+\tilde{q}^{102}+\tilde{q}^{125}+\tilde{q}^{132}+\ldots\right) \\
 I_{\left(\frac{101}{120}\right)}^{\left(\frac{59}{105}\right)} & = & \tilde{q}^{-\frac{47}{168}}\,\left(-\tilde{q}-\tilde{q}^2-\tilde{q}^4-\tilde{q}^6+\tilde{q}^{17}+\tilde{q}^{21}+\tilde{q}^{27}+\tilde{q}^{32}-\tilde{q}^{54}-\tilde{q}^{61}-\tilde{q}^{71}-\tilde{q}^{79}+\tilde{q}^{112}+\tilde{q}^{122}+\tilde{q}^{136}+\tilde{q}^{147}+\ldots\right) \\
 I_{\left(\frac{71}{120}\right)}^{\left(\frac{251}{420}\right)} & = & \tilde{q}^{\frac{1}{168}}\,\left(-1+\tilde{q}+\tilde{q}^5-\tilde{q}^{10}+\tilde{q}^{11}-\tilde{q}^{18}-\tilde{q}^{30}+\tilde{q}^{41}-\tilde{q}^{43}+\tilde{q}^{56}+\tilde{q}^{76}-\tilde{q}^{93}+\tilde{q}^{96}-\tilde{q}^{115}-\tilde{q}^{143}+\ldots\right) \\
 I_{\left(\frac{119}{120}\right)}^{\left(\frac{299}{420}\right)} & = & \tilde{q}^{-\frac{47}{168}}\,\left(-\tilde{q}-\tilde{q}^2-\tilde{q}^4-\tilde{q}^6+\tilde{q}^{17}+\tilde{q}^{21}+\tilde{q}^{27}+\tilde{q}^{32}-\tilde{q}^{54}-\tilde{q}^{61}-\tilde{q}^{71}-\tilde{q}^{79}+\tilde{q}^{112}+\tilde{q}^{122}+\tilde{q}^{136}+\tilde{q}^{147}+\ldots\right) \\
 I_{\left(\frac{71}{120}\right)}^{\left(\frac{311}{420}\right)} & = & \tilde{q}^{\frac{25}{168}}\,\left(-1-\tilde{q}^2-\tilde{q}^3-\tilde{q}^8+\tilde{q}^{13}+\tilde{q}^{22}+\tilde{q}^{25}+\tilde{q}^{37}-\tilde{q}^{47}-\tilde{q}^{63}-\tilde{q}^{68}-\tilde{q}^{87}+\tilde{q}^{102}+\tilde{q}^{125}+\tilde{q}^{132}+\ldots\right) \\
 I_{\left(\frac{101}{120}\right)}^{\left(\frac{89}{105}\right)} & = & \tilde{q}^{\frac{1}{168}}\,\left(-1+\tilde{q}+\tilde{q}^5-\tilde{q}^{10}+\tilde{q}^{11}-\tilde{q}^{18}-\tilde{q}^{30}+\tilde{q}^{41}-\tilde{q}^{43}+\tilde{q}^{56}+\tilde{q}^{76}-\tilde{q}^{93}+\tilde{q}^{96}-\tilde{q}^{115}-\tilde{q}^{143}+\ldots\right) \\
 I_{\left(\frac{29}{120}\right)}^{\left(\frac{101}{105}\right)} & = & \tilde{q}^{\frac{121}{168}}\,\left(-1-\tilde{q}-\tilde{q}^3-\tilde{q}^5+\tilde{q}^{16}+\tilde{q}^{20}+\tilde{q}^{26}+\tilde{q}^{31}-\tilde{q}^{53}-\tilde{q}^{60}-\tilde{q}^{70}-\tilde{q}^{78}+\tilde{q}^{111}+\tilde{q}^{121}+\tilde{q}^{135}+\tilde{q}^{146}+\ldots\right) \\
 I_{\left(\frac{101}{120}\right)}^{\left(\frac{104}{105}\right)} & = & \tilde{q}^{\frac{25}{168}}\,\left(-1-\tilde{q}^2-\tilde{q}^3-\tilde{q}^8+\tilde{q}^{13}+\tilde{q}^{22}+\tilde{q}^{25}+\tilde{q}^{37}-\tilde{q}^{47}-\tilde{q}^{63}-\tilde{q}^{68}-\tilde{q}^{87}+\tilde{q}^{102}+\tilde{q}^{125}+\tilde{q}^{132}+\ldots\right) \\
 I_{\left(\frac{119}{120}\right)}^{\left(\frac{419}{420}\right)} & = & \tilde{q}^{\frac{1}{168}}\,\left(-1+\tilde{q}+\tilde{q}^5-\tilde{q}^{10}+\tilde{q}^{11}-\tilde{q}^{18}-\tilde{q}^{30}+\tilde{q}^{41}-\tilde{q}^{43}+\tilde{q}^{56}+\tilde{q}^{76}-\tilde{q}^{93}+\tilde{q}^{96}-\tilde{q}^{115}-\tilde{q}^{143}+\ldots\right) \\
\end{array}
\end{equation}
with $I^{(S\mod 1)}_{(S\mod 1)}=\pm 1$ and the rest equal to zero. 

\subsection{General tree plumbings}
\label{sec:more-general-plumbings}

In this section we comment on modifications that are required in order to relax various simplifying assumptions made in Section \ref{sec:stokes-plumbed}.

\subsubsection{Resolving degeneracy of CS values}
\label{sec:deg-CS-values}

Without the assumption that in CS theory there is a single thimble for each critical value of CS functional, the relation (\ref{CS-fd-thimble-correspondence}) between the thimble in CS theory and the finite-dimensional model should be generalized to a non-trivial linear relation as follows.  We still assume that the distinguished CS thimble associated with the trivial flat connection corresponds to the thimble $\gamma^{(\emptyset)}$ in the finite-dimensional model, as in (\ref{CS-fd-zero-thimble-correspondence}). Consider all $\Z$-orbits $\alpha$ of the CS thimbles generated by the monodromies starting from the thimble associated with the trivial flat connection (see Section \ref{sec:single-thimble-info} for details). We can label them by critical values of the CS functional modulo 1 and an extra index $i$, running over a finite set: $\alpha\equiv (\CS_\bbalpha \mod 1,i)$.

Fix $(S \mod 1) \in \C/\Z$. For each finite-dimensional thimble $\gamma^{(H',\mathbb{n})}$ ($H'\neq \emptyset$) such that $(S^{(H',\mathbb{n})}_\ast\mod 1) = (S\mod 1)$ consider the corresponding perturbative expansion, with the dependence on the action factored out:
\begin{equation}
    2\,e^{-2\pi ikS^{(H',\mathbb{n})}_\ast}\int\limits_{\gamma^{(H',\mathbb{n})}}d^{|H|}v\,R(v)\,e^{2\pi ikS(v)}
    = 2\sum_{n\geq 0}\frac{a_n^{(H',\mathbb{n})}}{k^{n+\delta_{(H',n)}}}\,
    \label{fd-thimbe-expansion-no-action}
\end{equation}
where, as in (\ref{CS-fd-thimble-correspondence}), we introduced a factor of $2$ to take into account a 2-fold redundancy due to $\Z_2$ Weyl symmetry. For a generic enough plumbing, all integral combinations of such expansions are realized by the thimbles in the subgroup generated by the monodromies starting from the thimble $\gamma^{(\emptyset)}$. In what follows we assume this to be the case.

Since there is a finite number of $\Z$-orbits of thimbles in CS theory, the abelian subgroup of formal power series $\C[[k^{-\frac{1}{2}},k^{\frac{1}{2}}]$ generated by all perturbative expansions (\ref{fd-thimbe-expansion-no-action}) should be of finite rank. We can choose a basis of this subgroup to be $\tilde{I}^{(S\mod 1,i)}(k)$, defined as in equation (\ref{CS-I-tilde-def}). Then
\begin{equation}
    \sum_{i} N_{(S\mod 1,i)}^{(H',\mathbb{n})} \tilde{I}^{(S\mod 1,i)}(k)\pto 2\,e^{-2\pi ikS^{(H',\mathbb{n})}_\ast}\int\limits_{\gamma^{(H',\mathbb{n})}}d^{|H|}v\,R(v)\,e^{2\pi ikS(v)},
    \label{CS-fd-thimble-general-relation}
\end{equation}
for some $N_{(S\mod 1,i)}^{(H',\mathbb{n})}\in \Z$,  The proportionality factor is the same as in (\ref{CS-fd-zero-thimble-correspondence}). By construction, the relation can be inverted (over integers), so that we have:
\begin{equation}
    \tilde{I}^{(S\mod 1,i)}(k)\pto 2\sum_{\substack{(H',\mathbb{n}):\\ S^{(H',\mathbb{n})}_\ast=S\mod 1}}T^{(S\mod 1,i)}_{(H',\mathbb{n})}\,e^{-2\pi ikS^{(H',\mathbb{n})}_\ast}\int\limits_{\gamma^{(H',\mathbb{n})}}d^{|H|}v\,R(v)\,e^{2\pi ikS(v)},
    \label{CS-fd-thimble-general-inverse-relation}
\end{equation}
for some\footnote{In practice, the coefficients $N^{(H',\mathbb{n})}_{(S\mod 1,i)}$ and $T_{(H',\mathbb{n})}^{(S\mod 1,i)}$ can be obtained as follows. First, assuming some truncation ($\| \mathbb{n} \|<\Lambda\gg 1$) of the set of the tuples (that should correspond to lifts in CS theory), one finds a basis  $J^{(S\mod 1,i)}(k)$ over $\Q$ of the vector subspace in $\C[[k^{-\frac{1}{2}},k^{\frac{1}{2}}]$ generated by the series (\ref{fd-thimbe-expansion-no-action}). This can be done by choosing any maximal subset of the series linearly independent over $\Q$, using one of the integral relations algorithms (such as PSLQ or LLL). A basis over $\Q$ satisfies a relation of the form (\ref{CS-fd-thimble-general-relation}) but with some rational coefficients $\tilde{N}_{(S\mod 1,I)}^{(H',\mathbb{n})}\in\Q$ instead. Let $d$ be their common denominator. Consider then Hermite decomposition of the integral rectangular matrix (with the number of rows $\gg$ number of columns) $d\tilde{N}=UR$ where $U$ is a square integral unimodular and $R$ is rectangular integral upper triangular. Then the basis $\tilde{I}^{(S,i)}(k)$ over the integers can be obtained as the left block of $RJ/d$. The coefficients $N$ in the relations (\ref{CS-fd-thimble-general-relation}) is given by the left block of $U$ and the coefficients $T$ in (\ref{CS-fd-thimble-general-inverse-relation}) are given by the upper block of $U^{-1}$. } (non-unique) $T^{(S\mod 1,i)}_{(H',\mathbb{n})}\in\Z$. The generating function for the monodromy coefficients in CS theory then can be expressed through the Stokes coefficients of the finite-dimensional model as follows:
\begin{multline}
    I^{(S\mod 1,i)}_{(S'\mod 1,j)}(\tilde{q})=
    (-1)^{|H|-1}\times \\
    \sum_{\substack{(H',\mathbb{n}):\,S_\ast^{(H',\mathbb{n})}=S\mod 1\\ (K,\mathbb{m})}}
    T^{(S\mod 1,i)}_{(H',\mathbb{n})}M^{(H',\mathbb{n})}_{(K,\mathbb{m})}N^{(K,\mathbb{m})}_{(S',j)}\,
    \tilde{q}^{S_\ast^{(H',\mathbb{n})}-S_\ast^{(K,\mathbb{m})}}
    \times\left\{
    \begin{array}{rl}
        \frac{1}{2}, & H'=\emptyset,\,K\neq \emptyset \\
        1, & \text{otherwise},
    \end{array}
    \right.
    ,
    \label{CS-fd-monodromies-general}
\end{multline}
where the sign $(-1)^{|H|-1}$, as in (\ref{CS-fd-monodromies}), takes into account the monodromy of the proportionality coefficient in (\ref{CS-fd-thimble-general-relation}), which contains $k^{\frac{|H|-1}{2}}$.

\subsubsection{Non-weakly-negative-definite plumbings}
\label{sec:non-weakly-definite}

Consider now the situation when the plumbing is not weakly negative-definite, i.e. the condition $C<0$ is not satisfied. The case when $C>0$ can be related to the previously considered one by $k\rightarrow -k$ and therefore is already covered by the previous analysis. More generally, if $\det C\neq 0$ we conjecture that the relation (\ref{CS-fd-zero-thimble-correspondence}) still holds, with the finite-dimensional contour $\gamma^{(\emptyset)}$ now being the Lefschetz thimble in $\C^{|H|}$ with respect to $S(v)$, which is given by the same formula (\ref{S-expression}) as before. However, now $\gamma^{(\emptyset)}$ is not given by a Cartesian product of contours in individual $v_I$-planes. One can similarly consider contours associated to other pairs $(H'\subset H,\mathbb{n}\in \Z^{H'})$:
\begin{equation}
    \gamma^{(H',\mathbb{n})}=
    \left( \bigtimes_{I\in H'}\gamma^{(H',\mathbb{n})}_I\right)
    \times \gamma^{(H',\mathbb{n})}_{H''}
\end{equation}
where $\gamma^{(H',\mathbb{n})}_I,\,I\in H'$ is, as before, a circle contour in the $v_I$-plane (see Figure \ref{fig:contours-basis}),  and 
$\gamma^{(H',\mathbb{n})}_{H''}\subset \C^{|H''|}$ is the Lefschetz thimble in the space of all $v_J,\,J\in H''\equiv H\setminus H'$ integration variables with respect to the restricted action (\ref{S-restricted}). Since the restricted action is quadratic, the thimble is still an affine subspace $\cong \R^{|H''|}\subset \C^{|H''|}$, however it is not simply a product of lines in the individual $v_I$-planes. Therefore, the calculation of monodromies becomes more involved, as it cannot be reduced to the calculation of monodromies of contours in individual $v_I$-planes. Moreover, unlike before, the Stokes jumps will happen both at $k\in i\R_+$ and $k\in -i\R_+$ rays, so that both $\prescript{+}{}{I}^\alpha_\beta(\tilde{q})$ and $\prescript{-}{}{I}^\alpha_\beta(\tilde{q})$ (defined at the end of Section \ref{sec:CS-resurgence-review}) are non-trivial, with ${I}^\alpha_\beta(\tilde{q})=\prescript{+}{}{I}^\alpha_\beta(\tilde{q})+\prescript{-}{}{I}^\alpha_\beta(\tilde{q})+m^\alpha_\alpha\,\delta^\alpha_\beta$. It is still, in principle, possible to calculate monodromy coefficients in terms of intersection numbers, as explained in Section \ref{sec:fd-categorification} for the weakly negative-definite case $C<0$. 

Alternatively, one can analyze directly the singularities of the Borel transforms $B^{(H',\mathbb{n})}(\xi)$ of the perturbative expansions of the integrals along finite-dimensional thimble contours $\gamma^{(H',\mathbb{n})}$. We have
\begin{equation}
    \int_{\gamma^{(H',\mathbb{n})}}d^{|H|}v\,R(v)\,e^{2\pi ikS(v)}=\int\limits_{S^{(H',\mathbb{n})}_\ast+\frac{i}{k}\R_+}d\xi e^{2\pi ik\xi}B^{(H',\mathbb{n})}(\xi)    
    \label{fd-total-to-fiber-integration}
\end{equation}
where 
\begin{equation}
    B^{(H',\mathbb{n})}(\xi)=\int\limits_{\Omega^{(H',\mathbb{n})}\subset \{S(v)=\xi\}\setminus \{\text{poles of }R(v)\}}\,\frac{d^{|H|}v}{d\xi}\,R(v)
    \label{Borel-quadric-integral}
\end{equation}
and $\Omega^{(H',\mathbb{n})}$ is a compact contour of real dimension $|H|-1$ inside the quadric $\{S(v)=
\xi\}$ which, in turn, belongs to $\C^{|H|}$ with the hyperplanes at the poles of $R(v)$ removed. The contour is chosen such that the equality (\ref{fd-total-to-fiber-integration}) holds. In any case, since we are only interested in the subgroup of thimbles generated by the monodromies starting from $\gamma^{(\emptyset)}$ (that corresponds to the trivial flat connection in CS theory), in principle it suffices to describe $\Omega^{(\emptyset)}$. 
Assuming that there are no poles at $v_I=0$, $\Omega^{(\emptyset)}$ can be defined in the vicinity of $\xi=0$ as the standard vanishing contour. It then can be analytically continued to the rest of the $\xi$-plane. The monodromy coefficients can be deduced using (\ref{B-singularity}) and analyzing the behavior of the compact contours $\Omega^{(H',\mathbb{n})}$ as one varies $\xi$. We will follow this approach in one of the examples considered later in this section.  

The case when $\det C=0$ (still assuming that $|\det B|=1$) can be dealt as follows. For a moment, we continue to assume that $C<0$. Starting as in (\ref{Z-hat-to-LG-model}), we have:
\begin{multline}
Z(k)
    \pto
    \sum_{n\in\Z^V}
\prod_{i\in V} {F_{i,n_i}} \, 
q^{-\frac{n^TB^{-1}n}{4}}
\pto
\sum_{n\in \Z^V} \int\limits_{\R^{|V|}} d^{|V|}u
\prod_{I\in V} {F_{I,n_I}}\,
e^{\frac{\pi ik}{2}\,u^TBu}\,e^{\pi iu^Tn} \\
\pto \int\limits_{\tilde\gamma}d^{|V|}u\,\tilde{R}(u)\,e^{2\pi ik\tilde{S}(u)}
\label{Z-hat-to-LG-big}
\end{multline}
where now $\tilde{S}(u)$ and $\tilde{R}(u)$ are respectively holomorphic and meromorphic functions on the affine space $\C^{|V|}$ with the variables corresponding to all vertices $V$ of the plumbing graph, not just the subset $H\subset V$:
\begin{equation}
    \tilde{S}(u)=\frac{1}{4}\,u^TBu,
    \label{S-expression-big}
\end{equation}
\begin{equation}
    \tilde{R}(u)=\prod_{i\in V}\,
    \left(\sin\pi u_i\right)^{2-\deg(i)}.
    \label{R-expression-big}
\end{equation}
The integration contour $\tilde{\gamma}\subset \C^{|V|}$ is similar to $\gamma$ in (\ref{Z-hat-to-LG-model}). It is a Cartesian product 
\begin{equation}
    \tilde{\gamma}=\bigtimes_{i\in V}\tilde{\gamma}_{i}
\end{equation}
of the contours in $u_I$-planes corresponding to the vertices of the plumbing graph. For high-valency vertices $I\in H$ the contours are the same as before: $\tilde{\gamma}_I={\gamma}_I$ (see Figure \ref{fig:contours-original}, left panel), and for other vertices $i\in V\setminus H$ they are simply real lines: $\tilde{\gamma}_i=\R\subset \C$. 

This gives an alternative finite-dimensional model in a larger space. In particular, similarly to (\ref{CS-fd-zero-thimble-correspondence}) we have the following expression for CS thimble $\bbalpha_0$ corresponding to the trivial flat connection and the thimble $\tilde{\gamma}^{(\emptyset)}$  with respect to $\tilde{S}(u)$:
\begin{equation}
    I^{\bbalpha_0}(k)\pto \int_{\tilde{\gamma}^{(\emptyset)}}d^{|V|}u\,\tilde{R}(u)\,e^{2\pi ik\tilde{S}(u)}.
    \label{CS-big-fd-zero-thimble-correspondence}
\end{equation}
Even if the condition $C<0$ does not hold, the right-hand side of (\ref{CS-big-fd-zero-thimble-correspondence}) makes sense and is well defined for arbitrary $B$, with $|\det B|=1$, provided that  $\tilde{\gamma}^{(\emptyset)}$ is still defined as the Lefschetz thimble with respect to $\tilde{S}(u)$. We then conjecture that the relation (\ref{CS-big-fd-zero-thimble-correspondence}) holds in general.

\subsubsection*{A strongly indefinite example}

\begin{figure}
\centering
\begin{tikzpicture}[scale=0.4]

\draw[ultra thick] (5,0)  -- (9,0);
  \draw[ultra thick] (9,0) -- (12,3);
   \draw[ultra thick] (9,0) -- (12,-3);
    \draw[ultra thick] (5,0) -- (2,-3);
     \draw[ultra thick] (5,0) -- (2,3);
    \filldraw[black] (5,0) circle (8pt) node[left=5] {$[0,0]$};
    \draw[blue] (5,0) node[above right] {$v_1$};
    \filldraw[black] (2,3) circle (8pt) node[left=4] {$[2,0]$};
    \filldraw[black] (2,-3) circle (8pt) node[left=4] {$[3,0]$};
         \filldraw[black] (9,0) circle (8pt) node[right=5] {$[-1,0]$};
    \draw[blue] (9,0) node[above left] {$v_2$};
    \filldraw[black] (12,3) circle (8pt) node[right=4] {$[-7,0]$};
    \filldraw[black] (12,-3) circle (8pt) node[right=4] {$[3,0]$};
\end{tikzpicture}
 \caption{An example of a strongly indefinite plumbed integer homology sphere. The variables $v_{1,2}$ corresponding to the high-valency vertices are shown in blue.}
\label{fig:indef-example}
\end{figure}

Consider an example of a \textit{strongly indefinite} plumbing (i.e. not weakly definite) shown in Figure \ref{fig:indef-example}. 
The linking matrix, with its block decomposition corresponding to high- and low-valency vertices is the following:
\begin{equation}
    B=
\begin{array}{cccc|ccccc}
& & \multicolumn{2}{c}{H} & \multicolumn{4}{c}{L} & \\
\multirow{2}{*}{$H$} & \ldelim({6}{4mm} &0 & 1 & 1 & 1 & 0 & 0 & \rdelim){6}{4mm} \\
 & &1 & -1 & 0 & 0 & 1 & 1 & \\
 \cline{3-8}
\multirow{4}{*}{$L$} & &1 & 0 & 2 & 0 & 0 & 0 & \\
 & &1 & 0 & 0 & 3 & 0 & 0 & \\
 & &0 & 1 & 0 & 0 & 3 & 0 & \\
 & &0 & 1 & 0 & 0 & 0 & -7 & \\
\end{array}.
\end{equation}
Its inverse is
\begin{equation}
    B^{-1}=\begin{array}{cccc|ccccc}
& & \multicolumn{2}{c}{H} & \multicolumn{4}{c}{L} & \\
\multirow{2}{*}{$H$} & \ldelim({6}{4mm} 
 & 150 & 126 & -75 & -50 & -42 & 18 & \rdelim){6}{4mm} \\
 & &126 & 105 & -63 & -42 & -35 & 15 & \\
 \cline{3-8}
\multirow{4}{*}{$L$} & & -75 & -63 & 38 & 25 & 21 & -9 & \\
 & &-50 & -42 & 25 & 17 & 14 & -6 & \\
 & &-42 & -35 & 21 & 14 & 12 & -5 & \\
& & 18 & 15 & -9 & -6 & -5 & 2 & \\
\end{array}
\end{equation}
with the blocks given by the matrices $C$, $D$, $D^T$ and $A$, cf. (\ref{A-block-def})--(\ref{C-block-def}). Assuming labeling of the high-valency vertices by 1,2 as shown in Figure \ref{fig:indef-example}, the meromorphic function (\ref{R-expression}) in the integrand of the finite-dimensional model reads
\begin{equation}
    R(v)=-\frac{\sin \left(\frac{\pi  v_1}{3}\right) \sin \left(\frac{\pi  v_1}{2}\right) \sin \left(\frac{\pi  v_2}{7}\right) \sin \left(\frac{\pi  v_2}{3}\right)}{\sin \left(\pi  v_1\right) \sin \left(\pi  v_2\right)}.
    \label{R-indefinite-example}
\end{equation}
The action (\ref{S-expression}) is
\begin{equation}
    S(v)=-\frac{5v_1^2}{24}+\frac{v_2 v_1}{2}-\frac{25 v_2^2}{84},
\end{equation}
which is indefinite over real values of $v_I$. The contour $\Omega^{(\emptyset)}$ in the conic $\{S(v)=\xi\}\subset \C^2$ vanishing at $\xi=0$ can be explicitly described as follows. The following linear change of variables $(v_1,v_2)\rightarrow (s,t)$ with
\begin{align}
    s & = \left(\sqrt{\frac{6}{5}}-\frac{1}{\sqrt{105}}\right) v_2-\sqrt{\frac{5}{6}} v_1, \\
    t & = 
    \sqrt{\frac{5}{6}} v_1+\left(-\sqrt{\frac{6}{5}}-\frac{1}{\sqrt{105}}\right) v_2,
\end{align}
transforms the conic equation into $st=4\xi$. For $\xi\neq 0$ one can then globally parametrize the conic by $x\in \C^*$ with $s=2\sqrt{\xi}x$ and $t=2\sqrt{\xi}/x$. In terms of these, the original variables $v_{1,2}$ are expressed as follows:
\begin{align}
    v_1 & = \frac{\sqrt{\frac{3}{5}} \sqrt{\xi } \left(-\left(\left(\sqrt{2}+6 \sqrt{7}\right) x^2\right)+\sqrt{2}-6 \sqrt{7}\right)}{x}, \\
    v_2 & = -\frac{\sqrt{105} \sqrt{\xi } \left(x^2+1\right)}{x}.
    \label{indefinite-example-v-from-x}
\end{align}
The vanishing contour can be described in the neighborhood of $\xi=0$ as the unit circle: $\Omega^{(\emptyset)}=\{|x|=1\}$. It is unambiguously defined in this way, as (\ref{R-indefinite-example}) does not have poles in the vicinity of $v_1=0,v_2=0$. The positions of singularities of $R(v)$ in the $x$-plane can be obtained by solving in $x$ one of the equations (\ref{indefinite-example-v-from-x}), with  $v_I=\mathbb{n}_I\in \Z\setminus \{0\}$ being a pole of $R(v)$. With a generic value of $\xi$, there are two distinct solutions for a fixed $v_1=\mathbb{n}_1$ or $v_2=\mathbb{n}_2$. All the singularities tend to either $\infty$ or $0$ in the $x$-plane as $\xi\rightarrow 0$. 

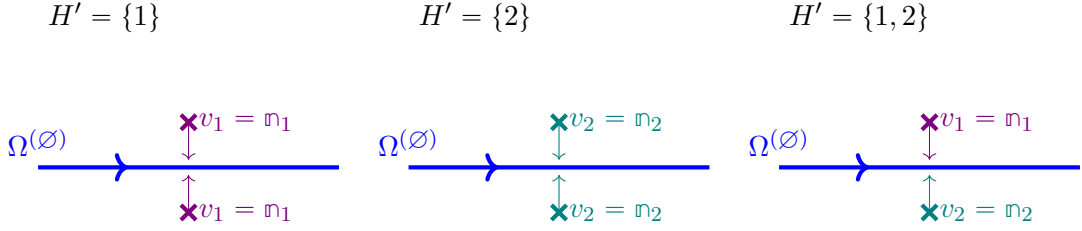
\begin{figure}
\centering
\begin{tikzpicture}
\begin{scope}[shift={(0,0.6)}]
\draw[violet,ultra thick] (-0.1,-0.1) -- (0.1,0.1);
\draw[violet,ultra thick] (-0.1,0.1) -- (0.1,-0.1);
\draw[violet,<-] (0,-0.5) -- (0,0) node[right] {$v_1=\mathbb{n}_1$}; 
\end{scope}

\draw (-2,2) node[right] {$H'=\{1\}$};

\begin{scope}[shift={(0,-0.6)}]
\draw[violet,ultra thick] (-0.1,-0.1) -- (0.1,0.1);
\draw[violet,ultra thick] (-0.1,0.1) -- (0.1,-0.1);
\draw[violet,<-] (0,0.5) -- (0,0) node[right] {$v_1=\mathbb{n}_1$};  
\end{scope}

\draw[blue,ultra thick,decoration={markings, mark=at position 0.3 with {\arrow{>}}}, postaction={decorate}] (-2,0) node[above] {$\Omega^{(\emptyset)}$} -- (2,0); 
\end{tikzpicture}
\;
\begin{tikzpicture}
\begin{scope}[shift={(0,0.6)}]
\draw[teal,ultra thick] (-0.1,-0.1) -- (0.1,0.1);
\draw[teal,ultra thick] (-0.1,0.1) -- (0.1,-0.1);
\draw[teal,<-] (0,-0.5) -- (0,0) node[right] {$v_2=\mathbb{n}_2$}; 
\end{scope}

\draw (-2,2) node[right] {$H'=\{2\}$};

\begin{scope}[shift={(0,-0.6)}]
\draw[teal,ultra thick] (-0.1,-0.1) -- (0.1,0.1);
\draw[teal,ultra thick] (-0.1,0.1) -- (0.1,-0.1);
\draw[teal,<-] (0,0.5) -- (0,0) node[right] {$v_2=\mathbb{n}_2$};  
\end{scope}

\draw[blue,ultra thick,decoration={markings, mark=at position 0.3 with {\arrow{>}}}, postaction={decorate}] (-2,0) node[above] {$\Omega^{(\emptyset)}$} -- (2,0); 
\end{tikzpicture}
\;
\begin{tikzpicture}
\begin{scope}[shift={(0,0.6)}]
\draw[violet,ultra thick] (-0.1,-0.1) -- (0.1,0.1);
\draw[violet,ultra thick] (-0.1,0.1) -- (0.1,-0.1);
\draw[violet,<-] (0,-0.5) -- (0,0) node[right] {$v_1=\mathbb{n}_1$}; 
\end{scope}

\draw (-2,2) node[right] {$H'=\{1,2\}$};

\begin{scope}[shift={(0,-0.6)}]
\draw[teal,ultra thick] (-0.1,-0.1) -- (0.1,0.1);
\draw[teal,ultra thick] (-0.1,0.1) -- (0.1,-0.1);
\draw[teal,<-] (0,0.5) -- (0,0) node[right] {$v_2=\mathbb{n}_2$};  
\end{scope}

\draw[blue,ultra thick,decoration={markings, mark=at position 0.3 with {\arrow{>}}}, postaction={decorate}] (-2,0) node[above] {$\Omega^{(\emptyset)}$} -- (2,0); 
\end{tikzpicture}

\caption{Decription of the singularties of $B^{(\emptyset)}(\xi)$ of different types corresponding to different subsets $H'\subset H$ in terms of ``pinching'' of the contour $\Omega^{(\emptyset)}$ in the quadric $S(v)=\xi$ by the sigularities of $R(v)$ at $v_I=\mathbb{n}_I\in \Z$, in the case of 2 high-valency vertices ($H=\{1,2\}$).}
\label{fig:contour-pinching}
\end{figure}

This provides an explicit description of the Borel transform $B^{(\emptyset)}(\xi)$ in terms of the integral (\ref{Borel-quadric-integral}) in the vicinity of $\xi=0$. In terms of the parametrization by $x$, the integration measure is simply $dv_1\wedge dv_2/d\xi\pto dx/x$, with a $\xi$-independent numerical factor. As one analytically continues $B^{(\emptyset)}(\xi)$ away from $\xi=0$, the positions of the singularities move around in the $x$-plane and can eventually reach the contour $|x|=1$. When a singularity reaches the contour, this, by itself does not lead to a singularity of $B^{(\emptyset)}(\xi)$, as one can deform the contour of integration away from it. The function $B^{(\emptyset)}(\xi)$ can become singular, however, when the contour is ``pinched'' between two singularities. There are 3 cases, illustrated in Figure \ref{fig:contour-pinching}:
\begin{itemize}

\item when 
the two colliding singularities correspond to two different solutions of $v_1=\mathbb{n}_1$ on the quadric, 

\item when they correspond to two different solutions of $v_2=\mathbb{n}_2$, and

\item when one of them corresponds to a solutions of $v_1=\mathbb{n}_1$ and the other to a solution of $v_2=\mathbb{n}_2$.

\end{itemize}
\noindent
Note that singularities corresponding to solutions of $v_I=\mathbb{n}_I$ and  $v_I=\mathbb{n}_I'$ for \textit{different} $\mathbb{n}_I$ and $\mathbb{n}_I'$ (with a fixed $I=1,2$) can never collide because $v_I$ is a well-defined function on the quadric. Such a classification of singularities is in direct correspondence with the classification of ``Lefschetz thimble'' contours in the finite-dimensional model by the pairs $(H'\subset H,\mathbb{n}\in \Z^{H'})$ considered before.

The values of $\xi$ for which the collisions of singularities happen are given by the corresponding critical values $S^{(H',\mathbb{n})}_*$ explicitly given by the formula (\ref{fd-S-extremal-value}). When two singularities collide, they do not necessarily pinch the contour $\Omega^{(\emptyset)}$. For $H'=\{1\}$ we have
\begin{equation}
    S^{(\{1\},\mathbb{n}_1)}=\frac{\mathbb{n}_1^2}{600}
\end{equation}
so that the singularities of $B^{(\emptyset)}$ can occur only on the positive real axis in the $\xi$-plane. A more detailed analysis shows that they are non-trivial for the following critical values of the action modulo 1:
\begin{multline}
    H'=\{1\}:\; S_*^{(H',\mathbb{n})} \in
    \left\{\frac{1}{600},\frac{1}{24},\frac{49}{600},\frac{121}{600},\frac{169}{600},\frac{241}{600},\frac{289}{600},\frac{361}{600},\frac{409}{600},\frac{481}{600},\frac{529}{600}\right\} \\ \mod 1.
\end{multline}
Unlike in the previously considered example, we will only list the generating function for Stokes coefficients between the trivial flat connection and one of the flat connections in the above class:
\begin{equation}
    I^{(0)}_{(\frac{1}{600})}(\tilde{q})=
    \tilde{q}^{-\frac{1}{600}}\left(
1-\tilde{q}^{-4}-\tilde{q}^{-17}
+\tilde{q}^{-37}-\tilde{q}^{-38}+\tilde{q}^{-66}+\ldots
    \right).
\end{equation}

Similarly, for $H'=\{2\}$ we have
\begin{equation}
    S^{(\{2\},\mathbb{n})}=\frac{\mathbb{n}_2^2}{420}.
\end{equation}
The corresponding singularities of $B^{(\emptyset)}$ again occur only on the positive real axis in the $\xi$-plane. The critical values of the action modulo 1 are the following:
\begin{multline}
    H'=\{2\}:\; S_*^{(H',\mathbb{n})} \in 
   \left\{\frac{1}{420},\frac{1}{105},\frac{4}{105},\frac{16}{105},\frac{109}{420},\frac{121}{420},\frac{169}{420},\frac{46}{105},\frac{64}{105},\frac{289}{420},\frac{79}{105},\frac{361}{420}\right\} \\ \mod 1.
\end{multline}
We give the following example of  a generating function for Stokes coefficients between the trivial flat connection and one of the flat connections in the above class:
\begin{equation}
    I^{(0)}_{(\frac{1}{420})}(\tilde{q})=
    \tilde{q}^{-\frac{1}{420}}\left(
-1-\tilde{q}^{-2}-\tilde{q}^{-4}
+\tilde{q}^{-12}-\tilde{q}^{-46}+\tilde{q}^{-68}+\tilde{q}^{-78}+\ldots
    \right).
\end{equation}

For $H'=\{1,2\}$ we have
\begin{equation}
    S^{(\{1,2\},\mathbb{n})}=-\frac{5\mathbb{n}_2^2}{24}
    +\frac{\mathbb{n}_1\mathbb{n}_2}{2}
    -\frac{25\mathbb{n}_1^2}{84}.
    \label{indef-example-action-n-indef}
\end{equation}
Due to the indefiniteness of this quadratic form, collisions of this type occur both on negative and positive real axes in $\xi$-plane. Moreover, an infinite number of collisions can happen simultaneously. This is because (\ref{indef-example-action-n-indef}), if considered as a Diophantine equation on $\mathbb{n}_1,\,\mathbb{n}_2$, with the fixed left-hand side, has an infinite number of solutions in general. However, only a finite number of colliding pairs can actually pinch a contour\footnote{This is different from what happens for the $q$-series $\hat{Z}(q)$ \cite{Gukov:2017kmk} associated to plumbings, where the infinite number of solutions to an analogous Diophantine equation does lead to an infinite number of contributions to a coefficient for given power of $q$, rendering $\hat{Z}(q)$ ill-defined (without an additional regularization). See \cite{Ri:2022bxf}.}. 

The set of critical values of the action modulo 1 is the following:
\begin{equation}
    H'=\{1,2\}:\; S_*^{(H',\mathbb{n})} \in 
    \left\{\frac{5}{168},\frac{47}{168},\frac{101}{168},\frac{125}{168},\frac{143}{168},\frac{167}{168}\right\} \mod 1.
\end{equation}
A detailed analysis shows that the contour pinching actually occurs only on the negative real axis in the $\xi$-plane.
We give the following example of  a generating function for Stokes coefficients:
\begin{equation}
    I^{(0)}_{(\frac{5}{168})}(\tilde{q})=
    \tilde{q}^{-\frac{5}{168}}\left(
-\tilde{q}^{7}+\tilde{q}^{20}
+\tilde{q}^{29}+\tilde{q}^{35}-\tilde{q}^{52}-\tilde{q}^{60}+\ldots
    \right).
\end{equation}

The singularities of $B^{(H',\mathbb{n})}(\xi)$ for $H'=\{1\}$ (similarly $H'=\{2\}$) can be analyzed in a similar way, by analyzing pinching of the corresponding contour $\Omega^{(\{1\},\mathbb{n}_1)}$ on the quadric, which is a difference of small circle contours surrounding the two solutions to $v_1=\mathbb{n}_1$. However, one can also proceed with the same analysis as for weakly negative plumbings because the thimble contours $\gamma^{(\{1\},\mathbb{n}_1)}$ factorize as before (since a 1-dimensional non-degenerate quadratic form is always definite).
We present the following example of a generating function for the Stokes coefficients:
\begin{equation}
    I^{(\frac{1}{600})}_{(\frac{5}{168})}(\tilde{q})=
    \tilde{q}^{\frac{1}{600}-\frac{5}{168}}\left(
-\tilde{q}^{3}-\tilde{q}^{7}
+\tilde{q}^{25}+\tilde{q}^{35}+\ldots
    \right).
\end{equation}

\subsubsection{Higher-genus plumbings}

Now consider the case of plumbings with general values of $g_i\geq 0,\,i\in V$ associated with the vertices. We conjecture that (\ref{Z-hat-to-LG-big}) generalizes directly with the same $\tilde{S}(u)$ as in (\ref{S-expression-big}), while $\tilde{R}(u)$ in (\ref{R-expression-big}) is modified as follows:
\begin{equation}
    \tilde{R}(u)=\prod_{i\in V}\,
    \left(\sin\pi u_i\right)^{2-\deg(i)-2g_i}.
    \label{R-expression-big-genus}
\end{equation}
This is motivated by the direct generalization of the naive analytic continuation of the WRT invariant for $g_i=0$ plumbings considered in \cite{Gukov:2017kmk} to the case of $g_i\geq 0$. Such a generalization leads to the relation as in (\ref{WRT-from-limit})--(\ref{Z-hat-from-F}), except that the coefficients $F_{i,n_i}$ are now defined as follows:
\begin{equation}
    \sum_{n\in\Z}F_{i,n}z^n  =
    \left.\frac{1}{2}\,(z-1/z)^{2-\deg(i)-2g_i}\right|_{\substack{\text{expansion at} \\ z=0}}+
    \left.\frac{1}{2}\,(z-1/z)^{2-\deg(i)-2g_i}\right|_{\substack{\text{expansion at} \\ z=\infty}}.
    \label{vertex-half-sum-expansion-genus}
\end{equation}

When $g_i=0$ for all low-valency vertices $i\in V\setminus H$, one can consider a generalization of the ``small'' version of the finite-dimensional model, where the integration is performed only over the variables corresponding to high-valency vertices. Namely, we conjecture that in this case (\ref{CS-fd-zero-thimble-correspondence}) holds with the same $S(v)$ as before but with $R(v)$ replaced by
\begin{equation}
    R(v)=\frac{ \prod_{a\in L} \sin \cfrac{\pi v_{h(a)}}{P_a}}{\prod_{I\in H}(\sin\pi v_I)^{2g_I+\deg(I)-2}}.
    \label{R-expression-high-val-g}
\end{equation}
When $|H|=1$, i.e. there is a single high-valency vertex $I\in H$, the plumbed manifold is a Seifert fibration over a genus $g_I$ surface. The formula (\ref{CS-fd-zero-thimble-correspondence}) with $R(v)$ as in (\ref{R-expression-high-val-g})  then becomes the formula obtained by Blau and Thompson via localization \cite{Blau:2006gh} (see also \cite{Gukov:2023srx} for an alternative approach via splicing).

\subsubsection*{Examples with $b_1 > 0$}

Let us elaborate on the family of examples encountered above, which provide a simple yet instructive supply of examples with $b_1 (Y) > 0$. Namely, let $Y$ be a Seifert fibration over genus-$g$ surface $\Sigma_g$ without singular fibers. For a degree-$p$ circle bundle over $\Sigma_g$, the 3-manifold corresponds to the plumbing graph consisting of a single vertex with label $[p,g]$. The fundamental group of $Y$ fits into the exact sequence
\begin{equation}
1 \to \pi_1 (S^1) \to \pi_1 (Y) \to \pi_1 (\Sigma_g) \to 1
\end{equation}
and can be explicitly described in terms of generators $(a_i, b_i)_{i=1, \ldots , g}$, $h$ and relations
\begin{equation}
\prod_{i=1}^{g} [a_i, b_i] = h^p \,, \qquad
[a_i, h] = 1 \,, \qquad
[b_i, h] = 1.
\end{equation}
The simplest members of this family are degree $p= \pm 1$ fibrations. In these cases, the above relations specialize to the following:
\begin{equation}
\prod_{i=1}^{g} [a_i, b_i] = h^{\pm 1} \,, \qquad
[a_i, h] = 1 \,, \qquad
[b_i, h] = 1.
\label{Sigmagh}
\end{equation}
We are interested in homomorphisms from this group into $SL(2,\C)$ (or, more generally, into $G_{\C}$). To avoid clutter, and with a small abuse of notations, we denote $SL(2,\C)$ holonomies by the same letters as the generators of $\pi_1 (Y)$. Then, for any $SL(2,\C)$ representation (i.e. both reducible and irreducible flat connections) one can show that $h \in SL (2, \C )$ is central. In the reducible case $h=1$ follows from the first relation above, whereas in the irreducible case it follows from the last two relations that $h = \pm 1$, where both choices of sign are allowed, for a given $p$. Then, we are only left with the first relation in \eqref{Sigmagh} which has a purely 2-dimensional interpretation in terms of gauge fields on $\Sigma_g$. In other words, we conclude
\begin{equation}
\mathcal{M} (Y) \; = \; \mathcal{M}_+ (\Sigma_g) \, \sqcup \, \mathcal{M}_- (\Sigma_g)
\label{MYMM}
\end{equation}
where
\begin{equation}
\mathcal{M}_{\pm} (\Sigma_g) : = \Big\{ a_i , b_j \in SL(2,\C) \; \Big\vert \; \prod_{i=1}^{g} [a_i, b_i] = \pm 1 \Big\} / \sim.
\end{equation}
These two moduli spaces, labeled by ``$+$'' and ``$-$'', respectively, can be understood as moduli spaces of complex flat connections on $\Sigma_g$ with one ramification point or, equivalently, as moduli spaces of rank-2 Higgs bundles on $\Sigma_g$ of even (resp. odd) degree \cite{MR0887284}. Both components, $\mathcal{M}_+ (\Sigma_g)$ and $\mathcal{M}_- (\Sigma_g)$, are non-compact because they are parametrized by holonomies valued in a non-compact group $SL(2,\C)$ or, in the language of Higgs bundles, because the norm of the Higgs field is unbounded. Both have real dimension
\begin{equation}
\dim \mathcal{M}_{\pm} (\Sigma_g) \; = \; 12g-12
\end{equation}
when $g>1$.
The second component, $\mathcal{M}_- (\Sigma_g)$, is smooth and its topology was studied by Hitchin in loc. cit. via the celebrated circle action on Higgs bundles; it has Betti numbers $b_i$ ranging from $i=0$ to the middle dimension of the moduli space, $i=6g-6$. The latter is exactly what we are interested in since it provides a bound on the number of independent Lefschetz thimbles. Specifically, according to \cite{MR0887284}, the Poincar\'e polynomial of $\mathcal{M}_- (\Sigma_g)$ is given by
\begin{multline}
\sum_{i=0}^{6g-6} t^i b_i (\mathcal{M}_- (\Sigma_g))
= \frac{(1+t^3)^{2g}}{(1 - t^2)(1 - t^4)}
- \frac{t^{4g-4}}{4 (1 - t^2) (1 - t^4)} \Big(
(1 + t^2)^2 (1 + t)^{2g} - (1 + t)^4 (1 - t)^{2g} \Big) \\
- (g-1) t^{4g-3} \frac{(1+t)^{2g-2}}{1-t}
+ 2^{2g-1} t^{4g-4} \Big( (1+t)^{2g-2} - (1-t)^{2g-2} \Big).
\label{MminusBetti}
\end{multline}
For the middle-dimensional homology it gives
\begin{equation}
b_{6g-6} (\mathcal{M}_- (\Sigma_g)) \; = \; g.
\label{MminusmidBetti}
\end{equation}
On the other hand, the Poincar\'e polynomial of $\mathcal{M}_+ (\Sigma_g)$ requires more care (and depends on the precise definition of $b_i$) since this component of the moduli space is not smooth. Following \cite{MR2755722}, we have
\begin{multline}
\sum_{i=0}^{6g-6} t^i b_i (\mathcal{M}_+ (\Sigma_g))
= P (\mathrm{Bun}_{SL(2)} (\Sigma_g))
- t^{4g-4} 
+ \frac{t^{2g+2} (1+t)^{2g}}{(1-t^2) (1-t^4)}
+ \frac{t^{4g-4} (1-t)^{2g}}{4 (1+t^2)} \\
+ \frac{t^{4g-4} (1+t)^{2g}}{2 (1-t^2)} \Big( \frac{5}{2} - \frac{2gt}{1+t} + \frac{1}{t^2-1} \Big)
+ \frac{1}{2} (2^{2g} - 1) t^{4g-2} \Big( (1+t)^{2g-2} (1-t)^{2g-2} -2 \Big)
\label{MplusBetti}
\end{multline}
where the first term is the Poincar\'e polynomial of the moduli space of semistable rank 2 bundles, $\mathrm{Bun}_{SL(2)} (\Sigma_g)$. Note, while $\mathrm{Bun}_{SL(2)} (\Sigma_g)$ is compact and has real dimension $6g-6$, the non-compact moduli space $\mathcal{M}_+ (\Sigma_g)$ has top non-zero betti number $b_{6g-4} (\mathcal{M}_+ (\Sigma_g)) \ne 0$, slightly above the middle dimension, which is also in contrast to \eqref{MminusBetti}. The Poincar\'e polynomial of $\mathrm{Bun}_{SL(2)} (\Sigma_g)$ is given by the celebrated the Harder-Narasimhan formula
\begin{equation}
P (\mathrm{Bun}_{SL(2)} (\Sigma_g)) \; = \; \frac{(1 + t^3)^{2g} - t^{2g} (1+t)^{2g}}{(1-t^2) (1-t^4)}.
\end{equation}
This latter expression contributes $+1$ to the Betti number $b_{6g-6} (\mathcal{M}_+ (\Sigma_g))$ for any value of $g>1$. On the other hand, the rest of the formula \eqref{MplusBetti} produces a contribution that grows very fast with the genus. Below we list the first few values for low genus:
$$
\begin{array}{c||c|c|c|c|c|c|c}
g & 2 & 3 & 4 & 5 & 6 & 7 & ~\cdots \\
\hline
~b_{6g-6} (\mathcal{M}_+ (\Sigma_g))~ & ~2~ & ~381~ & ~3829~ & ~28649~ & ~184281~ & ~1081285~ & ~\cdots 
\end{array}
$$
Numerically, we find that this growth is exponential, with the following rate
\begin{equation}
b_{6g-6} \left( \mathcal{M}_+ (\Sigma_g) \right) \; \simeq \; (4.72656 \ldots)^g.
\label{MplusmidBetti}
\end{equation}
This growth is much faster than \eqref{MminusmidBetti} and dominates
\begin{equation}
b_{6g-6} \left( \mathcal{M} (Y) \right) \; = \; b_{6g-6} \left( \mathcal{M}_+ (\Sigma_g) \right) \, + \, b_{6g-6} \left( \mathcal{M}_- (\Sigma_g) \right)
\end{equation}
for this class of 3-manifolds. As we will see shortly this provides a much larger set of potential Lefschetz thimbles than can be seen from the Borel plane for the perturbative expansion around the trivial flat connection. That is, not all the thimbles of complex Chern-Simons theory are in the subgroup generated by the thimble corresponding to the trivial flat connection in the sense described in Section \ref{sec:single-thimble-info}.

To compare with the Borel plane analysis, for concreteness let us focus on the case $g>1$ and $p=-1$. We have (with $v\equiv v_1$)
\begin{equation}
    S(v)=-\frac{v^2}{4},
\end{equation}
\begin{equation}
    R(v)=\frac{1}{(\sin\pi v)^{2g-2}}.
\end{equation}
We have two types of non-trivial contours in the $v$-plane: $\gamma^{(\emptyset)}$, the straight line Lefschetz thimble contour dodging the pole at $v=0$ (it does not matter from which side, as the residue at this pole is zero) and the circle contours $\gamma^{(\{1\},\mathbb{n})}$, $\mathbb{n}\in \Z\setminus \{0\}$ going around the poles at $v=\mathbb{n}$. For the second type, the perturbative expansion is \textit{finite} and has the following form:
\begin{equation}
    \int\limits_{\gamma^{(\{1\},\mathbb{n})}}dv\,R(v)\,e^{2\pi ik\,S(v)}=
    e^{2\pi ik\,S_*^{(\{1\},\mathbb{n})}}\left(\mathcal{A}_{g-1}(k)\mathbb{n}^{2g-3}
    +\mathcal{A}_{g-2}(k)\mathbb{n}^{2g-5}
    +\ldots+\mathcal{A}_{1}(k)\mathbb{n}\right)
\end{equation}
where $\mathcal{A}_{i}(k)$ are polynomials in $k$ of degree $g+i-2$ and $S_*^{(\{1\},\mathbb{n})}=-\mathbb{n}^2/4$. Modulo 1, there are just two distinct critical values of the action: $-1/4\in \C/\Z$ realized for $\gamma^{(\{1\},\mathbb{n})}$ with $\mathbb{n}$ odd, and $0\in \C/\Z$ realized for $\gamma^{(\{1\},\mathbb{n})}$ with $\mathbb{n}$ even as well as for $\gamma^{(\emptyset)}$. On the other hand, the integration variable $v$ can be identified with the holonomy of the $SL(2,\mathbb{C})$ connection over the circle fiber \cite{Blau:2006gh}, with $v=\mathbb{n}=0\mod 2$
corresponding to $h=1$, and $v=\mathbb{n}= 1\mod 2$ corresponding to $h=-1$. In other words, the two components in \eqref{MYMM} correspond to the different Chern-Simons values, respectively $0$ and $- \frac{1}{4}$, modulo 1.

Following the general method described in Section \ref{sec:deg-CS-values}, we can choose the basis of perturbative expansions in Chern-Simons theory (in the subgroup generated by Stokes jumps starting from the trivial flat connection, see Section \ref{sec:single-thimble-info}) $\tilde{I}^{(0,i)}(k),\,i=0,\ldots,g-1$ and $\tilde{I}^{(-\frac{1}{4},i)}(k),\,i=1,\ldots,g-1$  such that
\begin{equation}
    \tilde{I}^{(0,0)}(k)\pto \int\limits_{\gamma^{(H',\mathbb{n})}}d^{|H|}v\,R(v)\,e^{2\pi ikS(v)},
\end{equation}
i.e. 
\begin{equation}
    N^{(H',\mathbb{n})}_{(\emptyset)}=\left\{
        \begin{array}{ll}
             1, & (H',\mathbb{n})=(\emptyset), \\
             0, & \text{otherwise},
        \end{array}
    \right.
\end{equation}
in (\ref{CS-fd-thimble-general-relation})
and other fixed by taking
\begin{equation}
    N^{(\{1\},\mathbb{n})}_{(0,1)}=\frac{\mathbb{n}}{2},\;
    N^{(\{1\},\mathbb{n})}_{(0,2)}=\frac{\mathbb{n}^3-4\,\mathbb{n}}{48},\;
    N^{(\{1\},\mathbb{n})}_{(0,3)}=\frac{\mathbb{n}^5-20\,\mathbb{n}^3+64\,\mathbb{n}}{3840},\;\ldots
\end{equation}
for even $\mathbb{n}$ and 
\begin{equation}
    N^{(\{1\},\mathbb{n})}_{(0,1)}={\mathbb{n}},\;
    N^{(\{1\},\mathbb{n})}_{(0,2)}=\frac{\mathbb{n}^3-\mathbb{n}}{24},\;
    N^{(\{1\},\mathbb{n})}_{(0,3)}=\frac{\mathbb{n}^5-10\,\mathbb{n}^3+9\,\mathbb{n}}{1920},\;\ldots
\end{equation}
for odd $\mathbb{n}$. One can then take
\begin{equation}
    T_{(H',\mathbb{n})}^{(\emptyset)}=\left\{
        \begin{array}{ll}
             1, & (H',\mathbb{n})=(\emptyset), \\
             0, & \text{otherwise},
        \end{array}
    \right.
\end{equation}
in (\ref{CS-fd-thimble-general-inverse-relation}). The monodromy coefficients in the finite dimensional model are the following:
\begin{equation}
    M^{(\emptyset)}_{(\emptyset)}=-1,
\end{equation}
\begin{equation}
    M^{(\emptyset)}_{(\{1\},\mathbb{n})}=\left\{
        \begin{array}{ll}
             1, & \mathbb{n}\geq 0, \\
             -1, & \mathbb{n}<0,
        \end{array}
    \right.
\end{equation}
\begin{equation}
    M^{(\{1\},\mathbb{m})}_{(\{1\},\mathbb{n})}=0.
\end{equation}
From (\ref{CS-fd-monodromies-general}) we then get the following non-zero generating functions of Stokes coefficients:
\begin{equation}
    I^{(0,0)}_{(0,0)}(\tilde{q})=-1\;\in \Z[\tilde{q}],
\end{equation}
\begin{equation}
    I^{(0,0)}_{(0,i)}(\tilde{q})=
    \sum_{\substack{\mathbb{n}>0 \\ \mathbb{n}\text{ even}}}
    N^{(\{1\},\mathbb{n})}_{(0,i)}\;\tilde{q}^\frac{\mathbb{n}^2}{4}\;\in \Z[\tilde{q}],\qquad i=1,\ldots, g-1,
\end{equation}
and 
\begin{equation}
    I^{(0,0)}_{(-\frac{1}{4},i)}(\tilde{q})=
    \sum_{\substack{\mathbb{n}>0 \\ \mathbb{n}\text{ odd}}}
    N^{(\{1\},\mathbb{n})}_{(0,i)}\;\tilde{q}^\frac{\mathbb{n}^2}{4}\;\in \tilde{q}^\frac{1}{4}\Z[\tilde{q}],\qquad i=1,\ldots, g-1.
\end{equation}
In particular, the first non-trivial example is $g=2$. In this case, both $b_{6g-6} ( \mathcal{M}_+ (\Sigma_g))$ and $b_{6g-6} ( \mathcal{M}_- (\Sigma_g))$ are equal to 2. In the former case this agrees with the number of non-zero generating functions of Stokes coefficients:
\begin{eqnarray}
I^{(0,0)}_{(0,0)}(\tilde{q}) & = & -1, \\
I^{(0,0)}_{(0,1)}(\tilde{q})
& = & \tilde{q} +2 \tilde{q}^4 +3 \tilde{q}^9 +4 \tilde{q}^{16}
+5 \tilde{q}^{25} +6 \tilde{q}^{36} +7 \tilde{q}^{49}
+8 \tilde{q}^{64} + 9 \tilde{q}^{81} 
+ \ldots
\end{eqnarray}
whereas in the latter case we find only one non-trivial $\tilde q$-series from the Borel plane:
\begin{equation}
I^{(0,0)}_{(-\frac{1}{4},1)}(\tilde{q})
= \tilde{q}^{\frac{1}{4}} \left(
1 +3 \tilde{q}^2 +5 \tilde{q}^6 +7 \tilde{q}^{12}
+9 \tilde{q}^{20} +11 \tilde{q}^{30} +13 \tilde{q}^{42}
+15 \tilde{q}^{56} +17 \tilde{q}^{72} 
+ \ldots \right).
\end{equation}
This illustrates that, in general, not all homology classes are involved in Stokes jumps of the thimble for the trivial flat connection.

\subsubsection{Non-trivial torsion in homology}

Finally, consider yet another generalization to the case when $|\det B|>1$ and, therefore, there is a non-trivial torsion in homology: $\mathrm{Tor}\, H_1(Y,\Z)\cong \Z^V/B\Z^V\equiv \mathrm{Coker}\, B$. The conjecture of \cite{Gukov:2017kmk} about the analytic continuation of the WRT invariant for plumbed 3-manifolds was already formulated in this general case. It states that (\ref{WRT-from-limit}) holds in this more general case with
\begin{equation}
Z(k):=\frac{(-1)^{b_+}\,q^\frac{3\sigma-\Tr B}{4}}{2\sqrt{2k}|\mathrm{Tor} H_1(Y)|}\sum_{a\in \Z^V/B\Z^V}e^{-2\pi ik\,a^TB^{-1}a}\,Z_a(k),
\label{Z-hat-from-F-torsion}
\end{equation}
where
\begin{equation}
   Z_a(k)=\sum_{n\in \Z^V}\prod_{i\in V} F_{i,n_i}\,q^{-\frac{n^TB^{-1}n}{4}}\,e^{-2\pi ia^TB^{-1}n},
\end{equation}
and the coefficients $F_{i,n_i}$ are given by the same expression as before. Repeating the transformations in (\ref{Z-hat-to-LG-model}) and identifying $a^TB^{-1}a$ with the CS functional of a connected component of an abelian flat connection
\begin{multline}
a\in \pi_0\Hom(H_1(Y),U(1))/\{\pm 1\}\cong \Hom(\mathrm{Tor}\,H_1(Y),U(1))/\{\pm 1\} \\
\cong \mathrm{Tor}\,H_1(Y)/\{\pm 1\} \cong \Z^V/B\Z^V/\{\pm 1\}
\end{multline}
one obtains a generalization of (\ref{CS-big-fd-zero-thimble-correspondence}) to the CS thimbles $\bbalpha_a$ corresponding to abelian flat connections in the case of non-trivial torsion in homology:
\begin{equation}
    I^{\bbalpha_a}(k)\pto \int_{\tilde{\gamma}_a}d^{|V|}u\,\tilde{R}(u)\,e^{2\pi ik(\tilde{S}(u)+u^Ta)}
    \label{CS-big-fd-zero-thimble-correspondence-torsion}
\end{equation}
where $\tilde{\gamma}_a$ is the Lefschetz thimble with respect to $\tilde{S}(u)+u^Ta$ (cf. \cite{lawrence1999witten,Marino:2002fk,Beasley:2005vf,Blau:2006gh,Gukov:2023srx} for the case of Seifert manifolds).

\subsection{Relation to \texorpdfstring{$\hat{Z}$}{Z-hat}}
\label{sec:Z-hat}

In this section we comment further on the relationship between $\hat{Z}$-invariants of plumbed 3-manifolds introduced in \cite{Gukov:2017kmk} and Lefschetz thimble integrals. We will again restrict ourselves to the case of an integer homology sphere. In this case there is a single $q$-series invariant $\hat{Z}$ (more generally there is a family of $q$-series labeled by Spin$^c$-structures on the 3-manifold). Up to a simple factor, it is the same as the analytic continuation (\ref{Z-hat-from-F}) of the WRT invariant:
\begin{equation}
\hat{Z}(k):=(-1)^{b_+}\,q^\frac{3\sigma-\Tr B}{4}\sum_{n\in \Z^V}\prod_{i\in V} F_{i,n_i}\,q^{-\frac{n^TB^{-1}n}{4}}.
\end{equation}
As was shown in Section \ref{sec:stokes-weakly-negative}, formula (\ref{Z-hat-to-LG-model}), it is given by an integral over the contour $\gamma$ that factorizes into the contours shown in Figure \ref{fig:contours-original}:
\begin{equation}
    \hat{Z}(q)\pto
    \int_\gamma d^{|H|}v\,
    R(v)\,e^{2\pi ikS(v)}.
    \label{Z-hat-gamma-contour}
\end{equation}
As before, $\pto$ means equality up to a factor of the form $c\,k^{\frac{a}{2}}\,q^b$.
The contour can be expressed as a linear combination (with coefficients in $2^{-|H|}\Z$) of the ``thimble'' contours $\gamma^{(H',\mathbb{n})}$ by applying transformations (\ref{contour-shift-transformation}). Using the relations (\ref{CS-fd-thimble-general-relation}) between the thimble integrals in the finite-dimensional and Chern-Simons theory, one then can obtain the relation of the following form:
\begin{equation}
    \hat{Z}(q)\pto \sum_\bbalpha \hat{m}_\bbalpha\,I^{\bbalpha}(k),
    \label{Z-hat-to-thimbles}
\end{equation}
with certain coefficients\footnote{The coefficients are multiples of $2\times 2^{-|H|}$ because the integrand is an even function of $v$.} $\hat{m}_\bbalpha\in 2^{-|H|+1}\Z$. Using (\ref{CS-I-tilde-def}) instead, one can rewrite (\ref{Z-hat-to-thimbles}) as a \textit{finite} sum over the $\Z$-orbits of the CS thimbles:
\begin{equation}
     \hat{Z}(q)\pto \sum_\alpha \mathcal{S}_\alpha(\tilde{q})\,\tilde{I}^{\alpha}(k),
    \label{Z-hat-to-thimbles-finite}
\end{equation}
with $\tilde{q}=e^{-2\pi ik}$ and
\begin{equation}
    \mathcal{S}_\alpha(\tilde{q})=
    \sum_{\bbalpha\in \alpha}
    \hat{m}_\bbalpha \,
    \tilde{q}^{-\CS_\bbalpha}\qquad \in  \;\tilde{q}^{\{\CS_\alpha\}}\,\Z[[\tilde{q}]].
\end{equation}
As a specific example, consider again the plumbing from Figure \ref{fig:plumbing-example-2m3-5m3}. As before, we will label the $\Z$-orbits by the corresponding values of Chern-Simons functional modulo 1. We have:
\begin{equation}
    \hat{Z}(q)=q^{-\frac{1}{2}}\left(
    2-q+q^2+q^4-q^5+q^6-3 q^7-q^8+q^{10}-q^{12}-2 q^{13}+\ldots\right)
\end{equation}
and
\begin{equation}
\footnotesize
\begin{array}{rcl}
  {\mathcal{S}}_{\left(\frac{23}{312}\right)}  & = & -\tilde{q}^{-\frac{23}{312}}\,\left(-\tilde{q}-\tilde{q}^4-\tilde{q}^6-\tilde{q}^{12}+\tilde{q}^{29}+\tilde{q}^{41}+\tilde{q}^{47}+\tilde{q}^{62}-\tilde{q}^{96}-\tilde{q}^{117}-\tilde{q}^{127}+\ldots\right) \\
 {\mathcal{S}}_{\left(\frac{95}{312}\right)}  & = & -\tilde{q}^{-\frac{95}{312}}\,\left(-\tilde{q}^2-\tilde{q}^3-\tilde{q}^8-\tilde{q}^{10}+\tilde{q}^{33}+\tilde{q}^{37}+\tilde{q}^{52}+\tilde{q}^{57}-\tilde{q}^{103}-\tilde{q}^{110}-\tilde{q}^{135}-\tilde{q}^{143}+\ldots\right) \\
 {\mathcal{S}}_{\left(\frac{191}{312}\right)}  & = & -\tilde{q}^{-\frac{191}{312}}\,\left(\tilde{q}+\tilde{q}^5+\tilde{q}^6+\tilde{q}^{15}-\tilde{q}^{26}-\tilde{q}^{43}-\tilde{q}^{46}-\tilde{q}^{68}+\tilde{q}^{90}+\tilde{q}^{120}+\tilde{q}^{125}+\ldots\right) \\
 {\mathcal{S}}_{\left(\frac{263}{312}\right)}  & = & -\tilde{q}^{-\frac{263}{312}}\,\left(\tilde{q}-\tilde{q}^2-\tilde{q}^{12}+\tilde{q}^{17}-\tilde{q}^{24}+\tilde{q}^{31}+\tilde{q}^{61}-\tilde{q}^{72}+\tilde{q}^{86}-\tilde{q}^{99}-\tilde{q}^{149}+\ldots\right) \\
 {\mathcal{S}}_{\left(\frac{287}{312}\right)}  & = & -\tilde{q}^{-\frac{287}{312}}\,\left(\tilde{q}+\tilde{q}^4+\tilde{q}^8+\tilde{q}^{18}-\tilde{q}^{23}-\tilde{q}^{39}-\tilde{q}^{51}-\tilde{q}^{74}+\tilde{q}^{84}+\tilde{q}^{113}+\tilde{q}^{133}+\ldots\right) \\
 {\mathcal{S}}_{\left(\frac{311}{312}\right)}  & = & -\tilde{q}^{-\frac{311}{312}}\,\left(-\tilde{q}+\tilde{q}^3+\tilde{q}^{10}-\tilde{q}^{20}+\tilde{q}^{21}-\tilde{q}^{35}-\tilde{q}^{56}+\tilde{q}^{78}-\tilde{q}^{80}+\tilde{q}^{106}+\tilde{q}^{141}+\ldots\right) \\
\end{array}
\end{equation}

\begin{equation}
\footnotesize
\begin{array}{rcl}
 {\mathcal{S}}_{\left(\frac{59}{420}\right)}  & = & -\tilde{q}^{-\frac{59}{420}}\,\left(\tilde{q}-\tilde{q}^9-\tilde{q}^{15}-\tilde{q}^{19}+\tilde{q}^{35}+\tilde{q}^{41}+\tilde{q}^{53}-\tilde{q}^{87}+\tilde{q}^{125}+\ldots\right) \\
 {\mathcal{S}}_{\left(\frac{26}{105}\right)}  & = & -\tilde{q}^{-\frac{26}{105}}\,\left(\tilde{q}^3-\tilde{q}^{10}-\tilde{q}^{14}-\tilde{q}^{26}+\tilde{q}^{27}+\tilde{q}^{43}+\tilde{q}^{51}-\tilde{q}^{74}+\tilde{q}^{142}+\ldots\right) \\
 {\mathcal{S}}_{\left(\frac{131}{420}\right)}  & = & -\tilde{q}^{-\frac{131}{420}}\,\left(-\tilde{q}-\tilde{q}^7-\tilde{q}^{11}-\tilde{q}^{13}+\tilde{q}^{45}+\tilde{q}^{49}+\tilde{q}^{59}+\tilde{q}^{89}-\tilde{q}^{123}+\ldots\right) \\
 {\mathcal{S}}_{\left(\frac{41}{105}\right)}  & = & -\tilde{q}^{-\frac{41}{105}}\,\left(\tilde{q}+\tilde{q}^2+\tilde{q}^5-\tilde{q}^{18}+\tilde{q}^{37}-\tilde{q}^{66}-\tilde{q}^{81}-\tilde{q}^{90}+\tilde{q}^{122}+\tilde{q}^{133}+\ldots\right) \\
 {\mathcal{S}}_{\left(\frac{59}{105}\right)}  & = & -\tilde{q}^{-\frac{59}{105}}\,\left(\tilde{q}^3+\tilde{q}^4+\tilde{q}^7+\tilde{q}^{19}-\tilde{q}^{36}-\tilde{q}^{60}-\tilde{q}^{71}-\tilde{q}^{76}+\tilde{q}^{140}+\tilde{q}^{147}+\ldots\right) \\
 {\mathcal{S}}_{\left(\frac{251}{420}\right)}  & = & -\tilde{q}^{-\frac{251}{420}}\,\left(-\tilde{q}+\tilde{q}^5+\tilde{q}^{17}+\tilde{q}^{23}-\tilde{q}^{31}-\tilde{q}^{39}-\tilde{q}^{67}+\tilde{q}^{93}-\tilde{q}^{119}+\ldots\right) \\
 {\mathcal{S}}_{\left(\frac{299}{420}\right)}  & = & -\tilde{q}^{-\frac{299}{420}}\,\left(-\tilde{q}-\tilde{q}^3-\tilde{q}^9+\tilde{q}^{25}-\tilde{q}^{29}+\tilde{q}^{55}+\tilde{q}^{77}+\tilde{q}^{95}-\tilde{q}^{117}-\tilde{q}^{139}+\ldots\right) \\
 {\mathcal{S}}_{\left(\frac{311}{420}\right)}  & = & -\tilde{q}^{-\frac{311}{420}}\,\left(-\tilde{q}^2-\tilde{q}^4-\tilde{q}^6-\tilde{q}^{26}+\tilde{q}^{28}+\tilde{q}^{64}+\tilde{q}^{72}+\tilde{q}^{84}-\tilde{q}^{130}-\tilde{q}^{146}+\ldots\right) \\
 {\mathcal{S}}_{\left(\frac{89}{105}\right)}  & = & -\tilde{q}^{-\frac{89}{105}}\,\left(-\tilde{q}-\tilde{q}^2-\tilde{q}^{10}+\tilde{q}^{21}-\tilde{q}^{34}+\tilde{q}^{53}+\tilde{q}^{85}+\tilde{q}^{98}-\tilde{q}^{114}-\tilde{q}^{129}+\ldots\right) \\
 {\mathcal{S}}_{\left(\frac{101}{105}\right)}  & = & -\tilde{q}^{-\frac{101}{105}}\,\left(-\tilde{q}+\tilde{q}^6+\tilde{q}^{14}+\tilde{q}^{22}-\tilde{q}^{33}-\tilde{q}^{45}-\tilde{q}^{65}+\tilde{q}^{102}-\tilde{q}^{110}+\ldots\right) \\
 {\mathcal{S}}_{\left(\frac{104}{105}\right)}  & = & -\tilde{q}^{-\frac{104}{105}}\,\left(\tilde{q}+\tilde{q}^9+\tilde{q}^{12}+\tilde{q}^{17}-\tilde{q}^{40}-\tilde{q}^{49}-\tilde{q}^{56}-\tilde{q}^{104}+\tilde{q}^{108}+\ldots\right) \\
 {\mathcal{S}}_{\left(\frac{419}{420}\right)}  & = & -\tilde{q}^{-\frac{419}{420}}\,\left(\tilde{q}+\tilde{q}^3+\tilde{q}^5-\tilde{q}^{13}+\tilde{q}^{47}-\tilde{q}^{69}-\tilde{q}^{79}-\tilde{q}^{105}+\tilde{q}^{107}+\tilde{q}^{137}+\ldots\right) \\
 \end{array}
\end{equation}

\begin{equation}
\footnotesize
\begin{array}{rcl}
 {\mathcal{S}}_{\left(\frac{29}{120}\right)}  & = & \tilde{q}^{-\frac{29}{120}}\,\left(2 \tilde{q}-\tilde{q}^2+\tilde{q}^3+\tilde{q}^5-\tilde{q}^6+\tilde{q}^7-3 \tilde{q}^8-\tilde{q}^9+\tilde{q}^{11}-\tilde{q}^{13}-2 \tilde{q}^{14}-\tilde{q}^{17}-\tilde{q}^{18}+\tilde{q}^{19}-\tilde{q}^{20}+\ldots\right) \\
 {\mathcal{S}}_{\left(\frac{71}{120}\right)}  & = & \frac{1}{2}\tilde{q}^{-\frac{71}{120}}\,\left(-3 \tilde{q}-3 \tilde{q}^2+\tilde{q}^3-2 \tilde{q}^4+\tilde{q}^5+2 \tilde{q}^7+4 \tilde{q}^8-2 \tilde{q}^9+2 \tilde{q}^{10}+2 \tilde{q}^{11}+3 \tilde{q}^{12}-2 \tilde{q}^{13}-2 \tilde{q}^{14}+\ldots\right) \\
 {\mathcal{S}}_{\left(\frac{101}{120}\right)}  & = & \tilde{q}^{-\frac{101}{120}}\,\left(\tilde{q}+2 \tilde{q}^3-\tilde{q}^4+\tilde{q}^5-\tilde{q}^6-\tilde{q}^7+\tilde{q}^8-2 \tilde{q}^{10}-2 \tilde{q}^{12}-\tilde{q}^{13}-\tilde{q}^{14}+\tilde{q}^{15}-\tilde{q}^{17}-\tilde{q}^{19}+\tilde{q}^{20}+\ldots\right) \\
 {\mathcal{S}}_{\left(\frac{119}{120}\right)}  & = & \frac{1}{2}\tilde{q}^{-\frac{119}{120}}\,\left(-\tilde{q}-\tilde{q}^2-3 \tilde{q}^4+4 \tilde{q}^5+2 \tilde{q}^6-4 \tilde{q}^7+\tilde{q}^8+\tilde{q}^9+2 \tilde{q}^{11}+2 \tilde{q}^{12}-2 \tilde{q}^{13}+4 \tilde{q}^{14}+\ldots\right) \\
 \end{array}
\end{equation}
where we have grouped $\mathcal{S}_\alpha$ according to the subests $\{1\},\{2\},\{1,2\}\subset H$ to associated to flat connections $\alpha$. We see that, up to a sign, $\mathcal{S}_\alpha(\tilde{q})=I^{\alpha_0}_\alpha(\tilde{q})$ for flat connections $\alpha$ with corresponding subsets $H'\subset H$ consisting of a single element (i.e. $|H'|=1$) and that there exists a certain flat connection $\alpha_\ast$ (with $\CS_{\alpha_\ast}=29/120\mod 1$) with $H'=H$, such that up to a sign and a power of $q$,
\begin{equation}
    \hat{Z}(q)\pto \mathcal{S}_{\alpha_\ast}(q).
    \label{Z-hat-S-relation}
\end{equation}

We claim that those are general features. It has been previously conjectured (and proved in certain cases) 
\cite{Cheng:2018vpl,bringmann2020higher,Cheng:2019uzc,Mori:2021ost,Cheng:2022rqr,Cheng:2023row} that $\hat{Z}(q)$ for plumbed manifolds is a (component of a vector-valued) holomorphic quantum modular form of depth $|H|$. We can interpret (\ref{Z-hat-to-thimbles-finite}) as the modular $S$-transformation of such a holomorphic quantum modular form, so that $\mathcal{S}_\alpha$ appearing in the right-hand side are holomorphic quantum modular forms of depth $|H'|$, where $H'\subset H$ is the subset to which the flat connection is associated.

\begin{figure}[ht]
	\centering
	\includegraphics[trim={0.4in 4.9in 4.7in 0.6in},clip,width=3.0in]{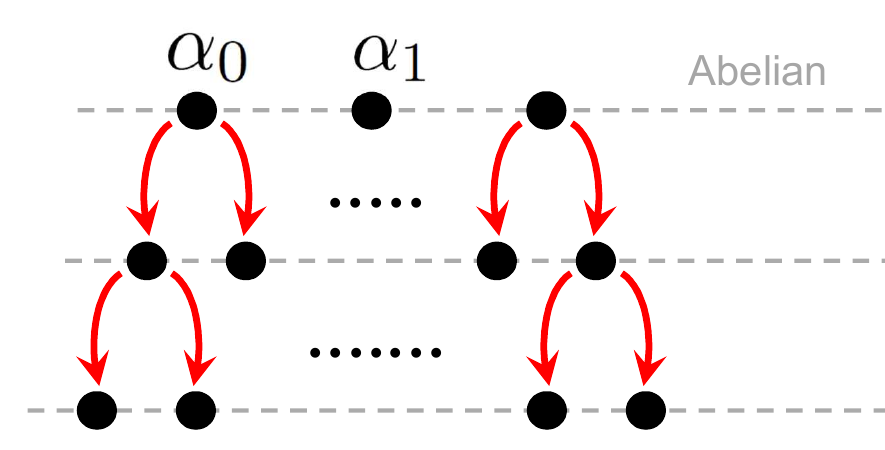}
	\caption{The cascading structure of flows. Abelian flat connections are at the top of the cascade. For a generic  plumbing (so that for each $0\leq|H'|\leq |H|$ there exists $\gamma^{(H',\mathbb{n})}$, such that the integral over it is non-zero) the number stages in the cascade is $|H|+1$.}
	\label{fig:cascade}
\end{figure}

As was previously explained, the structure of the Stokes phenomenon is such that the ``thimble'' contours $\gamma^{(H',\mathbb{n})}$ of the finite-dimensional model can only jump by the contours $\gamma^{(K,\mathbb{m})}$ such that $K\supset H'$ and $\mathbb{m}=\mathbb{n}\oplus \ldots \in \Z^{K}\supset \Z^{H'}$. Since the jumps occur due to flows between the flat connections, this gives the structure of flows schematically depicted in Figure \ref{fig:cascade}. One can identify the number of ``stages'' in the cascading structure of the flows with the depth of the quantum modular form plus one. When $|H|=1$, the plumbing manifold is Seifert-fibered. In this case the conditions $|H'|=1$ and $H'=H$ are equivalent so that there exists $\alpha$ such that simultaneously $\hat{Z}(q)\pto \mathcal{S}_{\alpha_\ast}(q)\pto I^{\alpha_0}_{\alpha_\ast}(q)$. This was already observed in \cite{Gukov:2016njj}.

One can also consider a version of $\hat{Z}(q)$ invariant, where instead of considering a symmetrized expansion as in (\ref{vertex-half-sum-expansion}) for $I\in H$, one uses the expansion in a particular ``chamber'', labeled by $\ch\in \{\pm 1\}^H$:
\begin{equation}
    \sum_{n\in\Z}F_{I,n}^{\ch} z^n  =
    \left.\,(z-1/z)^{2-\deg(I)}\right|_{\substack{\text{expansion at} \\ z^{\ch_I}\rightarrow 0}},\qquad I\in H.
    \label{vertex-chamber-expansion}
\end{equation}
One then proceeds as before, with 
\begin{equation}
2\sqrt{2k}\,Z^\ch(k)=\hat{Z}^\ch(q):=(-1)^{b_+}\,q^\frac{3\sigma-\Tr B}{4}\sum_{n\in \Z^V}\prod_{i\in V} F_{i,n_i}^\ch\,q^{-\frac{n^TB^{-1}n}{4}}.
\label{Z-hat-chamber-from-F}
\end{equation}
The original $\hat{Z}$ then can be recovered as 
\begin{equation}
    \hat{Z}(q)=\frac{1}{2^{|H|}}\sum_{\ch\in \{\pm 1\}^H}\hat{Z}^\ch(q).
\end{equation}
Note that one has identically $\hat{Z}^\ch= \hat{Z}^{-\ch}$, therefore generically there are $2^{|H|-1}$ linearly independent $q$-series $\hat{Z}^\ch(q)$. At least for $H$-shaped plumbings, the relation (\ref{WRT-from-limit}) to WRT invariant is satisfied with $Z(k)$ replaced with $Z^\ch(k)$, as follows from the analysis in \cite{Mori:2021ost}, and both $\hat{Z}^{\ch}(q)$ are depth two quantum modular forms \cite{bringmann2020higher}.

\begin{figure}
\centering
\begin{tikzpicture}
    \draw[->] (0,-4) -- (0,4) node[right=2] {$\re v_I$};
     \draw[->] (2,0) -- (-2,0) node[above=2] {$\im v_I$};
    \draw plot[only marks,mark=x,mark size=4pt,mark options={draw=red}] coordinates {(0,1) (0,2) (0,3) (0,-1) (0,-2) (0,-3)};

    \draw[blue,ultra thick, 
        decoration={markings, mark=at position 0.6 with {\arrow{>}}},
        postaction={decorate}
        ] (0.5,-3.5) -- (0.5,3.5) node[midway,above right=5] {$\gamma_I^\ch$};

\end{tikzpicture}
\qquad
\begin{tikzpicture}
    \draw[->] (0,-4) -- (0,4) node[right=2] {$\re v_I$};
     \draw[->] (2,0) -- (-2,0) node[above=2] {$\im v_I$};
    \draw plot[only marks,mark=x,mark size=4pt,mark options={draw=red}] coordinates {(0,1) (0,2) (0,3) (0,-1) (0,-2) (0,-3)};

     \draw[blue,ultra thick, 
        decoration={markings, mark=at position 0.7 with {\arrow{>}}},
        postaction={decorate}
        ] (-2,-3.5) -- (2,3.5);

     \draw[blue,ultra thick, 
        decoration={markings, mark=at position 0.7 with {\arrow{>}}},
        postaction={decorate}
        ] (0,-1) circle (0.4);

\draw[blue,ultra thick, 
        decoration={markings, mark=at position 0.7 with {\arrow{>}}},
        postaction={decorate}
        ] (0,-2) circle (0.4);

\draw[blue,ultra thick, 
        decoration={markings, mark=at position 0.7 with {\arrow{>}}},
        postaction={decorate}
        ] (0,-3) circle (0.4);
        
  \draw (1,0) arc (0:60:1) node[midway,right] {$\frac{1}{2}\mathrm{arg}\frac{i}{k}$};

\end{tikzpicture}
\caption{Left: the contour $\gamma_I^\ch$ in the $v_I$-plane corresponding to a chamber $\ch\in \{\pm 1\}^H$ with $\ch_I=+1$. Right: its deformation.}
\label{fig:contours-chamber}
\end{figure}
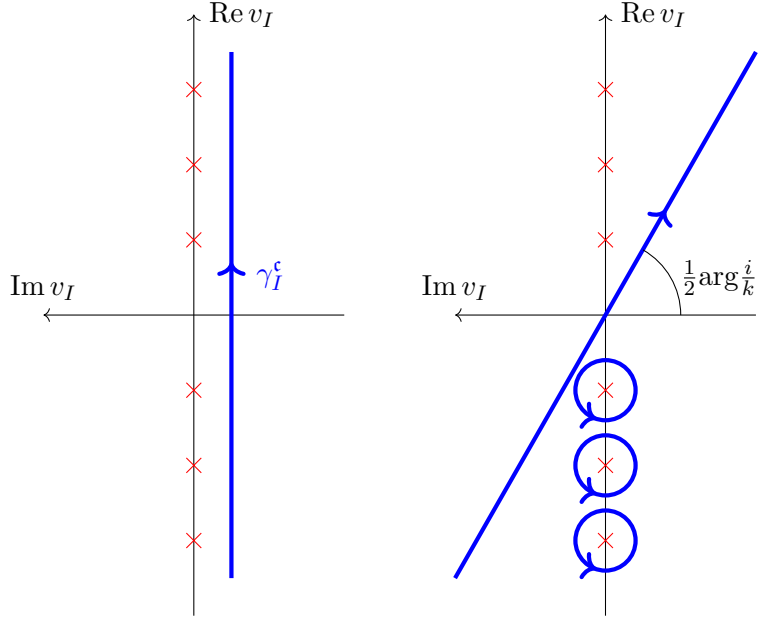

The sequence of equalities (\ref{Z-hat-to-LG-model}) can be repeated with the final expression being the same but with the contour of integration $\gamma^\ch=\bigtimes_{I\in H}\gamma_I^\ch$, where $\gamma_I^\ch =\R-i\ch_I\,\epsilon$. The factors can be similarly deformed, as shown in Figure \ref{fig:contours-chamber} for $\ch_I=+1$ case. Proceeding as before one then can obtain the analogue of (\ref{Z-hat-to-thimbles-finite}) for a particular chamber:
\begin{equation}
     \hat{Z}^\ch(q)\pto \sum_\alpha \mathcal{S}^\ch_\alpha(\tilde{q})\,\tilde{I}^{\alpha}(k).
    \label{Z-hat-chamber-to-thimbles-finite}
\end{equation}

For the same example, we have:
\begin{multline}
    \hat{Z}^{++}(q)\pto \mathcal{S}^{++}_{(\frac{29}{120})}(q)=\\
    q^{-\frac{29}{120}}\left(
4 q-2 q^2+2 q^3+2 q^5-2 q^6+2 q^7-6 q^8-2 q^9+q^{10}+2 q^{11}-2 q^{13}-4 q^{14}+\ldots
    \right),
\end{multline}
\begin{equation}
    \hat{Z}^{+-}(q)\pto \mathcal{S}^{+-}_{(\frac{29}{120})}(q)=
    q^{-\frac{29}{120}}\left(-q^{10}+q^{27}+q^{37}+q^{43}+q^{57}-q^{66}-q^{74}-q^{90}+\ldots
    \right).
\end{equation}

Finally, let us note that the coefficients $\hat{m}_\bbalpha$ (and their chamber analogues $\hat{m}_\bbalpha^\ch$) can be categorified similarly to the Stokes coefficients $m^\bbalpha_\bbbeta$ as done in Section \ref{sec:fd-categorification}. As for the Stokes coefficients, the first step is expressing $\hat{m}_\bbalpha$ via the intersection numbers (between $\gamma$ and the contours $\hat{\gamma}^{(H',\mathbb{n})}$ dual to the thimble contours). The categorification then will be provided by an appropriate version of Lagrangian Floer homology. Assuming the relation (\ref{Z-hat-S-relation}) this then provides a categorification for $\hat{Z}(q)$ itself.

\subsection{Behavior under cutting and gluing (surgery) operations}

For future generalizations of this work, it is important to understand how the invariants $I^{\alpha}_{\beta} (\tilde q)$ and $\mathcal{S}_\alpha(\tilde{q})$ behave under cutting and gluing (surgery) operations. A large class of plumbed 3-manifolds considered above provides a partial answer to this question, on which I can build generalizations to arbitrary 3-manifolds in the future work.

Since the physics motivation for this work involves a setup with a partial topological twist along $Y$, one can {\it a priori} expect surgery formulae for $I^{\alpha}_{\beta} (\tilde q)$ and $\mathcal{S}_\alpha(\tilde{q})$. However, such formulae may not have a simple form, as it happens e.g. in the case of Casson or Seiberg-Witten invariants of 3-manifolds. In particular, there are no {\it a priori} reasons to expect that either $I^{\alpha}_{\beta} (\tilde q)$ or $\mathcal{S}_\alpha(\tilde{q})$ come from a 3d TQFT in a sense of Atiyah. In fact, even the set of complex flat connections on $Y$ that labels these invariants behaves in a rather non-trivial way under cutting and gluing.

Certain plumbed 3-manifolds can be realized as knot surgeries. For a 3-manifold given by the $f$-surgery on a knot $K$, one can obtain the term $Z_0(k)$ in (\ref{Z-hat-from-F-torsion}) corresponding to the trivial flat connection by applying the linear ``Laplace transform'' to the two-variable series $F_K(x;q)$ considered in \cite{Gukov:2019mnk}: 
\begin{equation}
    Z_0(k)= F_K(x;q)(x-1/x)\Big|_{x^n\rightarrow q^{-\frac{n^2}{4f}}}.
\end{equation}
In fact, if there is a plumbing realization for an arbitrary $f$, this relation can be used to determine $F_K(x;q)$ uniquely.

\begin{figure}
\centering
\begin{tikzpicture}[scale=0.4]

\draw[ultra thick] (5,0)  -- (14,0);
\draw[ultra thick] (20,0)  -- (21,0);
    \draw[ultra thick] (5,0) -- (2,-3);
     \draw[ultra thick] (5,0) -- (2,3);
    \filldraw[black] (5,0) circle (8pt) node[left=8] {$[-1,0]$};
    \filldraw[black] (9,0) circle (8pt) node[above=5] {$[-3,0]$};
        \filldraw[black] (2,-3) circle (8pt) node[left=8] {$[-2,0]$};
    \filldraw[black] (2,3) circle (8pt) node[left=8] {$[f-2(2\ell+1),0]$};
\filldraw[black] (13,0) circle (8pt) node[above=5] {$[-2,0]$};

\draw (17,0) node {$\cdots$};

   \filldraw[black] (21,0) circle (8pt) node[above=5] {$[-2,0]$};

  \draw[thick, decorate,decoration={brace,amplitude=10pt,mirror}] (12.5,-0.5) -- (21.5,-0.5) node[midway,below=10] {$\ell-1 $ vertices};

\end{tikzpicture}
\caption{The plumbing that realizes $f$-surgery on $(2,2\ell+1)$ torus knot. In particular, for the trefoil ($\ell = 1$) we have no $[-2,0]$ vertices on the right leg.}
\label{fig:torus-knot-plumbing}
\end{figure}
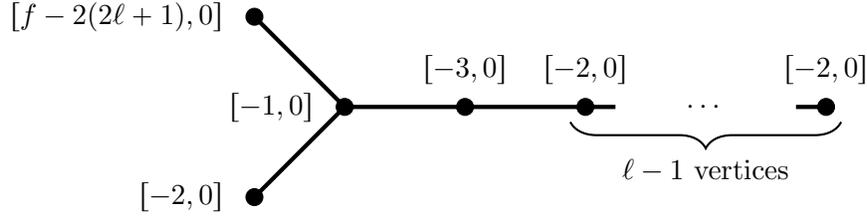

For example, the $f$-surgery on $(2,2\ell+1)$ torus knot $\forall f\in \Z$ is equivalent to the plumbed 3-manifold shown in Figure~\ref{fig:torus-knot-plumbing}. Consider the case $f=-1$ so that the 3-manifold is an integer homology sphere. Then we have $Z_0(k)=\hat{Z}(q)$. As was pointed out in Section \ref{sec:Z-hat} when $|H|=1$, we have 
\begin{equation}
  F_K(x;q)(x-1/x)\Big|_{x^n\rightarrow q^{\frac{n^2}{4}}} = \hat{Z}(q)\pto I^{\alpha_0}_{\alpha_\ast}(q)
\end{equation}
for a particular complex flat connection $\alpha_\ast$ with $|H'|=1$. We observe that in this family of examples, $I^{\alpha_0}_{\beta}(\tilde q)$ for other $\beta = (\mu_1, \mu_2, \mu_3)$ with $|H'|=1$ can also be expressed via $F^{(\mu_1,\mu_2)}_K(x; \tilde q)$ (up to a finite number of terms):
\begin{equation}
   I^{\alpha_0}_{\beta}(\tilde q)
\pto
   F^{(\mu_1,\mu_2)}_K(x;\tilde q)(x^{2\mu_3}-x^{-2\mu_3})\Big|_{x^n\rightarrow \tilde q^{\frac{n^2}{4}}}
  \label{FK-I-relation}
\end{equation}
for some $\mu_3 \in \Z$. Note that, although the set of values of $\mu_3$ is infinite, they produce only a finite collection of $\tilde q$-series (\ref{FK-I-relation}) up to an overall power of $\tilde q$. In general, for a start-shaped plumbing graph with $n$ legs (called ``leaves'' in the splicing terminology \cite{Gukov:2023srx}) the meaningful values of $\mu_i$, $i=1,\ldots,n$ that produce independent $\tilde q$-series are controlled by the denominators of the continued fraction expansions associated with legs (leaves) of the graph. In the present class of surgeries on $(2,2\ell+1)$ torus knots, we have $\mu_1 = 1$ and $\mu_2 = 1, \ldots, \ell$.

For the trefoil knot, $K = \kbf{3_1}$, there is only one knot complement invariant $F_K^{(1,1)} (x, \tilde q)$ obtained by \emph{not} integrating the vertex\footnote{In \cite{Gukov:2019mnk} such ``empty vertex'' was denoted by $\circ$.} with framing $f-2(2\ell+1)$ in Figure~\ref{fig:torus-knot-plumbing}:
\begin{equation}
F_{{{\kbf{3_1}}}}^{(1,1)} (x, \tilde q) =
\ldots + x - \tilde q x^5 - \tilde q^2 x^7 + \tilde q^5 x^{11} + \tilde q^7 x^{13} - \tilde q^{12} x^{17} - \tilde q^{15} x^{19} + \ldots
\label{F31mu1}
\end{equation}
Although we only display terms with positive powers of $x$ one can consider the \emph{symmetric completion} of this series as in \cite{Gukov:2019mnk}; this is what one set of ellipses on the right-hand side of \eqref{F31mu1} indicate. In the context of $\hat Z$ invariants, such symmetric completion is natural since it restores the gauge, i.e. the invariance under $\Z_2$ Weyl symmetry of $SL(2,\C)$. In the present context, this is less critical and one can also work with the asymmetric version of \eqref{F31mu1} which contains only positive powers of $x$. Both versions, symmetric and asymmetric, produce the same result for surgeries via the surgery formula \eqref{FK-I-relation}.

For the next case, $\ell = 2$, i.e. for the $(2,5)$ torus knot $K = \kbf{5_1}$, there are two choices of flat connections (Wilson lines) associated with a longer leg of the plumbing graph on Figure~\ref{fig:torus-knot-plumbing}. Up to an overall power of $\tilde q$, the corresponding $\tilde q$-series look like
\begin{equation}
F_{{\kbf{5_1}}}^{(1,1)} (x, \tilde q) =
\ldots + 
\tilde q x^3 - \tilde q^2 x^7 - \tilde q^5 x^{13} + \tilde q^8 x^{17} + \tilde q^{14} x^{23} - \tilde q^{19} x^{27}
- \tilde q^{28} x^{33}
+ \ldots
\label{F51mu1}
\end{equation}
and
\begin{equation}
F_{{ \kbf{5_1}}}^{(1,2)} (x, \tilde q) =
\ldots + 
\tilde q^2 x - \tilde q^4 x^9 - \tilde q^5 x^{11} + \tilde q^{11} x^{19} + \tilde q^{13} x^{21} - \tilde q^{23} x^{29} - \tilde q^{26} x^{31}
+ \ldots
\label{F51mu2}
\end{equation}
In line with the earlier remarks, both \eqref{F31mu1} and \eqref{F51mu1} coincide with $\hat Z (S^3 \setminus K)$ for the these knots, providing further support to the observation that the complex flat connection $\alpha_\ast$ with $(\mu_1 , \mu_2)= (1,1)$ plays a special tole. It would be interesting to study this further and demystify this chain of coincidences.

On the other hand, the series \eqref{F51mu2} is new in a sense that it does not coincide with any of the $\hat Z$ invariants associated with the knot $\kbf{5_1}$. To see this, and also as a step toward the interpretation of \eqref{F51mu2}, we can consider the ``classical limit'' $q \to 1$. In agreement with \cite{Gukov:2019mnk}, both \eqref{F31mu1} and \eqref{F51mu1} in the classical limit approach the inverse Alexander polynomial or, more precisely, $\frac{x-x^{-1}}{\Delta_K (x^2)}$, where the Alexander polynomial for a $(m,n)$ torus knot is given by the well known expression:
\begin{equation}
\Delta_{T_{m,n}} (x) = \frac{(x^{mn} - 1) (x-1)}{(x^m - 1) (x^n - 1)}.
\end{equation}
And for \eqref{F51mu2} the limit has a similar form, in particular also has the Alexander polynomial in the denominator,
\begin{equation}
\lim_{\tilde q \to 1} \, F_{{ \kbf{5_1}}}^{(1,2)} (x, \tilde q) \; = \; \frac{x^{-3} - x^{-1} + x - x^3}{\Delta_{{ \kbf{5_1}}} (x^2)}
\label{Alexlimit51}
\end{equation}
but the numerator is different: $x-x^{-1}$ is replaced by $x^{-3} - x^{-1} + x - x^3 = - \frac{x^{4} - x^{-4}}{x + x^{-1}}$. Continuing to larger values of $\ell$ we see that the pattern persists. Namely, for $(2,2\ell + 1)$ torus knots we have
\begin{equation}
\lim_{\tilde q \to 1} \, F_{T_{2,2\ell + 1}}^{(1,\mu_2)} (x, \tilde q) \; = \; \frac{x^{- 2 \mu_2} - x^{2 \mu_2}}{(x + x^{-1}) \; \Delta_{T_{2,2 \ell + 1}} (x^2)}.
\label{Alexlimit}
\end{equation}
For example, it is easy to verify that it holds true for the next example in this family, the case of $\ell = 3$:
\begin{equation}
F_{{ \kbf{7_1}}}^{(1,3)} (x, \tilde q) =
\ldots + 
\tilde q^4 x - \tilde q^7 x^{13} - \tilde q^8 x^{15} + \tilde q^{17} x^{27} + \tilde q^{19} x^{29} - \tilde q^{34} x^{41} - \tilde q^{37} x^{43} 
+ \ldots
\label{F71mu3}
\end{equation}
This, in turn, provides further support to the identification between complex flat connections on $Y$ and line operators in 3d theory $T[Y]$. Both correspond to topological boundary conditions in $T[Y]$ on $S^1 \times \Sigma$ and, therefore, are objects of $\text{MTC} [Y]$. In the class of examples considered here, they correspond to inserting $x$-dependent factors
\begin{equation}
x^{2\mu} - x^{- 2\mu} \quad \leftrightarrow \quad
\text{line operators in } T[Y].
\end{equation}
Depending on the conventions, this factor sometimes can be written as $x^{\mu} - x^{- \mu}$; with this choice of conventions half-integer powers of $x$ appear in $F_K (x;q)$ and in the integral formulae for plumbed manifolds, while some aspects can be more natural, e.g. the denominators in \eqref{Alexlimit51}--\eqref{Alexlimit} would involve $\Delta_K (x)$ rather than $\Delta_K (x^2)$.

An alternative way to produce invariants for knot complements is via the regularity in $\frac{1}{r}$ for the invarants of $-\frac{1}{r}$ (``small'') surgeries. As will be discussed in more detail further below, \emph{some} information about complex flat connections on $Y = S^3_{-1/r} (K)$ can be conveniently read off from the A-polynomial of the knot $K$. For the example, for the trefoil knot, $K = { \kbf{3_1}}$, the A-polynomial looks like
\begin{equation}
A(x,y) \; = \; (y-1) (y + x^{-6}).
\label{Atrefoil}
\end{equation}
Complex flat connections on a $-\frac{1}{r}$ surgery correspond to the intersection points with $y = x^{1/r}$, modulo the quotient by the Weyl symmetry $(x,y) \sim (x^{-1}, y^{-1})$ and discarding $(x,y) = (-1,-1)$. In our example, from the abelian branch of the A-polynomial, $y-1=0$, we quickly see that there is only one solution: the trivial flat connection with $(x,y)=(1,1)$. But from the non-abelain branch we have
\begin{equation}
x^{-6-\frac{1}{r}} = -1.
\end{equation}
Orbits of the Weyl symmetry can be characterized e.g. by solutions with $\text{Im} (y) < 0$. We quickly find a total of $3r$ irreducible complex flat connections:
\begin{equation}
(x,y) = ((-1)^r \omega^{nr}, -\omega^n) \,, \quad \omega = e^{\frac{2\pi i}{6r+1}},
\end{equation}
with $n = 1, \ldots 3r$.
Using a plumbing description of $Y = S^3_{-1/r} ({ \kbf{3_1}})$ we obtain
\begin{equation}
I^{\alpha_0}_{\alpha_\ast} (S^3_{-1/r} ({ \kbf{3_1}})) = \Phi_{36r+6}^{(6r-5)} (\tilde q) - \Phi_{36r+6}^{(6r+7)} (\tilde q)
\label{muone}
\end{equation}
where
\begin{gather}
\Phi^{(a)}_p (\tilde q) \;  :=  \; \sum_{n=0}^\infty \phi^{(a)}_{p}(n) \tilde q^{\frac{n^2}{4p}} \qquad \in \tilde q^\frac{a^2}{4p}\,\Z[[\tilde q]],
\label{falsetheta} \\
\phi^{(a)}_{p}(n)  =  \left\{
\begin{array}{cl}
(-1)^n, & n\equiv \pm a~\mod~ p\,, \\
0, & \text{otherwise}.
\end{array}\right. \nonumber
\end{gather}
It is a non-trivial fact that \eqref{muone} can be expressed as
\begin{equation}
I^{\alpha_0}_{\alpha_\ast} (S^3_{-1/r} ({ \kbf{3_1}})) =
\CL_{1/r} \Big[ \big( x^{\frac{1}{r}} - x^{-\frac{1}{r}} \big) F^{(1,1)}_{{ \kbf{3_1}}} (x, \tilde q) \Big]
\end{equation}
with the above $F^{(1,1)}_{{ \kbf{3_1}}} (x,\tilde q)$, thus providing yet another way to determine $F^{(1,1)}_{{ \kbf{3_1}}} (x,\tilde q)$. In other words, it says that the behavior in $1/r$ exhibits regularity that can be encoded in the $x$-dependence of this two-variable series.

To summarize, even though we do not expect $I^{\alpha}_{\beta}$ to form a TQFT --- after all, there isn't even a set of TQFT rules for the set of labels $\alpha$ and $\beta$ --- we find a lot of structure and regularity with respect to cutting and gluing at least for some $\alpha$ and $\beta$. Therefore, a natural problem for future work is to generalize these surgery formulae to more general classes of complex flat connections and to other types of knots and links.

\section{Flat connections at infinity}
\label{sec:flat-at-infinity}

We devote this section to a peculiar phenomenon that, to the best of our knowledge, has not been observed in the literature on Chern-Simons theory. The phenomenon can only arise in complex Chern-Simons theory, where moduli spaces are non-compact and have infinite asymptotic ends. What can happen is that flat connections at such infinite distance regions may still have finite action and, more importantly, show up on the Borel plane\footnote{A simple toy model for this phenomenon can be provided by a finite-dimensional integral $\int dxdydz\,e^{ikS(x,y,z)}$ with $S(x,y,z)=y^2/2-xyz+z-z^2/2$. The function $S$ (if considered as $\C^3\rightarrow \C$) has a unique critical point at $x=y=0,\,z=1$ with the critical value $1/2$. However, the Borel transform of the perturbative expansion around it has also a singularity corresponding to the critical value $0$. Its origin is the critical point at infinity with $z=0,y=\epsilon,x=1/\epsilon$, $S=\epsilon^2/2$, in the limit $\epsilon\rightarrow 0$. }.

We should note from the outset that there can be various formulations of the problem where such ``flat connections at infinity'' do not contribute and do not play a role. For example, they are not visible to the sheaf-theoretic model for $SL(2,\C)$ Floer homology introduced by Abouzaid and Manolescu \cite{MR4167016}. In our computations of the Stokes coefficients, though, such flat connections at infinity do show up and do play a role.

One implication of the discussion below is that it provides a counterexample to the folklore belief --- sometimes stated as a conjecture --- that the only singularities on the Borel plane in Chern-Simons theory correspond to complex flat connections.\\

\noindent
\textbf{Definition:} We call a complex flat connection $A_{\infty}$ a {\it flat connection at infinity} if there exists a family, $\{ A_{1/\epsilon} \}$, of $G_{\C}$-valued connections on $Y$ parametrized by $\epsilon$, such that the following two conditions are satisfied:
\begin{itemize}

\item all components of the curvature of $A_{1/\epsilon}$ approach zero as $\epsilon \to 0$, while this limit is singular for $A_{1/\epsilon}$ itself, and

\item the values $\text{CS} (A_{1/\epsilon})$ have a well-defined limit, denoted $\text{CS} (A_{\infty})$, as $\epsilon \to 0$.

\end{itemize}
Here, $\frac{1}{\epsilon}$ can go to $\infty$ either continuously or via an infinite discrete set of values. In other words, in this definition we allow the infinite family $\{ A_{1/\epsilon} \}$ to be either discrete or continuous.

Since, unlike its curvature, the gauge connection is not a gauge-invariant object, this requires some care. In particular, the presence of $\frac{1}{\epsilon}$ terms in the gauge connection might not mean that it is necessarily singular if one could remove these terms by a gauge transformation. However, as $\epsilon \to 0$ such gauge transformations would require larger and larger elements of the complex gauge group, which in the limit would mean that all components of $A_{\infty}$ can be made finite by an infinite gauge transformation. In practice, we will see that $G_{\C}$-valued holonomies of $A_{1/\epsilon}$ contain unipotent elements proportional to $\epsilon^{-1}$. For example, when $Y = S^3_p (K)$ is presented as a surgery on a knot $K$, one can consider a family of connections, such that the holonomy of the meridian is unipotent ($\epsilon \ne 0$):
\begin{equation}
\x \; = \;
\begin{pmatrix}
1 & \epsilon \\
0 & 1 
\end{pmatrix}.
\label{unipotentx}
\end{equation}
Below, we use this ansatz to search for flat connections at infinity for various knots and for $p=0,$ $\pm 1$. Note, \eqref{unipotentx} implies that the diagonal part of the meridian holonomy is $x=1$. And, from the surgery relation $y = x^p$ we learn that the diagonal part of the longitude holonomy is also $y=1$ for $p = \pm 1$. Therefore, a simple necessary criterion for $Y = S^3_{\pm 1} (K)$ to admit a flat connection at infinity of the form \eqref{unipotentx} is that the non-abelian part of the A-polynomial of the knot $K$ satisfies
\begin{equation}
A^{\text{irred}}_K (x,y) \big\vert_{x=1, \, y=1} \; = \; 0.
\label{Apoltest}
\end{equation}
By inspection, we find that the condition \eqref{Apoltest} is rarely satisfied among low-crossing prime knots, but is fairly common for composite knots or prime knots with larger crossing numbers.

While flat connections at infinity are a very interesting phenomenon, we have no other method of identifying them besides a direct search in each case. Just like the Borel plane can serve as a useful indicator of flat connections at infinity, it would be helpful to develop new variants of the $SL(2,\C)$ Casson invariant that count flat connections at infinity as well,\footnote{See e.g. \cite{Cu01,BC08,Gukov:2007ck,MR4167016} for prior work, though none of these variants accounts for flat connections at infinity.} something akin to the expression for the ordinary Casson invariant $\lambda (Y)$ in terms of the plumbing data:
\begin{equation}
- \frac{24}{|H|} \lambda (Y) \; = \; \sum_j e_j + 3s + \sum_j (2 - \delta_j) (B^{-1})_{jj}.
\end{equation}
Note, only when the group is complex, this subtle phenomenon is possible. It is also directly related to the inherent non-compactness of moduli spaces for surfaces and 3-manifolds with $b_1 (Y) > 0$.

\subsection{Surgeries on knots}

To describe complex flat connections, including flat connections at infinity, we need a fairly explicit description of $\pi_1 (Y)$. And, if $Y$ is a result of a surgery on a knot (or link) $K$, it means that we need the fundamental group of the knot complement as a starting point. The Wirtinger presentation of $\pi_1 (S^3 \setminus K)$ associates generators to arcs in the planar diagram of $K$ and relations to crossings. For a general knot or link $K$, this quickly becomes intractable as the crossing number of $K$ grows. However, for particular families of knots and links $\pi_1 (S^3 \setminus K)$ may still have a small number of generators and a small number of relations; as the complexity of $K$ grows within such a family, it affects the length of relations, but not their number.

This is precisely what happens in families of twist knots and torus knots from which we build some of the examples below.

\subsubsection*{The granny knot}

Based on the observation summarized below \eqref{Apoltest} it is natural to start with composite knots.
In general, under the connected sum operation,
\begin{equation}
K \; = \; K_1 \# K_2
\label{KKK}
\end{equation}
the meridians of $K_1$ and $K_2$ are identified, whereas the longitude of $K$ is a product of the longitudes of $K_1$ and $K_2$. As a result, it follows that the A-polynomial of $K$ is divisible by $A(K_1) A(K_2) / (x-1) = (x-1) A^{\text{irred}}_{K_1} A^{\text{irred}}_{K_2}$. In particular, $A^{\text{irred}}_K$ is divisible by $A^{\text{irred}}_{K_1} A^{\text{irred}}_{K_2}$.

The simplest examples of such knots are connected sums of left- and right-handed trefoil, namely the square (reef) knot ${ \kbf{3_1^r}} \# \kbf{3_1^{\ell}}$ and the granny knot $\kbf{3_1^r} \# \kbf{3_1^r}$. We have $A^{\text{irred}}_{\kbf{3_1^r}} = y + x^{-6}$ and $A^{\text{irred}}_{\kbf{3_1^{\ell}}} = y + x^6$, cf. \eqref{Atrefoil}. One can easily see that the suitable products of these factors divide the A-polynomial of the square knot
\begin{equation}
A^{\text{irred}}_{\kbf{3_1^r} \# \kbf{3_1^{\ell}}} \; = \; (y-1) \, (y + x^6) \, (y + x^{-6})
\label{Asquareknot}
\end{equation}
and the A-polynomial of the granny knot
\begin{equation}
A^{\text{irred}}_{\kbf{3_1^r} \# \kbf{3_1^r}} \; = \; (y + x^{-6})^2 \, (y - x^{-12}).
\label{Agrannyknot}
\end{equation}
We see that both meet the condition \eqref{Apoltest}. In fact, one can understand in detail how this happens. Both $\kbf{3_1^r}$ and $\kbf{3_1^{\ell}}$ have polynomial factors $A^{\text{irred}} (x,y)$ that vanish at $(x,y) = (1,-1)$,
\begin{equation}
A^{\text{irred}} (x,y) \big\vert_{x=1, \, y=-1} \; = \; 0.
\label{Aminustest}
\end{equation}
Since the longitudes are multiplied under the operation of connected sum, whereas meridians are identified, the condition \eqref{Aminustest} at the level of constituent knot is sufficient for the condition \eqref{Apoltest} at the level of the connected sum. This suggests that, in the case of composite knots \eqref{KKK}, one strategy for searching for flat connections at infinity could be based on finding flat connections at infinity for the constituents, with the meridian and longitude holonomies of the form
\begin{equation}
{\x} =
\begin{pmatrix}
1 & \epsilon \\
0 & 1 
\end{pmatrix}
\qquad , \qquad
\y =
\begin{pmatrix}
\pm 1 & * \\
0 & \pm 1 
\end{pmatrix}
\label{xyansatz}
\end{equation}
where the choice of sign has to be the same for both $K_1$ and $K_2$.

As we already mentioned earlier, surgeries on torus knots and on their connected sums can be represented by plumbing graphs. For example, small surgeries on the granny knot, $S^3_{- 1/r} (\kbf{3_1} \# \kbf{3_1})$, have the plumbing graph
\begin{equation}
    \begin{tikzpicture}[scale=0.4]
\draw[ultra thick] (5,0)  -- (9,0);
\draw[ultra thick] (9,0)  -- (9,3);
  \draw[ultra thick] (9,0) -- (13,0);
  \draw[ultra thick] (13,0) -- (16,3);
   \draw[ultra thick] (13,0) -- (16,-3);
    \draw[ultra thick] (5,0) -- (2,-3);
     \draw[ultra thick] (5,0) -- (2,3);
    \filldraw[black] (5,0) circle (8pt) node[left=5] {$[-1,0]$};
      \filldraw[black] (9,3) circle (8pt) node[above=5] {$[-r,0]$};
    \filldraw[black] (2,3) circle (8pt) node[left=4] {$[-3,0]$};
    \filldraw[black] (2,-3) circle (8pt) node[left=4] {$[-2,0]$};
    \filldraw[black] (9,0) circle (8pt) node[below=3] {$[-12,0]$};
    \filldraw[black] (13,0) circle (8pt) node[right=5] {$[-1,0]$};
    \filldraw[black] (16,3) circle (8pt) node[right=4] {$[-3,0]$};
    \filldraw[black] (16,-3) circle (8pt) node[right=4] {$[-2,0]$};
\end{tikzpicture}.
\label{twotrefsr}
\end{equation}

In particular, $S^3_{-1} (\kbf{3_1} \# \kbf{3_1})$ is equivalent to the following plumbed manifold
\begin{equation}
    \begin{tikzpicture}[scale=0.4]
\draw[ultra thick] (5,0)  -- (9,0);
  \draw[ultra thick] (9,0) -- (13,0);
  \draw[ultra thick] (13,0) -- (16,3);
   \draw[ultra thick] (13,0) -- (16,-3);
    \draw[ultra thick] (5,0) -- (2,-3);
     \draw[ultra thick] (5,0) -- (2,3);
    \filldraw[black] (5,0) circle (8pt) node[left=5] {$[-1,0]$};
    \filldraw[black] (2,3) circle (8pt) node[left=4] {$[-3,0]$};
    \filldraw[black] (2,-3) circle (8pt) node[left=4] {$[-2,0]$};
    \filldraw[black] (9,0) circle (8pt) node[below=3] {$[-11,0]$};
    \filldraw[black] (13,0) circle (8pt) node[right=5] {$[-1,0]$};
    \filldraw[black] (16,3) circle (8pt) node[right=4] {$[-3,0]$};
    \filldraw[black] (16,-3) circle (8pt) node[right=4] {$[-2,0]$};
\end{tikzpicture}.
\label{twotrefs}
\end{equation}
The Borel plane for this example shows  singularities at Chern-Simons values $\xi_*=-1/12\mod 1$ that do not correspond to a flat connection in the traditional sense, with finite values of its holonomies. Rather, it can be obtained as a limit of non-flat complex connections on $S^3_{-1} (\kbf{3_1} \# \kbf{3_1})$ with $-1$ diagonal values of the trefoil longitude in \eqref{xyansatz}. In the limit $\epsilon \to 0$ the connection becomes flat, but since some elements of the holonomies go to infinity, as $\frac{1}{\epsilon}$, we call these flat connections at infinity.

In the case of $S^3_{-1} (\kbf{3_1} \# \kbf{3_1})$ there is only one such flat connection at infinity. However, this example can be generalized in a number of different ways. For instance, one can consider surgeries on \eqref{KKK} where $K_i$, $i=1,2$, are torus knots or twist knots. The latter case will be considered in more detail later, while here we comment on the splicing of torus knots. As was noted earlier, they can be represented by plumbing graphs with trivalent vertices. According to the result of \cite{MR1002161}, a surgery on a composite knot $K = T_{m_1,n_1} \# T_{m_2,n_2}$ gives a Seifert manifold if and only if the surgery is integral, with the coefficient
$$
p = m_1 n_1 + m_2 n_2.
$$
Moreover, in this case the Seifert manifold has four singular fibers of order $m_1$, $n_1$, $m_2$, and $n_2$.
This agrees with (and can be understood via) the above splicing of the torus knots. For example, a generalization of 1-surgery to a connected sum of arbitrary torus knots has vertex labeled by $p - (m_1 n_1 + m_2 n_2) = 1 - (m_1 n_1 + m_2 n_2)$ in the middle. When this number vanishes, by the standard Kirby-Neumann move, we can pull the two trivalent vertices of the H-shaped graph together, to form a valency-4 vertex. (See the lower panel of Figure 10 in \cite{Gadde:2013sca}.) Under this move, the framing coefficients $a_1$ and $a_2$ of the two trivalent vertices add up to $a_1 + a_2$, which becomes the label of the new valency-4 vertex.

\subsubsection*{Connected sums of twist knots}

The trefoil knot $\kbf{3_1}$ is the first non-trivial knot in an infinite family of twist knots $K_n$. (Namely, $\kbf{3_1} = K_1$.) What makes this family particularly interesting for our analysis here is that the knot group has a very simple presentation.

From the Wirtinger presentation one finds that the knot group $\pi_1 (S^3 \setminus K_n)$ is generated by two elements, $a$ and $b$, which satisfy
\begin{equation}
(b a^{-1} b^{-1} a)^n a \; = \; b (b a^{-1} b^{-1} a)^n
\end{equation}
for $K_n$, and
\begin{equation}
(a b^{-1} a^{-1} b)^n a \; = \; b (a b^{-1} a^{-1} b)^n
\end{equation}
for $\bar K_{-n}$. In either case, $\x=a$ is the meridian and the longitude in the former case is $\y = (a b^{-1} a^{-1} b)^n (b a^{-1} b^{-1} a)^n$ and in the latter case is $\y = (b a^{-1} b^{-1} a)^n (a b^{-1} a^{-1} b)^n$. We look for solutions with
\begin{equation}
\x =
\begin{pmatrix}
1 & \epsilon \\
0 & 1 
\end{pmatrix}
\label{xtwist}
\end{equation}
such that the longitude holonomy $\y$ contains terms linear in $\epsilon^{-1}$. For all $n$, i.e. for the entire family of twist knots, the solutions are of the form
\begin{equation}
b =
\begin{pmatrix}
* & * \\
\frac{C}{\epsilon} & * 
\end{pmatrix}
\qquad , \qquad
\y =
\begin{pmatrix}
-1 & * \\
0 & -1 
\end{pmatrix}
\label{ytwist}
\end{equation}
where ``$*$'' represent expressions regular in the limit $\epsilon \to 0$. This form of the ansatz is motivated by looking at the A-polynomials of twist knots. Recall, that depending on which of the conditions \eqref{Apoltest} or \eqref{Aminustest} is satisfied for a given knot, it makes sense to search for solutions \eqref{xyansatz} with diagonal values of the longitude equal to $+1$ or $-1$, respectively. (When both conditions are satisfied, there can be solutions of both types; we will see such an example below.)

\begin{figure}[ht]
	\centering
	\includegraphics[trim={0.1in 0.1in 0.1in 0.1in},clip,width=3.0in]{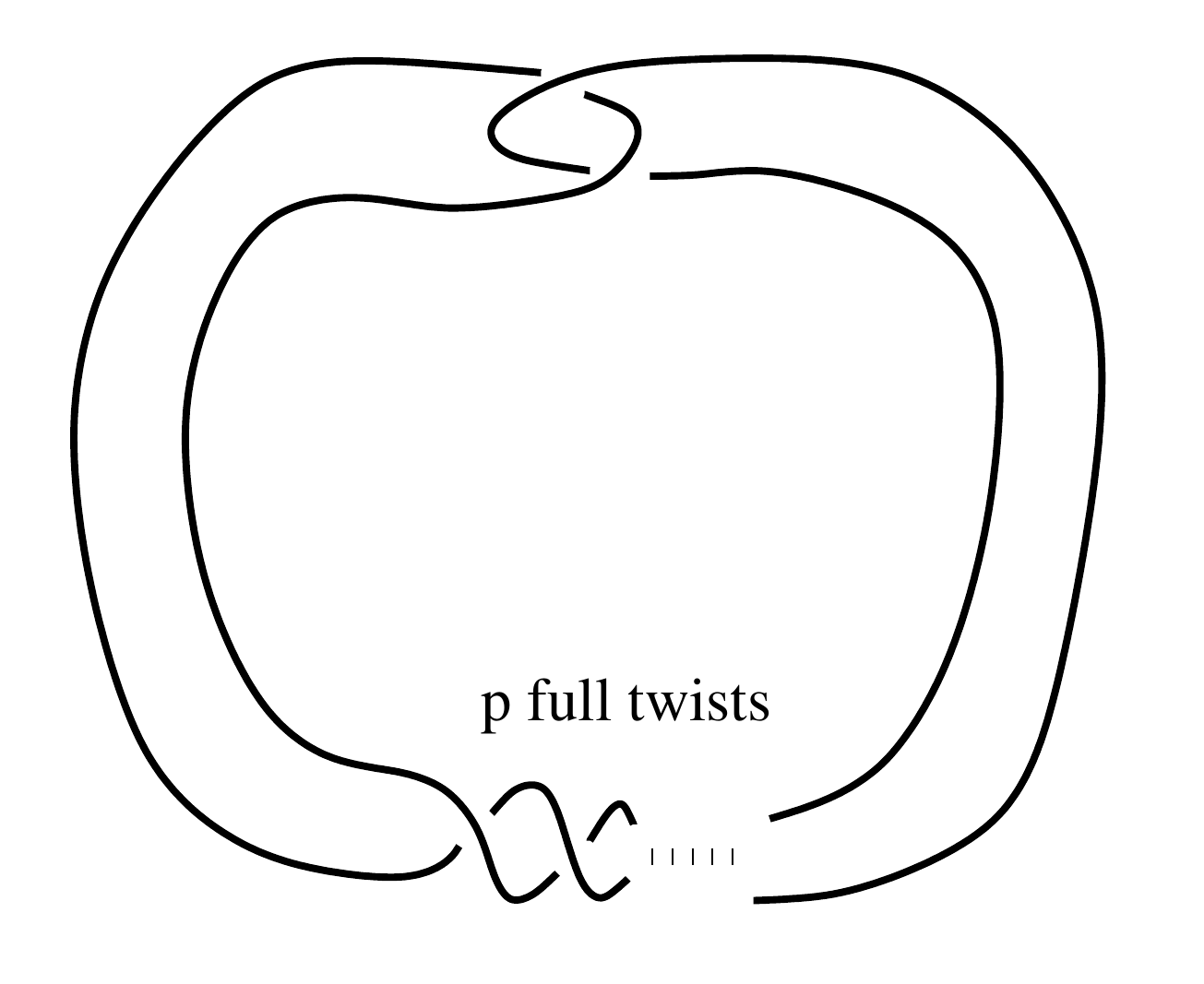}
	\caption{Twist knots $K_p$.}
	\label{fig:twistknots}
\end{figure}

In particular, in the above we already analyzed the trefoil knot and observed that its A-polynomial satisfies \eqref{Aminustest} but not \eqref{Apoltest}. The same is true for other examples of twist knots, e.g. for the figure-8 knot $\kbf{4_1} = K_{-1}$ and for the tweeny knot $\kbf{5_2} = K_2$. These knots have A-polynomials
\begin{equation}
A_{\kbf{4_1}}^{\text{irred}}(x, y) \; = \; -x^4 + (1 - x^2 - 2 x^4 - x^6 + x^8) y - x^4 y^2
\end{equation}
and
\begin{equation}
A_{\kbf{5_2}}^{\text{irred}}(x, y) = x^{14} + (x^4 - x^6 + 2 x^{10} + 2 x^{12} - x^{14}) y + (-1 + 2 x^2 + 2 x^4 - x^8 + x^{10}) y^2 + y^3
\end{equation}
respectively. And, it is easy to verify that they satisfy the condition \eqref{Aminustest} but not \eqref{Apoltest}. Therefore, to search for flat connections at infinity, we use the ansatz \eqref{xyansatz} with longitude diagonal elements $-1$. Indeed, we find that there are multiple such solutions, with the total number of solutions growing linearly with $|n|$ for twist knots $K_n$. Specifically, for the figure-8 knot $K_{-1} = {\kbf{4_1}}$ we find two Galois conjugate solutions of the form \eqref{xtwist}--\eqref{ytwist}. For the $K_2 = \kbf{5_2}$ knot, there are three such solutions with values of the constant $C$ in \eqref{ytwist} given by roots of the cubic polynomial
\begin{equation}
1 + 2C + C^2 + C^3 \; = \; 0.
\end{equation}
And for the twist knot $K_{-2} = 6_1$ we find a total of 4 solutions of the form \eqref{xtwist}--\eqref{ytwist}. Therefore, by taking a connected sum of two twist knots, $K_m \# K_n$, and performing $\pm 1$ surgery,
\begin{equation}
S^3_{\pm 1} \left( K_m \# K_n \right)
\label{twistmanifolds}
\end{equation}
we can construct a large infinite family of 3-manifolds that have many flat connections at infinity. Specifically, for a twist knot $K_n$ the number of solutions described here is expected to be $2n-1$ (resp. $2|n|$) for positive (resp. negative) values of $n$, and the total number of flat connections at infinity is a product of two such numbers, one for each knot in \eqref{twistmanifolds}.

\subsubsection*{Surgery on the pretzel knot $P(-2,3,7)$}

In all of the above examples we found flat connections at infinity on 3-manifolds produced via $\pm 1$ surgeries on composite knots. This invites an obvious question:\\

\noindent
\textbf{Question:} Is there a prime knot $K$, such that some surgery on $K$ admits flat connections at infinity?

\medskip
Below we answer this question in the affirmative by explicitly constructing flat connections at infinity on a surgery on the pretzel knot $P(-2,3,7) = \kbf{12n242}$. The irreducible component of the A-polynomial for this knot has the form
\begin{multline}
A_{\kbf{12n242}}^{\text{irred}} (x,y) = 1-x^{16} y + 2 x^{18} y - x^{20} y - 2 x^{36} y^2 - x^{38} y^2 + x^{72} y^4 \\
+ 2 x^{74} y^4 + x^{90} y^5 - 2 x^{92} y^5 + x^{94} y^5 - x^{110} y^6.
\end{multline}
It is easy to verify that it satisfies \emph{both} conditions \eqref{Apoltest} and \eqref{Aminustest}. This means that we should expect solutions of the form
\begin{equation}
\x =
\begin{pmatrix}
1 & \epsilon \\
0 & 1 
\end{pmatrix}
\qquad , \qquad
\y =
\begin{pmatrix}
\pm 1 & * \\
0 & \pm 1 
\end{pmatrix}
\label{pretzelxy}
\end{equation}
with both $+1$ and $-1$ diagonal values of the longitude holonomy. In order to construct the explicit solutions we need to write down the knot group.

Although the knot $P(-2,3,7) = \kbf{12n242}$ is pretty complicated in terms of the crossing number, its fundamental group has the same structure as that of twist knots: it has two generators and only one relation. Denoting the generators by $a$ and $b$, the relation is
\begin{equation}
a^2 b^{-1} a^2 = b^2 a b^2.
\end{equation}
In these notations, the meridian and the longitude are
\begin{align}
\x = a^{-1} b^2, \\
\y = a b^{-1} a^2 b^{-1} a^2.
\end{align}
As before, we look for solutions to the knot group relation where $\x$ and $\y$ are both upper triangular \eqref{pretzelxy} and some of the matrix elements of $a$ or $b$ involve $\epsilon^{-1}$ and, therefore, diverge in the limit $\epsilon \to 0$. The solutions that have $-1$ on the diagonal of the longitude holonomy can be used to produce flat connections at infinity for surgery on the connected sums, just like in the case of twist knots. We find 3 such solutions for the pretzel knot $P(-2,3,7)$.

On the other hand, the solutions with both $x$ and $y$ unipotent are more interesting; they can be used to produce flat connections at infinity for surgeries on the pretzel knot $P(-2,3,7)$ itself. We find a total of 4 such solutions for the pretzel knot $P(-2,3,7) = \kbf{12n242}$.

It would be interesting to find flat connections at infinity for surgeries on other prime knots, especially with smaller crossing number. By exhaustive search through the list of prime knots with low crossing number, we find that no prime knots with 8 crossings or less satisfy \eqref{Apoltest}. This means that there are no $\pm 1$ surgeries on low-crossing prime knots that have flat connections at infinity, at least constructed via the method described here. It still does not rule out a possibility that flat connections at infinity exist on surgeries on such knots with other surgery coefficients, and it would be interesting to look for them. Of particular interest are 0-surgeries that play a special role in complex Chern-Simons theory and in applications to low-dimensional topology (see e.g. \cite{Costin:2023kla} for a recent discussion) and it would be extremely helpful to study flat connections at infinity for 0-surgeries on knots. Although at present we do not find any flat connections at infinity on $S^3_0 (K)$ for prime knots with 5 crossings or less, we hope to continue this search in the future work.

Among knots with larger crossing number we found that the likelihood of \eqref{Apoltest} grows with the crossing number. In particular, this condition is obeyed for the following prime knots
\begin{multline}
\kbf{9_{22}} \,, \quad
\kbf{9_{25}} \,, \quad
\kbf{9_{29}} \,, \quad
\kbf{9_{36}} \,, \quad
\kbf{9_{37}} \,, \quad
\kbf{9_{38}} \,, \quad
\kbf{9_{41}} \,, \quad
\kbf{9_{42}} \,, \quad
\kbf{9_{43}} \,, \quad
\kbf{9_{44}} \,, \quad
\kbf{9_{45}} \,, \quad
\kbf{9_{47}} \,, \quad \\
\kbf{9_{48}} \,, \quad
\kbf{9_{49}} \,, \quad
\kbf{10_{46}} \,, \quad
\kbf{10_{47}} \,, \quad
\kbf{10_{48}} \,, \quad
\kbf{10_{54}} \,, \quad
\kbf{10_{61}} \,, \quad
\kbf{10_{124}} \,, \quad
\kbf{10_{125}} \,, \quad
\kbf{10_{126}} \,, \quad \\
\kbf{10_{127}} \,, \quad
\kbf{10_{128}} \,, \quad
\kbf{10_{129}} \,, \quad
\kbf{10_{130}} \,, \quad
\kbf{10_{131}} \,, \quad
\kbf{10_{132}} \,, \quad
\kbf{10_{133}} \,, \quad
\kbf{10_{134}} \,, \quad
\ldots
\end{multline}
It would be interesting to study more systematically which of these knots admit flat connections at infinity after $\pm 1$ surgery.

\section{Categorification}
\label{sec:categorification}

In this section we consider a somewhat naive categorification of the Stokes coefficients for plumbed 3-manifolds in terms of the finite-dimensional model presented in Section \ref{sec:stokes-weakly-negative}. The basic idea is to categorify the Stokes/monodromy coefficients of the finite-dimensional model (\ref{fd-monodromy-coefficients}) via a version of the Fukaya-Seidel category associated to the finite-dimensional integral (\ref{finte-dimensional-integral-general-contour}). By categorifying the relation (\ref{CS-fd-monodromies}), this provides some cateforification of the Stokes coefficients $m^\bbalpha_\bbbeta$ and, therefore, of their generating $\tilde{q}$-series $I^\alpha_\beta(\tilde{q})$.

It is tempting to conjecture that this categorification should agree with the categorification provided by the hypothetical ``Fukaya-Seidel category of CS functional'' or, maybe, a more realistic Fueter 1-category of the pair of holomorphic Lagrangians in the $SL(2,\C)$ character variety of a surface corresponding to a Heegaard splitting of the 3-manifold. However, we do not have any direct evidence for it, mainly because concrete computations in the infinite-dimensional setting of Chern-Simons theory or in the Fueter category are currently unavailable. There is only some indirect evidence that somewhat favors this possibility: the invariance of the cateforified Stokes coefficients under the $\Z$-action --- as in (\ref{CS-stokes-coefs-Z-invariance}) for the decategorified version --- and the invariance under the Kirby-Neumann moves on the plumbings. Put differently, the categorification proposed in this Section provides {\it some} homological invariant for a certain class of plumbed 3-manifolds.

Perhaps we should emphasize that it is actually not possible to have a complete equivalence between the Fukaya-Seidel category associated with the finite-dimensional model and that of Chern-Simons functional already on the level of objects. This can be already seen from the fact that, as explained in Section \ref{sec:single-thimble-info}, only a certain subset of Lefschetz thimbles in Chern-Simons theory is reproduced in the finite-dimensional model. Moreover, a Lefschetz thimble in Chern-Simons theory may correspond to a disjoint union of ``thimble'' contours in the finite-dimensional model (which, in the case $|H|>1$, are not related simply by the $\Z_2$ action $v\rightarrow -v$).

\subsection{Categorification of the finite-dimensional model}
\label{sec:fd-categorification}

In our approach to the categorification of the Stokes coefficients both the resulting homology and the underlying complex are considered to be 2-periodic, i.e. the homological grading is $\Z_2$. We leave the question of whether it can be canonically lifted to a  $\Z$-graded version out of the scope of this paper. Note that to have a well-defined \textit{absolute} $\Z$-grading in the Floer homology, one has to specify an additional structure on the Lagrangian submanifolds, namely their \textit{graded lifts} \cite{kontsevich1995homological,seidel2000graded}. In our setting Lagrangian submanifolds appear as integration contours and thus a priori only have orientation with no additional structure. This only allows us to define a $\Z_2$ absolute grading. We also cannot use consistently a \textit{relative} $\Z$-grading of Floer homology for a pair of Lagrangians in the finite-dimensional model, because in general (and as we will see explicitly in some examples) the categorification of the Stokes coefficients of Chern-Simons theory is obtained by taking a direct sum of Floer homologies of different pairs of Lagrangians. A good news is that the reduction to the 2-periodic version does not lead to infinite-dimensional vector spaces (for a categorification of a given Stokes coefficient). The categorification of the entire $\tilde{q}$-series $I^\alpha_\beta(\tilde{q})$ is then given by $\Z\times \Z_2$-graded vector spaces.

The first step of the categorification is the interpretation of the monodromy coefficients $M^{(H',\mathbb{n})}_{(K,\mathbb{m})}$ of the finite-dimensional model in terms of signed counts of intersection numbers between pairs of Lagrangians in the space of $v_I$ variables associated to the high-valency vertices:
\begin{equation}
 \mathcal{V}:=\left\{v_I\in \C,\;I\in H\;|\; v_I \text{ is not a pole/zero of }R(v)\right\}\subset \C^{|H|}.
 \label{fd-A-model-target}
\end{equation}
The interpretation will be given by a certain generalized version of the Picard-Lefschetz formula, cf. \cite{clemens1969picard,landman1973picard,Witten:2010cx}. The reason for excluding also zeros and not just poles in (\ref{fd-A-model-target}) is that one can interpret $d^{|H|}v\,R(v)$ as a holomorphic volume form on $\mathcal{V}$, see Section \ref{sec:A-model} for more details. In what follows we work under the same assumptions as in Section \ref{sec:stokes-plumbed}. In particular, we assume the plumbing to be weakly negative-definite.

\begin{figure}
\centering
\begin{tikzpicture}
    \draw[->] (0,-4) -- (0,4) node[right=2] {$\re v_I$};
     \draw[->] (3,0) -- (-3,0) node[above=2] {$\im v_I$};
    \draw plot[only marks,mark=x,mark size=4pt,mark options={draw=red}] coordinates {(0,1) (0,2) (0,3) (0,-1) (0,-2) (0,-3)};
  
    \draw[violet,ultra thick, 
        decoration={markings, mark=at position 0.7 with {\arrow{>}}},
        postaction={decorate}
        ] (0,2) circle (0.5)  node[left=20] {$\lambda_I^{(H',\mathbb{n})}$};

      \draw[orange,ultra thick, 
        decoration={markings, mark=at position 0.75 with {\arrow{>}}},
        postaction={decorate}
        ] (0,2) -- ++(50:3.5) node[above] {$\check{\lambda}^I_{(H',\mathbb{n})}$};

     \draw[magenta,ultra thick, 
        decoration={markings, mark=at position 0.75 with {\arrow{>}}},
        postaction={decorate}
        ] (0,2) -- ++(40:3.5) node[right] {$\hat{\lambda}^I_{(H',\mathbb{n})}$};

   \draw[dashed] (0,2) -- (3,2);

    \draw (1,2) arc (0:50:1) node[midway,right] {$\frac{1}{2}\mathrm{arg}\frac{i}{k}+\epsilon$};

     \draw (3,2) arc (0:40:3) node[midway,right] {$\frac{1}{2}\mathrm{arg}\frac{i}{k}-\epsilon$};

    \draw[violet,ultra thick, 
        decoration={markings, mark=at position 0.2 with {\arrow{>}}},
        postaction={decorate}
        ] (0,-1) circle (0.5);

      \draw[orange,ultra thick, 
        decoration={markings, mark=at position 0.75 with {\arrow{>}}},
        postaction={decorate}
        ] (0,-1) -- ++(230:3.5);

     \draw[magenta,ultra thick, 
        decoration={markings, mark=at position 0.75 with {\arrow{>}}},
        postaction={decorate}
        ] (0,-1) -- ++(220:3.5);

    \draw (2,1) node {$\mathbb{n}_I\geq 0$};

    \draw (2,-1) node {$\mathbb{n}_I<0$};

\end{tikzpicture}
\begin{tikzpicture}
    \draw[->] (0,-4) -- (0,4) node[right=2] {$\re v_I$};
     \draw[->] (3,0) -- (-3,0) node[above=2] {$\im v_I$};
    \draw plot[only marks,mark=x,mark size=4pt,mark options={draw=red}] coordinates {(0,1) (0,2) (0,3) (0,-1) (0,-2) (0,-3)};

     \draw[violet,ultra thick, 
        decoration={markings, mark=at position 0.9 with {\arrow{>}}},
        postaction={decorate}
        ] (0,0) -- ++(45:3) node[above right] {$\lambda_I^{(H',\mathbb{n})}$};
     \draw[violet,ultra thick] (0,0) -- ++(225:3);

      \draw[orange,ultra thick] (0,0) -- ++(50:3);
     \draw[orange,ultra thick, 
        decoration={markings, mark=at position 0.75 with {\arrow{>}}},
        postaction={decorate}
        ] (0,0) -- ++(230:3)  node[below] {$\check{\lambda}^I_{(H',\mathbb{n})}$};

     \draw[magenta,ultra thick] (0,0) -- ++(40:3);
     \draw[magenta,ultra thick, 
        decoration={markings, mark=at position 0.75 with {\arrow{>}}},
        postaction={decorate}
        ] (0,0) -- ++(220:3)  node[left] {$\hat{\lambda}^I_{(H',\mathbb{n})}$};

    \draw (1.5,0) arc (0:45:1.5) node[midway,right] {$\frac{1}{2}\mathrm{arg}\frac{i}{k}$};
\end{tikzpicture}
\caption{Factors $\lambda_I^{(H',\mathbb{n})}$ (violet), $\check{\lambda}^I_{(H',\mathbb{n})}$ (orange), and  $\hat{\lambda}^I_{(H',\mathbb{n})}$ (magenta) in the Cartesian product decomposition of contours $\lambda^{(H',\mathbb{n})}$, $\check{\lambda}_{(H',\mathbb{n})}$, and $\hat{\lambda}_{(H',\mathbb{n})}$, respectively.
Here, $\check{\lambda}_{(H',\mathbb{n})}$ is the contour dual to $\lambda^{(H',\mathbb{n})}$ for which the asymptotic slope is infinitesimally shifted in the negative direction relative to $\lambda^{(H',\mathbb{n})}$.
The contour $\hat{\lambda}_{(H',\mathbb{n})}$ is a copy of $\check{\lambda}_{(H',\mathbb{n})}$ with a rotation in the opposite direction. Top left: the case of $I\in H'$, $\mathbb{n}_I\geq0$. Bottom left: the case of $I\in H'$, $\mathbb{n}_I<0$. Right: the case of $I\notin H'$ (assuming $v_I=0$ is not a pole of  $R(v)$; otherwise, the contour should be infinitesimally shifted in the negative real direction).}
\label{fig:lambda-contours}
\end{figure}
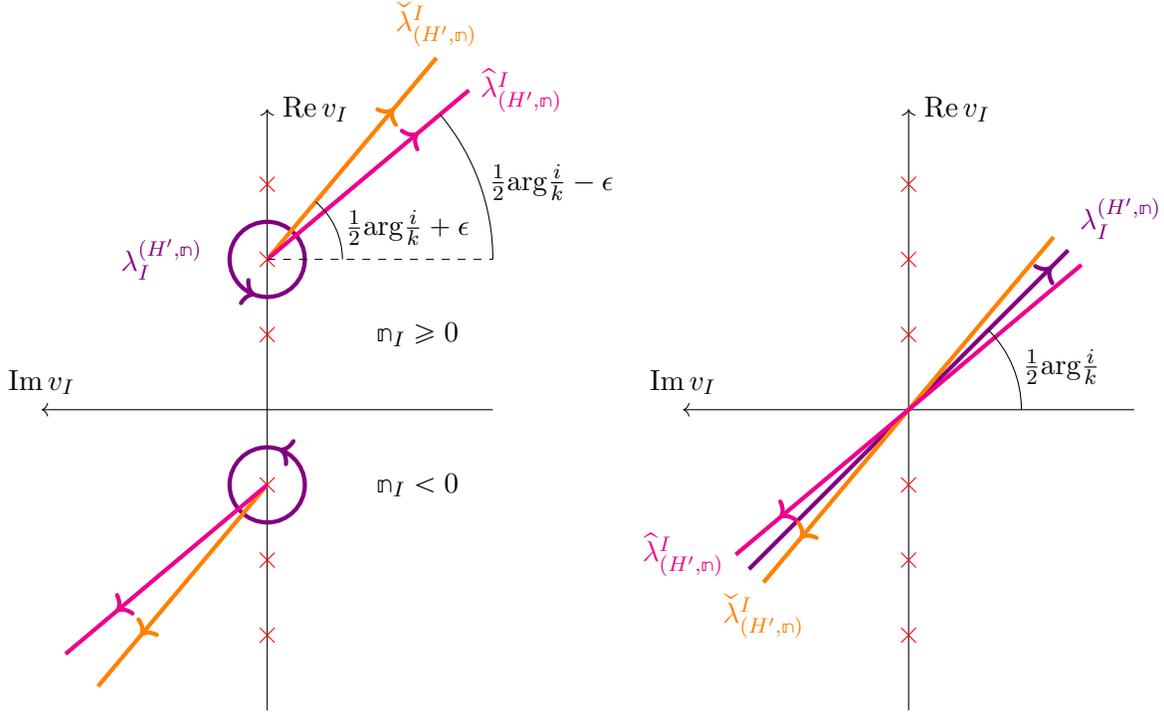

The relation between the monodromy coefficients and intersection numbers can be seen clearly in terms of the basis in the space of admissible contours which is slightly different from the ``Lefschetz thimble'' basis $\gamma^{(H',\mathbb{n})}$ considered in Section \ref{sec:stokes-plumbed}. We recall that, with our convention, admissible contours $\gamma\in \C^{|H|}$ are such that when $v$ goes to infinity along the contour, we have $ikS(v)\rightarrow -\infty$ along the real axis. This condition is essentially the same as the condition for a Lagrangian to be ``admissible'' in the formulation of the Fukaya-Seidel category associated to a holomorphic fibration $S:\mathcal{V}\rightarrow \C$ (with $S$ playing the role of the superpotential, ordinarily denoted by $W$) in terms of the infinitesimal Fukaya category of the total space. The categorification of intersection numbers between the Lagrangians will then be provided by the corresponding Hom-spaces in a version of infinitesimal Fukaya category of $\mathcal{V}$. We return to the details of Fukaya categories in Section~\ref{sec:wrappedstopped}. Here we provide a purely combinatorial description of the categorification, where such details do not play a role. It should be also possible to reformulate the categorification in terms of (partially) wrapped Fukaya category of $\mathcal{V}$ or in terms of the original Seidel's formulation of the Fukaya-Seidel category as the Fukaya category of a generic fiber of $S:\mathcal{V}\rightarrow\C$; although we will not attempt to do it in this paper, we present some initial steps in Section~\ref{sec:wrappedstopped}.

We also require the contours of integration to have boundary only at infinity of $\mathcal{V}$ (but not at the poles of $R(v)$). The new basis $\lambda^{(H',\mathbb{n})}$ that we wish to use here (shown in violet in Figure \ref{fig:lambda-contours}) is also indexed by pairs $(H'\subset H,\mathbb{n}\in \Z^{H'})$. As was the case for the basis $\gamma^{(H',\mathbb{n})}$, it has a Cartesian product decomposition
\begin{equation}
\lambda^{(H',\mathbb{n})} \; = \; \bigtimes_{I\in H}\lambda_I^{(H',\mathbb{n})}
\label{contourproduct}
\end{equation}
with factors being essentially the same as for $\gamma^{(H',\mathbb{n})}$, except that all line factors are required to pass through the origins of $v_I$-planes. The effect of the global monodromy (corresponding the increase of $\arg\frac{1}{k}$ by $2\pi$) on the factors of the $\lambda$-contours is also the same as the one for the $\gamma$-contours (shown in Figure \ref{fig:contour-monodromy}). However, in the present case, in order to express the result in terms of the $\lambda$-contours, no further shifts (shown in Figure \ref{fig:contour-shift}) are needed.

In order to obtain the same monodromy result in terms of the intersection theory, consider first the dual basis $\check{\lambda}_{(H',\mathbb{n})}$, in the sense that 
\begin{equation}
({\lambda}^{(H',\mathbb{n})},\check{\lambda}_{(K,\mathbb{m})})=\delta_{(K,\mathbb{m})}^{(H',\mathbb{n})}
\end{equation}
where $(\,\cdot\,,\,\cdot\,)$ denotes the intersection pairing, symmetric or skewsymmetric depending on the parity of the dimension. The dual basis is a basis in the space of the contours, dual to the space of the admissible contours. Those are the contours that satisfy an infinitesimally rotated condition at infinity: $ikS\rightarrow -\infty\,e^{i\epsilon}$ for $\epsilon>0$ and are instead \textit{relative} to the poles of $R(v)$, i.e. they are allowed to end on them (with the circle contours now becoming contractible). The dual contours $\check{\lambda}_{(H',\mathbb{n})}$ can also be decomposed into Cartesian products with factors either being lines (dual to lines) or rays originating from poles (dual to circles), as shown in Figure \ref{fig:lambda-contours} in orange. Note that the contours dual to $\gamma$-contours do not have such a simple factorization and have a more complicated description. This is the reason for using the $\lambda$-contours instead of the $\gamma$-contours. 

One can then consider the deformations $\hat{\lambda}_{(H',\mathbb{n})}$ of the dual contours $\check{\lambda}_{(H',\mathbb{n})}$ corresponding to the infinitesimal rotation at infinity in the opposite direction. Namely, at infinity the deformed contours should satisfy the condition $ikS\rightarrow -\infty\,e^{i\epsilon}$. They are shown in Figure \ref{fig:lambda-contours} in magenta. The analogue of the Picard-Lefschetz formula for the  monodromy transformation then reads:
\begin{equation}
\Gamma\longmapsto \sum_{(K,\mathbb{m})}(\Gamma,\hat{\lambda}_{(K,\mathbb{m})})\,{\lambda}^{(K,\mathbb{m})}.
\label{PL-lambda}
\end{equation}
It is easy to see explicitly that the formula also applies to the $\lambda$-contours, i.e. works with $\Gamma=\lambda^{(H',\mathbb{n})}$.

It is, of course, possible to express the $\lambda$-contours through the $\gamma$-contous:
\begin{equation}
    \lambda^{(K,\mathbb{m})}=
    \sum_{(L,\mathbb{p})}E_{(L,\mathbb{p})}^{(K,\mathbb{m})}\,\gamma^{(L,\mathbb{p})},
\end{equation}
with some integers $E_{(L,\mathbb{p})}^{(K,\mathbb{m})}\in\Z$ that can be algorithmically determined by applying the shifts as in Figure \ref{fig:contour-shift}. Namely, one can start by interpreting the line factors $\lambda_I^{(K,\mathbb{m})},\,I\in H\setminus K$ as the line factors $\gamma_I^{(\emptyset)}$ and the circle factors $\lambda_I^{(K,\mathbb{m})},\,I\in K$ simply as the circle contours $\sigma^{(\mathbb{m}_I)}_I$. Then one can apply the transformations (\ref{contour-shift-transformation}) recursively until the result is a linear combination of $\gamma$-contours. The monodromy transformation formula (\ref{PL-lambda}) then can be rewritten as
\begin{equation}
\Gamma\longmapsto \sum_{(L,\mathbb{p})}\left(
\Gamma,
\sum_{(K,\mathbb{m})}E_{(L,\mathbb{p})}^{(K,\mathbb{m})}
\hat{\lambda}_{(K,\mathbb{m})}
\right)\,{\gamma}^{(L,\mathbb{p})}.
\label{PL-gamma}
\end{equation}
In this form we can interpret 
\begin{equation}
\sum_{(K,\mathbb{m})}E_{(L,\mathbb{p})}^{(K,\mathbb{m})}\hat{\lambda}_{(K,\mathbb{m})}=\hat{\gamma}_{(K,\mathbb{m})}
\end{equation}
as the $\gamma$-analogs of the $\hat\lambda$-contours.

From (\ref{PL-gamma}) it follows that the momodromy coefficients (\ref{fd-monodromy-coefficients}) of the $\gamma$-contours  can be expressed as:
\begin{equation}
    M^{(H',\mathbb{n})}_{(K,\mathbb{m})}=
   \left(
   \gamma^{(H',\mathbb{n})},
\sum_{(L,\mathbb{p})}E^{(L,\mathbb{p})}_{(K,\mathbb{m})}\hat{\lambda}_{(L,\mathbb{p})}
\right).
\label{fd-monodromy-intersection}
\end{equation}
This, apart from providing an expression suitable for categorification in terms of Fukaya category, also provides an alternative way to calculate the monodromy coefficients. 

Let $\mathcal{C}^{(H',\mathbb{n})}_{(K,\mathbb{m})}=\prescript{}{0}{\mathcal{C}}^{(H',\mathbb{n})}_{(K,\mathbb{m})}\oplus \prescript{}{1}{\mathcal{C}}^{(H',\mathbb{n})}_{(K,\mathbb{m})}$ be the $\Z_2$-graded complex of vector spaces (over $\C$) generated by the intersection points $\gamma^{(H',\mathbb{n})}\cap \hat{\lambda}_{(L,\mathbb{p})}$ for all possible $(L,\mathbb{p})$ counted with multiplicities $E^{(L,\mathbb{p})}_{(K,\mathbb{m})}\in \Z$. More informally one can say that $\mathcal{C}^{(H',\mathbb{n})}_{(K,\mathbb{m})}$ is generated by $\gamma^{(H',\mathbb{n})}\cap \hat{\gamma}_{(K,\mathbb{m})}$. The grading of a generator is 0 if the intersection point (taking into account the multiplicity) contributes with a positive sign into (\ref{fd-monodromy-intersection}) and is 1 otherwise. The boundary morphism on the complex is the usual Fukaya boundary operator, given by counting holomorphic `bigons' with two edges lying on $\gamma^{(H',\mathbb{n})}$ and $\hat{\lambda}_{(L,\mathbb{p})}$, respectively. Denote by $H_\ast \big( \mathcal{C}^{(H',\mathbb{n})}_{(K,\mathbb{m})} \big)$ the corresponding $\Z_2$-graded homology. 

By construction we have
\begin{equation}
    M^{(H',\mathbb{n})}_{(K,\mathbb{m})}=
    \dim H_0 \big( \mathcal{C}^{(H',\mathbb{n})}_{(K,\mathbb{m})} \big)-
    \dim H_1 \big( \mathcal{C}^{(H',\mathbb{n})}_{(K,\mathbb{m})} \big)
    =
    \dim \prescript{}{0}{\mathcal{C}}^{(H',\mathbb{n})}_{(K,\mathbb{m})}-
    \dim \prescript{}{1}{\mathcal{C}}^{(H',\mathbb{n})}_{(K,\mathbb{m})}.
\end{equation}
Then one can propose a categorification of the CS monodromy coefficients $m^{(S)}_{(S')}$ by the $\Z_2$-graded vector space $\mathcal{H}^{(S)}_{(S')}$ obtained by categorifying the relation (\ref{CS-fd-monodromies}) in the most naive way. In the case when $H'\neq \emptyset$, we have:
\begin{equation}
    \mathcal{H}^{(S)}_{(S')}:=
\bigoplus_{(K,\mathbb{m}):\,S_\ast^{(K,\mathbb{m})}=S'} H_\ast \big( \mathcal{C}^{(H',\mathbb{n})}_{(K,\mathbb{m})} \big) [|H|-1]\otimes 
\left\{
\begin{array}{ll}
\C^{\frac{N_{(K,\mathbb{m})}}{N_{(H',\mathbb{n})}}}, & \frac{N_{(K,\mathbb{m})}}{N_{(H',\mathbb{n})}}\geq 0, \\
 \C^{-\frac{N_{(K,\mathbb{m})}}{N_{(H',\mathbb{n})}}}[1], & \frac{N_{(K,\mathbb{m})}}{N_{(H',\mathbb{n})}}<0, \end{array}
\right.
\label{CS-m-coefficient-categorification}
\end{equation}
where $(H',\mathbb{n})$ is such that $S=S_\ast^{(H',\mathbb{n})}$ and $[n]$ denotes the $\Z_2$-grading shift (with a copy of $\C$ assumed to have grading zero by default). Note that by construction ${N_{(K,\mathbb{m})}}/{N_{(H',\mathbb{n})}}\in\Z$ (see Section \ref{sec:stokes-plumbed}).
In the case $H'= \emptyset$ (with $S=0$), one has to take into account the 2-fold overcounting due to the Weyl symmetry (which in the decategorified version \eqref{CS-fd-monodromies} has been dealt with by introducing the factor $1/2$):
\begin{equation}
    \mathcal{H}^{(0)}_{(S')}:=
\bigoplus_{\substack{(K,\mathbb{m}):\,S_\ast^{(K,\mathbb{m})}=S' \\ \text{single representative} \\ \text{w.r.t. }(K,\mathbb{m})\sim(K,-\mathbb{m})}}H_\ast(\mathcal{C}^{(\emptyset)}_{(K,\mathbb{m})})\left[|H|-1\right]\otimes 
\left\{
\begin{array}{ll}
\C^{{N_{(K,\mathbb{m})}}}, & {N_{(K,\mathbb{m})}}\geq 0, \\
 \C^{-{N_{(K,\mathbb{m})}}}[1], & {N_{(K,\mathbb{m})}}<0. \end{array}
\right.
\label{CS-m-coefficient-categorification-from-zero}
\end{equation}
These vector spaces then can be combined into $(\Z+S-S')\times \Z_2$-graded vector spaces that categorify the $\tilde{q}$-series $I^{(S\mod 1)}_{(S'\mod 1)}(\tilde{q})$ (\ref{Stokes-series-from-fd}):
\begin{equation}
    \bigoplus_{S'\mod 1\text{ is fixed}}\mathcal{H}^{(S)}_{(S')}\llbracket S-S'\rrbracket,
    \label{Stokes-series-categorifiaction}
\end{equation}
where $\llbracket\ldots\rrbracket$ denotes the grading corresponding to the $\tilde{q}$ generating variable. Note that in order for $\mathcal{H}^{(S)}_{(S')}$ to be well-defined, the total vector space in the right-hand-side of (\ref{CS-m-coefficient-categorification}), up to an isomorphism, should not depend on the particular choice of $(H',\mathbb{m})$ such that $S=S_\ast^{(H',\mathbb{m})}$. And for \eqref{Stokes-series-categorifiaction} to be a function of $S$ and $S'$ only modulo 1, we must have $\mathcal{H}^{(S)}_{(S')}\cong \mathcal{H}^{(S+f)}_{(S'+f)},\;f\in\Z$. We do not provide a proof of these properties, but we have checked in many examples that they do hold.

The isomorphism class of the doubly-graded vector spaces (\ref{Stokes-series-categorifiaction}) can be incoded in the respective Poincar\'e series
\begin{multline}
    \mathcal{P}^{(S\mod 1)}_{(S'\mod 1)}(\tilde{q},t):=
    \\
    \sum_{S':\;S'\mod 1\text{ is fixed}}\left(\dim \prescript{}{0}{\mathcal{H}}^{(S)}_{(S')}+
    t\,\dim \prescript{}{1}{\mathcal{H}}^{(S)}_{(S')}
    \right)
    \tilde{q}^{S-S'}
    \;
    \in
    \;
    \tilde{q}^{S-S'}
    \Z[t][[\tilde{q},\tilde{q}^{-1}]]/(1-t^2)
    \label{Stokes-Poincare-series}
\end{multline}
such that
\begin{equation}
   I^{(S\mod 1)}_{(S'\mod 1)}(\tilde{q})= \mathcal{P}^{(S\mod 1)}_{(S'\mod 1)}(\tilde{q},-1).
\end{equation}

We observe that the homology categorifying  an individual monodromy coefficient in the finite-dimensional model turns out be always concentrated in a single degree, that is: 
\begin{equation}
H_\ast(\mathcal{C}^{(H',\mathbb{n})}_{(K,\mathbb{m})})\cong \left\{
\begin{array}{ll}
\C^{M^{(H',\mathbb{n})}_{(K,\mathbb{m})}}, & M^{(H',\mathbb{n})}_{(K,\mathbb{m})}\geq 0, \\
 \C^{-M^{(H',\mathbb{n})}_{(K,\mathbb{m})}}[1], & M^{(H',\mathbb{n})}_{(K,\mathbb{m})}<0. \end{array}
\right.
\end{equation}
Again, we do not provide a general proof of this statement. Nevertheless, as we will see in some examples below, the same is not always true for the vector spaces (\ref{CS-m-coefficient-categorification}) categorifying the monodromy coefficients in CS. 

Let us comment briefly on the topological invariance, i.e. the invariance under the Neumann moves \cite{neumann1981calculus}. If we restrict ourselves to the moves that keep the plumbing of the considered type (weakly negative-definite), we only need to consider blow-up/blow-down moves and 0-chain absorption/extrusion moves (R1 and R3 in loc. cit.). We can also then exclude 0-chain extrusions that increase the number of high-valency vertices, as it makes the matrix $C$ non-invertible. The inverse moves, 0-chain absorptions that decrease the number of high-valency vertices, therefore can be also ignored. Moves that do not introduce or remove high-valency vertices keep $S(v)$ and $R(v)$ unchanged.

It is then left to verify a blow-up move that creates a new high-valency vertex. This move introduces a single new variable associated with the new high-valency vertex. The function $R(v)$ is kept the same (and depends on the old variables only), but $S(v)$ is replaced with $S'(v')=S(v)+u^2$, where $u\in \C$ is a certain linear combination of the new variables $v'\in\C^{|H|+1}$ (which includes the new variable with a non-zero coefficient). In terms of the variables $(v,u)\in\C^{|H|+1}$ the set of $\gamma$-contours (``Lefschetz thimbles'') in the new finite-dimensional model is in a canonical one-to-one correspondence with the $\gamma$-contours in the old model. Namely, the new $\gamma$-contours are given by the products of the old $\gamma$-contours with the straight-line contour in the $u$-plane passing through the origin. The same holds for the $\lambda$-contours and their duals. Moreover, the factors of the $\gamma$- and $\hat{\lambda}$-contours in the $u$-plane always have a single intersection point (with multiplicity $-1$). It is then immediate to see that all the homology groups will remain the same, including the grading (the grading shift by $|H|-1$ in (\ref{CS-m-coefficient-categorification}) takes care of the flip of signs of the intersection numbers).

\subsubsection{Example 1}

\begin{figure}
\centering
$\mathcal{C}^{(\emptyset)}_{(\{1\},-17)}=\C$
\begin{tikzpicture}[scale=0.5,baseline=(current  bounding  box.center)]

\useasboundingbox (-3,-3) rectangle (3,3);

    \draw[->] (0,-3) -- (0,3) node[left=2] {$v_1$};
     \draw[->] (2.5,0) -- (-2.5,0);
    \draw plot[only marks,mark=x,mark size=4pt,mark options={draw=red}] coordinates {(0,-1)};

     \draw[blue,ultra thick, 
        decoration={markings, mark=at position 0.9 with {\arrow{>}}},
        postaction={decorate}
        ] (0,0) -- ++(55:3);
     \draw[blue,ultra thick] (0,0) -- ++(235:4);

     \draw[magenta,ultra thick, 
        decoration={markings, mark=at position 0.75 with {\arrow{>}}},
        postaction={decorate}
        ] (0,-1) -- ++(215:3);

 \draw[blue] (0,0) node[above left=5] {$\gamma_1^{(\emptyset)}$};

     \draw[magenta] (0,0) node[below right] {$\hat{\lambda}^1_{(\{1\},-17)}$};
        
\end{tikzpicture}      
                \begin{tikzpicture}[scale=0.5,baseline=(current  bounding  box.center)]
\useasboundingbox (-3,-3) rectangle (3,3);
                
    \draw[->] (0,-3) -- (0,3) node[left=2] {$v_2$};
     \draw[->] (2.5,0) -- (-2.5,0);

     \draw[blue,ultra thick, 
        decoration={markings, mark=at position 0.9 with {\arrow{>}}},
        postaction={decorate}
        ] (0,0) -- ++(55:3);
     \draw[blue,ultra thick] (0,0) -- ++(235:4);

     \draw[magenta,ultra thick] (0,0) -- ++(35:3);
     
     \draw[magenta,ultra thick, 
        decoration={markings, mark=at position 0.75 with {\arrow{>}}},
        postaction={decorate}
        ] (0,0) -- ++(215:3);
        
    \draw[blue] (0,0) node[above left=5] {$\gamma_2^{(\emptyset)}$};

     \draw[magenta] (0,0) node[below right] {$\hat{\lambda}^2_{(\{1\},-17)}$};

\end{tikzpicture}
\caption{A single point in the intersection of $\gamma^{(\emptyset)}$ with $\hat{\gamma}^{(\{1\},-17)}=\hat{\lambda}^{(\{1\},-17)}$   generating one-dimensional space $\mathcal{H}^{(0)}_{(1-\frac{23}{312})}=H_\ast(\mathcal{C}^{(\emptyset)}_{(\{1\},-17)})[1]=\mathcal{C}^{(\emptyset)}_{(\{1\},-17)}[1]\cong \C[1]$ corresponding to the $t\tilde{q}^1$ term in the Poincar\'e series (\ref{cat-example1-series1}). }
\label{fig:cat-example1-1}
\end{figure}
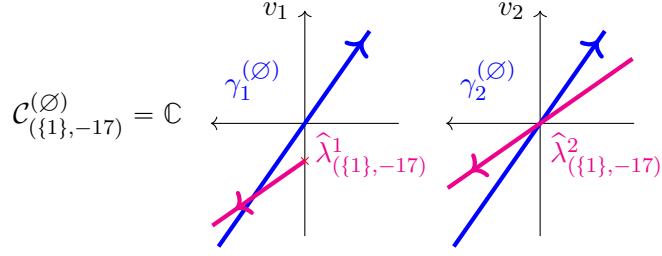

Let us first return to the example of the plumbing manifold depicted in Figure \ref{fig:plumbing-example-2m3-5m3} considered in Section \ref{sec:a-specific-example}. For the monodromy coefficients between pairs of Lefschetz thimbles in CS corresponding to the finite-dimensional thimbles labeled by $(H',\mathbb{n})$ and $(K,\mathbb{m})$, such $K$ has only one more element than $H'$, the categorification does not provide any additional information. In this case the vector spaces $\mathcal{H}^{(S)}_{(S')}$ categorifying individual coefficients are concentrated in a single degree, as the sums in (\ref{CS-m-coefficient-categorification})--(\ref{CS-m-coefficient-categorification-from-zero}) consist of a single term. For example,
\begin{multline}
      \mathcal{P}_{\left(\frac{23}{312}\right)}^{(0)}(\tilde{q},t)  = \\ \tilde{q}^{-\frac{23}{312}}\,\left(t\tilde{q}+t\tilde{q}^4+t\tilde{q}^6+t\tilde{q}^{12}+\tilde{q}^{29}+\tilde{q}^{41}+\tilde{q}^{47}+\tilde{q}^{62}+t\tilde{q}^{96}+t\tilde{q}^{117}+t\tilde{q}^{127}+\ldots\right).
      \label{cat-example1-series1}
\end{multline}
The vector space corresponding to each power of $\tilde{q}$ is generated by a single intersection point, with a trivial boundary operator. The intersection point generating the vector space corresponding to the $t\tilde{q}^1$ term is illustrated in Figure \ref{fig:cat-example1-1}.

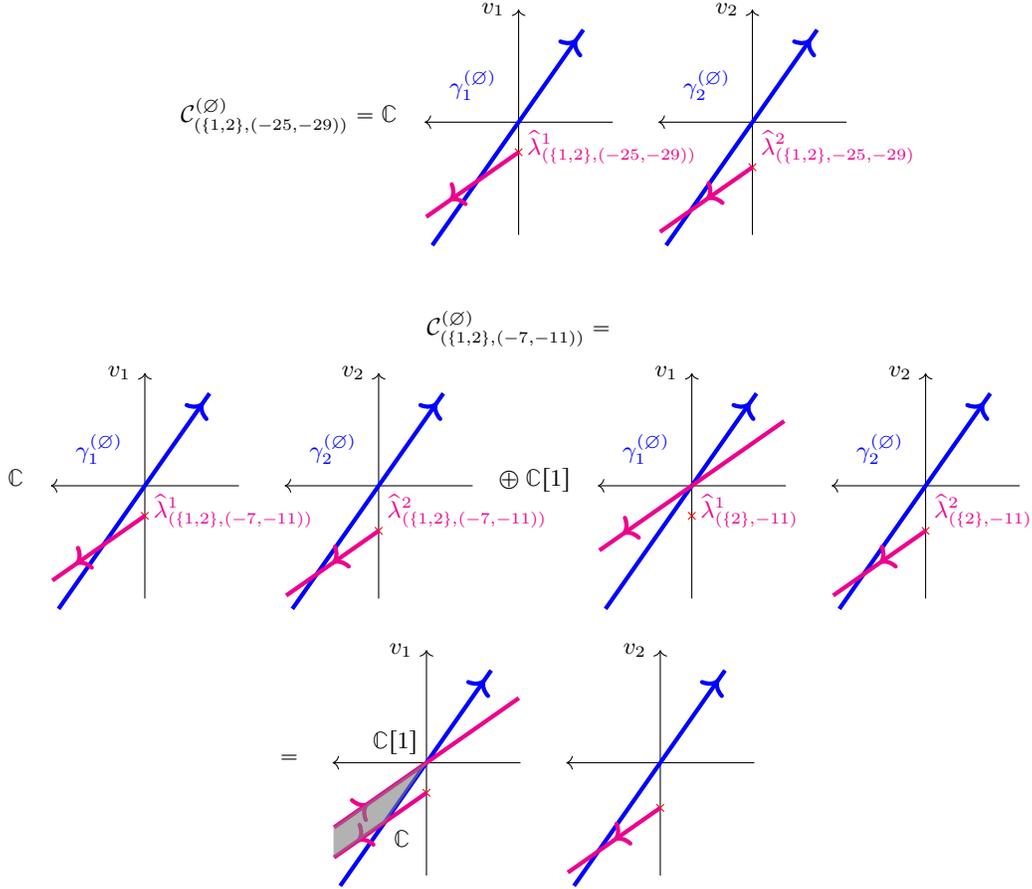
\begin{figure}
\centering
\footnotesize
$\mathcal{C}^{(\emptyset)}_{(\{1,2\},(-25,-29))}=\C$
\begin{tikzpicture}[scale=0.5,baseline=(current  bounding  box.center)]

\useasboundingbox (-3,-3) rectangle (3,3);
    \draw[->] (0,-3) -- (0,3) node[left=2] {$v_1$};
     \draw[->] (2.5,0) -- (-2.5,0);
    \draw plot[only marks,mark=x,mark size=4pt,mark options={draw=red}] coordinates {(0,-0.8)};

     \draw[blue,ultra thick, 
        decoration={markings, mark=at position 0.9 with {\arrow{>}}},
        postaction={decorate}
        ] (0,0) -- ++(55:3);
     \draw[blue,ultra thick] (0,0) -- ++(235:4);

     \draw[magenta,ultra thick, 
        decoration={markings, mark=at position 0.75 with {\arrow{>}}},
        postaction={decorate}
        ] (0,-0.8) -- ++(215:3);

 \draw[blue] (0,0) node[above left=5] {$\gamma_1^{(\emptyset)}$};

     \draw[magenta] (0,0) node[below right] {$\hat{\lambda}^1_{(\{1,2\},(-25,-29))}$};
        
\end{tikzpicture}   
\begin{tikzpicture}[scale=0.5,baseline=(current  bounding  box.center)]
\useasboundingbox (-3,-3) rectangle (3,3);
    \draw[->] (0,-3) -- (0,3) node[left=2] {$v_2$};
     \draw[->] (2.5,0) -- (-2.5,0);
    \draw plot[only marks,mark=x,mark size=4pt,mark options={draw=red}] coordinates {(0,-1.2)};

     \draw[blue,ultra thick, 
        decoration={markings, mark=at position 0.9 with {\arrow{>}}},
        postaction={decorate}
        ] (0,0) -- ++(55:3);
     \draw[blue,ultra thick] (0,0) -- ++(235:4);

     \draw[magenta,ultra thick, 
        decoration={markings, mark=at position 0.5 with {\arrow{>}}},
        postaction={decorate}
        ] (0,-1.2) -- ++(215:3);

 \draw[blue] (0,0) node[above left=5] {$\gamma_2^{(\emptyset)}$};

     \draw[magenta] (0,0) node[below right] {$\hat{\lambda}^2_{(\{1,2\},-25,-29)}$};
        
\end{tikzpicture}   

\vspace{3em}
$\mathcal{C}^{(\emptyset)}_{(\{1,2\},(-7,-11))}=$
\vspace{1em}

$\C$
\begin{tikzpicture}[scale=0.5,baseline=(current  bounding  box.center)]

\useasboundingbox (-3,-3) rectangle (3,3);

    \draw[->] (0,-3) -- (0,3) node[left=2] {$v_1$};
     \draw[->] (2.5,0) -- (-2.5,0);
    \draw plot[only marks,mark=x,mark size=4pt,mark options={draw=red}] coordinates {(0,-0.8)};

     \draw[blue,ultra thick, 
        decoration={markings, mark=at position 0.9 with {\arrow{>}}},
        postaction={decorate}
        ] (0,0) -- ++(55:3);
     \draw[blue,ultra thick] (0,0) -- ++(235:4);

     \draw[magenta,ultra thick, 
        decoration={markings, mark=at position 0.75 with {\arrow{>}}},
        postaction={decorate}
        ] (0,-0.8) -- ++(215:3);

 \draw[blue] (0,0) node[above left=5] {$\gamma_1^{(\emptyset)}$};

     \draw[magenta] (0,0) node[below right] {$\hat{\lambda}^1_{(\{1,2\},(-7,-11))}$};
        
\end{tikzpicture}   
\begin{tikzpicture}[scale=0.5,baseline=(current  bounding  box.center)]
\useasboundingbox (-3,-3) rectangle (3,3);
    \draw[->] (0,-3) -- (0,3) node[left=2] {$v_2$};
     \draw[->] (2.5,0) -- (-2.5,0);
    \draw plot[only marks,mark=x,mark size=4pt,mark options={draw=red}] coordinates {(0,-1.2)};

     \draw[blue,ultra thick, 
        decoration={markings, mark=at position 0.9 with {\arrow{>}}},
        postaction={decorate}
        ] (0,0) -- ++(55:3);
     \draw[blue,ultra thick] (0,0) -- ++(235:4);

     \draw[magenta,ultra thick, 
        decoration={markings, mark=at position 0.5 with {\arrow{>}}},
        postaction={decorate}
        ] (0,-1.2) -- ++(215:3);

 \draw[blue] (0,0) node[above left=5] {$\gamma_2^{(\emptyset)}$};

     \draw[magenta] (0,0) node[below right] {$\hat{\lambda}^2_{(\{1,2\},(-7,-11))}$};
        
\end{tikzpicture}   
$\oplus\;\C[1]$
\begin{tikzpicture}[scale=0.5,baseline=(current  bounding  box.center)]

\useasboundingbox (-3,-3) rectangle (3,3);
    \draw[->] (0,-3) -- (0,3) node[left=2] {$v_1$};
     \draw[->] (2.5,0) -- (-2.5,0);
    \draw plot[only marks,mark=x,mark size=4pt,mark options={draw=red}] coordinates {(0,-0.8)};

     \draw[blue,ultra thick, 
        decoration={markings, mark=at position 0.9 with {\arrow{>}}},
        postaction={decorate}
        ] (0,0) -- ++(55:3);
     \draw[blue,ultra thick] (0,0) -- ++(235:4);

     \draw[magenta,ultra thick, 
        decoration={markings, mark=at position 0.75 with {\arrow{>}}},
        postaction={decorate}
        ] (0,0) -- ++(215:3);

         \draw[magenta,ultra thick] (0,0) -- ++(35:3);

 \draw[blue] (0,0) node[above left=5] {$\gamma_1^{(\emptyset)}$};

     \draw[magenta] (0,0) node[below right] {$\hat{\lambda}^1_{(\{2\},-11)}$};
        
\end{tikzpicture}   
\begin{tikzpicture}[scale=0.5,baseline=(current  bounding  box.center)]

\useasboundingbox (-3,-3) rectangle (3,3);
    \draw[->] (0,-3) -- (0,3) node[left=2] {$v_2$};
     \draw[->] (2.5,0) -- (-2.5,0);
    \draw plot[only marks,mark=x,mark size=4pt,mark options={draw=red}] coordinates {(0,-1.2)};

     \draw[blue,ultra thick, 
        decoration={markings, mark=at position 0.9 with {\arrow{>}}},
        postaction={decorate}
        ] (0,0) -- ++(55:3);
     \draw[blue,ultra thick] (0,0) -- ++(235:4);

     \draw[magenta,ultra thick, 
        decoration={markings, mark=at position 0.5 with {\arrow{>}}},
        postaction={decorate}
        ] (0,-1.2) -- ++(215:3);

 \draw[blue] (0,0) node[above left=5] {$\gamma_2^{(\emptyset)}$};

     \draw[magenta] (0,0) node[below right] {$\hat{\lambda}^2_{(\{2\},-11)}$};
        
\end{tikzpicture}  

\vspace{2em}
$\;=\;$
\begin{tikzpicture}[scale=0.5,baseline=(current  bounding  box.center)]

\useasboundingbox (-3,-3) rectangle (3,3);
    \draw[->] (0,-3) -- (0,3) node[left=2] {$v_1$};
     \draw[->] (2.5,0) -- (-2.5,0);
    \draw plot[only marks,mark=x,mark size=4pt,mark options={draw=red}] coordinates {(0,-0.8)};

     \draw[blue,ultra thick, 
        decoration={markings, mark=at position 0.9 with {\arrow{>}}},
        postaction={decorate}
        ] (0,0) -- ++(55:3);
     \draw[blue,ultra thick,name path=PB] (0,0) -- ++(235:4);

     \draw[magenta,ultra thick, 
        decoration={markings, mark=at position 0.75 with {\arrow{<}}},
        postaction={decorate}
        ] (0,0) -- ++(215:3) node(d) {};

         \draw[magenta,ultra thick] (0,0) -- ++(35:3);

     \draw[magenta,ultra thick, 
        decoration={markings, mark=at position 0.75 with {\arrow{>}}},
        postaction={decorate},name path=PM] (0,-0.8) -- ++(215:3) node(c) {};

    \path[name intersections={of=PM and PB,by=b}];
    \filldraw[gray,opacity=0.6] (0,0) -- (b) -- (c.center) -- (d.center) -- cycle;

    \draw (0,0) node[above left] {$\C[1]$};

    \draw (b.center) node[below right] {$\C$};
    
\end{tikzpicture}   
\begin{tikzpicture}[scale=0.5,baseline=(current  bounding  box.center)]

\useasboundingbox (-3,-3) rectangle (3,3);
    \draw[->] (0,-3) -- (0,3) node[left=2] {$v_2$};
     \draw[->] (2.5,0) -- (-2.5,0);
    \draw plot[only marks,mark=x,mark size=4pt,mark options={draw=red}] coordinates {(0,-1.2)};

     \draw[blue,ultra thick, 
        decoration={markings, mark=at position 0.9 with {\arrow{>}}},
        postaction={decorate}
        ] (0,0) -- ++(55:3);
     \draw[blue,ultra thick] (0,0) -- ++(235:4);

     \draw[magenta,ultra thick, 
        decoration={markings, mark=at position 0.5 with {\arrow{>}}},
        postaction={decorate}
        ] (0,-1.2) -- ++(215:3);

\end{tikzpicture}  

\caption{The two complexes contributing to the vector space $\mathcal{H}^{(0)}_{(3-\frac{119}{120})}=H_\ast(\mathcal{C}^{(\emptyset)}_{(\{1,2\},(-25,-29))})[1]\oplus H_\ast(\mathcal{C}^{(\emptyset)}_{(\{1,2\},(-7,-11))})[1]\cong \C[1]\oplus 0$ corresponding to the $t\tilde{q}^3$ term in the Poincar\'e series (\ref{cat-example1-series2}). The complexes are generated by $\gamma^{(\emptyset)}\cap \hat{\gamma}_{(\{1,2\},(-25,-29))}$ 
 and $\gamma^{(\emptyset)}\cap \hat{\gamma}_{(\{1,2\},(-7,-11))}$ respectively, where $\hat{\gamma}_{(\{1,2\},(-25,-29))}=\hat{\lambda}_{(\{1,2\},(-25,-29))}$ 
 and $\hat{\gamma}_{(\{1,2\},(-7,-11))}=\hat{\lambda}_{(\{1,2\},(-7,-11))}-\hat{\lambda}_{(\{2\},-11)}$. The first complex, $\mathcal{C}^{(\emptyset)}_{(\{1,2\},(-25,-29))}$ consists of a single intersection point, with a trivial boundary operator. In the second complex, $\mathcal{C}^{(\emptyset)}_{(\{1,2\},(-7,-11))}$, using that $\hat{\lambda}_{(\{1,2\},(-7,-11))}^2=\hat{\lambda}_{(\{2\},-11)}^2$, one can combine the contours in the first plane together, as shown on the bottom. The shaded region depicts a holomorphic disk resulting in the boundary operator relating the two intersection points, rendering the homology trivial. }
\label{fig:cat-example1-2}
\end{figure}

In general, this is not the case: the boundary operator acting on the intersection points is non-trivial and the categorification of the coefficients of $\tilde{q}$-series is not concentrated in a single $\Z_2$-degree. For example, for the same 3-manifold we have
\begin{multline}
    \mathcal{P}_{\left(\frac{119}{120}\right)}^{(0)}(\tilde{q},t)  = \\ \tilde{q}^{-\frac{119}{120}}\,\left(\tilde{q}^2+t\tilde{q}^3+t\tilde{q}^{11}+\tilde{q}^{13}+t\tilde{q}^{19}+\tilde{q}^{20}+t\tilde{q}^{33}+\tilde{q}^{35}+(1+t)\tilde{q}^{42}
    \right.\\
 \left.
+\tilde{q}^{57}+\tilde{q}^{62}+t\tilde{q}^{63}+\tilde{q}^{69}+t\tilde{q}^{78}
 +\tilde{q}^{80}+t\tilde{q}^{89}+t\tilde{q}^{94}+t\tilde{q}^{106}+t\tilde{q}^{107}
    \right.\\
 \left.
 +t\tilde{q}^{116}+\tilde{q}^{118}+t\tilde{q}^{123}+t\tilde{q}^{126}+\tilde{q}^{134}+t\tilde{q}^{137}+t\tilde{q}^{148}+\ldots\right).
 \label{cat-example1-series2}
\end{multline}
The intersection points generating the complexes relevant for the $t\tilde{q}^3$ term are illustrated in Figure \ref{fig:cat-example1-2}.

\subsubsection{Example 2}

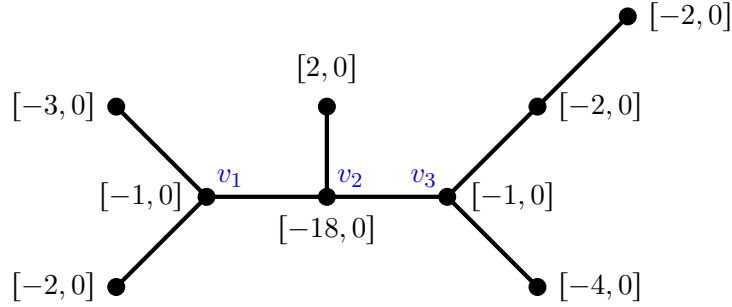
\begin{figure}
\centering
\begin{tikzpicture}[scale=0.4]

\draw[ultra thick] (5,0)  -- (9,0);
\draw[ultra thick] (9,0)  -- (9,3);
  \draw[ultra thick] (9,0) -- (13,0);
  \draw[ultra thick] (13,0) -- (16,3);
  \draw[ultra thick] (16,3) -- (19,6);
   \draw[ultra thick] (13,0) -- (16,-3);
    \draw[ultra thick] (5,0) -- (2,-3);
     \draw[ultra thick] (5,0) -- (2,3);
    \filldraw[black] (5,0) circle (8pt) node[left=5] {$[-1,0]$};
    \draw[blue] (5,0) node[above right] {$v_1$};
      \filldraw[black] (9,3) circle (8pt) node[above=5] {$[2,0]$};
    \filldraw[black] (2,3) circle (8pt) node[left=4] {$[-3,0]$};
    \filldraw[black] (2,-3) circle (8pt) node[left=4] {$[-2,0]$};
    \filldraw[black] (9,0) circle (8pt) node[below=3] {$[-18,0]$};
    \draw[blue] (9,0) node[above right] {$v_2$};
    \filldraw[black] (13,0) circle (8pt) node[right=5] {$[-1,0]$};
    \draw[blue] (13,0) node[above left] {$v_3$};
    \filldraw[black] (16,3) circle (8pt) node[right=4] {$[-2,0]$};
    \filldraw[black] (19,6) circle (8pt) node[right=4] {$[-2,0]$};
    \filldraw[black] (16,-3) circle (8pt) node[right=4] {$[-4,0]$};
\end{tikzpicture}
 \caption{An example of a weakly negative-definite plumbed integer homology sphere. The variables $v_{1,2,3}$ corresponding to the high-valency vertices are shown in blue.}
\label{fig:cat-example2-plumbing}
\end{figure}

To illustrate the independence of various choices made in the process of defining the vector spaces (\ref{Stokes-series-categorifiaction}) consider a slightly more involved example of plumbing, shown in Figure \ref{fig:cat-example2-plumbing}. The 3-manifold can be equivalently realized as $-1/2$-surgery on the connected sum of a trefoil and $T(4,3)$ torus knot, much as in the examples of Section~\ref{sec:flat-at-infinity}.

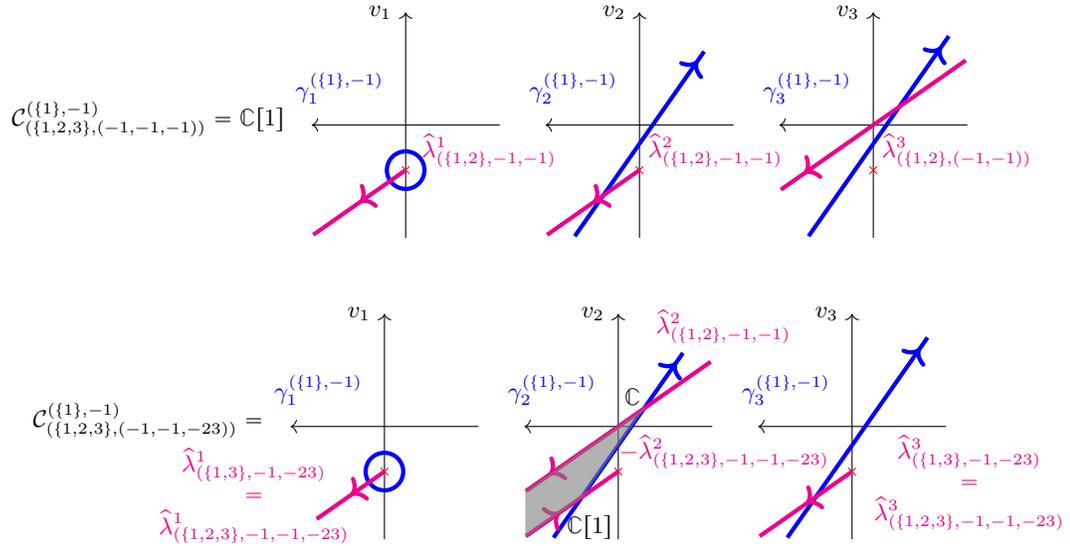
\begin{figure}
\centering
\footnotesize
$\mathcal{C}^{(\{1\},-1)}_{(\{1,2,3\},(-1,-1,-1))}=\C[1]$
\begin{tikzpicture}[scale=0.5,baseline=(current  bounding  box.center)]
\useasboundingbox (-3,-3) rectangle (3,3);
    \draw[->] (0,-3) -- (0,3) node[left=2] {$v_1$};
     \draw[->] (2.5,0) -- (-2.5,0);
    \draw plot[only marks,mark=x,mark size=4pt,mark options={draw=red}] coordinates {(0,-1.2)};

     \draw[blue,ultra thick] (0,-1.2) circle (0.5);

     \draw[magenta,ultra thick, 
        decoration={markings, mark=at position 0.5 with {\arrow{>}}},
        postaction={decorate}
        ] (0,-1.2) -- ++(215:3);

 \draw[blue] (0,0) node[above left=5] {$\gamma_1^{(\{1\},-1)}$};

     \draw[magenta] (0.2,0) node[below right] {$\hat{\lambda}^1_{(\{1,2\},-1,-1)}$};
        
\end{tikzpicture}   
\begin{tikzpicture}[scale=0.5,baseline=(current  bounding  box.center)]
\useasboundingbox (-3,-3) rectangle (3,3);
    \draw[->] (0,-3) -- (0,3) node[left=2] {$v_2$};
     \draw[->] (2.5,0) -- (-2.5,0);
    \draw plot[only marks,mark=x,mark size=4pt,mark options={draw=red}] coordinates {(0,-1.2)};

     \draw[blue,ultra thick, 
        decoration={markings, mark=at position 0.9 with {\arrow{>}}},
        postaction={decorate}
        ] (0,-0.5) -- ++(55:3);
     \draw[blue,ultra thick] (0,-0.5) -- ++(235:3);

     \draw[magenta,ultra thick, 
        decoration={markings, mark=at position 0.5 with {\arrow{>}}},
        postaction={decorate}
        ] (0,-1.2) -- ++(215:3);

 \draw[blue] (0,0) node[above left=5] {$\gamma_2^{(\{1\},-1)}$};

     \draw[magenta] (0,0) node[below right] {$\hat{\lambda}^2_{(\{1,2\},-1,-1)}$};
        
\end{tikzpicture}   
\begin{tikzpicture}[scale=0.5,baseline=(current  bounding  box.center)]

\useasboundingbox (-3,-3) rectangle (3,3);
    \draw[->] (0,-3) -- (0,3) node[left=2] {$v_3$};
     \draw[->] (2.5,0) -- (-2.5,0);
   \draw plot[only marks,mark=x,mark size=4pt,mark options={draw=red}] coordinates {(0,-1.2)};

     \draw[blue,ultra thick, 
        decoration={markings, mark=at position 0.9 with {\arrow{>}}},
        postaction={decorate}
        ] (0,-0.5) -- ++(55:3.5);
     \draw[blue,ultra thick] (0,-0.5) -- ++(235:3);

     \draw[magenta,ultra thick, 
        decoration={markings, mark=at position 0.75 with {\arrow{>}}},
        postaction={decorate}
        ] (0,0) -- ++(215:3);

          \draw[magenta,ultra thick] (0,0) -- ++(35:3);

 \draw[blue] (0,0) node[above left=5] {$\gamma_3^{(\{1\},-1)}$};

     \draw[magenta] (0,0) node[below right] {$\hat{\lambda}^3_{(\{1,2\},(-1,-1))}$};
        
\end{tikzpicture}   

\vspace{3em}
$\mathcal{C}^{(\{1\},-1)}_{(\{1,2,3\},(-1,-1,-23))}=$
\begin{tikzpicture}[scale=0.5,baseline=(current  bounding  box.center)]
\useasboundingbox (-3,-3) rectangle (3,3);
    \draw[->] (0,-3) -- (0,3) node[left=2] {$v_1$};
     \draw[->] (2.5,0) -- (-2.5,0);
    \draw plot[only marks,mark=x,mark size=4pt,mark options={draw=red}] coordinates {(0,-1.2)};

     \draw[blue,ultra thick] (0,-1.2) circle (0.5);

     \draw[magenta,ultra thick, 
        decoration={markings, mark=at position 0.5 with {\arrow{>}}},
        postaction={decorate}
        ] (0,-1.2) -- ++(215:2.2);

 \draw[blue] (0,0) node[above left=5] {$\gamma_1^{(\{1\},-1)}$};

     \draw[magenta] (0,0) node[below left=5] {$\begin{array}{c} \hat{\lambda}^1_{(\{1,3\},-1,-23)}\\ =\\ \hat{\lambda}^1_{(\{1,2,3\},-1,-1,-23)}\end{array}$};
        
\end{tikzpicture}   
\begin{tikzpicture}[scale=0.5,baseline=(current  bounding  box.center)]
\useasboundingbox (-3,-3) rectangle (3,3);
    \draw[->] (0,-3) -- (0,3) node[left=2] {$v_2$};
     \draw[->] (2.5,0) -- (-2.5,0);
    \draw plot[only marks,mark=x,mark size=4pt,mark options={draw=red}] coordinates {(0,-1.2)};

     \draw[blue,ultra thick, 
        decoration={markings, mark=at position 0.9 with {\arrow{>}}},
        postaction={decorate},name path=PBU
        ] (0,-0.5) -- ++(55:3);
     \draw[blue,ultra thick,name path=PBD] (0,-0.5) -- ++(235:3);

     \draw[magenta,ultra thick, 
        decoration={markings, mark=at position 0.75 with {\arrow{<}}},
        postaction={decorate},name path=PMR
        ] (0,-1.2) -- ++(215:3) node(d) {};

     \draw[magenta,ultra thick, 
        decoration={markings, mark=at position 0.75 with {\arrow{>}}},
        postaction={decorate}
        ] (0,0) -- ++(215:3) node(a) {};

          \draw[magenta,ultra thick,name path=PML] (0,0) -- ++(35:3);

 \draw[blue] (0,0) node[above left=5] {$\gamma_2^{(\{1\},-1)}$};

     \draw[magenta] (-0.2,0) node[below right] {$-\hat{\lambda}^2_{(\{1,2,3\},-1,-1,-23)}$};

        \draw[magenta] (2.8,2) node[above] {$\hat{\lambda}^2_{(\{1,2\},-1,-1)}$};

    \path[name intersections={of=PML and PBU,by=b}];
    
    \path[name intersections={of=PMR and PBD,by=c}];
    
    \filldraw[gray,opacity=0.6] (a.center) -- (b) -- (c) -- (d.center) -- cycle;

    \draw (b) node[above left=-2] {$\C$};

    \node[shift=({0.1,-0.4})] at (c) {$\C[1]$};

\end{tikzpicture}   
\begin{tikzpicture}[scale=0.5,baseline=(current  bounding  box.center)]

\useasboundingbox (-3,-3) rectangle (3,3);
    \draw[->] (0,-3) -- (0,3) node[left=2] {$v_3$};
     \draw[->] (2.5,0) -- (-2.5,0);
        \draw plot[only marks,mark=x,mark size=4pt,mark options={draw=red}] coordinates {(0,-1.2)};

     \draw[blue,ultra thick, 
        decoration={markings, mark=at position 0.9 with {\arrow{>}}},
        postaction={decorate}
        ] (0,-0.5) -- ++(55:3.5);
     \draw[blue,ultra thick] (0,-0.5) -- ++(235:3);

  \draw[magenta,ultra thick, 
        decoration={markings, mark=at position 0.5 with {\arrow{>}}},
        postaction={decorate}
        ] (0,-1.2) -- ++(215:3);

 \draw[blue] (0,0) node[above left=5] {$\gamma_3^{(\{1\},-1)}$};

     \draw[magenta] (0,0) node[below right] {$\begin{array}{c} \hat{\lambda}^3_{(\{1,3\},-1,-23)}\\ =\\ \hat{\lambda}^3_{(\{1,2,3\},-1,-1,-23)}\end{array}$};
        
\end{tikzpicture}

\caption{
The two complexes contributing to the vector space $\mathcal{H}^{(-\frac{1}{312})}_{(-\frac{59}{16})}=H_\ast(\mathcal{C}^{(\{1\},-1)}_{(\{1,2,3\},(-1,-1,-1))})[1]\oplus H_\ast(\mathcal{C}^{(\{1\},-1)}_{(\{1,2,3\},(-1,-1,-23))})[1]\cong \C \oplus 0$ corresponding to the $\tilde{q}^3$ term in the Poincar\'e series (\ref{cat-example2-series1}). The complexes are generated by $\gamma^{(\{1\},-1)}\cap \hat{\gamma}_{(\{1,2,3\},(-1,-1,-1))}$ 
 and $\gamma^{(\{1\},-1)}\cap \hat{\gamma}_{(\{1,2,3\},(-1,-1,-23))}$ respectively. 
For the first complex, we have $\hat{\gamma}_{(\{1,2,3\},(-1,-1,-1))}=\hat{\lambda}_{(\{1,2,3\},(-1,-1,-1))}-\hat{\lambda}_{(\{1,2\},(-1,-1))}-\hat{\lambda}_{(\{2,3\},(-1,-1))}+\hat{\lambda}_{(\{2\},-1)}$ where only the second term has a non-trivial intersection with $\gamma^{(\{1\},-1)}$. For the second complex, we have $\hat{\gamma}_{(\{1,2,3\},(-1,-1,-23))}=\hat{\lambda}_{(\{1,2,3\},(-1,-1,-23))}-\hat{\lambda}_{(\{1,3\},(-1,-23))}-\hat{\lambda}_{(\{2,3\},(-1,-23))}+\hat{\lambda}_{(\{3\},-23)}$ where only the first two terms have a non-trivial intersection with $\gamma^{(\{1\},-1)}$. The shaded region represents a holomorphic disk resulting in the boundary operator relating the two intersection points, which is responsible for the homology of the second complex being trivial.}
\label{fig:cat-example2-1}
\end{figure}

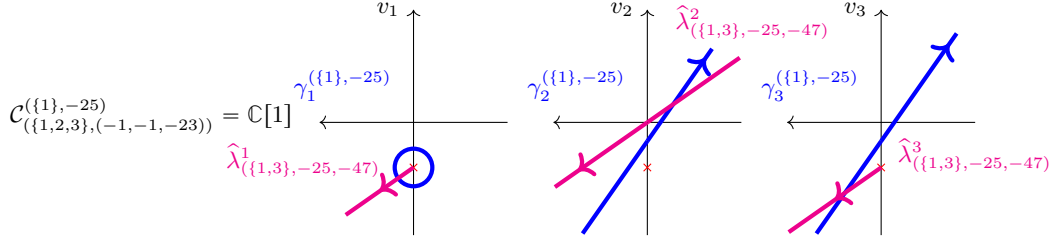
\begin{figure}
\centering
\footnotesize
$\mathcal{C}^{(\{1\},-25)}_{(\{1,2,3\},(-1,-1,-23))}=\C[1]$
\begin{tikzpicture}[scale=0.5,baseline=(current  bounding  box.center)]
\useasboundingbox (-3,-3) rectangle (3,3);
    \draw[->] (0,-3) -- (0,3) node[left=2] {$v_1$};
     \draw[->] (2.5,0) -- (-2.5,0);
    \draw plot[only marks,mark=x,mark size=4pt,mark options={draw=red}] coordinates {(0,-1.2)};

     \draw[blue,ultra thick] (0,-1.2) circle (0.5);

     \draw[magenta,ultra thick, 
        decoration={markings, mark=at position 0.5 with {\arrow{>}}},
        postaction={decorate}
        ] (0,-1.2) -- ++(215:2.2);

 \draw[blue] (0,0) node[above left=5] {$\gamma_1^{(\{1\},-25)}$};

     \draw[magenta] (-0.3,0) node[below left=5] {$\hat{\lambda}^1_{(\{1,3\},-25,-47)}$};
        
\end{tikzpicture}   
\begin{tikzpicture}[scale=0.5,baseline=(current  bounding  box.center)]
\useasboundingbox (-3,-3) rectangle (3,3);
    \draw[->] (0,-3) -- (0,3) node[left=2] {$v_2$};
     \draw[->] (2.5,0) -- (-2.5,0);
    \draw plot[only marks,mark=x,mark size=4pt,mark options={draw=red}] coordinates {(0,-1.2)};

     \draw[blue,ultra thick, 
        decoration={markings, mark=at position 0.9 with {\arrow{>}}},
        postaction={decorate},name path=PBU
        ] (0,-0.5) -- ++(55:3);
     \draw[blue,ultra thick,name path=PBD] (0,-0.5) -- ++(235:3);

     \draw[magenta,ultra thick, 
        decoration={markings, mark=at position 0.75 with {\arrow{>}}},
        postaction={decorate}
        ] (0,0) -- ++(215:3) node(a) {};

          \draw[magenta,ultra thick,name path=PML] (0,0) -- ++(35:3);

 \draw[blue] (0,0) node[above left=5] {$\gamma_2^{(\{1\},-25)}$};

        \draw[magenta] (2.8,2) node[above] {$\hat{\lambda}^2_{(\{1,3\},-25,-47)}$};

    \path[name intersections={of=PML and PBU,by=b}];

\end{tikzpicture}   
\begin{tikzpicture}[scale=0.5,baseline=(current  bounding  box.center)]

\useasboundingbox (-3,-3) rectangle (3,3);
    \draw[->] (0,-3) -- (0,3) node[left=2] {$v_3$};
     \draw[->] (2.5,0) -- (-2.5,0);
        \draw plot[only marks,mark=x,mark size=4pt,mark options={draw=red}] coordinates {(0,-1.2)};

     \draw[blue,ultra thick, 
        decoration={markings, mark=at position 0.9 with {\arrow{>}}},
        postaction={decorate}
        ] (0,-0.5) -- ++(55:3.5);
     \draw[blue,ultra thick] (0,-0.5) -- ++(235:3);

  \draw[magenta,ultra thick, 
        decoration={markings, mark=at position 0.5 with {\arrow{>}}},
        postaction={decorate}
        ] (0,-1.2) -- ++(215:3);

 \draw[blue] (0,0) node[above left=5] {$\gamma_3^{(\{1\},-25)}$};

     \draw[magenta] (0,0) node[below right=3] {$\hat{\lambda}^3_{(\{1,3\},-25,-47)}$};
        
\end{tikzpicture}

\caption{The complex contributing to the vector space $\mathcal{H}^{(-\frac{625}{312})}_{(-\frac{91}{16})}=H_\ast(\mathcal{C}^{(\{1\},-25)}_{(\{1,2,3\},(-25,-3,-47))})\cong \C$ corresponding to the $\tilde{q}^3$ term in the Poincar\'e series (\ref{cat-example2-series1}). The complex is generated by $\gamma^{(\{1\},-25)}\cap \hat{\gamma}_{(\{1,2,3\},(-25,-3,-47))}$, with $\hat{\gamma}_{(\{1,2,3\},(-25,-3,-47))}=\hat{\lambda}_{(\{1,2,3\},(-25,-3,-47))}-\hat{\lambda}_{(\{1,3\},(-25,-47))}$ where only the second term has a non-trivial intersection with $\gamma^{(\{1\},-25)}$.}
\label{fig:cat-example2-2}
\end{figure}

Consider Poincar\'e series of the vector space categorifying Stokes coefficients of jumps of the CS theory thimble with critical CS value $\frac{311}{312}\mod 1$ by thimbles with CS values $\frac{5}{16}\mod 1$:
\begin{multline}
   \mathcal{P}_{\left(\frac{5}{16}\right)}^{(\frac{311}{312})}(\tilde{q},t)  = \\ \tilde{q}^{\frac{427}{624}}\,\left(\tilde{q}+t\tilde{q}^2+\tilde{q}^{3}+\tilde{q}^{4}+t\tilde{q}^{8}+\tilde{q}^{9}+t\tilde{q}^{13}+\tilde{q}^{15}+2\tilde{q}^{16}
+2t\tilde{q}^{18}+(1+t)\tilde{q}^{19}+\ldots\right).
 \label{cat-example2-series1}
\end{multline}
The vector space corresponding to the $\tilde{q}^3$ term can be realized as $\mathcal{H}^{(-\frac{1}{312})}_{(-\frac{59}{16})}$ or as $\mathcal{H}^{(-\frac{625}{312})}_{(-\frac{91}{16})}$. In fact, there is an infinite number of ways to realize it. In the first case, we have 
\begin{equation}
    \mathcal{H}^{(-\frac{1}{312})}_{(-\frac{59}{16})}=H_\ast(\mathcal{C}^{(\{1\},-1)}_{(\{1,2,3\},(-1,-1,-1))})[1]\oplus H_\ast(\mathcal{C}^{(\{1\},-1)}_{(\{1,2,3\},(-1,-1,-23))})[1]\cong \C \oplus 0.
\end{equation}
The three intersection points that generate the relevant complexes are shown in Figure \ref{fig:cat-example2-1}.
In the second case, we have an isomorphic result:
\begin{equation}
    \mathcal{H}^{(-\frac{625}{312})}_{(-\frac{91}{16})}=H_\ast(\mathcal{C}^{(\{1\},-25)}_{(\{1,2,3\},(-25,-3,-47))})[1]\cong \C.
\end{equation} 
The single intersection point that generates the relevant complexes is shown in Figure \ref{fig:cat-example2-2}.

\subsection{Related perspectives}
\label{sec:wrappedstopped}

Although Fukaya categories are an active research area, their mathematical formulation is reasonably well developed when the target space is smooth and compact. In applications to Chern-Simons theory, even if we replace the infinite-dimensional space of complex gauge connections by a finite-dimensional model, the non-compactness (and, depending on the model, singularities) still requires attention and leads to alternative variants of the Fukaya category.

In the above we considered a finite-dimensional model based on the non-compact target space $\mathcal{V}$ and the potential function $S:\mathcal{V}\rightarrow\C$. There are several mathematical ways to approach this problem. First, one can study it directly \cite{kontsevich1998,MR1882336}, as a Landau-Ginzburg model of a pair $(\mathcal{V}, S)$. Objects of the resulting category $\text{FS} (\mathcal{V}, S)$ are supported on Lagrangian submanifolds in $\mathcal{V}$ that, outside of a compact subset, under $S$ project to $\frac{1}{ik} \R_+ + \text{const} \subset \C$. Alternatively, one can encode the information about the potential function $S$ in the ``stop'':
\begin{equation}
\mathfrak{f} \; := \; S^{-1} (- \infty)
\label{stop-def}
\end{equation}
and consider a Fukaya category of $\mathcal{V}$ where Lagrangian submanifolds are required to have a certain asymptotic behavior relative to the stop $\mathfrak{f}$. This also involves choices that lead to different versions of what one means by the ``Fukaya category of $(\mathcal{V} , \mathfrak{f})$.'' When Lagrangian submanifolds are required to avoid the stop \cite{MR3911570,MR4106794} we obtain a suitable partially wrapped Fukaya category $\mathcal{W} (\mathcal{V}, \mathfrak{f})$, whereas in the infinitesimal Fukaya category $\mathcal{F}_{\text{inf}} (\mathcal{V}, \mathfrak{f})$ Lagrangian submanifolds are required to tend to the stop at infinity \cite{MR2449059}.\footnote{In mirror symmetry, a choice of stop roughly corresponds to a choice of the compactification in the mirror B-model.} A categorification of the finite-dimensional model proposed in section \ref{sec:fd-categorification} was essentially based on the infinitesimal Fukaya category $\mathcal{F}_{\text{inf}} (\mathcal{V}, \mathfrak{f})$. In this subsection, we describe its alternative formulations, including $\text{FS} (\mathcal{V}, S)$ and $\mathcal{W} (\mathcal{V}, \mathfrak{f})$, which are all expected to be closely related:
\begin{equation}
\text{FS} (\mathcal{V}, S) \; \subseteq \;
\mathcal{W} (\mathcal{V}, \mathfrak{f}) \; \supseteq \;
\mathcal{F}_{\text{inf}} (\mathcal{V}, \mathfrak{f}).
\label{FS-categories}
\end{equation}
In addition, following Seidel, one can consider the directed category of vanishing cycles, i.e. the Fukaya category of a smooth fiber of $S:\mathcal{V}\rightarrow\C$. (See also \cite{MR3502098,MR3838112}.) This version works best in the setting of Lefschetz fibrations with a given choice of Lefschetz thimbles, which luckily happens to be the case of our interest.

\begin{figure}[ht]
	\centering
	\includegraphics[trim={0.3in 0.3in 0.3in 0.3in},clip,width=2.5in]{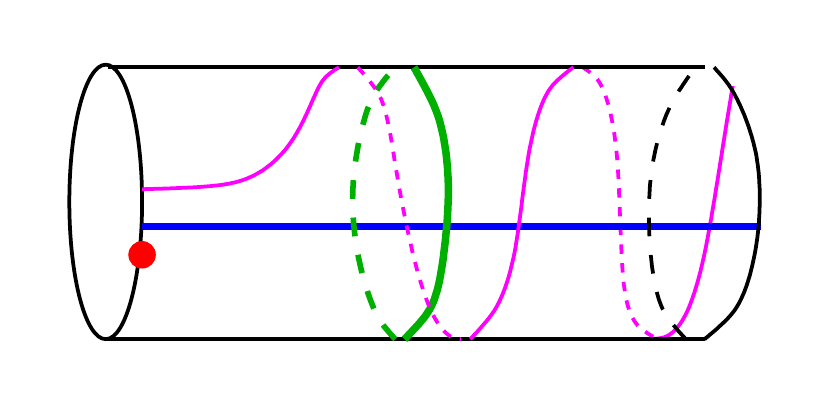}
	\caption{An illustration of wrapping in a simple example of $\mathcal{V} = \R \times S^1 \cong \C^*$. A stop $\mathfrak{f} = \textcolor{red}{\bullet}$ is shown in red. The core is shown in green and the cocore $L = \{ \R \times \text{pt} \}$ is shown in blue. The wrapped Fukaya category is generated by $L$ and, to compute morphisms, one needs to wrap $L$ until it runs to a stop (as shown in magenta).}
	\label{fig:cylinder4}
\end{figure}

In the partially wrapped Fukaya category $\mathcal{W} (\mathcal{V}, \mathfrak{f})$, the Hom-space between two Lagrangians $L_1$ and $L_2$ that avoid $\mathfrak{f}$ is defined to be the limit of the symplectic Floer homology of $(L_1^+, L_2)$, where $L_1^+$ is obtained from $L_1$ by applying a ``positive'' Hamiltonian isotopy. Equivalently, one can apply a negative Hamiltonian isotopy to $L_2$, or apply both transformations to $L_1$ and $L_2$. For example, with this definition it is easy to see that self-Homs of a Lagrangian object $L = \{ \R \times \text{pt} \}$ in $\R \times S^1 \cong \C^*$ with a linear potential give $\text{Hom} (L,L) = \C [x]$, in agreement with mirror symmetry that relates it to the category of coherent sheaves on $\C$. See Figure~\ref{fig:cylinder4}.

This simple example provides a good opportunity to illustrate an important difference between 
$\mathcal{W} (\mathcal{V}, \mathfrak{f})$ and $\mathcal{F}_{\text{inf}} (\mathcal{V}, \mathfrak{f})$ that we swept under the rug so far. Starting on the mirror side, there is clearly a difference between $\text{Coh} (\C)$ and $\text{Coh}_{\text{cpct}} (\C)$. Mirror symmetry relates these two versions of the category of coherent sheaves to $\mathcal{W} (\C^*, x)$ and $\mathcal{F}_{\text{inf}} (\C^*, x)$, respectively. Therefore, depending on the needs of a given problem, one must select the appropriate version accordingly. The former version, $\mathcal{W} (\C^*, x)$, is useful for accommodating infinite-dimensional spaces of morphisms that require infinitely many intersections.

When the target space is a Weinstein (exact symplectic) manifold, the partially wrapped Fukaya category is generated by cocores. And, in the presence of stops, one must also include linking disks (that can be thought of as cocores transverse to the cores that end on stops), so that altogether the partially wrapped Fukaya category is generated by cocores and linking disks \cite{CDGG17,MR4695507}. Stop removal corresponds to the quotient by cocores / linking disks.\footnote{Linking disks play an important role in knot-quiver correspondence \cite{Ekholm:2019lmb}, the geometry of which may have multiple connections to the subject of the present paper. We also point out that wrapped Fukaya categories of the type similar to the one studied here appear in recent works \cite{Lauda:2020tee} and \cite{LOT2021} on bordered Heegaard Floer homology. It would be interesting to explore all such connections further.}

We remind that a core (Lagrangian skeleton) of a Weinstein manifold $\mathcal{V}$ is the union of stable manifolds for the flow of the associated Liouville vector field $\xi$, where $\iota_\xi \omega = \theta$ is a Liouville form associated with the symplectic form $\omega$:
\begin{equation}
\text{Skel} (\mathcal{V}, \theta) \; = \; \text{core}_{\xi} (\mathcal{V}).
\end{equation}
Its tubular neighborhood encodes the symplectic topology of the ambient exact symplectic manifold. Locally, the Hamiltonian $H$ can be defined as the half distance squared from $\text{core}_{\xi} (\mathcal{V})$, and the retraction $\pi : \mathcal{V} \to \text{core}_{\xi} (\mathcal{V})$, induced by the metric gradient of $H$, has smooth Lagrangian fibers over generic points of $\text{core}_{\xi} (\mathcal{V})$. The non-compact ends of the skeleton play the role of stops since they represent directions along which wrapping is not allowed. Therefore, a natural generalization of a skeleton (core) of $(\mathcal{V}, \theta)$ equipped with a stop $\mathfrak{f}$ is the union with the ``conicalization'' of the stop:
\begin{equation}
\text{core}_{\xi} (\mathcal{V}) \; \cup \; \{ x \in \mathcal{V} ~\vert~ \phi^t_{\xi} (x) \in \mathfrak{f} \text{ for some } t \}
\label{coreunion}
\end{equation}
and admissible Lagrangian submanifolds must avoid it outside a compact set. Here, $\phi^t_{\xi}$ denotes a finite-$t$ transformation generated by vector field $\xi$.

\begin{figure}[ht]
	\centering
	\includegraphics[trim={0.3in 0.5in 0.3in 0.3in},clip,width=2.5in]{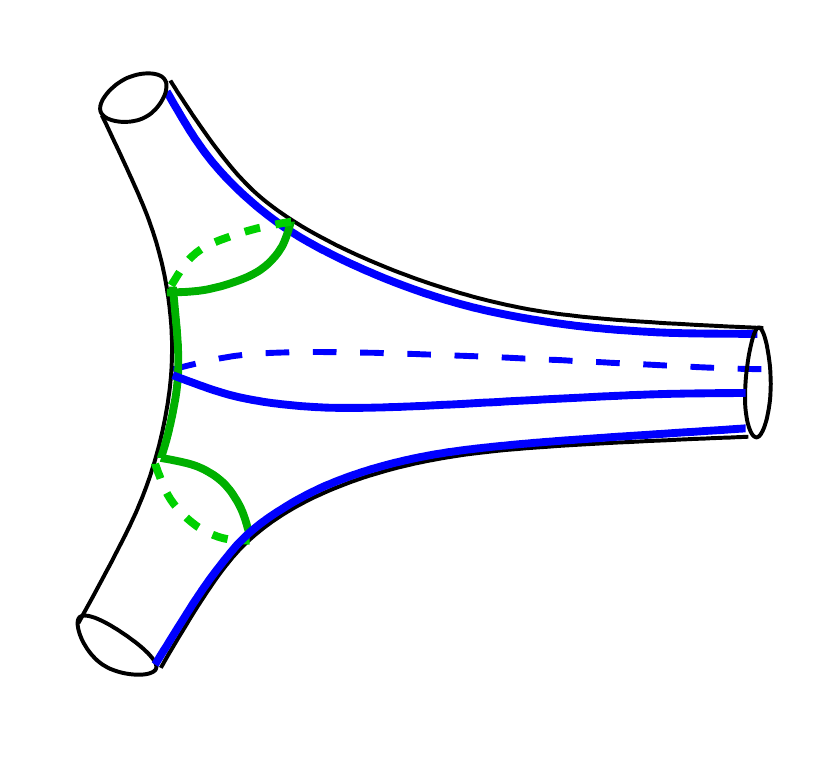}
	\caption{Another illustration of the core (shown in green) and cocores (shown in blue) for a 3-punctured sphere.}
	\label{fig:curve3}
\end{figure}

Note, in this general framework, a stop can be either a Weinstein hypersurface $F \subset \partial_{\infty} \mathcal{V}$ or a Legendrian submanifold $\Lambda \subset \partial_{\infty} \mathcal{V}$, cf. \cite{MR4634745}. The relation between these two versions is
\begin{equation}
\Lambda \; =  \;\text{core}_\xi (F)
\end{equation}
and the following equivalence holds \cite{MR4695507}:
\begin{equation}
\mathcal{W} (\mathcal{V}, F) \; \cong \; \mathcal{W} (\mathcal{V}, \text{core} (F)).
\end{equation}
Thanks to this equivalence, we can use the two versions of the ``stop'' interchangeably, sometimes referring to a Legendrian submanifold of the contact manifold $(\partial_{\infty} \mathcal{V}, \theta)$ and sometimes to a Weinstein hypersurface at infinity in $\mathcal{V}$.

\begin{figure}
\centering
\begin{tikzpicture}
    \draw[->] (0,-4) -- (0,4.5) node[right=2] {$\re v$};
     \draw[->] (3,0) -- (-3,0) node[above=3] {$\im v$};
    \draw plot[only marks,mark=x,mark size=4pt,mark options={draw=red}] coordinates {(0,1) (0,3) (0,-1) (0,-3)};

\draw[blue,ultra thick, 
        decoration={markings, mark=at position 0.7 with {\arrow{>}}},
        postaction={decorate}
        ] (-2.7,0.1) -- (2.7,0.1);

\draw[blue,ultra thick, 
        decoration={markings, mark=at position 0.7 with {\arrow{>}}},
        postaction={decorate}
        ] (-2.7,2) -- (2.7,2);

\draw[blue,ultra thick, 
        decoration={markings, mark=at position 0.7 with {\arrow{>}}},
        postaction={decorate}
        ] (-2.7,-2) -- (2.7,-2);

\draw[magenta,ultra thick, 
        decoration={markings, mark=at position 0.7 with {\arrow{>}}},
        postaction={decorate}
        ] (0.1,1) -- (2.7,1);

\draw[magenta,ultra thick, 
        decoration={markings, mark=at position 0.7 with {\arrow{>}}},
        postaction={decorate}
        ] (-0.1,1) -- (-2.7,1);

\draw[magenta,ultra thick, 
        decoration={markings, mark=at position 0.7 with {\arrow{>}}},
        postaction={decorate}
        ] (0.1,3) -- (2.7,3);

\draw[magenta,ultra thick, 
        decoration={markings, mark=at position 0.7 with {\arrow{>}}},
        postaction={decorate}
        ] (-0.1,3) -- (-2.7,3);

\draw[magenta,ultra thick, 
        decoration={markings, mark=at position 0.7 with {\arrow{>}}},
        postaction={decorate}
        ] (0.1,-1) -- (2.7,-1);

\draw[magenta,ultra thick, 
        decoration={markings, mark=at position 0.7 with {\arrow{>}}},
        postaction={decorate}
        ] (-0.1,-1) -- (-2.7,-1);

\draw[magenta,ultra thick, 
        decoration={markings, mark=at position 0.7 with {\arrow{>}}},
        postaction={decorate}
        ] (0.1,-3) -- (2.7,-3);

\draw[magenta,ultra thick, 
        decoration={markings, mark=at position 0.7 with {\arrow{>}}},
        postaction={decorate}
        ] (-0.1,-3) -- (-2.7,-3);

\draw[green,ultra thick] (0,1) circle (0.4)  node[left=9] {};

\draw[green,ultra thick] (0,-0.6) -- (0,0.6);

\draw[green,ultra thick] (0,1.4) -- (0,2.6);

\draw[green,ultra thick] (0,3.4) -- (0,4.0);

\draw[green,ultra thick] (0,-1.4) -- (0,-2.6);

\draw[green,ultra thick] (0,-3.4) -- (0,-4);

\draw[green,ultra thick] (0,3) circle (0.4)  node[left=9] {} node[left = 30] {};

\draw[green,ultra thick] (0,-1) circle (0.4)  node[right=9] {};

\draw[green,ultra thick] (0,-3) circle (0.4)  node[right=9] {};

\end{tikzpicture}
\caption{The core (shown in green) and cocores (shown in blue and magenta) for a complex $v$-plane with an infinite set of punctures (points removed).}
\label{fig:vcocores}
\end{figure}
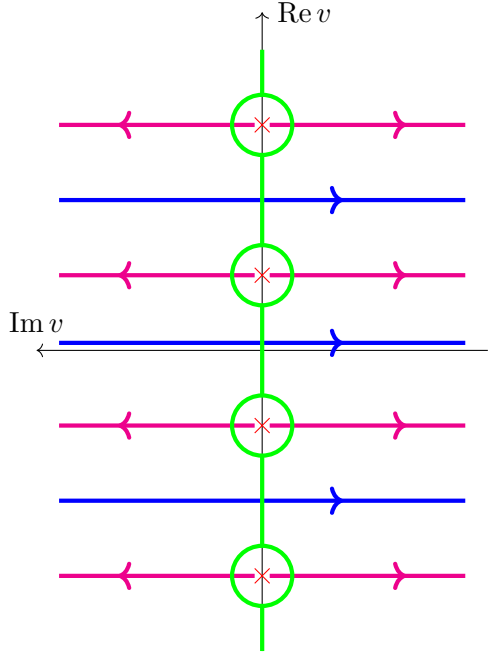

For a generic point $x$ that belongs to mid-dimensional strata of $\text{core}_{\xi} (\mathcal{V})$ or, more generally, to \eqref{coreunion}, the cocore is
\begin{equation}
U_x \; := \; \{ y \in \mathcal{V} ~\vert~ \lim_{t \to - \infty} \phi^t_v (y) = x \}.
\end{equation}
In the simple example above, illustrated in Figure~\ref{fig:cylinder4}, $\{ 0 \} \times S^1$ is the core of $T^* S^1 = \R \times S^1$, and $L = \R \times \{ \text{pt} \}$ is the cocore. A more interesting example of a 3-punctured sphere is shown in Figure~\ref{fig:curve3}. Here, the core consists of three components (two circles connected by an interval), and there are three corresponding cocores. This example can be generalized further to a sphere (or a complex line $\C$) with an infinite set of points removed, say at all integer values, $v \in \Z$. This generalization can be viewed as a limit of the wrapped Fukaya categories studied in \cite{MR3073884} and describes a model very close to our needs; it supports an infinite collection of cocores illustrated in Figure~\ref{fig:vcocores}.

An important property of the partially wrapped Fukaya category is that it is multiplicative under the product operation on target spaces (equipped with stops). This has direct relevance to our target space $\mathcal{V}$ defined as the product over high-valency vertices, with a collection of planes removed:
$$
\mathcal{V} \; = \; \Big\{ \, v_I\in \C,\;I\in H\;|\; v_I \text{ is not a pole/zero of }R(v) = \frac{ \prod_{a\in L} \sin \cfrac{\pi v_{h(a)}}{P_a}}{\prod_{I\in H}(\sin\pi v_I)^{\deg(I)-2}} \, \Big\} \; \subset \; \C^{|H|}.
$$
Due to this factorization structure, the Fukaya category for the problem at hand is generated by products of Lagrangian objects in each $v_I$-plane, making them easy to classify. This, in turn, leads to the product structure in \eqref{contourproduct}, modeling $\gamma^{(H',\mathbb{n})}=\bigtimes_{I\in H}\gamma_I^{(H',\mathbb{n})}$. Moreover, our target space $\mathcal{V}$ comes equipped with a quadratic potential function $S : \mathcal{V} \to \C$, which determines the stop(s) and the behavior of Lagrangians at infinity. If $\mathcal{V}$ was a single complex plane parametrized by $v$, there would be two stops restricting the wrapping of Lagrangian objects. With multiple $v_I$'s, the stop is connected, but still plays a similar role, restricting the wrapping and determining the directionality of Lagrangians in a way that agrees with what was illustrated in Figure~\ref{fig:lambda-contours}.

Comparing to Figure~\ref{fig:lambda-contours}, it is easy to see that cocores in Figure~\ref{fig:vcocores} indeed correspond to basic building blocks of the purely combinatorial model: $\lambda_I^{(H',\mathbb{n})}$, $\check{\lambda}^I_{(H',\mathbb{n})}$, and $\hat{\lambda}^I_{(H',\mathbb{n})}$. There are a few delicate points, though, that one should keep in mind. One subtlety is whether Hom-spaces $\mathcal{H}^{(S)}_{(S')}$ should be finite-dimensional or not when $S = S'$. From the perspective of the wrapped Fukaya category, infinite-dimensional spaces are more natural because the only stops are at infinity and nothing prevents wrapping near boundary components at finite values of $v_I$. Another subtle point is the grading assignment, cf. \cite{MR3073884,MR3838112}.

\section{Hemisphere partition function interpretation}
\label{sec:hemisphere}

In this section we show how one can formally interpret the integrals
\begin{equation}
\int_{\gamma^{(H',\mathbb{n})}} d^{|H|}v \, R(v) \,e^{2\pi i k\,S(v)}
\label{fd-integral-to-hemisphere}
\end{equation}
of finite-dimensional model, that appeared in Section \ref{sec:stokes-plumbed}, as a hemisphere partition function of 2d $\mathcal{N}=(2,2)$ theories \cite{Hori:2013ika}, both B-twisted with A-type boundary conditions and A-twisted with B-type boundary conditions. We also comment on why the two interpretations \textit{should not} be understood as mirror to each other. We provide interpretations using the notations of \cite{Hori:2013ika,Hori:2000ck} and point out similarities (and differences) to other candidates for finite-dimensional models where $\mathcal{M}_{\text{flat}} (G_{\mathbb{C}}, Y)$ is realized as (part of) the critical set, $\text{Crit} (W)$, of a superpotential function $W: X \to \mathbb{C}$.

\subsection{B-twist with A-type boundary conditions}
\label{sec:B-model}

Up to a simple overall factor (a power of the hemisphere radius and the UV cutoff parameter), the integral (\ref{fd-integral-to-hemisphere}) can be interpreted as the partition function preserving $\mathrm{B}_{(+)}^{\frac{\arg k}{2}}$-type supersymmetry on a hemisphere of radius $r=|k|$ of Landau-Ginzburg model with the target (\ref{fd-A-model-target}) equipped with the holomorphic volume form $\Omega=d^{|H|}v\,R(v)$ and the superpotential $W(v)=-S(v)$. The boundary condition should be considered as A-type, given by the Lagrangian submanifold $L_+=\gamma^{(H',\mathbb{n})}$.

\begin{figure}[ht]
	\centering
	\includegraphics[trim={0.1in 0.1in 0.1in 0.1in},clip,width=2.0in]{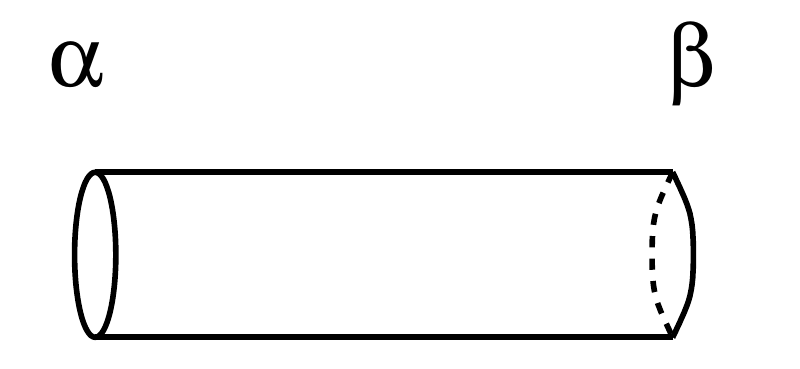}
	\caption{Two-dimensional topological theory realization of the Stokes coefficients.}
	\label{fig:cylinder2}
\end{figure}

As usual, the fact that A-type boundary conditions appear in the B-model comes from the topological twist and has to do with the curvature of the underlying geometry, namely the hemisphere in our problem at hand. If we instead consider a cylinder $S^1 \times [0,1]$ with a flat metric, the topological twist will have the same effect in all parts of the geometry and we instead obtain A-type supersymmetry preserved in the bulk and at the boundary. If, moreover, we choose the boundary conditions labeled by complex flat connections (including, possibly, flat connections at infinity), then this setup, illustrated in Figure~\ref{fig:cylinder2}, will be precisely a physical realization of \eqref{Stokes-series-definition}. This A-model setup has a natural categorification, which corresponds to replacing $S^1$ with $\R$ and leads to the categorification of the Stokes coefficients proposed in section \ref{sec:fd-categorification}. In other words, it is the A-model on a strip, $\R \times [0,1]$ with target space \eqref{fd-A-model-target} and the superpotential $S:\mathcal{V}\rightarrow\C$, much like the prototypical examples considered in \cite{Hori:2000ck}.

\begin{figure}[ht]
	\centering
	\includegraphics[trim={0.1in 0.1in 0.1in 0.1in},clip,width=2.0in]{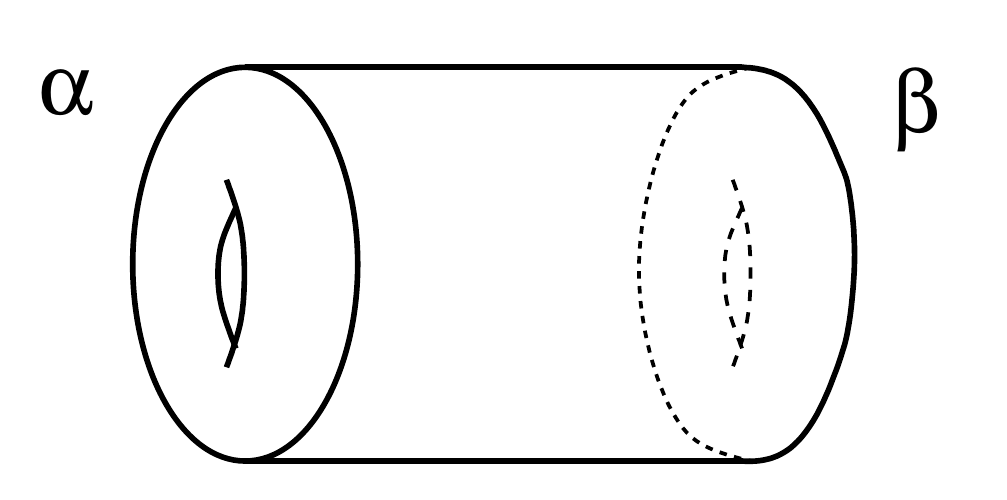}
	\caption{Thee-dimensional $\mathcal{N}=2$ theory on $T^2 \times [0,1]$ with a partial A-model twist along $S^1 \times [0,1]$.}
	\label{fig:cylinder3}
\end{figure}

It is natural to ask how this physical setup compares to other finite-dimensional models, in particular to the effective 3d $\mathcal{N}=2$ theory $T[Y]$ on $S^1$ or to 6d $(0,2)$ theory on $S^1 \times Y$. While the latter two are equivalent, and lead to a 2d theory with $\mathcal{N} = (2,2)$ supersymmetry very similar\footnote{\label{foot:allvacua} In particular, the space of vacua of the A-twisted theory $T[Y]$, denoted $K^0 (\text{MTC} [Y])$ in \cite{Gukov:2016gkn}, also has a basis indexed by $\alpha$, $\beta$, etc. that includes complex flat connections at infinity.} to the one discussed here, they are {\it not} the same as the A-model studied in this paper. One crucial difference is that the low-energy fields in 3d-3d correspondence (or in its 6d origin) are essentially holonomies of the complex flat connection in a surgery presentation of $Y$ (cf. Section \ref{sec:flat-at-infinity}), or in a Heegaard decomposition of $Y$, or in other ways of constructing $Y$. In the simplest non-trivial case of $G_{\mathbb{C}} = SL(2,\C)$, this leads to various finite-dimensional models of complex (a.k.a. analytically continued) Chern-Simons theory where the target space $X$ contains numerous $\C^*$ factors, which play an important role in reproducing generalized symmetries of the 6d theory on $Y$ and in other aspects of 3d-3d correspondence \cite{Chun:2019mal}. In these models, $X$ comes equipped with a potential function $W$ such that
\begin{equation}
\mathcal{M}_{\text{flat}} (G_{\mathbb{C}}, Y) \; \subsetneq \; \text{Crit} (W).
\label{vacua3d3d}
\end{equation}
Although our target space here looks similar, with $\C$ factors instead of $\C^*$, there is no simple relation between spaces $\mathcal{V}$ and $X$, in particular we can not say that one is a cover of the other. One reason is that, modulo $\Z$ action, an isolated flat connection corresponds to a \emph{single} point in multiplicative (or, K-theoretic) target spaces $X$ that arise in 3d $\mathcal{N}=2$ theories or other string constructions, whereas in general it corresponds to \emph{multiple} points in the finite-dimensional model considered here, based on the target space $\mathcal{V}$. There is also no localization argument (for $|H|>1$) that could justify the reduction of the infinite-dimensional gauge theory model to the finite-dimensional model with target $\mathcal{V}$.

Nevertheless, as explained in Section~\ref{sec:stokes-plumbed}, the Stokes coefficients in the finite-dimensional model of this paper should agree with those in infinite-dimensional QFT setting(s). It is also possible that the A-model considered here produces the same categorification of the Stokes coefficients as 3d theory $T[Y]$ on $\R \times S^1 \times [0,1]$ with two boundary conditions labeled by $\alpha$ and $\beta$. The fact that the space of vacua in \eqref{vacua3d3d} is strictly larger than the space of complex flat connections may, in fact, be favorable for accomplishing the equivalence with the A-model of this paper and, in particular, recovering flat connections at infinity (see Section \ref{sec:flat-at-infinity} and footnote \ref{foot:allvacua}).

\subsection{A-twist with B-type boundary conditions}
\label{sec:A-model}

The basic idea that we will follow is to interpret (\ref{fd-integral-to-hemisphere}) as a partition function on a hemisphere preserving A-type supersymmetry of a GLSM with gauge group $U(1)^{H}$, so that the integration variables $v_I$ are interpreted as vacuum expectation values of scalars in the vector multiplets. This strategy is frequently used in other applications of 3d-3d correspondence. Then, various factors in $R(v)$, given explicitly by \eqref{R-expression-high-val-g}, are interpreted as the contributions from charged matter multiplets in the bulk and the boundary Chan-Paton factor. The ``action'' $S(v)$ plays the role of the twisted superpotential. The boundary condition should be considered to be of B-type. The choice of the contour of integration $\gamma^{(H',\mathbb{n})}$ can be interpreted as the choice of the boundary conditions on the vector multiplets. However, it is more natural to have instead dependence only on the choice of boundary Chan-Paton vector space, and keep the integration contour fixed, similarly to what was considered in \cite{Hori:2013ika}.
In our setting we can recast the contour dependence into the dependence on the Chan-Paton factor only when $k\in \Z$. Such a restriction is not natural from the point of view of analytically continued Chern-Simons theory. Moreover, the resulting Chan-Paton vector spaces depend on the value of $k\in \Z$. All this makes the interpretation of the GLSM as a mirror of the LG model considered in Section \ref{sec:B-model} somewhat problematic.

To make the statements above more explicit, consider the case of weakly negative-definite plumbings. In this case, the contours factorize into the contours shown in Figure \ref{fig:contours-basis}. In each $v_I$-plane consider straight-line contours 
\begin{equation}
\rho_I^{(\mathbb{m}_I)}=-\frac{1}{2}+\mathbb{m}_I+\frac{1}{\sqrt{ik}}\R\;\subset \;\C,\qquad \mathbb{m}_I\in \Z.
\end{equation}
The difference between a pair of such contours is a disjoint union of the contours encircling the poles of $R(v)$ that lie on the real axis between the two contours (cf. Figure \ref{fig:contour-shift}). Therefore any thimble contour $\gamma^{(H',\mathbb{n})}_I$ can be realized as an integral linear combination of the contours 
\begin{equation}
\rho^{(\mathbb{m})}=\bigtimes_{I\in H}\rho^{(\mathbb{m}_I)}.    
\end{equation}
It is then enough to consider integrals over such contours. By a shift of the integration variables they all can be written as an integral over the contour $\rho^{(0)}$:
\begin{equation}
\int\limits_{\rho^{(\mathbb{m})}}R(v)\,e^{2\pi ikS(v)}
=    \int\limits_{\rho^{(0)}}R(v+\mathbb{m})\,e^{2\pi ikS(v+\mathbb{m})}.
\label{int-rho-contour-shift}
\end{equation}
Next, choose $b_I\in\Z_{>0}$, $I\in H$, such that
\begin{equation}
    \frac{b_{h(a)}}{P_a}\in\Z,\;\forall a\in L\qquad
    \text{and}\qquad    C^{-1}_{IJ}b_J\in\Z,\;\forall I,J\in H,
\end{equation}
and assume $k\in\Z$. After a change of integration variables $v_I=2irb_I\sigma_I$, where $r>0$ will play the role of the hemisphere radius, the integral (\ref{int-rho-contour-shift}), up to a $k$- and $\mathbb{m}$-independent constant, can be written in the following form:
\begin{equation}
  \int\limits_{\bigtimes\limits_{I\in H}\frac{i}{2b_I}+\sqrt{\frac{i}{k}}\R}
  d^{|H|}\sigma
\frac{e^{2\pi ikr^2\sum_{I,J}\sigma_{I}\sigma_{J}C^{-1}_{IJ}b_Ib_J}}{\prod_{I}(\sinh 2\pi rb_I\sigma_I)^{\deg(I)-2+2g_I}}\,\left(\sum_{i}e^{2\pi ir\mathfrak{q}_i^T\sigma}\,e^{\pi i\mathfrak{r}_i}\right)
\label{integral-over-sigma-vars}
\end{equation}
for some $\mathfrak{q}_i\in\Z^H,\mathfrak{r}_i\in\Q$ (possibly repeating). The integral (\ref{integral-over-sigma-vars}) then can be interpreted as a hemisphere partition function of GLSM, such that for each high-valency vertex $I\in H$ we have $U(1)$ gauge group and $\deg(I)-2+2g_I$ hypermultiplets with charge $b_I$ under this group. The model has twisted superpotential
\begin{equation}
    \tilde{W}(\sigma)=kr\sum_{I,J}\sigma_{I}\sigma_{J}C^{-1}_{IJ}b_Ib_J
\end{equation}
and the Chan-Paton vector space consisting of the states with $U(1)^H$ and R-symmetry charges $\mathfrak{q}_i$ and $\mathfrak{r}_i$ respectively.

Apart from the issues mentioned earlier, we see a few more. First, the overall coefficient in the twisted superpotential $\tilde{W}(\sigma)$ depends on the radius. Another important issue is that increasing the genus $g_I$ contributes two ordinary ($\C^2$ valued) hypermultiplets, whereas from compactification of 6d theory on $Y$ we expect $\C^* \times \C$ valued hypermultiplets \cite{Chun:2019mal}. The overall number of scalar 0-modes agrees, but the distinction between periodic (circle-valued) and non-compact ($\R$-valued) scalars is crucial for reproducing correctly the generalized global symmetries of the 3d theory $T[Y]$. This suggests that one should consider K-theoretic versions of the partition functions and moduli spaces.

\section*{Acknowledgments} We wish to thank Mohammed Abouzaid, Mina Aganagic, Denis Auroux, Joel Beimler, Pierrick Bousseau, Francesco Costantino, Ovidiu Costin, Gerald V. Dunne, Francesca Ferrari, Sheel Ganatra, Vasily Golyshev, Cagri Karakurt, Ludmil Katzarkov, William Mistegård, Du Pei, Mauricio Romo, Paul Seidel, Vivek Shende, Josef Svoboda, and Peng Zhou for useful discussions.
The work of SG is supported in part by a Simons Collaboration Grant on New Structures in Low-Dimensional Topology, by the NSF grant DMS-2245099, and by the U.S. Department of Energy, Office of Science, Office of High Energy Physics, under Award No. DE-SC0011632.

\appendix

\section{Borel transform technicalities}
\label{app:borel}

In this section we provide a short argument of why an overall coefficient of the form 
\begin{equation}
    c\,k^{a/2}e^{\frac{2\pi ib}{k}},\qquad c\in \C,\;b\in \Q,\;a\in \Z
\end{equation}
 in front of an oscillatory integral of the form (\ref{finte-dimensional-integral-general-contour}) does not affect the structure of the Stokes phenomenon and the Stokes coefficients in particular. 

First note that multiplication by $k^{a/2}$ for positive (negative) even $a\in 2\Z$ is equivalent to applying $2|a|$-th derivative (integration) to the Borel transforms of the perturbative expansions. This does not affect the structure of singularities (\ref{B-singularity}). 

Thus we can focus on the case of $a=-1/2$. The whole coefficient itself then can be realized as an oscillatory integral of the form
\begin{equation}
    \int dx\,e^{2\pi i \sqrt{b}\,x}\,e^{-\pi ikx^2/2}.
\end{equation}
One then can combine the corresponding complex plane of the $x$ integration variable with the $\C^{|H|}$ integration space of the $v_I$ variables, as considered in Section \ref{sec:stokes-plumbed}. The $R(v)$ and $S(v)$ functions will be modified accordingly:
\begin{equation}
    R(v)\longrightarrow R(v,x):=R(v)\,e^{2\pi i \sqrt{b}\,x},\qquad S(v)\longrightarrow S(v,x):=S(v)-x^2/4.
\end{equation}
In particular the new function $R(x,v)$ has the same poles as $R(v)$. The critical points of the new $S(v,x)$ (with $v$ possibly restricted to the poles of $R(v)$) are simply the critical points of $S(v)$ in the space of $v_I$ variables with $x=0$.
The new ``Lefschetz thimble contours'' are the Cartesian products of the thimble contours in the space of $v_I$ variables, same as before, and a line $\frac{1}{\sqrt{ik}}\R$ in the $x$-plane. Since $R(x,v)$ is holomorphic in $x$, it is clear the new finite-dimensional oscillatory integral has the same structure of the Stokes jumps as before.

\section{Another derivation of the finite-dimensional integral representation}
\label{app:another-derivation}

In this section we present a slightly alternative derivation of the final expression (\ref{Z-hat-to-LG-model}). Such a derivation, although being more involved, exhibits a direct relation between the integration variables $v_I$ and the variables $z_I$ in the integral representation of the $\hat{Z}$ invariants of plumbed 3-manifolds of \cite{Gukov:2017kmk}. We will restrict again ourselves to the case of weakly negative-definite plumbings that are integer homology spheres and use the same notations for the plumbing graph data as in Section \ref{sec:stokes-plumbed}. 

We now start with the following expression for $Z(k)\pto \hat{Z}(q)$ ($q=e^{\frac{2\pi i}{k}}$) \cite{Gukov:2017kmk}:
\begin{equation}
    Z(k)\pto \sum_{\pm}
    \int\limits_{|z_i|=1\pm \epsilon} \prod_{i\in V}\frac{dz_i}{z_i}
    \left(z_i-1/z_i\right)^{2-\deg (i)}\,
    \sum_{n\in \Z^V} q^{-\frac{n^TB^{-1}n}{4}}\,\prod_{i\in V}z_i^{n_i}
\end{equation}
where the signs $\pm$ are summed over independently for each integration variable. The integral over $z_i$ for $i\in V\setminus H$ can be easily performed\footnote{One can skip this first step and proceed similarly with the following steps to obtain instead (\ref{Z-hat-to-LG-big}) as the final expression.}:
\begin{multline}
    Z(k)\pto
    \sum_{s\in\{\pm 1\}^L}\sum_{m\in \Z^H}
    \prod_{a\in L}s_a
    \sum_{\pm}\int_{|z_I|=1\pm \epsilon}
    \prod_{I\in H}\frac{dz_I}{z_I}
    \left(z_I-1/z_I\right)^{2-\deg (I)}\,\times 
    \\
    \sum_{m\in \Z^H} q^{-\frac{m^TCm}{4}-\frac{m^TDs}{2}-\frac{s^TAs}{4}}\,\prod_{I\in H}z_I^{m_I}.
\end{multline}
Next, we make the change of integration variables $z_I=\exp \pi iv_I$:
\begin{multline}
    Z(k)\pto 
     \sum_{s\in\{\pm 1\}^L}\sum_{m\in \Z^H}
    \prod_{a\in L}s_a\,
    q^{-\frac{s^TAs}{4}}
    \sum_{\pm}\int\limits_{v_I\in [0,2]\pm i\epsilon}
    \prod_{I\in H}{dv_I}
    \left(\sin \pi v_I\right)^{2-\deg (I)}\,
    \times 
    \\
    \Theta\left(-\frac{1}{2k}\,C;\frac{1}{2}\,v-\frac{1}{2k}\,Ds\right)
    \label{alt-derivation-with-theta}
\end{multline}
where we have used the general notion of theta-function defined for a symmetric $N\times N$ matrix $\mathcal{T}$ with $\Im \mathcal{T}>0$ and a complex vector $y\in \C^N$:
\begin{equation}
    \Theta(\mathcal{T};y):=
    \sum_{m\in \Z^N}
    e^{\pi i m^T\mathcal{T}m+2\pi im^Ty}.
\end{equation}
It has the following modular transformation property:
\begin{equation}
   \Theta(\mathcal{T};y)=(\det \mathcal{T})^{-\frac{1}{2}}
   e^{-\pi iy^T\mathcal{T}y}\,\Theta(-\mathcal{T}^{-1};\mathcal{T}^{-1}y),
\end{equation}
up to a $\mathcal{T}$- and $y$-independent constant. Using it, we can rewrite (\ref{alt-derivation-with-theta}) as follows:
\begin{multline}
Z(k)       \pto 
     \sum_{s\in\{\pm 1\}^L}\sum_{m\in \Z^H}
    \prod_{a\in L}s_a\,
    q^{-\frac{s^TAs}{4}}
    \sum_{\pm}\int\limits_{v_I\in [0,2]\pm i\epsilon}
    \prod_{I\in H}{dv_I}
    \left(\sin \pi v_I\right)^{2-\deg (I)}\,
    \times 
    \\
    e^{2\pi ik(v/2-Ds/2k)^TC^{-1}(v/2-Ds/2k)}
\,  \Theta\left(2k\,C^{-1};k\,C^{-1}v-C^{-1}Ds\right).
\end{multline}
As in Section \ref{sec:stokes-weakly-negative} we can use the fact that the matrix $A-D^TC^{-1}D$ is diagonal to write the expression in the following form:
\begin{multline}
Z(k)       \pto 
     \sum_{s\in\{\pm 1\}^L}\sum_{m\in \Z^H}
    \prod_{a\in L}s_a\,
    \sum_{\pm}\int\limits_{v_I\in [0,2]\pm i\epsilon}
    \prod_{I\in H}{dv_I}
    \left(\sin \pi v_I\right)^{2-\deg (I)}\,
    \times 
    \\
\sum_{m\in \Z^H}e^{\frac{\pi ik}{2}
(v-2m)^TC^{-1}(v-2m)
}\,e^{-\pi i(v-2m)^TC^{-1}Ds}
\\
\pto
\sum_{m\in \Z^H}
        \sum_{\pm}\int\limits_{v_I\in [0,2]\pm i\epsilon}
    \prod_{I\in H}{dv_I}
\sum_{m\in \Z^H} R(v-2m)\,e^{2\pi ikS(v-2m)}.
\end{multline}
Finally, we can recast the sum over $m\in \Z^H$ into a sum over shifts of the interval integration contours:
\begin{multline}
Z(k)       
\pto
\sum_{m\in \Z^H}
        \sum_{\pm}\int\limits_{v_I\in [0,2]-2m\pm i\epsilon}
    \prod_{I\in H}{dv_I}
\sum_{m\in \Z^H} R(v)\,e^{2\pi ikS(v)}=\\
=        \sum_{\pm}\int\limits_{v_I\in \R\pm i\epsilon}
    \prod_{I\in H}{dv_I}
\sum_{m\in \Z^H} R(v)\,e^{2\pi ikS(v)}.
\end{multline}

\vspace{10ex}
\bibliography{CS}

\providecommand{\href}[2]{#2}\begingroup\raggedright\begin{thebibliography}{10}

\bibitem{Gukov:2004hz}
S.~Gukov, A.S.~Schwarz and C.~Vafa, \emph{{Khovanov-Rozansky homology and
  topological strings}},
  \href{https://doi.org/10.1007/s11005-005-0008-8}{\emph{Lett. Math. Phys.}
  {\bfseries 74} (2005) 53}
  [\href{https://arxiv.org/abs/hep-th/0412243}{{\ttfamily hep-th/0412243}}].

\bibitem{Gukov:2007ck}
S.~Gukov, \emph{{Gauge theory and knot homologies}},
  \href{https://doi.org/10.1002/prop.200610385}{\emph{Fortsch. Phys.}
  {\bfseries 55} (2007) 473} [\href{https://arxiv.org/abs/0706.2369}{{\ttfamily
  0706.2369}}].

\bibitem{witten2011fivebranes}
E.~Witten, \emph{Fivebranes and knots}, {\emph{Quantum Topology} {\bfseries 3}
  (2011) 1}.

\bibitem{khovanov2000}
M.~Khovanov, \emph{{A categorification of the Jones polynomial}},
  \href{https://doi.org/10.1215/S0012-7094-00-10131-7}{\emph{Duke Mathematical
  Journal} {\bfseries 101} (2000) 359 }.

\bibitem{khovanov2008matrix}
M.~Khovanov and L.~Rozansky, \emph{Matrix factorizations and link homology},
  {\emph{Fundamenta Mathematicae} {\bfseries 199} (2008) 1}.

\bibitem{ozsvath2004holomorphic}
P.~Ozsv{\'a}th and Z.~Szab{\'o}, \emph{Holomorphic disks and topological
  invariants for closed three-manifolds}, {\emph{Annals of Mathematics} (2004)
  1027}.

\bibitem{Kronheimer_Mrowka_2007}
P.~Kronheimer and T.~Mrowka, \emph{Monopoles and Three-Manifolds}, New
  Mathematical Monographs, Cambridge University Press (2007).

\bibitem{floer1988instanton}
A.~Floer, \emph{An instanton-invariant for 3-manifolds}, {\emph{Communications
  in mathematical physics} {\bfseries 118} (1988) 215}.

\bibitem{hutchings2009embedded}
M.~Hutchings, \emph{The embedded contact homology index revisited}, {\emph{New
  perspectives and challenges in symplectic field theory} {\bfseries 49} (2009)
  263}.

\bibitem{Witten:2010cx}
E.~Witten, \emph{{Analytic Continuation Of Chern-Simons Theory}}, {\emph{AMS/IP
  Stud. Adv. Math.} {\bfseries 50} (2011) 347}
  [\href{https://arxiv.org/abs/1001.2933}{{\ttfamily 1001.2933}}].

\bibitem{Haydys:2010dv}
A.~Haydys, \emph{{Fukaya-Seidel category and gauge theory}},
  \href{https://doi.org/10.4310/JSG.2015.v13.n1.a5}{\emph{J. Sympl. Geom.}
  {\bfseries 13} (2015) 151} [\href{https://arxiv.org/abs/1010.2353}{{\ttfamily
  1010.2353}}].

\bibitem{doan2022holomorphic}
A.~Doan and S.~Rezchikov, \emph{{Holomorphic Floer theory and the Fueter
  equation}}, {\emph{arXiv preprint arXiv:2210.12047} (2022) }.

\bibitem{Bousseau:2022jaf}
P.~Bousseau, \emph{{Holomorphic Floer theory and Donaldson-Thomas invariants}},
   \href{https://arxiv.org/abs/2210.17001}{{\ttfamily 2210.17001}}.

\bibitem{Kontsevich}
M.~Kontsevich, ``{Resurgence from the path integral perspective (Perimeter
  Institute, 2012); Exponential integrals (SCGP and at IHES, 2014 and 2015);
  Resurgence and wall-crossing via complexified path integral (TFC Sendai,
  2016)}.''.

\bibitem{Gukov:2016njj}
S.~Gukov, M.~Marino and P.~Putrov, \emph{{Resurgence in complex Chern-Simons
  theory}},  \href{https://arxiv.org/abs/1605.07615}{{\ttfamily 1605.07615}}.

\bibitem{Gukov:2003na}
S.~Gukov, \emph{{Three-dimensional quantum gravity, Chern-Simons theory, and
  the A polynomial}},
  \href{https://doi.org/10.1007/s00220-005-1312-y}{\emph{Commun. Math. Phys.}
  {\bfseries 255} (2005) 577}
  [\href{https://arxiv.org/abs/hep-th/0306165}{{\ttfamily hep-th/0306165}}].

\bibitem{Dimofte:2009yn}
T.~Dimofte, S.~Gukov, J.~Lenells and D.~Zagier, \emph{{Exact Results for
  Perturbative Chern-Simons Theory with Complex Gauge Group}},
  \href{https://doi.org/10.4310/CNTP.2009.v3.n2.a4}{\emph{Commun. Num. Theor.
  Phys.} {\bfseries 3} (2009) 363}
  [\href{https://arxiv.org/abs/0903.2472}{{\ttfamily 0903.2472}}].

\bibitem{Witten:1988hf}
E.~Witten, \emph{{Quantum Field Theory and the Jones Polynomial}},
  \href{https://doi.org/10.1007/BF01217730}{\emph{Commun. Math. Phys.}
  {\bfseries 121} (1989) 351}.

\bibitem{Freed:1991wd}
D.S.~Freed and R.E.~Gompf, \emph{{Computer calculation of Witten's three
  manifold invariant}}, \href{https://doi.org/10.1007/BF02100006}{\emph{Commun.
  Math. Phys.} {\bfseries 141} (1991) 79}.

\bibitem{Jeffrey:1992tk}
L.C.~Jeffrey, \emph{{Chern-Simons-Witten invariants of lens spaces and torus
  bundles, and the semiclassical approximation}},
  \href{https://doi.org/10.1007/BF02097243}{\emph{Commun. Math. Phys.}
  {\bfseries 147} (1992) 563}.

\bibitem{ohtsuki2002problems}
T.~Ohtsuki et~al., \emph{Problems on invariants of knots and 3-manifolds},
  \href{https://doi.org/10.48550/arXiv.math/0406190}{\emph{Geom. Topol. Monogr}
  {\bfseries 4} (2002) 377}.

\bibitem{Gukov:2006ze}
S.~Gukov and H.~Murakami, \emph{{SL(2,C) Chern-Simons theory and the asymptotic
  behavior of the colored Jones polynomial}},
  \href{https://doi.org/10.1007/s11005-008-0282-3}{\emph{Lett. Math. Phys.}
  {\bfseries 86} (2008) 79}
  [\href{https://arxiv.org/abs/math/0608324}{{\ttfamily math/0608324}}].

\bibitem{Axelrod:1991vq}
S.~Axelrod and I.M.~Singer, \emph{{Chern-Simons perturbation theory}},  in
  \emph{{International Conference on Differential Geometric Methods in
  Theoretical Physics}}, pp.~3--45, 1991
  [\href{https://arxiv.org/abs/hep-th/9110056}{{\ttfamily hep-th/9110056}}].

\bibitem{Axelrod:1993wr}
S.~Axelrod and I.M.~Singer, \emph{{Chern-Simons perturbation theory. II}},
  {\emph{J. Diff. Geom.} {\bfseries 39} (1994) 173}
  [\href{https://arxiv.org/abs/hep-th/9304087}{{\ttfamily hep-th/9304087}}].

\bibitem{kontsevich1994feynman}
M.~Kontsevich, \emph{Feynman diagrams and low-dimensional topology},  in
  \emph{First European Congress of Mathematics Paris, July 6--10, 1992: Vol.
  II: Invited Lectures (Part 2)}, pp.~97--121, Springer, 1994.

\bibitem{Dimofte:2012qj}
T.D.~Dimofte and S.~Garoufalidis, \emph{{The Quantum content of the gluing
  equations}}, {\emph{Geom. Topol.} {\bfseries 17} (2013) 1253}
  [\href{https://arxiv.org/abs/1202.6268}{{\ttfamily 1202.6268}}].

\bibitem{ohtsuki1996polynomial}
T.~Ohtsuki, \emph{A polynomial invariant of rational homology 3-spheres},
  \href{https://doi.org/10.1007/s002220050025}{\emph{Inventiones mathematicae}
  {\bfseries 123} (1996) 241}.

\bibitem{pham1983}
F.~Pham, \emph{Vanishing homologies and the {$n$} variable saddlepoint method},
   in \emph{Singularities, {P}art 2 ({A}rcata, {C}alif., 1981)}, vol.~40 of
  \emph{Proc. Sympos. Pure Math.}, pp.~319--333, Amer. Math. Soc., Providence,
  RI (1983), \href{https://doi.org/10.1090/pspum/040.2/713258}{DOI}.

\bibitem{Garoufalidis:2020nut}
S.~Garoufalidis, J.~Gu and M.~Marino, \emph{{The Resurgent Structure of Quantum
  Knot Invariants}},
  \href{https://doi.org/10.1007/s00220-021-04076-0}{\emph{Commun. Math. Phys.}
  {\bfseries 386} (2021) 469}
  [\href{https://arxiv.org/abs/2007.10190}{{\ttfamily 2007.10190}}].

\bibitem{Garoufalidis:2020xec}
S.~Garoufalidis, J.~Gu and M.~Marino, \emph{{Peacock patterns and resurgence in
  complex Chern-Simons theory}},
  \href{https://arxiv.org/abs/2012.00062}{{\ttfamily 2012.00062}}.

\bibitem{Garoufalidis:2021osl}
S.~Garoufalidis, J.~Gu, M.~Marino and C.~Wheeler, \emph{{Resurgence of
  Chern-Simons theory at the trivial flat connection}},
  \href{https://arxiv.org/abs/2111.04763}{{\ttfamily 2111.04763}}.

\bibitem{Wheeler:2023cht}
C.~Wheeler, \emph{{Quantum modularity for a closed hyperbolic 3-manifold}},
  \href{https://arxiv.org/abs/2308.03265}{{\ttfamily 2308.03265}}.

\bibitem{neumann1981calculus}
W.D.~Neumann, \emph{A calculus for plumbing applied to the topology of complex
  surface singularities and degenerating complex curves}, {\emph{Transactions
  of the American Mathematical Society} {\bfseries 268} (1981) 299}.

\bibitem{Costin:2023kla}
O.~Costin, G.V.~Dunne, A.~Gruen and S.~Gukov, \emph{{Going to the Other Side
  via the Resurgent Bridge}},
  \href{https://arxiv.org/abs/2310.12317}{{\ttfamily 2310.12317}}.

\bibitem{Gukov:2019mnk}
S.~Gukov and C.~Manolescu, \emph{{A two-variable series for knot complements}},
  \href{https://doi.org/10.4171/qt/145}{\emph{Quantum Topol.} {\bfseries 12}
  (2021) 1} [\href{https://arxiv.org/abs/1904.06057}{{\ttfamily 1904.06057}}].

\bibitem{reshetikhin1991invariants}
N.~Reshetikhin and V.G.~Turaev, \emph{Invariants of 3-manifolds via link
  polynomials and quantum groups}, {\emph{Inventiones mathematicae} {\bfseries
  103} (1991) 547}.

\bibitem{Gukov:2017kmk}
S.~Gukov, D.~Pei, P.~Putrov and C.~Vafa, \emph{{BPS spectra and 3-manifold
  invariants}}, \href{https://doi.org/10.1142/S0218216520400039}{\emph{J. Knot
  Theor. Ramifications} {\bfseries 29} (2020) 2040003}
  [\href{https://arxiv.org/abs/1701.06567}{{\ttfamily 1701.06567}}].

\bibitem{Gopakumar:1998ii}
R.~Gopakumar and C.~Vafa, \emph{{M theory and topological strings. 1.}},
  \href{https://arxiv.org/abs/hep-th/9809187}{{\ttfamily hep-th/9809187}}.

\bibitem{lawrence1999modular}
R.~Lawrence and D.~Zagier, \emph{Modular forms and quantum invariants of
  3-manifolds}, {\emph{Asian Journal of Mathematics} {\bfseries 3} (1999) 93}.

\bibitem{deHaro:2004id}
S.~de~Haro and M.~Tierz, \emph{{Brownian motion, Chern-Simons theory, and 2-D
  Yang-Mills}},
  \href{https://doi.org/10.1016/j.physletb.2004.09.033}{\emph{Phys. Lett. B}
  {\bfseries 601} (2004) 201}
  [\href{https://arxiv.org/abs/hep-th/0406093}{{\ttfamily hep-th/0406093}}].

\bibitem{deHaro:2004wn}
S.~de~Haro, \emph{{Chern-Simons theory, 2d Yang-Mills, and Lie algebra
  wanderers}},
  \href{https://doi.org/10.1016/j.nuclphysb.2005.09.009}{\emph{Nucl. Phys. B}
  {\bfseries 730} (2005) 312}
  [\href{https://arxiv.org/abs/hep-th/0412110}{{\ttfamily hep-th/0412110}}].

\bibitem{hikami2005quantum}
K.~Hikami, \emph{{On the quantum invariant for the Brieskorn homology
  spheres}}, {\emph{International Journal of Mathematics} {\bfseries 16} (2005)
  661}.

\bibitem{Blau:2006gh}
M.~Blau and G.~Thompson, \emph{{Chern-Simons theory on S1-bundles:
  Abelianisation and q-deformed Yang-Mills theory}},
  \href{https://doi.org/10.1088/1126-6708/2006/05/003}{\emph{JHEP} {\bfseries
  05} (2006) 003} [\href{https://arxiv.org/abs/hep-th/0601068}{{\ttfamily
  hep-th/0601068}}].

\bibitem{hikami2006quantum}
K.~Hikami, \emph{{On the quantum invariants for the spherical Seifert
  manifolds}}, {\emph{Communications in mathematical physics} {\bfseries 268}
  (2006) 285}.

\bibitem{hikami2011decomposition}
K.~Hikami, \emph{{Decomposition of Witten-Reshetikhin-Turaev invariant: linking
  pairing and modular forms}}, {\emph{AMS/IP Stud. Adv. Math} {\bfseries 50}
  (2011) 131}.

\bibitem{Gukov:2016gkn}
S.~Gukov, P.~Putrov and C.~Vafa, \emph{{Fivebranes and 3-manifold homology}},
  \href{https://doi.org/10.1007/JHEP07(2017)071}{\emph{JHEP} {\bfseries 07}
  (2017) 071} [\href{https://arxiv.org/abs/1602.05302}{{\ttfamily
  1602.05302}}].

\bibitem{Murakami:2023oam}
Y.~Murakami, \emph{{A proof of a conjecture of Gukov-Pei-Putrov-Vafa}},
  \href{https://arxiv.org/abs/2302.13526}{{\ttfamily 2302.13526}}.

\bibitem{Murakami:2023csv}
Y.~Murakami and Y.~Terashima, \emph{{Homological blocks with simple Lie
  algebras and Witten--Reshetikhin--Turaev invariants}},
  \href{https://arxiv.org/abs/2308.04010}{{\ttfamily 2308.04010}}.

\bibitem{Ri:2022bxf}
S.J.~Ri, \emph{{Refined and Generalized $\hat{Z}$ Invariants for Plumbed
  3-Manifolds}}, \href{https://doi.org/10.3842/SIGMA.2023.011}{\emph{SIGMA}
  {\bfseries 19} (2023) 011}
  [\href{https://arxiv.org/abs/2205.08197}{{\ttfamily 2205.08197}}].

\bibitem{lawrence1999witten}
R.~Lawrence and L.~Rozansky, \emph{{Witten--Reshetikhin--Turaev Invariants of
  Seifert Manifolds}}, {\emph{Communications in mathematical physics}
  {\bfseries 205} (1999) 287}.

\bibitem{Marino:2002fk}
M.~Marino, \emph{{Chern-Simons theory, matrix integrals, and perturbative three
  manifold invariants}},
  \href{https://doi.org/10.1007/s00220-004-1194-4}{\emph{Commun. Math. Phys.}
  {\bfseries 253} (2004) 25}
  [\href{https://arxiv.org/abs/hep-th/0207096}{{\ttfamily hep-th/0207096}}].

\bibitem{Beasley:2005vf}
C.~Beasley and E.~Witten, \emph{{Non-Abelian localization for Chern-Simons
  theory}}, {\emph{J. Diff. Geom.} {\bfseries 70} (2005) 183}
  [\href{https://arxiv.org/abs/hep-th/0503126}{{\ttfamily hep-th/0503126}}].

\bibitem{Gukov:2023srx}
S.~Gukov, L.~Katzarkov and J.~Svoboda, \emph{{$\hat{Z}_b$ for plumbed manifolds
  and splice diagrams}},  \href{https://arxiv.org/abs/2304.00699}{{\ttfamily
  2304.00699}}.

\bibitem{MR0887284}
N.J.~Hitchin, \emph{The self-duality equations on a {R}iemann surface},
  {\emph{Proc. London Math. Soc. (3)} {\bfseries 55} (1987) 59}.

\bibitem{MR2755722}
G.D.~Daskalopoulos, R.A.~Wentworth and G.~Wilkin, \emph{Cohomology of {${\rm
  SL}(2,\mathbb{C})$} character varieties of surface groups and the action of
  the {T}orelli group}, {\emph{Asian J. Math.} {\bfseries 14} (2010) 359}.

\bibitem{Cheng:2018vpl}
M.C.N.~Cheng, S.~Chun, F.~Ferrari, S.~Gukov and S.M.~Harrison, \emph{{3d
  Modularity}}, \href{https://doi.org/10.1007/JHEP10(2019)010}{\emph{JHEP}
  {\bfseries 10} (2019) 010}
  [\href{https://arxiv.org/abs/1809.10148}{{\ttfamily 1809.10148}}].

\bibitem{bringmann2020higher}
K.~Bringmann, K.~Mahlburg and A.~Milas, \emph{Higher depth quantum modular
  forms and plumbed 3-manifolds}, {\emph{Letters in Mathematical Physics}
  {\bfseries 110} (2020) 2675}.

\bibitem{Cheng:2019uzc}
M.C.N.~Cheng, F.~Ferrari and G.~Sgroi, \emph{{Three-Manifold Quantum Invariants
  and Mock Theta Functions}},
  \href{https://doi.org/10.1098/rsta.2018.0439}{\emph{Phil. Trans. Roy. Soc.
  Lond.} {\bfseries 378} (2019) 20180439}
  [\href{https://arxiv.org/abs/1912.07997}{{\ttfamily 1912.07997}}].

\bibitem{Mori:2021ost}
A.~Mori and Y.~Murakami, \emph{{Witten-Reshetikhin-Turaev Invariants,
  Homological Blocks, and Quantum Modular Forms for Unimodular Plumbing
  H-Graphs}}, \href{https://doi.org/10.3842/SIGMA.2022.034}{\emph{SIGMA}
  {\bfseries 18} (2022) 034}
  [\href{https://arxiv.org/abs/2110.10958}{{\ttfamily 2110.10958}}].

\bibitem{Cheng:2022rqr}
M.C.N.~Cheng, S.~Chun, B.~Feigin, F.~Ferrari, S.~Gukov, S.M.~Harrison et~al.,
  \emph{{3-Manifolds and VOA Characters}},
  \href{https://arxiv.org/abs/2201.04640}{{\ttfamily 2201.04640}}.

\bibitem{Cheng:2023row}
M.C.N.~Cheng, I.~Coman, D.~Passaro and G.~Sgroi, \emph{{Quantum Modular $\hat
  Z^G$-Invariants}},  \href{https://arxiv.org/abs/2304.03934}{{\ttfamily
  2304.03934}}.

\bibitem{MR4167016}
M.~Abouzaid and C.~Manolescu, \emph{A sheaf-theoretic model for {${\rm
  SL}(2,\mathbb{C})$} {F}loer homology}, {\emph{J. Eur. Math. Soc. (JEMS)}
  {\bfseries 22} (2020) 3641}.

\bibitem{Cu01}
C.L.~Curtis, \emph{An intersection theory count of the
  {SL(2,C)}-representations of the fundamental group of a 3-manifold},
  \href{https://doi.org/https://doi.org/10.1016/S0040-9383(99)00083-X}{\emph{Topology}
  {\bfseries 40} (2001) 773}.

\bibitem{BC08}
H.U.~Boden and C.L.~Curtis, \emph{Splicing and the {SL}(2,{C}) {Casson}
  invariant}, {\emph{Proc. Amer. Math. Soc.} {\bfseries 136} (2008) 2615}.

\bibitem{MR1002161}
J.~Kalliongis and C.M.~Tsau, \emph{Seifert fibered surgery manifolds of
  composite knots}, {\emph{Proc. Amer. Math. Soc.} {\bfseries 108} (1990)
  1047}.

\bibitem{Gadde:2013sca}
A.~Gadde, S.~Gukov and P.~Putrov, \emph{{Fivebranes and 4-manifolds}},
  \href{https://doi.org/10.1007/978-3-319-43648-7_7}{\emph{Prog. Math.}
  {\bfseries 319} (2016) 155}
  [\href{https://arxiv.org/abs/1306.4320}{{\ttfamily 1306.4320}}].

\bibitem{kontsevich1995homological}
M.~Kontsevich, \emph{Homological algebra of mirror symmetry},  in
  \emph{Proceedings of the International Congress of Mathematicians: August
  3--11, 1994 Z{\"u}rich, Switzerland}, pp.~120--139, Springer, 1995.

\bibitem{seidel2000graded}
P.~Seidel, \emph{{Graded lagrangian submanifolds}}, {\emph{Bulletin de la
  Soci{\'e}t{\'e} Math{\'e}matique de France} {\bfseries 128} (2000) 103}.

\bibitem{clemens1969picard}
C.~Clemens, \emph{{Picard-Lefschetz theorem for families of nonsingular
  algebraic varieties acquiring ordinary singularities}}, {\emph{Transactions
  of the American Mathematical Society} {\bfseries 136} (1969) 93}.

\bibitem{landman1973picard}
A.~Landman, \emph{{On the Picard-Lefschetz transformation for algebraic
  manifolds acquiring general singularities}}, {\emph{Transactions of the
  American Mathematical Society} {\bfseries 181} (1973) 89}.

\bibitem{kontsevich1998}
M.~Kontsevich, \emph{{Lectures at ENS Paris}},  in \emph{Set of notes taken by
  J. Bellaiche, J.-F. Dat, I. Marin, G. Racinet and H. Randriambololona.},
  1998.

\bibitem{MR1882336}
P.~Seidel, \emph{More about vanishing cycles and mutation},  in
  \emph{Symplectic geometry and mirror symmetry ({S}eoul, 2000)}, pp.~429--465,
  World Sci. Publ., River Edge, NJ (2001).

\bibitem{MR3911570}
Z.~Sylvan, \emph{On partially wrapped {F}ukaya categories}, {\emph{J. Topol.}
  {\bfseries 12} (2019) 372}.

\bibitem{MR4106794}
S.~Ganatra, J.~Pardon and V.~Shende, \emph{Covariantly functorial wrapped
  {F}loer theory on {L}iouville sectors}, {\emph{Publ. Math. Inst. Hautes
  \'{E}tudes Sci.} {\bfseries 131} (2020) 73}.

\bibitem{MR2449059}
D.~Nadler and E.~Zaslow, \emph{Constructible sheaves and the {F}ukaya
  category}, {\emph{J. Amer. Math. Soc.} {\bfseries 22} (2009) 233}.

\bibitem{MR3502098}
M.~Abouzaid and D.~Auroux, \emph{Homological mirror symmetry for hypersurfaces
  in {$(\mathbb{C}^\ast)^n$}},
  \href{https://arxiv.org/abs/2111.06543}{{\ttfamily 2111.06543}}.

\bibitem{MR3838112}
D.~Auroux, \emph{Speculations on homological mirror symmetry for hypersurfaces
  in {$(\mathbb{C}^\ast)^n$}},  in \emph{Surveys in differential geometry 2017.
  {C}elebrating the 50th anniversary of the {J}ournal of {D}ifferential
  {G}eometry}, vol.~22 of \emph{Surv. Differ. Geom.}, pp.~1--47, Int. Press,
  Somerville, MA (2018).

\bibitem{CDGG17}
B.~Chantraine, G.~Dimitroglou~Rizell, P.~Ghiggini and R.~Golovko,
  \emph{{Geometric generation of the wrapped Fukaya category of Weinstein
  manifolds and sectors}},  \href{https://arxiv.org/abs/1712.09126}{{\ttfamily
  1712.09126}}.

\bibitem{MR4695507}
S.~Ganatra, J.~Pardon and V.~Shende, \emph{Sectorial descent for wrapped
  {F}ukaya categories}, {\emph{J. Amer. Math. Soc.} {\bfseries 37} (2024) 499}.

\bibitem{Ekholm:2019lmb}
T.~Ekholm, P.~Kucharski and P.~Longhi, \emph{{Multi-cover skeins, quivers, and
  3d $\mathcal{N}=2$ dualities}},
  \href{https://doi.org/10.1007/JHEP02(2020)018}{\emph{JHEP} {\bfseries 02}
  (2020) 018} [\href{https://arxiv.org/abs/1910.06193}{{\ttfamily
  1910.06193}}].

\bibitem{Lauda:2020tee}
A.D.~Lauda, A.M.~Licata and A.~Manion, \emph{{From hypertoric geometry to
  bordered Floer homology via the $m=1$ amplituhedron}},
  \href{https://arxiv.org/abs/2009.03981}{{\ttfamily 2009.03981}}.

\bibitem{LOT2021}
R.~Lipshitz, P.S.~Ozsv\'{a}th and D.P.~Thurston, \emph{{A bordered HF- algebra
  for the torus}},  \href{https://arxiv.org/abs/2108.12488}{{\ttfamily
  2108.12488}}.

\bibitem{MR4634745}
T.~Ekholm and Y.~Lekili, \emph{Duality between {L}agrangian and {L}egendrian
  invariants}, {\emph{Geom. Topol.} {\bfseries 27} (2023) 2049}.

\bibitem{MR3073884}
M.~Abouzaid, D.~Auroux, A.I.~Efimov, L.~Katzarkov and D.~Orlov,
  \emph{Homological mirror symmetry for punctured spheres}, {\emph{J. Amer.
  Math. Soc.} {\bfseries 26} (2013) 1051}.

\bibitem{Hori:2013ika}
K.~Hori and M.~Romo, \emph{{Exact Results In Two-Dimensional (2,2)
  Supersymmetric Gauge Theories With Boundary}},
  \href{https://arxiv.org/abs/1308.2438}{{\ttfamily 1308.2438}}.

\bibitem{Hori:2000ck}
K.~Hori, A.~Iqbal and C.~Vafa, \emph{{D-branes and mirror symmetry}},
  \href{https://arxiv.org/abs/hep-th/0005247}{{\ttfamily hep-th/0005247}}.

\bibitem{Chun:2019mal}
S.~Chun, S.~Gukov, S.~Park and N.~Sopenko, \emph{{3d-3d correspondence for
  mapping tori}}, \href{https://doi.org/10.1007/JHEP09(2020)152}{\emph{JHEP}
  {\bfseries 09} (2020) 152}
  [\href{https://arxiv.org/abs/1911.08456}{{\ttfamily 1911.08456}}].

\end{thebibliography}\endgroup
\bibliographystyle{JHEP}

\end{document}